\documentclass[format=manuscript, review=false, screen=true]{acmart}

\usepackage{algorithm}%
\usepackage[noend]{algpseudocode}
\makeatletter
\def\BState{\State\hskip-\ALG@thistlm}
\makeatother

\usepackage{multirow}%
\usepackage{pst-node}%
\usepackage{pst-rel-points}%
\usepackage{xspace}%
\usepackage{paralist}%
\usepackage{calligra} %
\usepackage{amsthm}
\usepackage{stmaryrd}
\usepackage{verbatim}

\usepackage[framemethod=PSTricks]{mdframed}
\usepackage{wrapfig}
\usepackage{fmtcount}
\usepackage{amsfonts}
\usepackage{amsmath}
\usepackage{amssymb}
\usepackage{isomath}
\usepackage{colortbl}

\DeclareMathAlphabet{\mathcalligra}{T1}{calligra}{m}{n}
\DeclareFontShape{T1}{calligra}{m}{n}{<->s*[2.2]callig15}{}

\acmJournal{TOPLAS}

\definecolor{blue}{named}{blue}
\definecolor{brown}{named}{brown}
\newrgbcolor{lightblue}{.75 0.85 1}

\newcommand{\remText}{\protect\green}
\newcommand{\bitem}{\item[$\bullet$]}

\psset{unit=1mm}
\newrgbcolor{brown}{.65 0.15 .0}
\newrgbcolor{violet}{.84 0 .96}
\newrgbcolor{mygreen}{.24 .84 .72}

\newcommand{\Start}[1]{\text{\sf Start$_{#1}$}}
\newcommand{\Startscriptsize}[1]{\text{\scriptsize\sf Start$_{#1}$}}
\newcommand{\End}[1]{\text{\sf End$_{#1}$}}
\newcommand{\Endscriptsize}[1]{\text{\scriptsize\sf End$_{#1}$}}

\newcommand{\vars}{\text{\sf\em L\/}\xspace}

\newcommand{\ptrs}{\text{\sf\em P\/}\xspace}

\newcommand{\gpredsubscript}{\text{\sf\em pred\/}\xspace}
\newcommand{\flow}{\text{$R$}\xspace}  % dangerous name
\newcommand{\flowbar}{\text{\protect$\overbar{\protect\protect\rule{0em}{.8em}\flow}$}\xspace}%%{\text{$R$}\xspace}  % dangerous name

\newcounter{linectr}
\DeclareRobustCommand{\printlinectr}{%
\darkgray\footnotesize\sf\padzeroes[2]{\decimal{linectr}} 
}

\protect{\newlength{\BRACEL}}
\newcommand{\OB}{\lcb\hspace*{\BRACEL}}
\newcommand{\CB}{\rcb}
\renewcommand{\OB}{\{\hspace*{\BRACEL}}
\renewcommand{\CB}{\}}
\protect{\newlength{\TAL}} 
\newcommand{\NL}[1]{\hspace*{#1\TAL}}
\newlength{\codeLineLength}
\DeclareRobustCommand{\codeLine}[3]{\refstepcounter{linectr}\printlinectr
&%
\psframebox[framesep=0,fillstyle=solid,fillcolor=#3,
	linestyle=none]{\makebox[\codeLineLength][l]{%
	\rule[-.25em]{0em}{1.15em}{%\ttfamily%
	 \NL{#1}{\text{#2}}}}}
\\ }

\newcommand{\codeLineOne}[4]{\setcounter{linectr}{#1}\printlinectr&%
\psframebox[framesep=0,fillstyle=solid,fillcolor=#4,
	linestyle=none]{\makebox[\codeLineLength][l]{%
	\rule[-.25em]{0em}{1.15em}{%\ttfamily%
	 \NL{#2}{\text{#3}}}}}
\\ }
\newcommand{\codeLineNoNumber}[3]{&%
\psframebox[framesep=0,fillstyle=solid,fillcolor=#3,
	linestyle=none]{\makebox[\codeLineLength][l]{%
	\rule[-.25em]{-2.5em}{1.15em}{%\ttfamily%
	 \NL{#1}{\text{#2}}}}}
\\ }

\newcommand{\ptsG}{\text{\sf PTG\/}\xspace} 

\newcommand{\Gen}{\text{\sf Gen}\xspace}
\newcommand{\Kill}{\text{\sf Kill}\xspace}
\newcommand{\blocked}{\text{\sf Blocked\/}\xspace}
\newcommand{\relgpus}{\text{\sf Prospective\_Producer\_\gpus}\xspace}
\newcommand{\IGPUs}{\text{\sf Ind\gpus\/}\xspace}

\newcommand{\overbar}[1]{\protect\mkern 2.5mu\protect\overline{\protect\mkern-2.5mu\protect#1\protect\mkern-1.5mu}\protect\mkern 2.5mu}

\newcommand{\EBIn}[1]{\text{{\sf EBIn}$_{#1}$}\xspace}
\newcommand{\EBOut}[1]{\text{{\sf EBOut}$_{#1}$}\xspace}
\newcommand{\RIn}[1]{\text{{\sf RGIn}$_{#1}$}\xspace}
\newcommand{\ROut}[1]{\text{{\sf RGOut}$_{#1}$}\xspace}
\newcommand{\RGen}[1]{\text{{\sf RGGen}$_{#1}$}\xspace}
\newcommand{\BRGen}[1]{\text{$\overline{\text{\sf RGGen}}_{#1}$}\xspace}
\newcommand{\BRKill}[1]{\text{$\overline{\text{\sf RGKill}}_{#1}$}\xspace}
\newcommand{\RKill}[1]{\text{{\sf RGKill}$_{#1}$}\xspace}
\newcommand{\BRIn}[1]{\text{$\overline{\text{\sf RGIn}}_{#1}$}\xspace}
\newcommand{\BROut}[1]{\text{$\overline{\text{\sf RGOut}}_{#1}$}\xspace}

\newcommand{\PIn}[1]{\text{{\sf PIn}$_{#1}$}\xspace}
\newcommand{\POut}[1]{\text{{\sf POut}$_{#1}$}\xspace}
\newcommand{\PKill}[1]{\text{{\sf PKill}$_{#1}$}\xspace}

\newcommand{\tscomp}{\text{\sf\em TS\/}\xspace}
\newcommand{\sscomp}{\text{\sf\em SS\/}\xspace}
\newcommand{\ttcomp}{\text{\sf\em TT\/}\xspace}
\newcommand{\stcomp}{\text{\sf\em ST\/}\xspace}
\newcommand{\ecomptype}{\text{${\tau}$}\xspace}
\newcommand{\ecompwt}{\text{$\circ^{\ecomptype}$}\xspace}
\newcommand{\ecomp}{\text{$\hspace*{1pt}\circ\hspace*{1pt}$}\xspace}
\newcommand{\rcomp}{\text{$\hspace*{1pt}\circ\hspace*{1pt}$}\xspace}

\newcommand{\boundary}{\text{\sf\em BI\/}\xspace}

\newcommand{\singledef}[2]{\text{{\sf Singledef}$\,({#1},\; {#2})$}}

\newcommand{\match}{\text{\sf Match}\xspace}
\newcommand{\conskill}{\text{\sf Kill}\xspace}
\newcommand{\unblockall}{\text{\sf UbAll}\xspace}
\newcommand{\unblockindirect}{\text{\sf UbInd}\xspace}
\newcommand{\consgen}{\text{\sf Gen}\xspace}
\newcommand{\memkill}{\text{\sf Memkill}\xspace}
\newcommand{\Def}{\text{\sf Def}\xspace}

\newcommand{\must}{\text{\sf\em must}\xspace}
\newcommand{\may}{\text{\sf\em may}\xspace}

\newcommand{\normalmay}{\text{\sf\em may}\xspace}

\newcommand{\candedge}{\text{$\mathsfbfit{c}$}\xspace}
\newcommand{\prevedge}{\text{$\mathsfbfit{p}$}\xspace}
\newcommand{\baredge}{\text{$\mathsfbfit{b}$}\xspace}
\newcommand{\rededge}{\text{$\mathsfbfit{r}$}\xspace}
\newcommand{\usenode}{\text{$\mathsfbfit{u}$}\xspace}

\newcommand{\flab}{\text{${s}$}\xspace}
\newcommand{\slab}{\text{${t}$}\xspace}
\newcommand{\tlab}{\text{${u}$}\xspace}
\newcommand{\plab}{\text{${v}$}\xspace}

\newcounter{exmpctr}%[section]
\newcommand{\mem}{\text{$M$}\xspace}
\newcommand{\mempp}{\text{$\mem_{\!\pp}$}\xspace}

\newcommand{\indlev}{\text{\sf\em indlev}\xspace}
\newcommand{\indlevs}{\text{{\sf\em indlev\hspace*{1pt}}s}\xspace}
\newcommand{\sindlevs}{\text{{\sf\em indlev\hspace*{.5pt}}s}\xspace}
\newcommand{\sindlev}{\text{\sf\em indlev}\xspace}
\newcommand{\indlist}{\text{\sf\em indlist}\xspace}
\newcommand{\indlists}{\text{{\sf\em indlist\hspace*{1pt}}s}\xspace}

\newcommand{\sindlist}{\text{\sf\em indlist}\xspace}

\newcommand{\rem}{\text{\sf Remainder}\xspace}
\newcommand{\srem}{\text{\sf sRemainder}\xspace}

\newcommand{\ssa}{\text{\sf\em ss1}\xspace}
\newcommand{\ssb}{\text{\sf\em ss2}\xspace}
\newcommand{\ssc}{\text{\sf\em ss3}\xspace}

\newcommand{\tsa}{\text{\sf\em ts1}\xspace}
\newcommand{\tsb}{\text{\sf\em ts2}\xspace}
\newcommand{\tsc}{\text{\sf\em ts3}\xspace}

\newcounter{pqr}

\newcommand{\smallblock}[1]{%
	\begin{pspicture}(0,0)(#1,6)
	\putnode{plla}{origin}{0}{0}{}
	\putnode{prla}{origin}{#1}{0}{}
	\putnode{prta}{origin}{#1}{6}{}
	\putnode{plta}{origin}{0}{6}{}

	\psframe[linecolor=white,fillstyle=solid,fillcolor=lightgray](0,0)(#1,6)

	\setcounter{pqr}{#1}
	\addtocounter{pqr}{7}
	
	\putnode{pllb}{plla}{12}{6}{}
	\putnode{prlb}{prla}{12}{6}{}
	\putnode{prtb}{prta}{12}{6}{}
	\putnode{pltb}{plta}{12}{6}{}

	\psline[linecolor=white,fillstyle=solid,fillcolor=lightgray]%
		(\x{prta},\y{prta})
		(\x{prtb},\y{prtb})
		(\x{prlb},\y{prlb})
		(\x{prla},\y{prla})
	\psline[linecolor=white,fillstyle=solid,fillcolor=lightgray]%
		(\x{plta},\y{plta})
		(\x{pltb},\y{pltb})
		(\x{prtb},\y{prtb})
		(\x{prta},\y{prta})
	\end{pspicture}
	}

\newcommand{\bigblock}[1]{%
	\begin{pspicture}(0,0)(#1,9)
	\putnode{plla}{origin}{0}{0}{}
	\putnode{prla}{origin}{#1}{0}{}
	\putnode{prta}{origin}{#1}{9}{}
	\putnode{plta}{origin}{0}{9}{}

	\psframe[linecolor=white,fillstyle=solid,fillcolor=lightgray](0,0)(#1,9)

	\putnode{pllb}{plla}{10}{5}{}
	\putnode{prlb}{prla}{10}{5}{}
	\putnode{prtb}{prta}{10}{5}{}
	\putnode{pltb}{plta}{10}{5}{}

	\psline[linecolor=white,fillstyle=solid,fillcolor=lightgray]%
		(\x{prta},\y{prta})
		(\x{prtb},\y{prtb})
		(\x{prlb},\y{prlb})
		(\x{prla},\y{prla})
	\psline[linecolor=white,fillstyle=solid,fillcolor=lightgray]%
		(\x{plta},\y{plta})
		(\x{pltb},\y{pltb})
		(\x{prtb},\y{prtb})
		(\x{prta},\y{prta})
	\end{pspicture}
	}

\newcommand{\stscomp}{\text{\sf\em TS\/}\xspace}
\newcommand{\ssscomp}{\text{\sf\em SS\/}\xspace}

\newcommand{\admissible}{\text{\sf\em admissible}\xspace}
\newcommand{\inadmissible}{\text{\sf\em inadmissible}\xspace}
\newcommand{\desirable}{\text{\sf\em desirable}\xspace}
\newcommand{\desirability}{\text{\sf\em desirability}\xspace}

\newcommand{\admissibility}{\text{\sf\em admissibility}\xspace}
\newcommand{\undesirable}{\text{\sf\em undesirable}\xspace}

\newcommand{\valid}{\text{\sf\em valid}\xspace}
\newcommand{\invalid}{\text{\sf\em invalid}\xspace}
\newcommand{\validity}{\text{\sf\em validity}\xspace}
\newcommand{\Validity}{\text{\sf\em Validity}\xspace}
\newcommand{\svalidity}{\text{\sf\em validity}\xspace}
\newcommand{\fisd}{\text{$\dagger$}\xspace}

\newlength{\arrowsubscriptheight}
\newlength{\arrowsubscriptmove}

\newcommand{\de}[3]{\text{${#1}\!\xrightarrow%
		{\settoheight{\arrowsubscriptheight}{$\scriptstyle#2$}\raisebox{-.5pt}[{\arrowsubscriptheight}][0pt]{$\scriptstyle#2$}}%
		\!{#3}$}}

\newcommand{\denew}[4]{\text{${#1}\!\xrightarrow%
		[\settoheight{\arrowsubscriptheight}{$\scriptstyle#4$}%
		\ifthenelse{\lengthtest{\arrowsubscriptheight < 3.5pt}}%
		 	{\setlength{\arrowsubscriptmove}{1.1\arrowsubscriptheight}}%
		 	{\setlength{\arrowsubscriptmove}{0.4\arrowsubscriptheight}}%
		{\raisebox{\arrowsubscriptmove}[0.5\arrowsubscriptheight][0pt]{$\scriptstyle#4$}}]%
		{\settoheight{\arrowsubscriptheight}{$\scriptstyle#2$}\raisebox{-.5pt}[{0.9\arrowsubscriptheight}][0pt]{$\scriptstyle#2$}}%
		\!{#3}$}}

\newcommand{\gpbsym}{\text{$\delta$}\xspace}
\newcommand{\mtsym}{\text{$\Delta$}\xspace}
\newcommand{\cfg}{\text{CFG}\xspace}
\newcommand{\cfgs}{\text{CFGs}\xspace}
\newcommand{\edge}{\text{$\gamma$}\xspace}

\newcommand{\gpu}{\text{GPU}\xspace}

\newcommand{\gpus}{\text{GPUs}\xspace}
\newcommand{\scriptgpu}{\text{\footnotesize GPU}\xspace}
\newcommand{\scriptgpus}{\text{\footnotesize GPUs}\xspace}
\newcommand{\gpb}{\text{GPB}\xspace}
\newcommand{\gpbs}{\text{GPBs}\xspace}

\newcommand{\undesComp}{\text{\sf Undes\_comp}\xspace}
\newcommand{\gpg}{\text{GPG}\xspace}
\newcommand{\gpgs}{\text{GPGs}\xspace}
\newcommand{\queued}{\text{\sf Queued}\xspace}

\newcommand{\FALSE}{\text{\bf false}\xspace}%
\newcommand{\TRUE}{\text{\bf true}\xspace}%
\newcommand{\GENEDGES}{\text{\sf Red}\xspace}%
\newcommand{\WL}{\text{\em W}\xspace}%

\newcommand{\WCompose}{\text{\em composed }}%
\newcommand{\UNDes}{\text{\em undes }}%
\newcommand{\succeeds}{\text{\em succeeds }}%
\newcommand{\POSTPONE}{\text{\em postpone}\xspace}%
\newcommand{\compress}{\text{\sf Compress}\xspace}

\newcommand{\filter}{\text{\sf Filter}\xspace}
\newcommand{\eflow}{\text{\sf Edgeflow}\xspace}

\newcommand{\Rprev}{\text{\em Rprev}\xspace}
\newcommand{\RprevBar}{\text{$\overline{R}$\em prev}\xspace}
\newcommand{\Rcur}{\text{\em Rcurr}\xspace}
\newcommand{\RcurBar}{\text{$\overline{R}$\em curr}\xspace}

\floatstyle{plain}
\newfloat{Definition}{t}{def}

\mdfdefinestyle{toplasexample}{
everyline=false,
leftline=true,
topline=true,
bottomline=true,
rightline=true,
innerleftmargin=4pt,
innerrightmargin=4pt,
innertopmargin=2pt,
innerbottommargin=2pt,
splittopskip=0pt,
splitbottomskip=0pt,
skipbelow=4pt,
skipabove=4pt,
}

\newenvironment{example}[1]{%
\begin{mdframed}[style=toplasexample]
	\noindent\text{\bfseries Example \refstepcounter{exmpctr}\theexmpctr.\label{#1} 
	}
}{\end{mdframed}\vskip5pt}
\newcommand{\edgeReduction}{\text{\sf \gpu\!\!\_reduction}\xspace}%

 \newcommand{\InT}[1]{\text{{\sf GIn}$_{#1}$}\xspace}
\newcommand{\OutT}[1]{\text{{\sf GOut}$_{#1}$}\xspace}
 \newcommand{\InB}[1]{\text{{\sf CIn}$_{#1}$}\xspace}
\newcommand{\OutB}[1]{\text{{\sf COut}$_{#1}$}\xspace}
\newcommand{\LT}{\text{\sf TDef}\xspace}
\newcommand{\RT}{\text{\sf TRef}\xspace}
\newcommand{\bDdep}{\text{$\overline{\text{\sf DDep}}$}\xspace}
\newcommand{\Ddep}{\text{\sf DDep}\xspace}
\newcommand{\deref}{\text{\sf deref}\xspace}
\newcommand{\Type}{\text{\sf typeof}\xspace}

\newcommand{\TypeFlow}{\text{\sf gpuFlow}\xspace}
\newcommand{\coalescePred}{\text{\sf coalesce}\xspace}

\newcommand{\partition}{\text{$\Pi$}\xspace}
\newcommand{\group}{\text{$\pi$}\xspace}
\newcommand{\Ggroup}{\text{$G(\group_i, n)$}\xspace}
\newcommand{\Ggroupp}{\text{$G(\group_i, p)$}\xspace}

\newcommand{\true}{\text{\sf\em true}\xspace}
\newcommand{\false}{\text{\sf\em false}\xspace}
\newcommand{\RaW}{\text{RaW}\xspace}
\newcommand{\WaW}{\text{WaW}\xspace}
\newcommand{\WaR}{\text{WaR}\xspace}
\newcommand{\RaR}{\text{RaR}\xspace}

\newrgbcolor{pink}{1 .5 .6}
\newrgbcolor{darkgreen}{0 .5 0}

\newcommand{\revOne}{\protect\blue}
\newcommand{\revTwo}{\protect\red}
\title{Generalized Points-to Graphs: A New Abstraction of Memory in the Presence of Pointers}

\author{Pritam M. Gharat}
\email{pritamg@cse.iitb.ac.in}
\author{Uday P. Khedker}
\email{uday@cse.iitb.ac.in}
\affiliation{%
  \institution{Indian Institute of Technology Bombay}
  \country{India}}
\author{Alan Mycroft}
\affiliation{%
 \institution{University of Cambridge}
 \country{UK}}
\email{Alan.Mycroft@cl.cam.ac.uk}

\newcommand{\AMcomment}[1]{{\color{magenta}[\protect#1]}}

\newcommand{\seechange}{N}

\ifthenelse{\equal{\seechange}{Y}}{
	
	}{
	
	}

\global\mdfdefinestyle{exampledefault}{%
pstricksappsetting={\addtopsstyle{mdfmiddlelinestyle}{%
doubleline=true,doublesep=1.5pt}}}

\begin{abstract}
Computing precise (fully flow- and context-sensitive) and exhaustive (as against demand-driven)
points-to information is known to be computationally expensive. Prior approaches to
flow- and context-sensitive points-to analysis (FCPA) have not scaled;
for top-down approaches, the problem centers on repeated analysis of the same procedure;
for bottom-up approaches, the abstractions used to represent procedure summaries have not scaled while preserving
precision.
Bottom-up approaches for points-to analysis require modelling unknown pointees accessed
indirectly through pointers that may be defined in the callers. 
We propose a novel abstraction called the Generalized Points-to Graph (\gpg)
which views points-to relations as memory updates and generalizes them
using the counts of indirection levels
leaving the unknown pointees implicit. This allows us to construct \gpgs\
as compact representations of bottom-up procedure summaries 
in terms of memory updates and control flow between them.
Their compactness is ensured by the following optimizations:
strength reduction reduces the indirection levels, redundancy elimination removes redundant memory updates and
minimizes control flow (without over-approximating data dependence between memory updates), 
and call inlining 
enhances the opportunities of these optimizations.
We devise novel operations and data flow analyses for
these optimizations. 
Our quest for scalability of points-to analysis leads to the following insight: The 
real killer of scalability in program analysis is not the amount of data
but the amount of control flow that it may be subjected to in search of precision.
The effectiveness of \gpgs lies in the fact that they 
discard as much control flow as possible without losing precision (i.e., by
preserving data dependence without over-approximation).
This is the reason why the \gpgs are very small even for main procedures that contain the effect of the entire
program. This allows our implementation to scale to 158kLoC for C programs.
At a more general level,
\gpgs provide a convenient
abstraction of memory and memory transformers in the presence of pointers. 
Future investigations can try to combine it with other abstractions for
static analyses that can benefit from points-to information.
\end{abstract}


	\allowdisplaybreaks[1]
	
\begin{document}
	\maketitle

\newcommand{\used}{\text{\small\sf Used\/}\xspace}
\newcounter{fmdefinition}

\newlength{\boxwidth}
\setlength{\boxwidth}{45.5mm}

\newcommand{\defbeg}[2]{\noindent\text{{\em\bfseries Definition }\refstepcounter{fmdefinition}\thefmdefinition: {\small\sf\em #1}\label{#2}}}
\newcommand{\defcaption}[2]{\caption{\small\sf\em #1}\label{#2}}

\newcommand{\BEGDEF}[2]{%
\refstepcounter{fmdefinition}%
%\vskip5pt
\noindent
\begin{tabular}{|l@{}r@{\ }c@{\ }l@{\ }|}
	\hline%
	\multicolumn{4}{|@{}l@{\ }|}{%
	\label{#2}
	\rule[-1.mm]{0mm}{3.5mm}%
	\makebox{\text{\small\bfseries\em Definition \thefmdefinition:\sf\em\protect\ #1}}}%
	 \\	
        \hline
}

\newcommand{\BEGDEFWITHTEXT}[2]{%
\refstepcounter{fmdefinition}%
%\vskip5pt
\noindent
\begin{tabular}{|l|l|}
	\hline%
	\multicolumn{2}{|@{}l|}{%
	\label{#2}
	\rule[-1.mm]{0mm}{3.5mm}%
	\makebox{\text{\small\bfseries\em Definition \thefmdefinition:\sf\em\protect\ #1}}}%
	 \\	
        \hline
\begin{tabular}{@{}l@{}r@{\ }c@{\ }l}
}

\newcommand{\CONTDEF}[1]{%
%\vskip5pt
\noindent
\begin{tabular}{|l@{}r@{\ }c@{\ }l@{\ }|}
	\multicolumn{4}{|l@{\ }|}{%
	\rule[-1.mm]{0mm}{3.5mm}%
	\makebox[\boxwidth]{\text{\sf\em\protect#1 (Continued)}}}%
	 \\	
        \hline
}

\newcommand{\ENDDEF}{%
%\hline
\end{tabular}
%\vskip5pt
}

\newcommand{\ENDDEFWITHTEXT}[2]{%
\end{tabular}
&
\begin{tabular}{c@{}}
\begin{minipage}{#1mm}
\protect#2
\end{minipage}
\end{tabular}
\\ \hline
\end{tabular}
%\vskip5pt
}

\newcommand{\oneColumn}[1]{
	\multicolumn{4}{l@{\ }}{%
	$\protect#1$}% 
        \\
	}
\newcommand{\oneColumnWITHTEXT}[1]{
	\multicolumn{4}{@{}l@{\ }}{%
	$\protect#1$}% 
        \\
	}
\newcommand{\threeColumns}[4]{
	\hspace*{#1mm}
	& $\protect{#2}$
	& $\protect{#3}$
	& $\protect{#4}$
	\\
	}
\newcommand{\twoColumns}[2]{
        \hspace*{#1mm}
	& 
	\multicolumn{3}{@{}l@{\ }|}{%
	$\protect#2$}% 
        \\
	}

\newcommand{\DEFRule}{%
\arrayrulecolor{lightgray}\hline\arrayrulecolor{black}
}
\newcommand{\CASERule}{%
\arrayrulecolor{lightgray}\cline{1-2}\arrayrulecolor{black}
}

\newcommand{\orderComputationRevised}{\hspace*{-2mm}\text{$
\begin{array}{|c|c|c|c|c|}
\hline
& \text{Desirable} & \text{Undesirable} & \begin{tabular}{c} \text{Semantically} \\ \text{Invalid} \end{tabular} & \text{No Composition}
\\ \hline \hline
	\ttcomp 
	& \labsrcp \leq \labtgtp \leq \labtgtn
	& \labtgtp > \labtgtn
	& -
	& -
\\ \hline
	\stcomp
	& \labsrcp \leq \labtgtp < \labsrcn
	& -
	& \labtgtp > \labsrcn
	& \labtgtp = \labsrcn
\\ \hline
	\tscomp
	& \labtgtp \leq \labsrcp \leq \labtgtn
	& \labsrcp > \labtgtn
	& -
	& -
\\ \hline
	\sscomp
	& \labtgtp \leq \labsrcp \leq \labsrcn
	& -
	& \labsrcp > \labsrcn
	& \labsrcp = \labsrcn
\\ \hline
\end{array}\!\!\!$}}

\newcommand{\orderComputation}{\hspace*{-2mm}\text{$
\begin{array}{c|c|c|c}
\multicolumn{2}{c|}{\text{\tscomp Composition} \; (\tgtn=\srcp)\rule{0em}{1em}} 
	& \multicolumn{2}{c}{\text{\sscomp Composition} \; (\srcn=\srcp)}
	\\ \hline
%%	\\ \hline\hline
\text{Cond.} 
	& \langle\labsrcr \;,\; \labtgtr \rangle
	&  \text{Cond.} 
	& \langle\labsrcr \;,\; \labtgtr \rangle
	\rule[-.5em]{0em}{1.5em}
	\\ \hline%\hline
\rule[-.6em]{0em}{1.7em}
\labtgtn<\labsrcp
	& \langle
	 \labsrcn + \labsrcp - \labtgtn \;,\; \labtgtp
	 \rangle
	 & \labsrcn < \labsrcp
	& \langle
	 \labtgtp \;,\; \labtgtn + \labsrcp - \labsrcn
	 \rangle
	\\ \hline
\rule[-.6em]{0em}{1.7em}
\labtgtn>\labsrcp
	& \langle
         \labsrcn \;,\; \labtgtp+\labtgtn - \labsrcp
	  \rangle
	& \labsrcn > \labsrcp
	& \langle
          \labtgtp + \labsrcn -  \labsrcp \;,\; \labtgtn
	   \rangle
	\\ \hline
\rule[-.6em]{0em}{1.6em}
\labtgtn=\labsrcp 
	& \langle
	\labsrcn \;,\; \labtgtp
	\rangle
	& \labsrcn = \labsrcp & \text{
		No composition
		}
\end{array}\!\!\!$}}

\newcommand{\gpuDef}{
%\defbeg{Generalized Points-to Update.}{def:gpu}
Given variables $x$ and $y$ and $i > 0$, $j \geq 0$, a \emph{generalized points-to update} (\gpu) $\denew{x}{i|j}{y}{\flab}$
represents a memory transformer in which %%that
%%the statement \flab which defines
all locations reached by $i-1$ indirections from $x$ in the abstract memory
are defined by the pointer assignment labelled \flab,
to hold the address of all locations
reached by $j$ indirections from $y$.
The pair \text{$i|j$} represents indirection levels and is called the \indlev of the \gpu
($i$ is the \indlev of $x$, and $j$ is the \indlev  of $y$).
The letter $\edge$ is used to denote a \gpu{} unless named otherwise.
}

\newcommand{\gpgDef}{
%\defbeg{A Generalized Points-to Block and a Generalized Points-to Graph.}{def:gpg}
A \emph{generalized points-to block} (\gpb), denoted \gpbsym, is a set of \gpus abstracting memory updates.
A \emph{generalized points-to graph} (\gpg) of a procedure, denoted \mtsym, is a graph
 $(N,E)$ whose nodes in $N$ are labelled with
\gpbs and edges in $E$ abstract the control flow of the procedure.
By common abuse of notation, we often conflate nodes and their
\gpb labellings.
}

\newcommand{\gpuReductionDef}{
\renewcommand{\arraystretch}{.8}%
\setlength{\codeLineLength}{100mm}%
\newcommand{\WHILE}{\text{\bf while }}%
\newcommand{\IF}{\text{\bf if }}%
\newcommand{\ELSE}{\text{\bf else }}%
\newcommand{\FOR}{\text{\bf for }}%
\newcommand{\RETURN}{\text{\bf return }}%
\newcommand{\EACH}{\text{\bf each }}%
\newcommand{\INPUT}{\text{\bf Input}}%
\newcommand{\OUTPUT}{\text{\bf Output}}%
\newcommand{\EXTRACT}{\text{extract }}%
\newcommand{\SUCCEEDS}{\text{ succeeds }}%
\newcommand{\WCOMPOSE}{\WCompose}
\newcommand{\FROM}{\text{ from }}%
\newcommand{\FILTER}{\text{\em filter}}%
\newcommand{\FAIL}{\text{\em fail}}%
\newcommand{\CHECK}{\text{\em no\_barrier\_between\_edges}}%
\begin{tabular}{rc}
\codeLineNoNumber{0}{\INPUT: \candedge \rule{15.5mm}{0mm} \slash\slash \; The consumer \gpu to be simplified}{white}
\codeLineNoNumber{1}{\rule{4mm}{0mm} \flow \rule{15mm}{0mm} \slash\slash \; The context (set of \gpus) in which \candedge is to be simplified}{white}
%%\codeLineNoNumber{1}{\rule{4mm}{0mm} \flow \rule{20mm}{0mm} \slash\slash \; The set of \gpus using which \candedge is to be simplified  }{white}
\codeLineNoNumber{0}{\OUTPUT: \GENEDGES \hfill \rule{9mm}{0mm} \slash\slash \;  The set of simplified \gpus equivalent to \candedge}{white}
\codeLineOne{1}{0}{\edgeReduction(\candedge, \flow)}{white}
\codeLine{0}{\OB $\GENEDGES = \emptyset$}{white}
\codeLine{1}{$\WL = \{\candedge\}$}{white}
\codeLine{1}{$\WHILE(\WL \neq \emptyset)$}{white}
\codeLine{1}{\OB $\EXTRACT  w  \FROM \WL$}{white}
\codeLine{2}{$\WCOMPOSE = \FALSE$}{white}
\codeLine{2}{$\FOR \EACH \edge \in \flow$}{white}
%%\codeLine{2}{\OB $\IF \big(\CHECK(w, \edge) = \TRUE\big)$}{white}
\codeLine{2}{\OB $\IF (r = w \ecomp^{\textrm{ts}} \edge) \SUCCEEDS$}{white}
\codeLine{3}{\OB $\WL = \WL \cup \{r\}$}{white}
\codeLine{4}{$\WCOMPOSE = \TRUE$}{white}
\codeLine{3}{\CB}{white}
\codeLine{3}{$\ELSE \IF (r = w \ecomp^{\textrm{ss}} \edge) \SUCCEEDS$}{white}
\codeLine{3}{\OB $\WL = \WL \cup \{r\}$}{white}
\codeLine{4}{$\WCOMPOSE = \TRUE$}{white}
\codeLine{3}{\CB}{white}
\codeLine{2}{\CB}{white}
\codeLine{2}{$\IF (\neg\,\WCOMPOSE)$}{white}
\codeLine{3}{$\GENEDGES = \GENEDGES \cup \{w\}$}{white}
\codeLine{1}{\CB}{white}
\codeLine{1}{\RETURN \GENEDGES}{white}
\codeLine{0}{\CB}{white}
\end{tabular}
}

\newcommand{\reachingGPUdfaDef}{
\begin{tabular}{@{}l@{}r@{\ }c@{\ }l@{}}
\threeColumns{0}%
	{
		\rule[-6.7mm]{0mm}{15.5mm}%
		\RIn{\flab}}%
	{\coloneq}%
	{	
		\begin{cases}
		\big\{ \denew{x}{\ell|\ell}{x'}{\flab} \mid x\in \ptrs, 0 < \ell \leq \kappa \big\}	
				& \flab = \Start{}, \text{$\kappa$ is the largest \indlev} 	
		\\
		\bigcup\limits_{p\; \in\; \gpredsubscript(\flab)} \ROut{p}
				&  \text{otherwise}
		\end{cases}
	}
\DEFRule
\threeColumns{0}%
	{
		\rule[-1.7mm]{0mm}{5.5mm}%
		\ROut{\flab}}%
	{\coloneq}%
	{\left(\RIn{\flab} - \RKill{\flab}\right) \cup \RGen{\flab}}
\DEFRule
\threeColumns{0}%
	{
		\rule[-1.7mm]{0mm}{5.5mm}%
		\RGen{\flab}}%
	{\coloneq}%
	{\consgen\left(\gpbsym_{\flab},\; \RIn{\flab}\right)}
\DEFRule
\threeColumns{0}%
	{
		\rule[-1.7mm]{0mm}{5.5mm}%
		\RKill{\flab}}%
	{\coloneq}%
	{\conskill\left(\RGen{\flab},\; \RIn{\flab}\right)}
\DEFRule
\threeColumns{0}%
	{
		\rule[-2.5mm]{0mm}{6mm}%
		\consgen(X,\flow)}%
	{\coloneq}%
	{\bigcup\limits_{\edge\; \in X} \edge \rcomp \flow}
\DEFRule
\threeColumns{0}%
	{
		\rule[-1.7mm]{0mm}{5.5mm}%
		\conskill(X,\flow)}%
	{\coloneq}%
	{ 
%	  \big\{ \edge_1 \mid \edge_1 \in \!\match(\edge, \flow), \edge \in X,
%		|\Def(X, \edge)| \!=\! 1 \big\}
          \big\{ \edge_1 \mid \exists\, \edge \in X \mbox{ such that }
		|\Def(X, \edge)| \!=\! 1 \wedge
                \edge_1 \in \!\match(\edge, \flow)
          \big\}
	}
%%\DEFRule
%%\threeColumns{0}%
%%	{
%%		\rule[-1.7mm]{0mm}{5mm}%
%%		\ppoint(\denew{x}{i|j}{y}{\flab})}
%%	{\coloneq}%
%%	{\flab}
\DEFRule
\threeColumns{0}%
	{
		\rule[-1.7mm]{0mm}{5mm}%
		\match(\denew{x}{i|j}{y}{\flab}, \flow)}%
	{\coloneq}%
	{
%          \{\denew{u}{k|l}{v}{\slab} \mid \denew{u}{k|l}{v}{\slab} \in \flow,\; x = u,\; i = k\}
	 \big\{\edge \in \flow \mid
            \edge = \denew{u}{k|l}{v}{\slab},\; x = u,\; i = k \big\}
        }
\DEFRule
\threeColumns{0}%
	{ \rule[-2mm]{0mm}{5mm}%
		\Def\big(X, \denew{w}{k|l}{z}{\flab}\big)%
	}%
	{\coloneq}%
	{\big\{(x, i) \mid \denew{x}{i|j}{y}{\flab} \in X\big\}
}
\ENDDEF
}

\newcommand{\reachingGPUWBdfaDef}{
\begin{tabular}{@{}l@{}r@{\ }c@{\ }l@{}}
\threeColumns{2}%
	{
		\rule[-6.7mm]{0mm}{15.5mm}%
		\BRIn{\flab}}%
	{\coloneq}%
	{	
		\begin{cases}
		\big\{ \denew{x}{\ell|\ell}{x'}{\flab} \mid x\in \ptrs, 0 < \ell \leq \kappa \big\}	
				& \flab = \Start{}, \text{$\kappa$ is the largest \indlev} 	
		\\
		\bigcup\limits_{p\; \in\; \gpredsubscript(\flab)} \BROut{p}
				&  \text{otherwise}
		\end{cases}
	}
\DEFRule
\threeColumns{2}%
	{
		\rule[-2.2mm]{0mm}{6.mm}%
		\BROut{\flab}}%
	{\coloneq}%
	{\left(\BRIn{\flab} - \left(\BRKill{\flab} \cup \blocked(\BRIn{\flab},\BRGen{\flab})\right)\right) \cup \BRGen{\flab}}
\DEFRule
\threeColumns{2}%
	{
		\rule[-7.5mm]{0mm}{16.5mm}%
		\blocked\,(I,G)
	}%
	{\coloneq}%
	{
	\begin{cases}
	\emptyset
		& G = \emptyset  
			 %\rule{24.5mm}{0mm}\text{(Case 1)}
	\\
		\rule[-2.2mm]{0mm}{3.5mm}%
	\{\edge \mid \edge \in I, \bDdep(\IGPUs(G), \{\edge\})\}
			& |\IGPUs(G)| > 1 %%\exists\, \denew{x}{i|j}{y}{\slab} \in G, i > 1
			 %\rule{5mm}{0mm}\text{(Case 2)}
	\\
	\{\edge \mid \edge \in \IGPUs(I), \bDdep(G, \{\edge\})\} & \text{otherwise}
	\end{cases}
	}
\DEFRule
\threeColumns{2}%
	{
		\rule{0mm}{4.5mm}%
		\IGPUs\,(X)
	}%
	{\coloneq}%
	{
	\{\denew{x}{i|j}{y}{\flab} \mid \denew{x}{i|j}{y}{\flab} \in X, i > 1\}
	}
\DEFRule
\threeColumns{2}%
	{
		\rule[-1.7mm]{0mm}{5.5mm}%
		\BRGen{\flab}}%
	{\coloneq}%
	{\consgen\left(\gpbsym_{\flab},\; \BRIn{\flab}\right)}
\DEFRule
\threeColumns{2}%
	{
		\rule[-1.7mm]{0mm}{5.5mm}%
		\BRKill{\flab}}%
	{\coloneq}%
	{\conskill\left(\BRGen{\flab},\; \BRIn{\flab}\right)}
\DEFRule
\threeColumns{2}%
	{
		\rule[-1.7mm]{0mm}{5.5mm}%
		\bDdep(B, I)}%
	{\Leftrightarrow}%
	{
		\begin{array}[t]{@{}l}
		\LT(B) \cap 
		\left(\LT(I) \cup \RT(I)\right) \neq \emptyset
		\end{array}
	}
\DEFRule
\threeColumns{2}%
	{
		\rule[-1.7mm]{0mm}{5.5mm}%
		\LT(X)}%
	{\coloneq}%
	{ 
		\big\{ \Type(x, i) \mid  \denew{x}{i|j}{y}{\flab} \, \in\, X \big\}
	}
\DEFRule
\threeColumns{2}%
	{
		%\rule[-1.7mm]{0mm}{5.5mm}%
		\RT(X)}%
	{\coloneq}%
	{ 	
		\begin{array}[t]{@{}l}
			\big\{ \Type(x, k) \mid  1 \leq k < i, \denew{x}{i|j}{y}{\flab} \in X \big\} \; \cup
			\\
			\big\{ \Type(y, k) \mid  1 \leq k < j, \denew{x}{i|j}{y}{\flab} \in X \big\}
		\end{array}
	}
\DEFRule
\oneColumn{
	\text{Note: The definitions of \consgen and \conskill are
	\rule[-.75em]{0em}{2em}
         same as in Definition~\protect\ref{def:dfaReachingGPUAnalysis}}
	}
%%\DEFRule
%%\threeColumns{2}%
%%	{
%%		\rule[-2.5mm]{0mm}{6mm}%
%%		\consgen(X,\flow)}%
%%	{\coloneq}%
%%	{\bigcup\limits_{\edge\; \in X} \edge \rcomp \flow}
%%\DEFRule
%%\threeColumns{2}%
%%	{
%%		\rule[-1.7mm]{0mm}{5.5mm}%
%%		\conskill(X,\flow)}%
%%	{\coloneq}%
%%	{ 
%%%	  \big\{ \edge_1 \mid \edge_1 \in \!\match(\edge, \flow), \edge \in X,
%%%		|\Def(X, \edge)| \!=\! 1 \big\}
%%          \big\{ \edge_1 \mid \exists\, \edge \in X \mbox{ such that }
%%		|\Def(X, \edge)| \!=\! 1 \wedge
%%                \edge_1 \in \!\match(\edge, \flow)
%%          \big\}
%%	}
%%%%\DEFRule
%%%%\threeColumns{2}%
%%%%	{
%%%%		\rule[-1.7mm]{0mm}{5mm}%
%%%%		\ppoint(\denew{x}{i|j}{y}{\flab})}
%%%%	{\coloneq}%
%%%%	{\flab}
%%\DEFRule
%%\threeColumns{2}%
%%	{
%%		\rule[-1.7mm]{0mm}{5mm}%
%%		\match(\denew{x}{i|j}{y}{\flab}, \flow)}%
%%	{\coloneq}%
%%	{
%%%          \{\denew{u}{k|l}{v}{\slab} \mid \denew{u}{k|l}{v}{\slab} \in \flow,\; x = u,\; i = k\}
%%	 \big\{\edge \in \flow \mid
%%            \edge = \denew{u}{k|l}{v}{\slab},\; x = u,\; i = k \big\}
%%        }
%%\DEFRule
%%\threeColumns{2}%
%%	{ \rule[-2mm]{0mm}{5mm}%
%%		\Def\big(X, \denew{w}{\cdot|\cdot}{z}{\flab}\big)%
%%	}%
%%	{\coloneq}%
%%	{\big\{(x, i) \mid \denew{x}{i|j}{y}{\flab} \in X\big\}}
\ENDDEF
}

\newcommand{\ptaDefA}{
\begin{tabular}{@{}l@{}r@{\ }c@{\ }l@{}}
\threeColumns{2}%
	{
		\rule[-6.7mm]{0mm}{15.5mm}%
		\PIn{\flab}}%
	{\coloneq}%
	{	
		\begin{cases}
		\boundary 
				& \flab = \Start{}	
		\\
		\bigcup\limits_{p\; \in\; \gpredsubscript(\flab)} \POut{p}
				&  \text{otherwise}
		\end{cases}
	}
\DEFRule
\threeColumns{2}%
	{
		\rule[-1.7mm]{0mm}{5.5mm}%
		\POut{\flab}}%
	{\coloneq}%
	{\left(\PIn{\flab} - \PKill{\flab}\right) \cup \llbracket\gpbsym_{\flab}\rrbracket \PIn{\flab}}
\DEFRule
\threeColumns{2}%
	{
		\rule[-1.7mm]{0mm}{5.5mm}%
		\PKill{\flab}}%
	{\coloneq}%
	{\memkill\left(\llbracket\gpbsym_{\flab}\rrbracket \PIn{\flab},\;\PIn{\flab}\right)}
\DEFRule
\threeColumns{2}%
	{
		\rule[-1.7mm]{0mm}{5.5mm}%
		\memkill(\mem, P)}%
	{\coloneq}%
	{ \big\{ \edge_1 \mid \edge_1 \in \!\match(\edge, P), \edge \in \mem,
		\singledef{\mem}{\edge}\big\}
	}
\DEFRule
\threeColumns{2}%
	{ \rule[-2mm]{0mm}{5mm}%
		\singledef{\mem}{\denew{x}{i|j}{y}{\flab}}%
	}%
	{\coloneq}%
	{|\mem\,^{i-1}\{x\}| = 1\; \wedge\; \mem\,^{i-1}\{x\} \neq \{?\}}
%%	{\big\{(x, i) \mid \denew{x}{i|j}{y}{\flab} \in X\big\}}
\ENDDEF
}

\newcommand{\edgeReductionDefAugmented}{
\renewcommand{\arraystretch}{.8}%
\setlength{\codeLineLength}{114mm}%
\newcommand{\WHILE}{\text{\bf while }}%
\newcommand{\IF}{\text{\bf if }}%
\newcommand{\ELSE}{\text{\bf else }}%
\newcommand{\FOR}{\text{\bf for }}%
\newcommand{\RETURN}{\text{\bf return }}%
\newcommand{\EACH}{\text{\bf each }}%
\newcommand{\INPUT}{\text{\bf Input}}%
\newcommand{\OUTPUT}{\text{\bf Output}}%
\newcommand{\EXTRACT}{\text{extract }}%
\newcommand{\SUCCEEDS}{\succeeds}%
\newcommand{\FROM}{\text{ from }}%
\newcommand{\WCOMPOSE}{\text{\em composed}}%
\newcommand{\WCOMPOSEA}{\text{\em tscomp}}%
\newcommand{\WCOMPOSEB}{\text{\em sscomp}}%
\newcommand{\POSTA}{\text{\em tspost}}%
\newcommand{\POSTB}{\text{\em sspost}}%
\newcommand{\UNDES}{\UNDes}%
\newcommand{\FILTER}{\text{\em filter}}%
\newcommand{\FAIL}{\text{\em fail}}%
\newcommand{\IRED}{\ired}
\newcommand{\OR}{\text{\bf or}\xspace}
\newcommand{\AND}{\text{\bf and}\xspace}
\newcommand{\BLOCKED}{\text{\em blocked}\xspace}%
\newcommand{\COMPOSE}{\text{\small\sf Compose\_\gpus}\xspace}%
\newcommand{\ADD}{\queued}
\newcommand{\augmentedEdgeReduction}{\text{\small\sf Augmented\_\gpu\!\!\_reduction }}%
\begin{tabular}{@{}rc@{}}
\codeLineNoNumber{0}{\INPUT: \rule{2mm}{0mm} \candedge \rule{12.5mm}{0mm} \slash\slash \; The consumer \gpu to be simplified}{white}
\codeLineNoNumber{1}{\rule{7mm}{0mm} \flow \rule{12mm}{0mm} \slash\slash \; The set of \gpus using which \candedge is to be simplified  }{white}
\codeLineNoNumber{1}{\rule{7mm}{0mm} \flowbar \rule{12mm}{0mm} \slash\slash \; The set of \gpus that have been blocked by a
barrier  }{white}
%%\codeLineNoNumber{1}{\rule{4mm}{0mm} \IRED \rule{4mm}{0mm} \slash\slash \; The flag which checks $\flow \neq \flowbar$}{white}
%%\codeLineNoNumber{0}{\rule{9mm}{0mm} \ADD\hfill \rule{1mm}{0mm} \slash\slash \;  The set of \gpus for which \gpu reduction is postponed}{white}
\codeLineNoNumber{0}{\OUTPUT: \GENEDGES \hfill \rule{9mm}{0mm} \slash\slash \;  The set of simplified \gpus equivalent to \candedge}{white}
\codeLineNoNumber{0}{\rule{12mm}{0mm} \ADD\hfill \rule{3mm}{0mm} \slash\slash \;  The set of \gpus which may be used later}{white}
\codeLineOne{1}{0}{\augmentedEdgeReduction(\candedge, \flow, \flowbar)}{white}
\codeLine{0}{\OB $\GENEDGES = \ADD = \emptyset$}{white}
\codeLine{1}{$\WL = \{\candedge\}$}{white}
\codeLine{1}{$\WHILE(\WL \neq \emptyset)$}{white}
\codeLine{1}{\OB $\EXTRACT  w  \FROM \WL$}{white}
%%\codeLine{2}{${\text{\protect$\overbar{\protect\protect\rule{0em}{.8em}\flow}$}\xspace} = \FILTER(w, \flow)$}{white}
\codeLine{2}{$\FOR \EACH \edge \in \flow$}{white}
\codeLine{2}{\OB $\langle \WL, \WCOMPOSEA, \POSTA\rangle = \COMPOSE(ts,w,\edge,\WL,\flowbar)$}{white}
\codeLine{3}{$\langle \WL, \WCOMPOSEB, \POSTB\rangle = \COMPOSE(ss,w,\edge,\WL,\flowbar)$}{white}
\codeLine{3}{$\IF (\POSTA \; \OR \; \POSTB)$}{white}
\codeLine{4}{$\queued = \queued\, \cup \{\edge\}$}{white}
\codeLine{2}{\CB}{white}
\codeLine{2}{$\IF (\neg\,(\WCOMPOSEA \; \OR\;  \WCOMPOSEB))$}{white}
\codeLine{3}{$\GENEDGES = \GENEDGES \; \cup \{w\}$}{white}
%%\codeLine{3}{$\IF \big(\neg\,\PTEDGE(w)) \; \vee \; \POSTPONE \big)$}{white}
%%\codeLine{4}{$\ADD = \ADD \; \cup \{w\}$}{white}
%%\codeLine{2}{\CB}{white}
\codeLine{1}{\CB}{white}
\codeLine{1}{\RETURN (\GENEDGES\hspace*{.01pt}, \ADD)}{white}
\codeLine{0}{\CB}{white}

\codeLine{0}{$\COMPOSE(\ecomptype,w,\edge,\WL,\BLOCKED)$}{white}
\codeLine{0}{\OB $\WCOMPOSE=\POSTPONE=\FALSE$}{white}
\codeLine{1}{$\IF (r = w\, \ecomp \edge) \;\SUCCEEDS$}{white}
\codeLine{1}{\OB $\IF (\edge \notin \BLOCKED)$}{white}
\codeLine{2}{\OB $\WL = \WL \cup \{r\}$}{white}
\codeLine{3}{$\WCOMPOSE = \TRUE$}{white}
\codeLine{2}{\CB}{white}
\codeLine{2}{$\ELSE \POSTPONE = \TRUE$}{white}
\codeLine{1}{\CB}{white}
\codeLine{1}{$\ELSE \IF (\undesComp(\ecomptype,w,\edge))$}{white}
\codeLine{2}{$\POSTPONE = \TRUE$}{white}
\codeLine{1}{$\RETURN \langle \WL, \WCOMPOSE, \POSTPONE\rangle$}{white}
\codeLine{0}{\CB}{white}
\end{tabular}
}

\newcommand{\blockedGPUdfaDef}{
\begin{tabular}{@{}l@{}r@{\ }c@{\ }l@{}}
\threeColumns{0}%
	{
		\rule[-6.1mm]{0mm}{14mm}%
		\BRIn{\flab}}%
	{\coloneq}%
	{	
		\begin{cases}
		\qquad\qquad \emptyset & \flab = \Start{}	
		\\
		\bigcup\limits_{p\in\gpredsubscript(\flab)}\!\! \BROut{p}
				&  \text{otherwise}
		\end{cases}
	}
%%	{\bigcup\limits_{p\; \in\; \gpredsubscript(\flab)} \BROut{p}}
\DEFRule
%%\threeColumns{0}%
%%	{
%%		\rule[-4.4mm]{0mm}{8.5mm}%
%%		\RGen{\flab}}%
%%	{\coloneq}%
%%	{\bigcup\limits_{\edge \in \gpbsym_{\flab}} \first\big(\augmentedEdgeReduction(\edge, \RIn{\flab}, \overbar{\RIn{\flab}})\big)}
\DEFRule
\threeColumns{0}%
	{
		\rule[-8.2mm]{0mm}{18.5mm}%
		\BROut{\flab}}%
	{\coloneq}%
	{
	\begin{cases}
	\BRIn{\flab} 
		& \RGen{\flab} = \emptyset  
			 \rule{24.5mm}{0mm}\text{(Case 1)}
	\\
	\BRIn{\flab} \; \cup \; \unblockall(\RIn{\flab}, \RGen{\flab})
			& \exists\, \denew{x}{i|j}{y}{\slab} \in \RGen{\flab}, i > 1
			 \rule{5mm}{0mm}\text{(Case 2)}
	\\
	\BRIn{\flab} \; \cup \; \unblockindirect(\RIn{\flab}, \RGen{\flab})
			%\big\{ \denew{x}{i|j}{y}{\slab} \in \RIn{\flab}, i > 1 \big\}
		& \text{otherwise}
			 \rule{29.5mm}{0mm}\text{(Case 3)}
	\end{cases}
	}
\DEFRule
\threeColumns{0}%
	{
		\rule[-4.5mm]{0mm}{10.5mm}%
		\unblockall(I,G)}%
	{\coloneq}%
	{	
		\left\{
		\denew{x}{i|j}{y}{\slab} \; \middle\vert \; 
		\denew{x}{i|j}{y}{\slab} \in I, 
		\denew{x}{k|l}{z}{\tlab} \notin  G
		\right\}
	}
\DEFRule
\threeColumns{0}%
	{
		\rule[-4.5mm]{0mm}{10.5mm}%
		\unblockindirect(I,G)}%
	{\coloneq}%
	{	
		\left\{
		\denew{x}{i|j}{y}{\slab} \; \middle\vert \;
		\denew{x}{i|j}{y}{\slab} \in I, i>1,
		\denew{x}{k|l}{z}{\tlab} \notin  G
		\right\}
	}
%%\DEFRule
%%\threeColumns{0}%
%%	{
%%		\rule[-1.7mm]{0mm}{5.5mm}%
%%		\first((A, B))}%
%%	{\coloneq}%
%%	{A}
%%\DEFRule
%%\threeColumns{0}%
%%	{
%%		\rule[-1.7mm]{0mm}{5.5mm}%
%%		\second((A, B))}%
%%	{\coloneq}%
%%	{B}
\ENDDEF
}

\newcommand{\handlingRecursionDefNew}{
\renewcommand{\arraystretch}{.8}%
\setlength{\codeLineLength}{120mm}%
\newcommand{\IF}{\text{\bf if }}%
\newcommand{\RETURN}{\text{\bf return }}%
\newcommand{\INPUT}{\text{\bf Input}}%
\newcommand{\OUTPUT}{\text{\bf Output}}%
\newcommand{\refineGPG}{\text{\small\sf Refine\_\gpg}\xspace}%
\begin{tabular}{rc@{}}
\codeLineNoNumber{0}{\INPUT: \rule{2mm}{0mm} $p, \mtsym^1_p$, $\mtsym^i_p$ \rule{3.5mm}{0mm} \slash\slash \; A recursive procedure, 
         its first incomplete \gpg containing only}{white}
\codeLineNoNumber{0}{\rule{31mm}{0mm} \slash\slash \; recursive calls, and its $i^{th}$ \gpg in the fixed-point computation}{white}
\codeLineNoNumber{0}{\OUTPUT: $\mtsym^{i+1}_p$ \rule{12mm}{0mm} \slash\slash \; Optimized $(i+1)^{th}$ \gpg for procedure $p$}{white}
\codeLineOne{1}{0}{\refineGPG $(p, \mtsym^1_p, \mtsym^i_p)$}{white}
\codeLine{0}{\OB}{white}
\codeLine{1}{$\Rprev = \ROut{\Endscriptsize{}}(\mtsym^i_p)$}{white}
\codeLine{1}{$\RprevBar = \BROut{\Endscriptsize{}}(\mtsym^i_p)$}{white}
\codeLine{1}{Compute $\mtsym^{i+1}_p$ by inlining recursive calls in $\mtsym^{1}_p$ with their
		latest \gpgs}
		{white}
\codeLine{1}{Perform both variants of reaching \gpus analysis over $\mtsym^{i+1}_p$}{white}
\codeLine{1}{$\Rcur = \ROut{\Endscriptsize{}}(\mtsym^{i+1}_p)$}{white}
\codeLine{1}{$\RcurBar = \BROut{\Endscriptsize{}}(\mtsym^{i+1}_p)$}{white}
\codeLine{1}{$\IF\left((\Rcur \neq \Rprev) \vee (\RcurBar \neq \RprevBar)\right)$}{white}
\codeLine{3}{Push callers of $p$ on the worklist}{white}
%%\codeLine{2}{$\IF\left(p \text{ is the source of back edge} \right)$}{white}
%%\codeLine{3}{Push $p$ on the worklist}{white}
%%\codeLine{1}{\CB}{white}
\codeLine{1}{$\text{Perform strength reduction and redundancy elimination optimizations over }\mtsym^{i+1}_p$}{white}
\codeLine{1}{$\RETURN \mtsym^{i+1}_p$}{white}
\codeLine{0}{\CB}{white}
\end{tabular}
}

\newcommand{\edgeReductionDefHeap}{
\renewcommand{\arraystretch}{.8}%
\setlength{\codeLineLength}{103mm}%
\newcommand{\WHILE}{\text{\bf while }}%
\newcommand{\IF}{\text{\bf if }}%
\newcommand{\FOR}{\text{\bf for }}%
\newcommand{\RETURN}{\text{\bf return }}%
\newcommand{\EACH}{\text{\bf each }}%
\newcommand{\INPUT}{\text{\bf Input}}%
\newcommand{\OUTPUT}{\text{\bf Output}}%
\newcommand{\EXTRACT}{\text{extract }}%
\newcommand{\SUCCEEDS}{\text{ succeeds }}%
\newcommand{\WCOMPOSE}{\WCompose}
\newcommand{\FROM}{\text{ from }}%
\newcommand{\FILTER}{\text{\em filter}}%
\newcommand{\FAIL}{\text{\em fail}}%
\newcommand{\CHECK}{\text{\em no\_barrier\_between\_edges}}%
\begin{tabular}{rc@{}}
\codeLineNoNumber{0}{\INPUT: \candedge \rule{15.5mm}{0mm} \slash\slash \; The consumer \gpu to be simplified}{white}
\codeLineNoNumber{1}{\rule{4mm}{0mm} \flow \rule{15mm}{0mm} \slash\slash \; The context (set of \gpus) in which \candedge is to be simplified}{white}
\codeLineNoNumber{1}{\rule{4mm}{0mm} \used \rule{10mm}{0mm} \slash\slash \; The set of \gpus used for \gpu reduction for a \gpu}{white}
\codeLineNoNumber{0}{\OUTPUT: \GENEDGES \hfill \rule{8mm}{0mm} \slash\slash \;  The set of simplified \gpus equivalent to \candedge}{white}
\codeLineOne{1}{0}{\edgeReduction(\candedge, \flow, \used)}{white}
\codeLine{0}{\OB $\GENEDGES = \emptyset$}{white}
%%\codeLine{1}{$\WL = \{\candedge\}$}{white}
%%\codeLine{1}{$\WHILE(\WL \neq \emptyset)$}{white}
%%\codeLine{1}{\OB $\EXTRACT  w  \FROM \WL$}{white}
\codeLine{1}{$\WCOMPOSE = \FALSE$}{white}
\codeLine{1}{$\FOR \EACH \edge \in (\flow - \used)$}{white}
%%\codeLine{2}{\OB $\IF \big(\CHECK(w, \edge) = \TRUE\big)$}{white}
\codeLine{1}{\OB $\FOR \EACH r \in (\candedge \ecomp^{\textrm{ts}} \edge)$}{white}
\codeLine{2}{\OB $\GENEDGES = \GENEDGES\, \cup \edgeReduction\;(r, \flow, \used\,\cup\,\{\edge\})$}{white}
\codeLine{3}{$\WCOMPOSE = \TRUE$}{white}
\codeLine{2}{\CB}{white}
\codeLine{2}{$\FOR \EACH r \in (\candedge \ecomp^{\textrm{ss}} \edge)$}{white}
\codeLine{2}{\OB $\GENEDGES = \GENEDGES\, \cup \edgeReduction\;(r, \flow, \used\,\cup\,\{\edge\})$}{white}
\codeLine{3}{$\WCOMPOSE = \TRUE$}{white}
\codeLine{2}{\CB}{white}
\codeLine{1}{\CB}{white}
\codeLine{1}{$\IF (\neg\,\WCOMPOSE)$}{white}
\codeLine{2}{$\GENEDGES = \GENEDGES\, \cup \{\candedge\}$}{white}
%%\codeLine{1}{\CB}{white}
\codeLine{1}{\RETURN \GENEDGES}{white}
\codeLine{0}{\CB}{white}
\end{tabular}
}

\newcommand{\backedgesDefA}{
\begin{tabular}{@{}l@{}r@{\ }c@{\ }l@{}}
\threeColumns{2}%
	{
		\rule[-5mm]{0mm}{12mm}%
		\EBIn{\flab}}%
	{\coloneq}%
	{
		\compress\left(
	\bigcup\limits_{\slab\in\gpredsubscript(\flab)} \eflow\left(\slab\rightarrow \flab,\EBOut{\slab}\right)
	\right)
	}
\DEFRule
\threeColumns{2}%
	{
		\rule[-1.7mm]{0mm}{5.5mm}%
		\EBOut{\flab}}%	
	{\coloneq}%
	{
	\Gen_{\flab} \; \cup  \; \filter\left(\flab,\; \EBIn{\flab} - \Kill_{\flab}\right)
	}
\DEFRule
\threeColumns{2}%
	{
		\rule[-1.7mm]{0mm}{5.5mm}%
		\Gen_{\flab}}%
	{\coloneq}%
	{
	\left\{
	\langle \edge, \emptyset \rangle \mid \edge \in \RGen{\flab} \cap \relgpus
	\right\}
	}
\DEFRule
\threeColumns{2}%
	{
		\rule[-1.7mm]{0mm}{5.5mm}%
		\Kill_{\flab}}%
	{\coloneq}%
	{
	\left\{
	\langle \edge, \beta \rangle \mid \edge \in \RKill{\flab}
	\right\}
	}
\ENDDEF
}

\newcommand{\gpucompIndlevDef}{
\begin{tabular}{@{}l@{}r@{\ }c@{\ }l@{}}
\threeColumns{2}%
	{
		\rule[-6mm]{0mm}{14mm}%
		\big(\denew{z}{i|j}{x}{\slab}\big) \; \ecomp^{\textrm{ts}} \big(\denew{v}{k|l}{y}{\flab}\big)
	}%
	{\coloneq}%
	{
		\begin{cases}
		\denew{z}{i|(l+j-k)}{y}{\slab} & (v = x) \wedge (l \leq k \leq j) \;\;\;
			\\
		\text{fail} & 	\text{otherwise}	
		\end{cases}
	}
\DEFRule
\threeColumns{2}%
	{
		\rule[-6mm]{0mm}{14mm}%
		\big(\denew{x}{i|j}{z}{\slab}\big) \; \ecomp^{\textrm{ss}} \big(\denew{v}{k|l}{y}{\flab}\big)
	}%
	{\coloneq}%
	{
		\begin{cases}
		\denew{y}{(l+i-k)|j}{z}{\slab} & (v = x) \wedge (l \leq k < i) \;\;\;
			\\
		\text{fail} & 	\text{otherwise}	
		\end{cases}
	}	
\ENDDEF
}

\newcommand{\gpucompIndlistDef}{
\begin{tabular}{@{}l@{}r@{\ }c@{\ }l@{}}
\threeColumns{2}%
	{
		\rule[-6mm]{0mm}{14mm}%
		\big(\denew{z}{il_1|il_2}{x}{\slab}\big) \; \ecomp^{\textrm{ts}} \big(\denew{v}{il_3|il_4}{y}{\tlab}\big)
	}%
	{\coloneq}%
	{
		\begin{cases}
		\denew{z}{il_1|il_5}{y}{\slab} 
			& (v = x) \wedge (il_2=il_3@il_6) 
				  \wedge (il_5=il_4@il_6) 
			\\
		\text{fail} & 	\text{otherwise}	
		\end{cases}
	}
\DEFRule
\threeColumns{2}%
	{
		\rule[-6mm]{0mm}{14mm}%
		\big(\denew{x}{il_1|il_2}{z}{\slab}\big) \; \ecomp^{\textrm{ss}} \big(\denew{v}{il_3|il_4}{y}{\tlab}\big) 
	}%
	{\coloneq}%
	{
		\begin{cases}
		\denew{y}{il_5|il_2}{z}{\slab} 
			& \begin{array}[t]{@{}l}
				(v = x) \wedge  (il_1=il_3@il_6) 
				  \wedge  (il_5=il_4@il_6) 
				\\
				  \wedge \;  il_6 \neq [\,]
		          \end{array}	
			\\
		\text{fail} & 	\text{otherwise}	
		\end{cases}
	}	
\ENDDEF
}

\newcommand{\coalescingdfaDef}{
\begin{tabular}{@{}l@{}r@{\ }c@{\ }l@{}}
\threeColumns{2}%
	{
		\rule[-6.7mm]{0mm}{15.25mm}%
		\InB{n} 
	}%
	{\coloneq}%
	{	
		\begin{cases}
		\false & \text{$n$ is \Start{}}
		\\
 		\displaystyle\bigwedge_{p\, \in\, pred(n)}\coalescePred(p, n) & \text{otherwise}
		\end{cases}
	}
\DEFRule
\threeColumns{2}%
	{
		\rule[-6.7mm]{0mm}{15.25mm}%
		\OutB{n} 
	}%
	{\coloneq}%
	{	
		\begin{cases}
		\false & \text{$n$ is \End{}}
			\\
		\displaystyle\bigwedge_{s \in succ(n)} \InB{s} 
		\hspace*{24mm}
			& \text{otherwise}
		\end{cases}
	}
\DEFRule
\threeColumns{2}%
	{
		\rule[-1.7mm]{0mm}{5.5mm}%
		\coalescePred(p, n) 
	}%
	{\Leftrightarrow}%
	{
		\OutB{p} \wedge  \big(\OutT{p} = \emptyset \vee \TypeFlow(p, n) \neq \emptyset\big)
	}
\DEFRule
\threeColumns{2}%
	{
		\rule[-6.2mm]{0mm}{14.25mm}%
		\InT{n} 
	}%
	{\coloneq}%
	{	
		\begin{cases}
		\emptyset & \text{$n$ is \Start{}}
		\\
		 \bigcup\limits_{p\, \in\, pred(n)}\TypeFlow(p, n)
		\hspace*{8mm}
			& \text{otherwise}
		\end{cases}
	}
\DEFRule
\threeColumns{2}%
	{
		\rule[-4.4mm]{0mm}{10.5mm}%
		\OutT{n} 
	}%
	{\coloneq}%
	{
		\begin{cases}
		\InT{n} \cup \gpbsym_n & \text{$\InB{n} = \true$}
		\\ 
		\gpbsym_n & \text{otherwise}
		\end{cases}
	}
\DEFRule
\threeColumns{2}%
	{
		\rule[-4.4mm]{0mm}{10.5mm}%
		\TypeFlow(p, n) 
	}%
	{\coloneq}%
	{
		\begin{cases}
		\emptyset & 
		\neg\InB{n}
		\wedge \Ddep(\OutT{p}, \gpbsym_n) 
		\\
		\OutT{p} & \text{otherwise}
		\end{cases}
	}
\DEFRule
\threeColumns{2}%
	{
		\rule[-7.mm]{0mm}{10.6mm}%
		\Ddep(X, Y)
	}%
	{\Leftrightarrow}%
	{
		\begin{array}[t]{@{}l}
		\big(\deref(X) \vee \deref(Y)\big) \wedge
		\rule[-.5em]{0em}{1em}
		\\
		%\TypeGPB(\gpbsym_s) \cap 
		\big(\LT(Y) \cup \RT(Y)\big) \cap
		\LT(X - Y) \neq \emptyset
		\end{array}
	}
\DEFRule
\threeColumns{2}%
	{
		\rule{0mm}{4.5mm}%
		\deref(X)
	}%
	{\Leftrightarrow}%
	{
		\exists\; \denew{x}{i|j}{y}{\flab} \in X \; \text{ s.t. } \;
		(i > 1) \vee (j > 1)  
	}
\ENDDEF
}

\section{Introduction}
\label{sec:intro}

Points-to analysis discovers  information  about indirect  accesses in  a
program. Its  precision influences  the precision and scalability of client program analyses significantly.
Computationally intensive  analyses such as model  checking are noted as being ineffective
on  programs  containing  pointers, partly  because  of  imprecision of points-to 
%analyses~\cite{Ball:2002:SLP:503272.503274, Beyer:2007:CSV:1770351.1770419, Clarke04atool, Fischer:2005:JDP:1081706.1081742, ivanvcic2005model, Jhala:2009:SMC:1592434.1592438}. 
analysis~\cite{Ball:2002:SLP:503272.503274}.

\subsection{The Context of this Work}

We focus on exhaustive as against demand-driven~\cite{Dillig:2008:SCS:1375581.1375615, demand.driven.1, demand.driven.2,ecoop16boomerang} points-to
analysis. A demand-driven points-to analysis computes points-to information that is relevant to a query raised by a client
analysis; for a different query, the points-to analysis needs to be repeated. An exhaustive analysis, on the other hand, computes all
points-to information which can be queried later by a client analysis; multiple queries do not require 
points-to analysis to be repeated.  For precision of points-to information, we are interested in 
full flow- and context-sensitive points-to analysis. 
A flow-sensitive analysis  respects  the control flow  
and  computes  separate data flow information at each program point.  
This matters because a pointer could have different pointees at different program points
because of redefinitions.
Hence, a flow-sensitive analysis provides more precise results than a flow-insensitive 
analysis
but can become inefficient at the interprocedural level.
A context-sensitive
analysis distinguishes between different  calling contexts of procedures
and restricts the analysis to interprocedurally valid control flow paths (i.e.\ control flow paths from program entry to program
exit in which every return from a procedure is matched with a call to the procedure such that all call-return matchings are
properly nested).  A fully context-sensitive analysis does not lose precision  even in the presence of recursion.
%%(relative to the abstraction chosen to create a decidable version of the analysis).
Both flow- and context-sensitivity enhance precision and we aim to achieve this without compromising
efficiency.

A  top-down   approach  to   interprocedural context-sensitive analysis  propagates
information  from  callers to  callees~\cite{summ2}  
effectively traversing the call graph
top-down. In the process, it 
analyzes   a   procedure   each   time    a   new   data   flow   value
reaches it from some call.   Several   popular   approaches   fall   in
this  category:  the call-strings  method~\cite{sharir.pnueli},  its  
value-based  variants~\cite{call_string.vbt,vasco}  and the  tabulation-based
functional method~\cite{graph_reach,sharir.pnueli}. By contrast, bottom-up
approaches~\cite{Chatterjee:1999:RCI:292540.292554,DBLP:conf/aplas/FengWDD15,Saturn,Yu:2010:LLM:1772954.1772985,Kahlon:2008:BTS:1375581.1375613,summ1,reps.ide,sharir.pnueli,purity1,Whaley,ptf,Yan:2012:RSS:2259051.2259053,yorsh.ipdfa,summ2} 
avoid  analyzing a procedure multiple  times by  constructing its {\em procedure summary
\/} which  is used
to incorporate the effect of calls to the procedure.
Effectively, this approach traverses the call graph bottom-up.\footnote%
{We use the terms top-down and bottom-up for traversals over a call graph; 
traversals over a control flow graph are termed forward and backward. Hence these terms are orthogonal.
Thus, both a forward data flow analysis (e.g. available expressions analysis) and a backward data flow analysis
(e.g. live variables analysis) could be implemented as a top-down or a bottom-up analysis at the interprocedural level.
}
A flow- and context-sensitive interprocedural analysis using procedure summaries is 
performed in two phases: the first phase constructs the procedure summaries and
the second phase applies them at the call sites
to compute the desired  information.
%%\change{}{which, in our case, is the classical points-to information.}
%%
%%\change{}{
%%A top-down approach to interprocedural analysis may not scale well 
%%but can be more precise than a bottom-up approach if
%%the latter's procedure summaries fail to capture all relevant details
%%of a procedure valid for all possible calls to it.
%%}

\newcommand{\maybe}{\text{\em may be\/}\xspace}
\newcommand{\yes}{\text{\em yes\/}\xspace}
\newcommand{\no}{\text{\em no\/}\xspace}

\subsection{Our Contributions}
\label{sec:contribs}

This paper advocates a new form of bottom-up procedure summaries, called the \emph{generalized points-to graphs} (\gpgs)
for flow- and context-sensitive points-to analysis.
\gpgs represent memory transformers (summarizing the effect of a procedure)
and contain \gpus (generalized points-to updates)
representing individual memory updates along with the control flow between them.
\gpgs are compact---their compactness is achieved by a careful choice of a suitable representation and 
a series of optimizations as described below.
	\begin{enumerate}
	\item  Our representation of memory updates, called the \emph{generalized
	       points-to update} (\gpu)
               leaves accesses of unknown pointees implicit
		without losing precision.
	\item  \gpgs undergo aggressive optimizations 
 		that are applied repeatedly to improve the compactness of \gpgs incrementally.
                These optimizations are
		similar to the optimizations performed by compilers and
\label{li:sec.1.step.2}
		are governed by the following possibilities of data dependence 
		between two memory updates (illustrated in Example~\ref{exmp:caller.dependence} in
		Section~\ref{sec:bottom.up.nomenclature})
		\begin{itemize}
		\item {\bf Case A.} The memory updates have a data dependence between them. It 
			could be 
			\begin{itemize}
				\item {\bf Case 1.} a read-after-write (\RaW) dependence,
				\item {\bf Case 2.} a write-after-read (\WaR) dependence, or
				\item {\bf Case 3.} a write-after-write (\WaW) dependence.
			\end{itemize}
			A read-after-read (\RaR) dependence is irrelevant.
		\item {\bf Case B.} The memory updates do not have a data dependence between them.
		\item {\bf Case C.} More information is needed to find out
			whether the memory updates have a data dependence between them.
		\end{itemize}
		These cases are exploited by the optimizations described below:

	\begin{itemize}
	\item \emph{Strength reduction} optimization exploits case A1. It 
		simplifies memory updates by using the information from other memory updates 
		to eliminate data dependence between them.
	\item \emph{Redundancy elimination} optimizations handle cases A2, A3, and B. They 
	remove redundant memory updates (case A3) and
               minimize control flow (case B). Case A2 is an anti-dependence and is modelled
	by eliminating control flow and ensuring that it is not viewed as
		a \RaW dependence (Example~\ref{eg:war-dep} in Section~\ref{sec:gpg.def}).
%%		 It is based on exploiting (lack of) data dependence between memory updates. 
%%              These opportunities are enhanced by strength reduction optimization.
%%		In case C, we use type information to rule out data dependence to eliminate 
%%		redundant control flow.
	\item \emph{Call inlining} optimization handles case C by progressively 
		providing more information. It 
		inlines the summaries of the callees of a procedure. 
		%%at the call sites in the summary of the procedure 
		This enhances the opportunities of strength reduction and redundancy elimination 
		and enables context-sensitive analysis.
	\item \emph{Type-based non-aliasing}.
		We use the types specified in the program 
                to resolve some additional instances of case C into case B.
	\end{itemize}
Our measurements suggest that the real killer of scalability in program analysis is not the amount of data
but the amount of control flow that it may be subjected to in search of precision.
Our optimizations are effective because they eliminate data dependence wherever possible 
and discard irrelevant control flow without losing precision.
Flow and context insensitivity discard control flow but over-approximate data dependence 
         causing imprecision.

\item Interleaving call inlining and strength reduction of \gpgs 
	facilitates  a novel optimization that computes
	flow- and context-sensitive points-to information 
	in the first phase of a bottom-up approach. This obviates the need for the usual second phase.
\end{enumerate}
In order to perform these optimizations:
	\begin{itemize}
	\item We define operations of \emph{\gpu composition} (to create new \gpus by eliminating data dependence between two
              \gpus), and \emph{\gpu reduction} (to eliminate the data dependence 
		of a \gpu with the \gpus in a given set).
	\item We propose  novel data flow analyses such as two variants of 
	      \emph{reaching \gpus analysis} (to identify the effects of memory updates reaching a given statement) and
	\emph{coalescing analysis} (to eliminate the redundant control flow in the \gpg).
        \item We handle recursive calls by refining the \gpgs through a fixed-point computation. 
%%		eliminate  recursive calls  by a bounded inlining of callee \gpgs without over-approximation.
	Calls through function pointers are proposed to be handled through delayed inlining.
	\end{itemize}
% Apart from interesting optimizations on the \gpgs that we construct, we also employ optimizations when
%%In addition to these techniques, we employ the following existing optimizations:
%%%% when constructing \gpgs and when applying them to compute classical points-to information at program points:
%%{\revOne CHECK: second part may have to change}
%%\begin{itemize}
%%\item For optimizing the process of construction of \gpgs, we traverse the 
%%	def-use chains of SSA for efficiently simplifying  memory updates involving SSA variables. 
%%{\revTwo
%%\item An interleaving of strength reduction and call inlining helps to compute 
%%	points-to information in the first phase of the bottom-up approach 
%%	(during \gpg construction), thereby rendering the second phase redundant.
%%}
%%\deleted{
%%\item For optimizing the process of using \gpgs for computing points-to information within a procedure,
%%	we extend the concept of bypassing~\cite{hakjoo2,hakjoo1}
%%        to pointers thereby filtering out the points-to information that is 
%%	not accessed in the procedure.
%%}
%%\end{itemize}
At a practical level, our main contribution is a method of flow-sensitive, field-sensitive, 
and context-sensitive exhaustive points-to analysis of C programs that scales to large real-life programs.

The core ideas of \gpgs have been presented before~\cite{gpg.sas.16}. 
This paper provides a complete treatment and
enhances the core ideas significantly. 
We describe our formulations for a C-like language.

\subsection{The Organization of the Paper}

Section~\ref{sec:background.motivation.keyideas} 
describes the limitations of past approaches as a background to motivate our key ideas that
overcome them.
%%provides background and illustrates our key ideas through a motivating example. 
Section~\ref{sec:intro.our.mt} introduces the concept of
generalized points-to updates (\gpus) that form the basis of \gpgs and provides
a brief overview of \gpg construction through a motivating example.
Section~\ref{sec:strength-reduction-optimization} describes 
the strength-reduction optimization
performed on \gpgs by formalizing
the operations such as \gpu composition and \gpu reduction and defining
data flow equations for reaching  \gpus analyses. 
Section~\ref{sec:structural-optimizations} describes redundancy elimination optimizations
performed on \gpgs. 
Section~\ref{sec:interprocedural.extensions} explains the interprocedural use of \gpgs
by defining call inlining
and shows how recursion 
is handled. 
Section~\ref{sec:dfv_compute} shows how \gpgs are used for performing points-to analysis. 
Section~\ref{sec:heap} describes the handling of structures, unions and the heap.  
Section~\ref{sec:level_3} describes the handling of function pointers.
Section~\ref{sec:emp-eval} presents empirical evaluation on SPEC benchmarks and
Section~\ref{sec:big-picture-pta} describes related work. 
Section~\ref{sec:conclusions} concludes the paper. 

%%\change{}{
%%We have described our formulations for a C-like language
%% and have handled all pointer-related features of C.
%%For simplicity of exposition, 
%%the paper is organized around the features as follows:
%%\begin{itemize}
%%\item Pointers to scalars  (intra-procedural analysis): 
%%	Sections~\ref{sec:intro.our.mt}, \ref{sec:strength-reduction-optimization},
%%	and~\ref{sec:structural-optimizations}.
%%\item Function calls, recursion, and function pointers:
%%	Section~\ref{sec:interprocedural.extensions}.
%%\item Pointers to structures, unions, and the heap: Section~\ref{sec:heap}.
%%\item Pointer arithmetic, pointers to arrays, address-escaping locals:
%%	Section~\ref{sec:emp-eval}.
%%\end{itemize}
%%}

\section{Existing Approaches and Their \protect Limitations}
\label{sec:background.motivation.keyideas}

This section begins by reviewing some basic concepts and then describes the challenges
in constructing procedure summaries for points-to analysis. 
It concludes by describing the limitations of the past approaches and 
outlining our key ideas. For further details of related work, see 
Section~\ref{sec:big-picture-pta}. 

\subsection{Basic Concepts}

In this section we describe the nature of memory, memory updates, and memory transformers.

\subsubsection{Abstract and Concrete Memory}
There are two views of memory and operations on it. Firstly we have the concrete memory view
(or semantic view) corresponding to run-time operations where a memory location always points to exactly one
memory location or NULL (which is a distinguished memory location).
Unfortunately this is, in general, statically uncomputable.
Secondly, as is traditional
in program analysis, we can consider an abstract view of memory where
an abstract location represents one or more concrete locations; this conflation and the uncertainty of
conditional branches means that abstract memory locations can point to multiple other locations---as in
the classical points-to graph.
These views are not independent and abstract operations must over-approximate concrete operations to ensure
soundness.
Formally, let \vars and \text{$\ptrs \subseteq \vars$} denote the sets of locations
and pointers respectively. 
The \emph{concrete memory} after a pointer assignment is a
function \text{$\mem: \ptrs \to \vars$}.
 %where ``?'' denotes an undefined location.
%\AMcomment{Do we really use `?'?  Could we not merely say about `sets of locations (including NULL)' as
%we have done elsewhere?  Sometimes it's convenient to assume `?' arrives via definition-free paths, but
%we're not doing this here.  Should we?  Anyway, either we should remove `?' here, or replace definition-free
%path with a flow of `?'}
The \emph{abstract memory} after a pointer assignment is a
relation \text{$\mem \subseteq \ptrs \times \vars$}. 
In either case, we view \mem as a graph
with \vars as the set of nodes.
An edge $x \rightarrow y$ in \mem is a \emph{points-to edge} indicating
that $x \in \ptrs$ contains the address of $y \in \vars$.
Unless noted explicitly, all subsequent references to memory locations and transformers refer to the abstract view.

The (abstract) memory
associated with a statement \flab is
an over-approximation of the concrete memory
associated with every occurrence of \flab in the same or different control flow paths.

\subsubsection{Memory Transformer}

A procedure summary for points-to analysis should
represent memory updates in terms of copying locations, loading from locations, 
or storing to locations.
It is called a \emph{memory transformer} because it
updates the memory before a call
to the procedure to compute the memory after the call.
Given a memory \mem and a memory transformer $\mtsym$, the updated memory $\mem'$ is computed 
by \text{$\mem'=\mtsym(\mem)$} as illustrated in Example~\ref{exmp:ph.illustration} (Section~\ref{sec:limitations-past-work}).

\subsubsection{Strong and Weak Updates}
\label{sec:memory.updates}

In concrete memory, every assignment 
overwrites the contents of the memory location corresponding to the LHS of the assignment.
However, in abstract memory, we may be uncertain as to which of several locations a 
variable (say $p$) points to.
Hence an indirect assignment such as $*p=\&x$  does not
overwrite any of its pointees, but merely \emph{adds} $x$ to the possible pointees.
This is a \emph{weak update}.  Sometimes however, there is only one possible abstract location described
by the LHS of an assignment, and in this case we may, in general, \emph{replace} the contents
of this location.  This is a \emph{strong update}.  There is just one subtlety which we return to later:
prior to the above assignment we may only have one assignment to \emph{p} (say $p=\&a$).
If this latter assignment dominates the former, then a strong update is appropriate.
But if the latter assignment only appears on some control flow paths to the former, then we say that
the read of $p$ in $*p=\&x$ is \emph{upwards exposed} (live on entry to the current procedure)
and therefore may have additional pointees unknown to the current procedure.
Thus, the criterion for a strong update in an assignment is that its LHS references a
single location \emph{and} the location referenced is not
%%(there can be more than one in statements like \texttt{****q = \&a}) 
 upwards exposed (for more details, see Section~\ref{sec:may.must.xxp.edges}).  
An important special case is that a direct assignment to a variable
(e.g.\ $p = \&x$) is always a strong update. 

When a value is stored in a location, we say that the location is \emph{defined}
without specifying whether the update is strong or weak and
make the distinction only where required.

\subsection{Challenges in Constructing Procedure Summaries for Points-to Analysis}
\label{sec:diff.pta}
\label{sec:bottom.up.nomenclature}

In the absence of pointers, data dependence between memory updates within a procedure can be 
inferred by using variable names without requiring any information from the callers. In such a
situation, procedure summaries for some analyses, including various bit-vector
data flow analyses (such as live variables analysis), 
can be precisely represented by constant \emph{gen} and \emph{kill} sets or
graph paths discovered using reachability~\cite{dfa_book}. In the presence of pointers, 
these (bit-vector)
summaries can be constructed using externally supplied points-to information.

Procedure summaries for points-to analysis, however, cannot be represented in terms of constant \emph{gen} and 
\emph{kill} sets because the association between pointer variables and their pointee locations
could change in the procedure and may depend on the aliases 
between pointer variables established in the callers of the procedure.
%%\change{}{\footnote{%
%%Unlike other analyses, a points-to analysis 
%%cannot rely on externally supplied points-to information.
%%}}
%%\subsubsection{Nomenclature for Bottom-Up Analyses}
%%In bottom-up program analysis, we analyze a procedure to obtain a \emph{procedure summary}
%%which is in turn used to create the procedure summaries for procedures which call it.
Often, and particularly for points-to analysis, we have a situation where a procedure summary
must either lose information or retain internal details which can only be resolved when its
caller is known.  

\begin{example}{exmp:caller.dependence}
Consider procedure $f$ on the right.
For many calls, $f()$ 
simply returns $\&a$ but until
\setlength{\intextsep}{-.8mm}%
\setlength{\columnsep}{2mm}%
\begin{wrapfigure}{r}{34mm}
\setlength{\codeLineLength}{25mm}%
\renewcommand{\arraystretch}{.7}%
	\begin{tabular}{|rc}
	%\hline
   	\codeLineOne{1}{0}{$\tt int\; a, b, *x, *\!*\!p;$}{white}
	\codeLine{0}{$\tt int *f() \; \OB$}{white} 
	\codeLine{1}{$\tt x =  \&a;$}{white}
	\codeLine{1}{$\tt *p =  \&b;$   }{white}
	\codeLine{1}{$\tt return \; x; $   }{white}
	\codeLine{0}{$\CB$\rule[-.5em]{0em}{1em}}{white}
	 %\hline
	\end{tabular}
\end{wrapfigure}
we are certain that $*p$
does not alias with $x$, we cannot perform this constant-propagation optimization.
We say that the assignment 04 \emph{blocks} this optimization.
%(We also use the word `\emph{barrier}' for a blocking assignment.)
There are four possibilities:
\begin{itemize}
\item If it is known that $*p$ and $x$ \emph{always} alias then we can optimize $f$ to return $\&b$.
\item If it is known that $*p$ and $x$ alias on some control flow paths containing a call to $f$
but not on all, then the procedure returns $\&a$ in some cases and $\&b$ in 
	other cases. While procedure $f$ cannot be optimized to do this, a static analysis
	can compute such a summary.\\[-16pt]
\end{itemize}

\begin{itemize}
\item If  it is known that they \emph{never} alias we can optimize this code to return $\&a$.
\item If  nothing is known about the alias information, then to preserve precision,
 we must retain this blocking assignment in the procedure summary for $f$.
\end{itemize}
The first two situations correspond to case (A1) 
in item~(\ref{li:sec.1.step.2}) in Section~\ref{sec:contribs}.
The third and the fourth
situations correspond to cases (B) and (C) respectively. 

The key idea is that information from the calling context(s) can 
determine whether a potentially 
blocking assignment really blocks an optimization or not.
As such we say that we \emph{postpone} optimizations that we would like to do until it is safe to do them.
\end{example}

%\change{Thus, three challenges in constructing flow-sensitive memory transformers are:}{
The above example illustrates the following challenges in constructing
flow-sensitive memory transformers:
%}
\begin{inparaenum}[(a)]
\item representing indirectly accessed unknown pointees,
\item identifying blocking assignments and postponing some optimizations, and
\item recording control flow between memory updates so that potential
	data dependence between them is neither violated nor over-approximated.
\end{inparaenum}

Thus, the main problem in constructing flow-sensitive memory transformers for points-to analysis is to find a representation
that is compact and yet captures memory updates and the minimal control flow between them succinctly.
%%The main challenge in achieving this is:
%%\begin{quote}
%%\em
%%The memory transformers need to handle indirectly accessed unknown pointees and preserve data 
%%dependence between them without introducing any imprecision. 
%%\end{quote}
%%\change{}{
%%This is what distinguishes  procedure summaries  of points-to
%%analysis from those of other analyses.
%%}

\subsection{Limitations of Existing Procedure Summaries for Points-to Analysis}
\label{sec:limitations-past-work}

A common solution for modelling indirect accesses of unknown pointees in a 
memory transformer is to use \emph{placeholders}\footnote{Placeholders have also been known as
external variables~\cite{purity1,Whaley,summ1} and extended parameters~\cite{ptf}. 
They are parameters of the procedure summary 
and not necessarily of the procedure for which the summary is constructed.} which are pattern-matched
against the input memory to compute the output memory.   
Here we describe two broad approaches that use placeholders.

The first approach, which we call a \emph{multiple transfer functions} (MTF) approach, proposed a  
precise representation of a procedure summary for points-to analysis
as a collection of \emph{partial transfer functions} 
(PTFs)~\cite{Chatterjee:1999:RCI:292540.292554,ptf,Yu:2010:LLM:1772954.1772985,Kahlon:2008:BTS:1375581.1375613}.\footnote{In level-by-level 
analysis~\cite{Yu:2010:LLM:1772954.1772985}, multiple PTFs are combined into a
single function with a series of condition checks for different points-to information
occurring in the calling contexts.}
 Each PTF
corresponds to a combination of aliases that might occur in the callers of a procedure.
%Unfortunately, this representation fails to scale because of a possibility of 
%combinatorially many aliases.
Our work is inspired by the second approach, which we call a \emph{single transfer function} (STF) 
approach~\cite{purity1,Whaley,summ1}. This approach does not customize procedure summaries for combinations of aliases.
However, the existing STF approach fails to be precise. We illustrate this approach and its limitations to
motivate our key ideas using Figure~\ref{fig:mt.exmp}. It shows a procedure and two memory transformers 
($\mtsym'$ and $\mtsym''$) for it and the associated input and output memories.
The effect of $\mtsym'$ is explained in Example~\ref{exmp:ph.illustration} and
that of $\mtsym''$, in Example~\ref{exmp:ph.improvement}.

\begin{figure}[t]
\centering
\renewcommand{\tabcolsep}{2pt}
\begin{tabular}{c|c||c|c|}
%\cline{2-4}
\multicolumn{1}{c}{\rule{0em}{1em}}
&
\multicolumn{1}{c}{Procedure $f$}
&
\multicolumn{1}{c}{Example 1}
&
\multicolumn{1}{c}{Example 2}
\\ \cline{2-4}\cline{2-4}
&
\rule{0em}{1em}
Control flow graph
&
Input Memory $\mem_{1}$
&
Input Memory $\mem_{2}$
\\ \cline{2-4}\cline{2-4}
&
\begin{tabular}{@{}c@{}}
\psset{unit=.25mm}
\begin{pspicture}(4,0)(80,120)
%\psframe(4,0)(80,120)
	\small
	\putnode{n0}{origin}{44}{108}{\psframebox[framesep=3]{$\Startscriptsize{f}$}}
	\putnode{n1}{n0}{0}{-48}{\psframebox{$\renewcommand{\arraystretch}{.8}%
		\begin{array}{@{}l@{}}
		\tt p = *y \\
			\tt *x = q \\
			\tt q = *y \\
			\end{array}$}}
			\putnode{w}{n1}{-32}{13}{1}
			\putnode{w}{n1}{-32}{1}{2}
			\putnode{w}{n1}{-32}{-12}{3}
			\putnode{n2}{n1}{0}{-48}{\psframebox[framesep=3]{$\Endscriptsize{f}$}}
			\ncline{->}{n0}{n1}
			\ncline{->}{n1}{n2}
			\end{pspicture}
			\end{tabular}
			&
			\begin{tabular}{@{}c@{}}
			\newcommand{\curvestrikeoff}{%
				\ncput[npos=.29]{/}
				\ncput[npos=.35]{/}
				\ncput[npos=.41]{/}
				\ncput[npos=.47]{/}
				\ncput[npos=.53]{/}
				\ncput[npos=.59]{/}
				\ncput[npos=.65]{/}
				\ncput[npos=.72]{/}
			}%
\begin{pspicture}(-4,-2)(26,20)
%\psframe(0,-2)(22,20)
	\putnode{x1}{origin}{3}{25}{}
	\putnode{y1}{x1}{0}{-8}{\pscirclebox[framesep=.77]{$y$}}
	\putnode{p1}{y1}{8}{-8}{\pscirclebox[framesep=.77]{$q$}}
	\putnode{z1}{y1}{0}{-16}{\pscirclebox[framesep=1]{$x$}}
	\putnode{g1}{y1}{8}{0}{\pscirclebox[framesep=1]{$r$}}
	\putnode{g2}{g1}{8}{0}{\pscirclebox[framesep=.77]{$a$}}
	\putnode{g3}{z1}{8}{0}{\pscirclebox[framesep=.77]{$s$}}
	\putnode{g4}{g3}{8}{0}{\pscirclebox[framesep=.77]{$c$}}
	\putnode{g5}{p1}{8}{0}{\pscirclebox[framesep=.77]{$b$}}
	\ncline[nodesepA=-.2,nodesepB=-.3]{->}{y1}{g1}
	\ncline[nodesepA=-.2,nodesepB=-.3]{->}{p1}{g5}
	\ncline{->}{g1}{g2}
	\ncline{->}{z1}{g3}
	\ncline{->}{g3}{g4}
	\end{pspicture}
	\end{tabular}
	&
	\begin{tabular}{@{}c@{}}
	\newcommand{\curvestrikeoff}{%
		\ncput[npos=.29]{/}
		\ncput[npos=.35]{/}
		\ncput[npos=.41]{/}
		\ncput[npos=.47]{/}
		\ncput[npos=.53]{/}
		\ncput[npos=.59]{/}
		\ncput[npos=.65]{/}
		\ncput[npos=.72]{/}
	}%
\newcommand{\smallstrikeoff}{%
	%\ncput[npos=.26,nrot=20]{/}
	\ncput[npos=.3,nrot=20]{/}
	\ncput[npos=.5,nrot=20]{/}
	\ncput[npos=.7,nrot=20]{/}
	%\ncput[npos=.75,nrot=20]{/}
}%
\begin{pspicture}(-4,-2)(26,20)
%\psframe(-4,-2)(26,20)
	\putnode{x1}{origin}{3}{25}{}
	\putnode{y1}{x1}{0}{-8}{\pscirclebox[framesep=.77]{$y$}}
	\putnode{z1}{y1}{0}{-8}{\pscirclebox[framesep=1]{$x$}}
	\putnode{p1}{z1}{8}{-8}{\pscirclebox[framesep=.77]{$q$}}
	\putnode{g1}{y1}{8}{-4}{\pscirclebox[framesep=.77]{$r$}}
	\putnode{g2}{g1}{8}{0}{\pscirclebox[framesep=1]{$a$}}
	\putnode{g5}{p1}{8}{0}{\pscirclebox[framesep=.77]{$b$}}
	\ncline[nodesepA=-.2,nodesepB=-.3]{->}{y1}{g1}
	\ncline[nodesepA=-.2,nodesepB=-.3]{->}{z1}{g1}
	\ncline[nodesepA=-.2,nodesepB=-.3]{->}{p1}{g5}
	\ncline{->}{g1}{g2}
	\end{pspicture}
	\end{tabular}
	\\ \cline{2-4}  \cline{2-4}
	\rule[-.4em]{0em}{1.5em}
	&  
	Memory Transformer $\mtsym'$
	& \begin{tabular}{@{}l@{}}
		\rule{0em}{1em}Output Memory \\
		$\mem_1' = \mtsym'(\mem_1)$
	  \end{tabular}
	& \begin{tabular}{@{}l@{}}
		\rule{0em}{1em}Output Memory \\
		$\mem_2' = \mtsym'(\mem_2)$
	  \end{tabular}
	\\ \cline{2-4}\cline{2-4}
	\begin{minipage}{32mm}
	\small\raggedright
	The memory transformer $\mtsym'$ is compact but imprecise because it uses the same placeholder for
every access of a pointee. Thus it
	over-approximates the memory.
	\end{minipage}
	\;
	&
	\begin{tabular}{@{}c@{}}
	\newcommand{\curvestrikeoff}{%
		\ncput[npos=.29]{/}
		\ncput[npos=.35]{/}
		\ncput[npos=.41]{/}
		\ncput[npos=.47]{/}
		\ncput[npos=.53]{/}
		\ncput[npos=.59]{/}
		\ncput[npos=.65]{/}
		\ncput[npos=.72]{/}
	}%
\begin{pspicture}(0,-2)(22,29)
%\psframe(0,-2)(22,29)
	\putnode{x1}{origin}{3}{25}{\pscirclebox[framesep=.77]{$p$}}
	\putnode{y1}{x1}{0}{-8}{\pscirclebox[framesep=.77]{$y$}}
	\putnode{p1}{y1}{0}{-8}{\pscirclebox[framesep=.77]{$q$}}
	\putnode{z1}{p1}{0}{-8}{\pscirclebox[framesep=1]{$x$}}
	\putnode{g1}{y1}{8}{4}{\pscirclebox[framesep=-.1]{\small$\phi_1$}}
	\putnode{g2}{g1}{8}{0}{\pscirclebox[framesep=-.1]{\small$\phi_2$}}
	\putnode{g3}{z1}{8}{0}{\pscirclebox[framesep=-.1]{\small$\phi_3$}}
	\putnode{g4}{g3}{8}{0}{\pscirclebox[framesep=-.1]{\small$\phi_4$}}
	\ncline[nodesepA=-.2,nodesepB=-.3]{->}{y1}{g1}
	\ncline{->}{g1}{g2}
	\ncline{->}{g5}{g6}
	\ncline{->}{z1}{g3}
	\ncline[linewidth=0.6,arrowsize=1.5]{->}{g3}{g4}
	\nccurve[angleA=0,angleB=140,nodesepB=-.6,linewidth=0.6,arrowsize=1.5]{->}{x1}{g2}
	\nccurve[angleA=0,angleB=220,nodesepB=-.6,linewidth=0.6,arrowsize=1.5]{->}{p1}{g2}
	%%\nccurve[angleA=-30,angleB=135,nodesepB=-.8,nodesepA=-.6]{->}{p1}{g4}
	%%\curvestrikeoff
	\end{pspicture}
	\end{tabular}
	&
	\begin{tabular}{@{}c@{}}
	\newcommand{\curvestrikeoff}{%
		\ncput[npos=.29]{/}
		\ncput[npos=.35]{/}
		\ncput[npos=.41]{/}
		\ncput[npos=.47]{/}
		\ncput[npos=.53]{/}
		\ncput[npos=.59]{/}
		\ncput[npos=.65]{/}
		\ncput[npos=.72]{/}
	}%
\newcommand{\smallstrikeoff}{%
	%\ncput[npos=.26,nrot=20]{/}
	\ncput[npos=.3,nrot=20]{/}
	\ncput[npos=.5,nrot=20]{/}
	\ncput[npos=.7,nrot=20]{/}
	%\ncput[npos=.75,nrot=20]{/}
}%
\begin{pspicture}(0,-2)(22,29)
%\psframe(0,0)(22,30)
	\putnode{x1}{origin}{11}{25}{\pscirclebox[framesep=.67]{$p$}}
	\putnode{y1}{x1}{-8}{-8}{\pscirclebox[framesep=.77]{$y$}}
	\putnode{p1}{y1}{8}{-8}{\pscirclebox[framesep=.77]{$q$}}
	\putnode{z1}{y1}{0}{-16}{\pscirclebox[framesep=1]{$x$}}
	\putnode{g1}{y1}{8}{0}{\pscirclebox[framesep=1]{$r$}}
	\putnode{g2}{g1}{8}{0}{\pscirclebox[framesep=.77]{$a$}}
	\putnode{g3}{z1}{8}{0}{\pscirclebox[framesep=.77]{$s$}}
	\putnode{g4}{g3}{8}{0}{\pscirclebox[framesep=.77]{$c$}}
	\putnode{g5}{p1}{8}{0}{\pscirclebox[framesep=.77]{$b$}}
	\ncline[nodesepA=-.2,nodesepB=-.3]{->}{y1}{g1}
	%%\ncline[nodesepA=-.2,nodesepB=-.3]{->}{p1}{g5}
	%%\smallstrikeoff
	\ncline{->}{g1}{g2}
	\ncline{->}{z1}{g3}
	\ncline{->}{g3}{g4}
	\ncline[nodesepA=-.7,nodesepB=-.8]{->}{g3}{g5}
	\ncline[nodesepA=-.7,nodesepB=-.8]{->}{x1}{g2}
	%%\nccurve[angleA=0,angleB=140,nodesepB=-.6]{->}{x1}{g2}
	\nccurve[angleA=45,angleB=225,nodesepA=-.8,nodesepB=-.8]{->}{p1}{g2}
	\end{pspicture}
	\end{tabular}
	&
	\begin{tabular}{@{}c@{}}
	\newcommand{\curvestrikeoff}{%
		\ncput[npos=.29]{/}
		\ncput[npos=.35]{/}
		\ncput[npos=.41]{/}
		\ncput[npos=.47]{/}
		\ncput[npos=.53]{/}
		\ncput[npos=.59]{/}
		\ncput[npos=.65]{/}
		\ncput[npos=.72]{/}
	}%
\newcommand{\smallstrikeoff}{%
	%\ncput[npos=.26,nrot=20]{/}
	\ncput[npos=.3,nrot=20]{/}
	\ncput[npos=.5,nrot=20]{/}
	\ncput[npos=.7,nrot=20]{/}
	%\ncput[npos=.75,nrot=20]{/}
}%
\begin{pspicture}(-4,-2)(26,29)
%\psframe(-4,-2)(26,30)
	\putnode{x1}{origin}{9}{25}{\pscirclebox[framesep=1]{$p$}}
	\putnode{y1}{x1}{-8}{-8}{\pscirclebox[framesep=.77]{$y$}}
	\putnode{z1}{y1}{0}{-8}{\pscirclebox[framesep=1]{$x$}}
	\putnode{p1}{z1}{8}{-8}{\pscirclebox[framesep=.77]{$q$}}
	\putnode{g1}{y1}{8}{-4}{\pscirclebox[framesep=.77]{$r$}}
	\putnode{g2}{g1}{8}{0}{\pscirclebox[framesep=1]{$a$}}
	\putnode{g5}{p1}{8}{0}{\pscirclebox[framesep=.77]{$b$}}
	\ncline[nodesepA=-.2,nodesepB=-.3]{->}{y1}{g1}
	\ncline[nodesepA=-.2,nodesepB=-.3]{->}{z1}{g1}
	\ncline[nodesepA=-.2,nodesepB=-.3]{->}{p1}{g5}
	\ncline[nodesepA=-.5,nodesepB=-.6]{->}{g1}{g5}
	\ncline{->}{g1}{g2}
	\ncline[nodesepA=-.5,nodesepB=-.6]{->}{x1}{g2}
	\ncline[nodesepA=-.5,nodesepB=-.6]{->}{p1}{g2}
	\nccurve[angleA=0,angleB=30,nodesepB=-.6]{->}{x1}{g5}
	\end{pspicture}
	\end{tabular}
	\\ \cline{2-4}  \cline{2-4}
	\rule[-.4em]{0em}{1.5em}
	&
	Memory Transformer $\mtsym''$
	& \begin{tabular}{@{}l@{}}
		\rule{0em}{1em}Output Memory \\
		$\mem''_1 = \mtsym''(\mem_1)$
	  \end{tabular}
	& \begin{tabular}{@{}l@{}}
		\rule{0em}{1em}Output Memory \\
		$\mem''_2 = \mtsym''(\mem_2)$
	  \end{tabular}
	\\ \cline{2-4}\cline{2-4}
	\begin{minipage}{32mm}
	\small\raggedright
	The memory transformer $\mtsym''$ shows that precision can be improved by using a separate placeholder for every access of a pointee. However, the size of the memory
	transformer increases.
	\end{minipage}
	\;
	&
	\begin{tabular}{@{}c@{}}
	\newcommand{\curvestrikeoff}{%
		\ncput[npos=.29]{/}
		\ncput[npos=.35]{/}
		\ncput[npos=.41]{/}
		\ncput[npos=.47]{/}
		\ncput[npos=.53]{/}
		\ncput[npos=.59]{/}
		\ncput[npos=.65]{/}
		\ncput[npos=.72]{/}
	}%
\begin{pspicture}(0,-3)(22,30)
%\psframe(0,0)(22,30)
	\putnode{x1}{origin}{3}{25}{\pscirclebox[framesep=.77]{$p$}}
	\putnode{y1}{x1}{0}{-8}{\pscirclebox[framesep=.77]{$y$}}
	\putnode{p1}{y1}{0}{-8}{\pscirclebox[framesep=.77]{$q$}}
	\putnode{z1}{p1}{0}{-8}{\pscirclebox[framesep=1]{$x$}}
	\putnode{g1}{y1}{8}{4}{\pscirclebox[framesep=-.1]{\small$\phi_1$}}
	\putnode{g2}{g1}{8}{0}{\pscirclebox[framesep=-.1]{\small$\phi_2$}}
	\putnode{g3}{z1}{8}{0}{\pscirclebox[framesep=-.1]{\small$\phi_3$}}
	\putnode{g4}{g3}{8}{0}{\pscirclebox[framesep=-.1]{\small$\phi_4$}}
	\putnode{g5}{y1}{8}{-4}{\pscirclebox[framesep=-.1]{\small$\phi_5$}}
	\putnode{g6}{g5}{8}{0}{\pscirclebox[framesep=-.1]{\small$\phi_6$}}
	\ncline[nodesepA=-.2,nodesepB=-.3]{->}{y1}{g1}
	\ncline[nodesepA=-.2,nodesepB=-.3]{->}{y1}{g5}
	\ncline{->}{g1}{g2}
	\ncline{->}{g5}{g6}
	\ncline{->}{z1}{g3}
	\ncline[linewidth=0.6,arrowsize=1.5]{->}{g3}{g4}
	\nbput[npos=.5,labelsep=.25]{\small\psframebox[linestyle=none,fillstyle=solid,fillcolor=lightgray,framesep=.5]{2}}
	\nccurve[angleA=0,angleB=140,nodesepB=-.6,linewidth=0.6,arrowsize=1.5]{->}{x1}{g2}
	\naput[npos=.5,labelsep=.25]{\small\psframebox[linestyle=none,fillstyle=solid,fillcolor=lightgray,framesep=.5]{1}}
	\nccurve[angleA=0,angleB=220,nodesepB=-.6,linewidth=0.6,arrowsize=1.5]{->}{p1}{g6}
	\nbput[npos=.85,labelsep=0]{\small\psframebox[linestyle=none,fillstyle=solid,fillcolor=lightgray,framesep=.5]{3}}
	%%\nccurve[angleA=-30,angleB=135,nodesepB=-.8,nodesepA=-.6]{->}{p1}{g4}
	%%\curvestrikeoff
	\end{pspicture}
	\end{tabular}
	&
	\begin{tabular}{@{}c@{}}
	\newcommand{\curvestrikeoff}{%
		\ncput[npos=.29]{/}
		\ncput[npos=.35]{/}
		\ncput[npos=.41]{/}
		\ncput[npos=.47]{/}
		\ncput[npos=.53]{/}
		\ncput[npos=.59]{/}
		\ncput[npos=.65]{/}
		\ncput[npos=.72]{/}
	}%
\newcommand{\smallstrikeoff}{%
	%\ncput[npos=.26,nrot=20]{/}
	\ncput[npos=.3,nrot=20]{/}
	\ncput[npos=.5,nrot=20]{/}
	\ncput[npos=.7,nrot=20]{/}
	%\ncput[npos=.75,nrot=20]{/}
}%
\begin{pspicture}(0,-2)(22,29)
%\psframe(0,0)(22,30)
	\putnode{x1}{origin}{11}{25}{\pscirclebox[framesep=.67]{$p$}}
	\putnode{y1}{x1}{-8}{-8}{\pscirclebox[framesep=.77]{$y$}}
	\putnode{p1}{y1}{8}{-8}{\pscirclebox[framesep=.77]{$q$}}
	\putnode{z1}{y1}{0}{-16}{\pscirclebox[framesep=1]{$x$}}
	\putnode{g1}{y1}{8}{0}{\pscirclebox[framesep=1]{$r$}}
	\putnode{g2}{g1}{8}{0}{\pscirclebox[framesep=.77]{$a$}}
	\putnode{g3}{z1}{8}{0}{\pscirclebox[framesep=.77]{$s$}}
	\putnode{g4}{g3}{8}{0}{\pscirclebox[framesep=.77]{$c$}}
	\putnode{g5}{p1}{8}{0}{\pscirclebox[framesep=.77]{$b$}}
	%%\putnode{x1}{origin}{3}{25}{\pscirclebox[framesep=1]{$x$}}
	%%\putnode{y1}{x1}{0}{-8}{\pscirclebox[framesep=.77]{$y$}}
	%%\putnode{p1}{y1}{0}{-8}{\pscirclebox[framesep=.77]{$p$}}
	%%\putnode{z1}{p1}{0}{-8}{\pscirclebox[framesep=1]{$z$}}
	%%%
	%%\putnode{g1}{y1}{8}{0}{\pscirclebox[framesep=1]{$a$}}
	%%\putnode{g2}{g1}{8}{0}{\pscirclebox[framesep=.77]{$b$}}
	%%\putnode{g3}{z1}{8}{0}{\pscirclebox[framesep=.97]{$c$}}
	%%\putnode{g4}{g3}{8}{0}{\pscirclebox[framesep=.5]{$d$}}
	%%\putnode{g5}{p1}{8}{0}{\pscirclebox[framesep=.77]{$q$}}
	%
	\ncline[nodesepA=-.2,nodesepB=-.3]{->}{y1}{g1}
	%%\ncline[nodesepA=-.2,nodesepB=-.3]{->}{p1}{g5}
	%%\smallstrikeoff
	\ncline{->}{g1}{g2}
	\ncline{->}{z1}{g3}
	\ncline{->}{g3}{g4}
	\smallstrikeoff
	\ncline[nodesepA=-.7,nodesepB=-.8]{->}{g3}{g5}
	\ncline[nodesepA=-.7,nodesepB=-.8]{->}{x1}{g2}
	%%\nccurve[angleA=0,angleB=140,nodesepB=-.6]{->}{x1}{g2}
	\nccurve[angleA=45,angleB=225,nodesepA=-.8,nodesepB=-.8]{->}{p1}{g2}
	\end{pspicture}
	\end{tabular}
	&
\begin{tabular}{@{}c@{}}
\newcommand{\curvestrikeoff}{%
	\ncput[npos=.29]{/}
	\ncput[npos=.35]{/}
	\ncput[npos=.41]{/}
	\ncput[npos=.47]{/}
	\ncput[npos=.53]{/}
	\ncput[npos=.59]{/}
	\ncput[npos=.65]{/}
	\ncput[npos=.72]{/}
	}%
\newcommand{\smallstrikeoff}{%
	%\ncput[npos=.26,nrot=20]{/}
	\ncput[npos=.3,nrot=20]{/}
	\ncput[npos=.5,nrot=20]{/}
	\ncput[npos=.7,nrot=20]{/}
	%\ncput[npos=.75,nrot=20]{/}
}%
\begin{pspicture}(0,-2)(22,30)
%\psframe(0,0)(22,30)
\putnode{x1}{origin}{11}{25}{\pscirclebox[framesep=1]{$p$}}
\putnode{y1}{x1}{-8}{-8}{\pscirclebox[framesep=.77]{$y$}}
\putnode{z1}{y1}{0}{-8}{\pscirclebox[framesep=1]{$x$}}
\putnode{p1}{z1}{8}{-8}{\pscirclebox[framesep=.77]{$q$}}
\putnode{g1}{y1}{8}{-4}{\pscirclebox[framesep=.77]{$r$}}
\putnode{g2}{g1}{8}{0}{\pscirclebox[framesep=1]{$a$}}
\putnode{g5}{p1}{8}{0}{\pscirclebox[framesep=.77]{$b$}}
\ncline[nodesepA=-.2,nodesepB=-.3]{->}{y1}{g1}
\ncline[nodesepA=-.2,nodesepB=-.3]{->}{z1}{g1}
\ncline[nodesepA=-.2,nodesepB=-.3]{->}{p1}{g5}
\ncline[nodesepA=-.5,nodesepB=-.6]{->}{g1}{g5}
\ncline{->}{g1}{g2}
\smallstrikeoff
\ncline[nodesepA=-.5,nodesepB=-.6]{->}{x1}{g2}
%%\nccurve[angleA=0,angleB=140,nodesepB=-.6]{->}{x1}{g2}
\end{pspicture}
\end{tabular}
\\ \cline{2-4}
\end{tabular}

\caption{An STF-style  memory transformer $\mtsym'$ 
and its associated transformations. 
$\mtsym''$ is its flow-sensitive version.
Unknown pointees are denoted by
placeholders $\phi_i$. Thick edges in a memory transformer represent the
points-to edges \emph{generated} by it, other edges are carried forward from the input memory.
Labels of the points-to edges in $\mtsym''$ 
correspond to the statements indicating the sequencing of edges.
Edges that are \emph{killed} in the memory are 
struck off.} 
\label{fig:mt.exmp}
\end{figure}
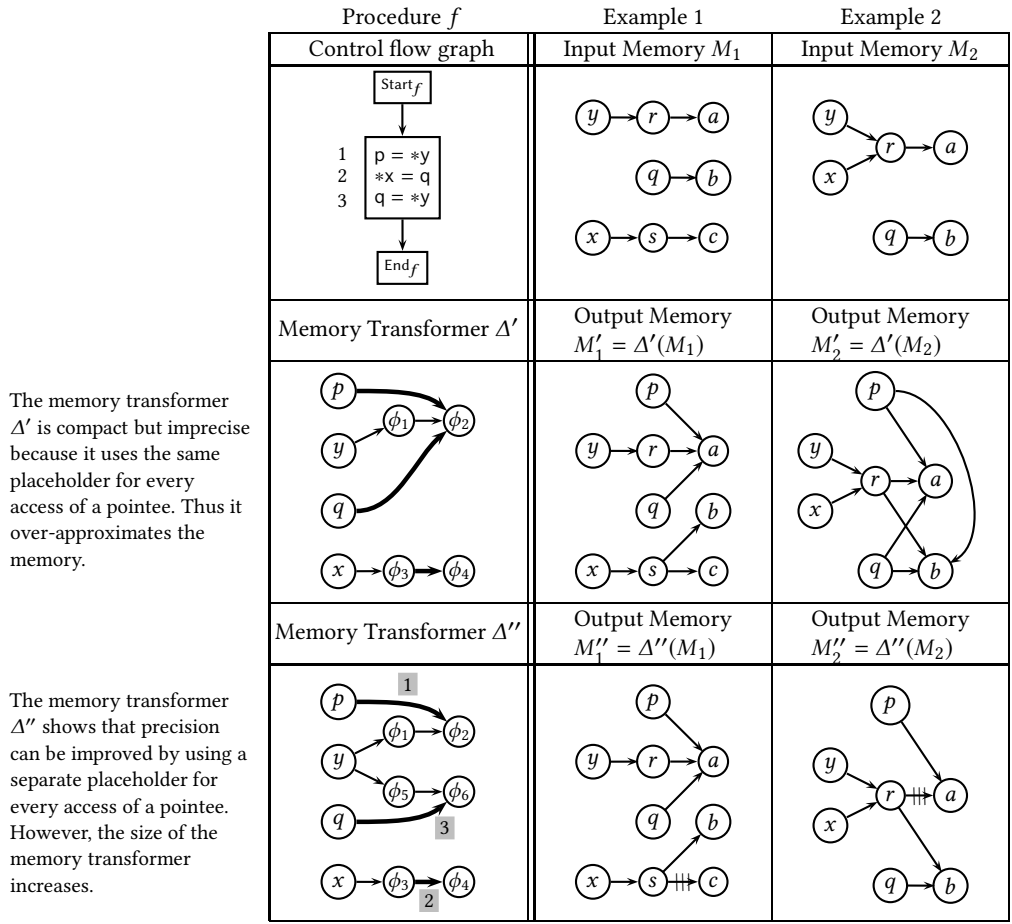

\begin{example}{exmp:ph.illustration}
Transformer $\mtsym'$ is constructed by the STF approach~\cite{purity1,Whaley,summ1}.
It can be viewed as an abstract points-to graph
containing placeholders $\phi_i$ for modelling
unknown pointees of the pointers appearing in $\mtsym'$. For example, $\phi_1$
represents the pointees of $y$ and $\phi_2$ represents the pointees of pointees of $y$, both
of which are not known in the procedure.
The placeholders are pattern matched against the input memory (e.g. $\mem_1$ or 
$\mem_2$) to compute the corresponding output memory ($\mem_1'$ and $\mem_2'$ respectively). 
A crucial difference between a memory and a memory transformer is: a memory is a snapshot
of points-to edges whereas a memory transformer needs to distinguish 
the points-to edges that are generated by it 
(shown by thick edges) from those that are carried forward from the input 
memory (shown by thin edges). 
%%However, in the memory, there is no such distinction.

The two accesses of $y$ in statements 1 and 3 may or may not refer
to the same location because of a possible side-effect of the intervening assignment in
statement 2. 
If $x$ and $y$ are aliased in the input memory (e.g. in $\mem_2$), statement 2 redefines 
the pointee of $y$ and hence $p$ and $q$ will not be aliased in the output memory. 
However, $\mtsym'$ 
uses the same placeholder for all accesses of a pointee. 
Further, $\mtsym'$ also
suppresses strong updates because the control flow ordering between memory updates is not recorded.
Hence, points-to edge \denew{s}{}{c}{} in $\mem_1'$ is not deleted. Similarly,  
points-to edge \denew{r}{}{a}{} in $\mem_2'$ is not deleted and $q$ spuriously points to $a$.
Additionally, $p$ spuriously points-to $b$.
Hence, $p$
and $q$ appear to be aliased in the output memory $\mem_2'$.
\end{example}

The use of control flow ordering between the points-to edges that are
\emph{generated} by a memory transformer can improve its precision
as shown by the following example.

\begin{example}{exmp:ph.improvement}
In Figure~\ref{fig:mt.exmp}, memory transformer $\mtsym''$  differs from $\mtsym'$
in two ways.  Firstly it uses a separate placeholder for every access 
of a pointee to avoid an over-approximation of memory (e.g. placeholders $\phi_1$ and $\phi_2$ to represent $*y$ in statement 1,
and $\phi_5$ and $\phi_6$ to represent 
$*y$ in statement 3).
This, along with control flow, allows strong updates thereby killing the points-to edge \denew{r}{}{a}{}  
and hence $q$ does not point to $a$ (as shown in $\mem_2''$).
Secondly, the points-to edges generated by the memory transformer are
ordered based on the control flow of a procedure, thereby adding some form
of flow-sensitivity which $\mtsym'$ lacks. 
To see the role of control flow, 
observe that if the points-to edge corresponding to statement 2
is considered first, then $p$ and $q$ will always be aliased because the possible
side-effect of statement 2 will be ignored.

The output memories $\mem_1''$  and $\mem_2''$ computed using $\mtsym''$
are more precise than the corresponding output memories $\mem_1'$ and $\mem_2'$ computed using $\mtsym'$.
\end{example}

Observe that, although $\mtsym''$ is more precise than $\mtsym'$, it uses
a larger number of placeholders and also requires control flow information. This affects
the scalability of points-to analysis.

A fundamental problem with placeholders is that they use a low-level representation of 
memory expressed in terms of classical points-to edges. Hence a placeholder-based approach is forced to
explicate unknown pointees by naming them, resulting in either a large number of
placeholders (in the STF approach) or multiple PTFs (in the MTF 
approach).
The need of control flow ordering further increases the number of placeholders in the former approach. The latter
approach obviates the need of ordering because the PTFs
are customized for combinations of aliases.

\begin{figure}[t]
%%\centering
\small
\begin{center}
\begin{tabular}{ccc}
\begin{tabular}{@{}c@{}}
\psset{unit=.25mm}
\begin{pspicture}(-8,0)(128,194)
%\psframe(0,0)(120,194)
\small
\putnode{n0}{origin}{50}{180}{\psframebox[framesep=3]{$\Startscriptsize{g}$}}
\putnode{n1}{n0}{0}{-42}{\psframebox{$\renewcommand{\arraystretch}{.8}%
\begin{array}{@{}l@{}}
\tt r = \&a
\\
\tt *q = \&m 
\end{array}$}}
\putnode{w}{n1}{-36}{7}{01}
\putnode{w}{n1}{-36}{-7}{02}
\putnode{n2}{n1}{40}{-40}{\psframebox{$\tt q = \&b$}}
\putnode{w}{n2}{-35}{0}{03}
\putnode{n3}{n2}{-40}{-42}{\psframebox{$\renewcommand{\arraystretch}{.8}%
	\begin{array}{@{}l@{}}
	\tt e = *p \\
	\tt q = \&e
	\end{array}$}}
\putnode{w}{n3}{-38}{7}{04}
\putnode{w}{n3}{-38}{-7}{05}
\putnode{n4}{n3}{0}{-44}{\psframebox[framesep=3]{$\Endscriptsize{g}$}}
\ncline{->}{n0}{n1}
\ncline{->}{n1}{n2}
\nccurve[angleA=240,angleB=110]{->}{n1}{n3}
\ncloop[arm=8,offsetB=6,linearc=2,angleA=270,angleB=90,loopsize=-36]{->}{n2}{n1}
\ncline{->}{n3}{n4}
\end{pspicture}
\end{tabular}
&
\begin{tabular}{@{}c@{}}
\psset{unit=.25mm}
\begin{pspicture}(-8,0)(128,194)
%\psframe(0,0)(120,194)
\small
\putnode{n0}{origin}{50}{184}{\psframebox[framesep=3]{$\Startscriptsize{f}$}}
\putnode{n1}{n0}{0}{-48}{\psframebox{$\renewcommand{\arraystretch}{.8}%
		\begin{array}{@{}l@{}}
		\tt p = \&c \\
		\tt q = \&d \\
		\tt d = \&n \\
		\end{array}$}}
\putnode{w}{n1}{-38}{13}{06}
\putnode{w}{n1}{-38}{1}{07}
\putnode{w}{n1}{-38}{-12}{08}
\putnode{n2}{n1}{0}{-52}{\psframebox{$\tt call\; g()$}}
\putnode{w}{n2}{-40}{0}{09}
\putnode{n3}{n2}{0}{-38}{\psframebox{$ \tt *q = \&o $}}
\putnode{w}{n3}{-40}{0}{10}
\putnode{n4}{n3}{0}{-34}{\psframebox[framesep=3]{$\Endscriptsize{f}$}}
\ncline{->}{n0}{n1}
\ncline{->}{n1}{n2}
\ncline{->}{n2}{n3}
\ncline{->}{n3}{n4}
\end{pspicture}
\end{tabular}
&
\begin{tabular}{c}
$
\begin{array}{|l|l|}
\hline
\rule[-.5em]{0em}{1.5em}%
\text{Variables} & \text{Types}
\\ \hline\hline
{\tt m}, {\tt n}, {\tt o} 
& \tt int 
\\ \hline
{\tt a}, {\tt b}, {\tt c}, {\tt d}, {\tt e}
& \tt int* 
\\ \hline
{\tt p}, {\tt q}, {\tt r}
& \tt int\!*\!*
\\ \hline
\end{array}
$

\\ \\
All variables are global
\end{tabular}

\end{tabular}

\caption{A motivating example. Procedures are represented by their control flow graphs (\cfgs). 
}
\label{fig:mot_eg.cfgs}
\end{center}
\end{figure}
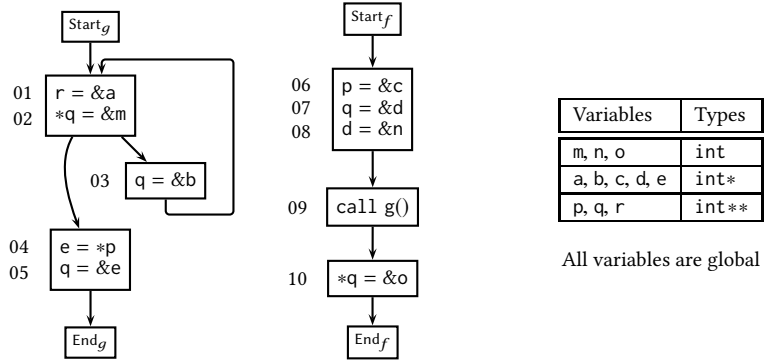

\subsection{Our Key Ideas}

We propose a \emph{generalized points-to graph} (\gpg) as a
representation for a memory transformer of a procedure;
special cases of \gpgs also represent memory as a points-to relation.
A \gpg
is characterized by the following key ideas that overcome the two limitations described in
Section~\ref{sec:limitations-past-work}.

\begin{itemize}
\item A \gpg leaves the placeholders implicit by using the counts of indirection levels. Simple arithmetic
on the counts allows us to combine the effects of multiple memory updates.
\item A \gpg uses a flow relation to order memory updates. 
An interesting property of the flow relation is that it can be compressed
dramatically without losing precision and can be
transformed into a compact acyclic flow relation in most cases, even if the
procedure it represents has loops or recursive calls.
%%\change{}{
%%Table~\ref{tab:stats2} shows that very few procedures (the number is in single digit) 
%%have back edges in their optimized \gpgs.
%%}
\end{itemize} 
Section~\ref{sec:intro.our.mt} illustrates them using a motivating example
and gives a big-picture view.

%\section{Overall approach leading to \gpgs as memory transformers}
\section{The Generalized Points-to Graphs and an Overview of their Construction}
\label{sec:intro.our.mt}

\begin{figure}[t]
\centering
%
%
%%%%%\begin{tabular}{cc}
%%%%%\begin{pspicture}(0,0)(68,52)
%%%%%\psframe(0,0)(68,52)
%%%%%%\small
%%%%%\psset{linestyle=none,fillstyle=solid,framearc=.07}
%%%%%
%%%%%\putnode{n0}{origin}{34}{26}{%
%%%%%	\psframebox[fillcolor=lightgray]{%
%%%%%	\;\;\,
%%%%%	\begin{tabular}{c}
%%%%%	\rule[-.75em]{0em}{1.25em}\gpg Optimizations
%%%%%		\\
%%%%%		\psframebox[fillcolor=white]{%
%%%%%	\;\;
%%%%%		\begin{tabular}{@{\ }c@{\ }}
%%%%%		\rule[-.75em]{0em}{1.25em}Data Flow Analyses over \gpgs
%%%%%			\\
%%%%%			\psframebox[fillcolor=lightgray]{%
%%%%%	\;\;
%%%%%			\begin{tabular}{c}
%%%%%			\rule[-.75em]{0em}{1.25em}\gpu Operations
%%%%%				\\
%%%%%				\psframebox[fillcolor=white]{%
%%%%%				\begin{tabular}{c}
%%%%%				Abstractions\rule[-.75em]{0em}{2.25em}
%%%%%				\end{tabular}
%%%%%				}
%%%%%			\\
%%%%%			\rule[-.75em]{0em}{1.25em}\lightgray Operations
%%%%%			\end{tabular}
%%%%%	\;\;
%%%%%			}
%%%%%			\\
%%%%%		\rule[-.75em]{0em}{1.25em}\white Data Flow Analysis
%%%%%		\end{tabular}
%%%%%	\;\;
%%%%%		}
%%%%%		\\
%%%%%	\rule[-.75em]{0em}{1.35em}\lightgray \gpg Optimizations
%%%%%	\end{tabular}
%%%%%	\;\;\,
%%%%%	}
%%%%%}
%%%%%%
%%%%%\end{pspicture}
%%%%%&
\begin{pspicture}(0,0)(68,52)
%\psframe(0,0)(68,52)
%\small
\psset{linestyle=none,fillstyle=solid}

\putnode{n0}{origin}{34}{43}{%
	\psframebox[fillcolor=lightgray]{%
	\makebox[80mm]{\rule[-1.em]{0em}{2.6em}%
	\gpg Optimizations%
	}}}
\putnode{n1}{n0}{0}{-12}{%
	\psframebox[fillcolor=white]{%
	\makebox[80mm]{%\rule[-.9em]{0em}{2.3em}%
	Data Flow Analyses over \gpgs%
	}}}
\putnode{n2}{n1}{0}{-11}{%
	\psframebox[fillcolor=lightgray]{%
	\makebox[80mm]{\rule[-1.em]{0em}{2.5em}%
	\gpu Operations%
	}}}
\putnode{n3}{n2}{0}{-11}{%
	\psframebox[fillcolor=white]{%
	\makebox[80mm]{%\rule[-1.em]{0em}{2.5em}%
	Abstractions%
	}}}
\putnode{w}{n0}{0}{-17}{\psframebox[doubleline=true,linecolor=gray,linestyle=solid,fillstyle=none]{\makebox[80mm]{\rule{0mm}{43.5mm}}}}
\end{pspicture}
%%%%%%%\end{tabular}
\caption{Inter-relationships between ideas and algorithms for defining and computing \gpus, \gpbs, and \gpgs.
Each layer is defined in terms of the layers below it.
Figure~\protect\ref{fig:analyses.optimizations}
fleshes out this picture by listing specific abstractions, operations,
data flow analyses, and optimizations.}
\label{fig:overview.of.the.big.picture}
\end{figure}
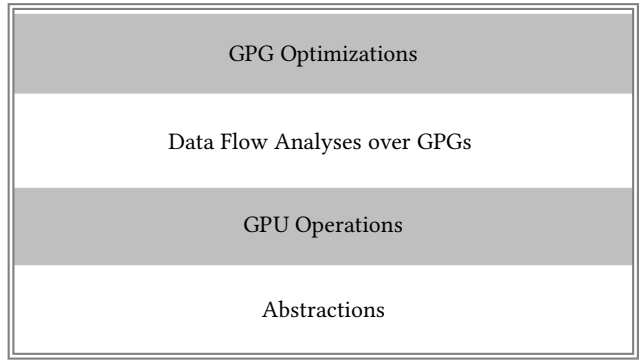

In this section, we define a \emph{generalized points-to graph} (\gpg) 
which serves as our memory transformer. 
It is a graph with \emph{generalized points-to blocks} (\gpbs) as nodes which contain 
\emph{generalized points-to updates} (\gpus). The ideas and algorithms for
defining and computing these three representations of memory transformers
can be seen as a collection of abstractions, operations, data flow analyses, and optimizations.
Their relationships are shown in Figure~\ref{fig:overview.of.the.big.picture}.
A choice of key abstractions enables us to define \gpu operations which
are used for performing three data flow analyses. The information computed by these analyses enables optimizations 
over \gpgs.

This section presents an overview of 
our approach in a limited
setting of our motivating example of Figure~\ref{fig:mot_eg.cfgs}. Towards the
end of this section, Figure~\ref{fig:analyses.optimizations}
fleshes out Figure~\ref{fig:overview.of.the.big.picture} to list specific abstractions, operations,
analyses, and optimizations. 

\subsection{Defining a Generalized Points-to Graph (\gpg)}
\label{sec:gpg.def}

We model the effect of a pointer assignment on an abstract memory 
by defining the concept of \emph{generalized points-to update} (\gpu) in
Definition~\ref{def:gpu}.
% A \gpu leaves the placeholders implicit.  [Confusing here and done well below]
We use the statement label \flab to capture weak versus strong updates and
for computing points-to information.\footnote{We omit the statement labels in \gpus at some
places when they are not required.} 
Definition~\ref{def:gpu} gives the abstract semantics of a \gpu. 
The concrete semantics of a \gpu \denew{x}{i|j}{y}{\flab} can be viewed as
the following C-style pointer assignment with $i-1$ dereferences of $x$\footnote{Alternatively, $i$ dereferences of $\&x$.
We choose $i-1$ dereference from $x$ because the left-hand side cannot be $\&x$.} and 
$j$ dereferences of $\&y$:\\[-4mm]
\[
\rule[-1.5em]{0em}{1em}
{\tt \rnode{a}{*}*\ldots\rnode{b}{*}\,x = \rnode{c}{*}*\ldots\rnode{d}{*}\&y}
\ncbar[angleA=270, angleB=270, arm=2mm, nodesep=.3mm, linearc=.5]{a}{b}\nbput[labelsep=2pt]{(i-1)}
\ncbar[angleA=270, angleB=270, arm=2mm, nodesep=.3mm, linearc=.5]{c}{d}\nbput[labelsep=2pt]{j}
\]

\begin{Definition}
\psframebox[framesep=5pt,doubleline=true,doublesep=1.5pt,linewidth=.2mm]{%
\begin{minipage}{132mm}
\gpuDef
\end{minipage}
}
\defcaption{Generalized Points-to Update.}{def:gpu}
\end{Definition}

A \gpu
\text{$\edge:\denew{x}{i|j}{y}{\flab}$} generalizes a points-to edge\footnote{Although 
a \gpu can be drawn as an arrow just like a points-to edge, we avoid
the term `edge' for a \gpu because of the risk of confusion with a
`control flow edge' in a \gpg.
}
from $x$ to $y$ with the following properties:
\begin{itemize}
\item The direction indicates that the source $x$ with \indlev $i$ identifies the locations
     being defined and the target $y$ with \indlev $j$ identifies the locations whose addresses are read.
\item The \gpu \edge abstracts away \text{$i-1+j$} placeholders. 
\item The \gpu \edge represents \may information because different locations may be reached from $x$ and $y$ 
	along different control flow paths reaching the statement \flab in the procedure.  
\end{itemize}
We refer to a \gpu with $i=1$ and $j=0$ as a \emph{classical points-to edge} as
it encodes the same information as edges in classical points-to graphs.

\begin{example}{}
The pointer assignment in statement 01 in Figure~\ref{fig:mot_eg.cfgs} 
is represented by a \gpu \denew{r}{1|0}{a}{01}
where the indirection levels ($1|0$) appear above the arrow and the statement 
number (01) appears below the arrow.
The indirection level 1 in ``$1|0$'' indicates that $r$ is defined 
by the assignment and the indirection level 0 in ``$1|0$'' 
indicates that the address of $a$ is read.
Similarly, statement 02 is represented by a \gpu \denew{q}{2|0}{m}{02}.
The indirection level 2 for $q$ indicates that some pointee of $q$ is
being defined and the indirection level 0
indicates that the address of $m$ is read.
%%; since $a$ is read, its indirection level is 1.
\end{example}

\begin{figure}[t]
\centering\small
\setlength{\tabcolsep}{1.25mm}
\psset{arrowsize=1.5}
\begin{tabular}{|l|c|c|}
\hline
\multicolumn{1}{|l|}{Pointer \rule[-.1em]{0em}{1em}}
	& \multirow{2}{*}{\scriptgpu}
	& \multicolumn{1}{l|}{Relevant memory graph }
\\
assignment
	& 
	& \multicolumn{1}{l|}{after the assignment}
	\\ \hline\hline
$\flab\!:\tt x = \&y$
	& \denew{x}{1|0}{y}{\flab}
	& 
	\begin{tabular}{@{}c@{}}
	\psset{unit=.1mm}
	\begin{pspicture}(30,50)(300,90)
	  \putnode{x}{origin}{60}{65}{\pscirclebox[fillstyle=solid,fillcolor=black,framesep=8.5]{}}
	  \putnode{w}{x}{-20}{0}{$x\;$}
	  \putnode{z}{x}{70}{0}{\pscirclebox[doubleline=true,fillstyle=solid,fillcolor=black,framesep=5]{}}
	  \putnode{w}{z}{30}{0}{$y$}
	  \ncline[linewidth=7,arrowsize=18,doublesep=.]{->}{x}{z}
          \end{pspicture}
        \end{tabular}
	\\ \hline
$\flab\!:\tt x = y$
	& \denew{x}{1|1}{y}{\flab}
	& 
	\begin{tabular}{@{}c@{}}
	\psset{unit=.1mm}
	\begin{pspicture}(30,50)(300,90)
	  \putnode{x}{origin}{60}{65}{\pscirclebox[fillstyle=solid,fillcolor=black,framesep=8.5]{}}
	  \putnode{w}{x}{-20}{0}{$x\;$}

	  \putnode{y}{x}{140}{0}{\pscirclebox[fillstyle=solid,fillcolor=black,framesep=8.5]{}}
	  \putnode{w}{y}{20}{0}{$\;y$}

	  \putnode{z}{x}{70}{0}{\pscirclebox[doubleline=true,fillstyle=solid,fillcolor=black,framesep=5]{}}

	  \ncline{->}{y}{z}
	  \ncline[linewidth=7,arrowsize=18,doublesep=2]{->}{x}{z}
          \end{pspicture}
        \end{tabular}
	\\ \hline
$\flab\!:\tt x = *y$
	& \denew{x}{1|2}{y}{\flab}
	& 
	\begin{tabular}{@{}c@{}}
	\psset{unit=.1mm}
	\begin{pspicture}(30,50)(300,90)
	  \putnode{x}{origin}{60}{65}{\pscirclebox[fillstyle=solid,fillcolor=black,framesep=8.5]{}}
	  \putnode{w}{x}{-20}{0}{$x\;$}

	  \putnode{y}{x}{210}{0}{\pscirclebox[fillstyle=solid,fillcolor=black,framesep=8.5]{}}
	  \putnode{w}{y}{20}{0}{$\;y$}

	  \putnode{z}{x}{70}{0}{\pscirclebox[doubleline=true,fillstyle=solid,fillcolor=black,framesep=5]{}}
	  \putnode{p}{x}{140}{0}{\pscirclebox[fillstyle=solid,fillcolor=black,framesep=8.5]{}}

	  \ncline{->}{y}{p}
	  \ncline{<-}{z}{p}
	  \ncline[linewidth=7,arrowsize=18,doublesep=2]{->}{x}{z}
          \end{pspicture}
        \end{tabular}
	\\ \hline
$\flab\!:\tt *x = y$
	& \denew{x}{2|1}{y}{\flab}
	& 
	\begin{tabular}{@{}c@{}}
	\psset{unit=.1mm}
	\begin{pspicture}(30,50)(300,90)
	  \putnode{x}{origin}{60}{65}{\pscirclebox[fillstyle=solid,fillcolor=black,framesep=8.5]{}}
	  \putnode{w}{x}{-20}{0}{$x\;$}

	  \putnode{y}{x}{210}{0}{\pscirclebox[fillstyle=solid,fillcolor=black,framesep=8.5]{}}
	  \putnode{w}{y}{20}{0}{$\;y$}

	  \putnode{z}{x}{70}{0}{\pscirclebox[fillstyle=solid,fillcolor=black,framesep=8.5]{}}
	  \putnode{p}{x}{140}{0}{\pscirclebox[doubleline=true,fillstyle=solid,fillcolor=black,framesep=5]{}}

	  \ncline{->}{y}{p}
	  \ncline[linewidth=7,arrowsize=18,doublesep=2]{->}{z}{p}
	  \ncline{->}{x}{z}
          \end{pspicture}
        \end{tabular}
	\\ \hline
\end{tabular}
\caption{\scriptgpus for basic pointer assignments in C\@. 
In the memory graphs, a double circle indicates
the location whose address is being assigned, a thick arrow shows the 
generated edges. Unnamed nodes may represent multiple pointees (implicitly representing placeholders).}
\label{fig:basic.gpg.edges}
\end{figure}

Figure~\ref{fig:basic.gpg.edges} presents the \gpus for basic pointer assignments in C\@.
(To deal with C structs and unions, \gpus are augmented to encode lists of field names---for details see
Figure~\ref{fig:basic.gpg.edges.heap}).

\gpus are useful rubrics of our abstractions because they
can be composed to construct new \gpus with smaller indirection levels
whenever possible thereby converting them progressively to classical points-to edges.
The composition between \gpus eliminates the data dependence between them and thereby,  
the need for control flow ordering between them.
Section~\ref{sec:gpg-operations} briefly describes the operations of
\emph{\gpu composition} and \emph{\gpu reduction} which are used for the purpose;
they are defined formally in later sections.

\begin{Definition}
\psframebox[framesep=5pt,doubleline=true,doublesep=1.5pt,linewidth=.2mm]{%
\begin{minipage}{132mm}
\gpgDef
\end{minipage}
}
\defcaption{Generalized Points-to Blocks and Generalized Points-to Graphs.}{def:gpg}
\end{Definition}

A \gpu can be seen as a atomic transformer which
is used as a building block for the 
\emph{generalized points-to graph} (\gpg) as a 
memory transformer
for a procedure (Definition~\ref{def:gpg}).
The \gpg for a procedure differs from its control flow graph (\cfg) in the following way:
\begin{itemize}
\item The \cfg could have procedure calls whereas the \gpg does not.\footnote{In the presence of
recursion and calls through function pointers (Sections~\ref{sec:handling_recur} 
and~\ref{sec:handling_fp}), 
we need an intermediate form of \gpg called an \emph{incomplete} \gpg containing 
unresolved calls that are resolved when more information becomes available.}
	Besides, a \gpg is acyclic in almost all cases, even if the
         procedure it represents has loops or recursive calls.
%%\change{}{
%%Table~\ref{tab:stats2} shows that very few procedures (the number is in single digits) out of 
%%hundreds of procedures
%%have back edges in their optimized \gpgs.
%%}
\item The \gpbs which form the nodes in a \gpg are analogous to the basic blocks of a \cfg except 
	that the
	basic blocks are sequences of statements but \gpbs are (unordered) sets of \gpus.
\end{itemize}

A concrete semantic reading of a \gpb \gpbsym is defined in terms of the semantics of 
executing a \gpu (Definition~\ref{def:gpu}).  
Execution of \gpbsym implies that the \gpus in \gpbsym are executed non-deterministically in any order.
This gives a correct abstract reading of a \gpb as a \emph{may} property.
But a stronger concrete semantic reading also holds as a \emph{must} property: 
Let \gpbsym contain \gpus corresponding to some statement \flab.  Define
\text{$X_\flab \subseteq \gpbsym$} by
\text{$X_\flab = \{ \denew{x}{i|j}{y}{\flab} \in \gpbsym \}$},  
\text{$X_\flab \neq \emptyset$}. Then, whenever statement \flab is reached in
any execution, at least one \gpu in $X_\flab$ \emph{must} be executed.
This semantics corresponds to that of the points-to information generated for a
statement in the classical points-to analysis.
This gives \gpbs their expressive power---multiple \gpus arising from a single
statement, produced by \gpu-reduction (see later), represent
\emph{may}-alternative updates, but one of these \emph{must} be
executed.\footnote{%
A subtlety is that a
\gpb \gpbsym may contain a spurious \gpu that can never be executed because 
the flow functions of points-to analysis are non-distributive~\cite{dfa_book}.
%This is consequence of introducing over-approximation to compute
%a decidable version of an undecidable analysis.
}

\begin{example}{}
Consider a \gpb \text{$\{\edge_1\!:\!\denew{x}{1|0}{a}{11},
\edge_2\!:\!\denew{x}{1|0}{b}{11},
\edge_3\!:\!\denew{y}{1|0}{c}{12},
\edge_4\!:\!\denew{z}{1|0}{d}{13},
\edge_5\!:\!\denew{t}{1|0}{d}{13},
\}$}.
After executing this \gpb (abstractly or concretely) we know that
the points-to sets of $x$ is overwritten to become $\{a,b\}$ (i.e.\ $x$ definitely points to
one of $a$ and $b$) because \gpus $\edge_1$ and $\edge_2$ both represent statement 11 and define a single location $x$. 
Similarly, the points-to set of $y$ is overwritten to become $\{c\}$
because $\edge_3$ defines a single location $c$ in statement 12.
However, this $\gpb$ causes the points-to sets of $z$ and $t$ to 
\emph{include} $\{d\}$ (without removing the existing pointees) because $\edge_4$ and $\edge_5$ both represent  statement $13$ but
define separate locations.
Thus, $x$ and $y$ are strongly updated
(their previous pointees are removed)
but $z$ and $t$ are weakly updated (their previous pointees are augmented).
\end{example}
The above example also illustrates how \gpu statement labels capture the distinction between strong and weak updates. 

The \may property of the absence of control flow between the \gpus in a \gpb 
allows us to model a \WaR dependence
	as illustrated in the following example:

\begin{example}{eg:war-dep}
Consider the code snippet on the right.
There is a \WaR data dependence
between
\setlength{\intextsep}{-.8mm}%
\setlength{\columnsep}{2mm}%
\begin{wrapfigure}{r}{21mm}
\setlength{\codeLineLength}{13mm}%
\renewcommand{\arraystretch}{.7}%
	\begin{tabular}{|rc@{}}
	%\hline
   	\codeLineOne{1}{0}{$\tt y = x;$}{white}
	\codeLine{0}{$\tt x = \&a;$}{white} 
	 %\hline
	\end{tabular}
\end{wrapfigure}
statements 01 and 02. If the control flow
is not maintained, the statements could be executed
in the reverse order and $y$ could erroneously point to $a$.

We construct a \gpb \text{$\{\denew{y}{1|1}{x}{01}, \denew{x}{1|0}{a}{02}\}$}
for the code snippet. The \may property
of this \gpb ensures that there is no data dependence between these \gpus.
The execution of this \gpb in the context of the memory represented by 
the \gpu \denew{x}{1|0}{b}{12}, computes the
points-to information 
$\{\denew{y}{}{b}{}, \denew{x}{}{a}{}\}$.
It does not compute the erroneous points-to information \denew{y}{}{a}{} thereby
preserving the \WaR dependence.
Thus, \WaR dependence can be handled without maintaining control flow.
\end{example}

\begin{figure}[t]
\small
\begin{center}
\psset{xunit=.99mm}
\begin{pspicture}(-38,10)(85,40)
%%\psframe(-38,10)(85,40)
\putnode{n1}{origin}{18}{13}{\smallblock{88}}
\putnode{n1}{origin}{18}{13}{
	\renewcommand{\arraystretch}{.7}
	\begin{tabular}{c}
	A generalized points-to update (\scriptgpu) $\edge\!:\! \denew{x}{i|j}{y}{\flab}$ %in memory \mem
	\end{tabular}
	}
\putnode{w}{n1}{49}{3}{
	\rotatebox{30}{
	\; 
	Sec. \ref{sec:gpg.def}
	}
}

\putnode{n2}{n1}{3}{8}{\bigblock{52}}
\putnode{n2}{n1}{3}{8}{
	\begin{tabular}{c}
	\gpu composition $\edge_1 \ecompwt \edge_2$ 
	%(Fig.~\ref{fig:edge.composition})
		\\
	$\ecompwt:\edge \times \edge \to \edge$ (partial function)
	\end{tabular}
	}
\putnode{w}{n2}{30}{2}{
	\rotatebox{30}{
	\; Sec.~\ref{sec:edge.composition}
	}
}
\putnode{n3}{n2}{0}{9}{\bigblock{42}}
\putnode{n3}{n2}{0}{9}{
	\begin{tabular}{c}
	\gpu reduction $\edge \rcomp \flow$ 
		\\
	$\rcomp:\edge \times \flow \to 2^{\edge}$
	\end{tabular}
	}
\putnode{w}{n3}{25}{2}{
	\rotatebox{30}{
	\; 
	Sec. \ref{sec:edge.reduction}
	}
}
%%%%%%%%%%%%%%%%%%%%%%%%%%%%%%%%%%%%%%%%%%%%%5555555
\putnode{m1}{n1}{28}{9}{}
\putnode{m2}{m1}{0}{38}{}
\putnode{m3}{m1}{0}{20}{}
%\ncline{m1}{m2}
%%\putnode[l]{u}{m3}{-5}{2}{\psframebox[linestyle=none]{
%%	\begin{tabular}{@{}l@{}}
%%		Computing points-to 
%%		\\
%%		information using \scriptgpb 
%%		\gpbsym
%%		\end{tabular}}}
%%\putnode[r]{u}{m3}{0}{2}{\psframebox[linestyle=none]{
%%	\begin{tabular}{@{}l@{}}
%%	 Construction of \scriptgpb \gpbsym 
%%		\end{tabular}}}
%%%%%%%%%%%%%%%%%%%%%%%%%%%%%%%%%%%%%%%%%%%%%5555555

%%\putnode{n6}{n1}{33}{8}{\bigblock{46}}
%%\putnode{n6}{n1}{33}{8}{
%%	\begin{tabular}{c}
%%	%\renewcommand{\arraystretch}{.9}
%%	\gpu evaluation $\llbracket \edge \rrbracket \mem$
%%		\\
%%	$\llbracket \; \rrbracket:
%%	\edge \times \mem \to \mem$
%%	\end{tabular}
%%	}
%%\putnode{w}{n6}{27}{2}{
%%	\rotatebox{30}{
%%	\; 
%%	Sec. \ref{sec:dfv_compute}
%%	}
%%}
%%\putnode{n7}{n6}{0}{9}{\bigblock{36}}
%%\putnode{n7}{n6}{0}{9}{
%%	\begin{tabular}{c}
%%	%\renewcommand{\arraystretch}{.9}
%%	\scriptgpb evaluation $\llbracket \gpbsym \rrbracket \mem$
%%		\\
%%	$\llbracket \; \rrbracket:
%%	\gpbsym \times \mem \to \mem$
%%	\end{tabular}
%%	}
%%\putnode{w}{n7}{22}{2}{
%%	\rotatebox{30}{
%%	\; 
%%	Sec. \ref{sec:dfv_compute}
%%	}
%%}
\end{pspicture}
\end{center}
\caption{A hierarchy of core operations involving \gpus. Each operation is defined in terms of 
the layers below it.
The set of \gpus reaching a \gpu \edge (computed using the reaching \gpus analyses of 
Sections~\ref{sec:reach.gpus.analysis} and~\ref{sec:blocked.gpus.analysis})
is denoted by \flow. 
By abuse of notation, we use \edge, \gpbsym, and \flow  also as types to indicate the
signatures of the operations. The operator ``$\rcomp$''  is overloaded and can be disambiguated
using the types of the operands.
}
\label{fig:overview.gpg.pta}
\end{figure}
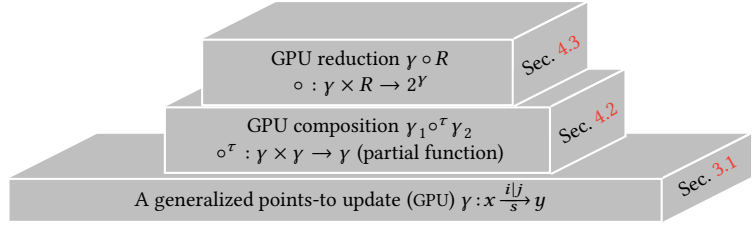

\subsection{An Overview of \gpg Operations}
\label{sec:gpg-operations}

Figure~\ref{fig:overview.gpg.pta} lists the \gpg operations based on the concept of generalized 
points-to updates (\gpus). 
Each layer is defined in terms of the layers below it.  
For each operation, Figure~\ref{fig:overview.gpg.pta} describes the types of its
operands and result, and lists the section in which the operation is defined.

\subsubsection{\gpu Composition}
In a compiler, the sequence \text{$p=\&a; *p=x$} is usually simplified
to \text{$p=\&a; a=x$} to facilitate further optimizations.
Similarly, the sequence \text{$p=\&a; q=p$} is usually simplified
to \text{$p=\&a; q=\&a$}.
While both simplifications are forms of constant propagation,
they play rather different roles, and in the \gpg framework,
are instances of (respectively) \sscomp and \tscomp variants
of \emph{\gpu composition} (Section~\ref{sec:edge.composition}).

Suppose a \gpu  $\edge_1$ precedes $\edge_2$ on some control flow path.
If there is a \RaW dependence between $\edge_1$  and $\edge_2$ 
then, a \gpu composition \text{$\edge_2\, \ecompwt \edge_1$} computes a new \gpu
where \ecomptype is \sscomp or \tscomp.
The resulting \gpu $\edge_3$ is a 
simplified version of the \emph{consumer} \gpu $\edge_2$ obtained by
using the points-to information in the \emph{producer}
\gpu $\edge_1$ such that:
\begin{itemize}
\item The \indlev of $\edge_3$ (say $i|j$) does not \emph{exceed} that  of $\edge_2$ (say 
	$i'|j'$), i.e.\ \text{$i\le i'$} and \text{$j\le j'$}.
	The two \gpus $\edge_2$ and $\edge_3$ are equivalent in the context of \gpu $\edge_1$. 
	%i.e.\ $\edge_3$ is a simplified form of $\edge_2$.
\item The type of \gpu composition (denoted $\ecomptype$) is governed by the 
	role of the common node (later called the `pivot') between $\edge_1$ and $\edge_2$.
The forms of \gpu composition important here are \tscomp and \sscomp compositions. In \tscomp composition, the pivot is the 
target of \gpu
$\edge_2$ and the source of $\edge_1$, whereas in \sscomp composition, the pivot is the source of both $\edge_1$ and $\edge_2$.
\end{itemize}
Both forms of \gpu composition are partial functions---either succeeding
with a simplified \gpu or signalling failure. A comparison of
\indlevs allow us to determine whether a \gpu composition is possible; if so,
simple arithmetic on \indlevs allows us to compute the \indlev of the resulting \gpu.

\begin{example}{}
For statement sequence \text{$p=\&a; *p=x$}, 
the consumer \gpu $\edge_2\!:\!\denew{p}{2|1}{x}{2}$ (statement 2) is simplified to $\edge_3\!:\!\denew{a}{1|1}{x}{2}$
by replacing the source $p$ of $\edge_2$ using the producer \gpu $\edge_1\!:\!\denew{p}{1|0}{a}{1}$ (statement 1).
\gpu $\edge_3$ can be further
simplified to one or more points-to edges (i.e. \gpus with \indlev $1|0$)
when \gpus representing the pointees of $x$ (the target of $\edge_3$)
become available.
% by simplifying it target $x$ when its pointees become available.
\end{example}

The above example illustrates the following:
\begin{itemize} 
\item Multiple \gpu compositions may be required to reduce the \indlev of a \gpu to
	convert it to an equivalent \gpu with \indlev $1|0$ (a classical points-to edge).
%%\AMcomment{Need to clarify somewhere that \indlev $1|0$ is essentially a
%%classical points-to edge.}
%%\UKcomment{The first list in Section~\ref{sec:gpg.def} already says so.}
\item \sscomp and \tscomp variants of \gpu composition respectively
allow a source or target
	to be resolved into a simpler form.
%%\item \deleted{A composition may create multiple \gpus. \AMcomment{Untrue! It's a partial function. Edge reduction?}}
\end{itemize}

\subsubsection{\gpu Reduction}
We generalize the above operation as follows.
If we have a set $\RIn{\flab}$ of \gpus (representing
generalized-points-to knowledge from previous statements and obtained
from the \emph{reaching \gpus analyses} of Sections~\ref{sec:reach.gpus.analysis} and~\ref{sec:blocked.gpus.analysis})
and a single \gpu $\edge_s \in \gpbsym_s$,
representing a \gpu statement $s$,
then \emph{\gpu reduction} \text{$\edge_\flab \rcomp \RIn{\flab}$} constructs a set of one or more \gpus,
all of which correspond to statement $s$. 
This is considered as the information generated for statement \flab and is denoted by
$\RGen{\flab}$. It is a union of all such sets created for every \gpu 
$\edge_s \in \gpbsym_s$ and is semantically equivalent to 
$\gpbsym_{\flab}$ in the context of
$\RIn{\flab}$ and, as suggested above, may beneficially replace 
$\gpbsym_{\flab}$.

%%\deleted{
%%The reaching \gpus analysis is an intraprocedural forward
%%data flow analysis in the spirit of the 
%%classical reaching definitions analysis except that it is not a bit-vector framework because it computes sets of
%%\gpus by processing pointer assignments. Like any other data flow analysis,
%%it may require a fixed-point computation in the case of loops. 
%%}

\gpu reduction plays a vital role in
constructing \gpgs in two ways. First, inlining the \gpg of a callee procedure and  performing
\gpu reduction eliminates procedure calls.  Further, 
%%%and this is more remarkable, \AMcomment{it was more remarkable when it always removed back edges}
\gpu reduction helps in removing redundant control flow
 wherever possible and resolving recursive calls. In particular,  
a \gpu reduction \text{$\edge_\flab \rcomp \RIn{\flab}$} eliminates the \RaW data dependence of 
$\edge_\flab$ on \RIn{\flab} thereby eliminating
the need for a control flow between $\edge_\flab$ and the \gpus in \RIn{\flab}.

%%\deleted{
%%\subsubsection{Operations for Computing Points-to Information}
%%
%%\mbox{ }
%%{\revOne CHECK: may have to be removed.}
%%For computing points-to information, we define
%%\emph{\gpu evaluation} \text{$\llbracket \edge \rrbracket \mem$} as an operation that computes 
%%the set of points-to edges 
%%that would be created by executing the \gpu \edge in memory \mem. This operation 
%%is then used to
%%define \emph{\gpb evaluation} \text{$\llbracket\gpbsym\rrbracket \mem$} which 
%%computes a set of points-to edges $\mem'$ using \gpu evaluation iteratively.
%%Let \mtsym be the \gpg of a procedure. Given a memory \mem before a call to the procedure, we compute the memory after the call
%%using \gpg \emph{application} \text{$\mtsym\left(\mem\right)$}.
%%%%\text{$\llbracket\mtsym\rrbracket \mem$};
%%This is 
%%computed using \text{$\llbracket\gpbsym\rrbracket \mem$} for \gpbs \gpbsym of \mtsym iterated over
%%the control flow of \mtsym.
%%}

\subsection{An Overview of \gpg Construction}
\label{sec:gpg-construction}

Recall that a \gpg of procedure $f$ (denoted $\mtsym_f$) 
 is a graph whose nodes are \gpbs (denoted \gpbsym) abstracting sets of memory updates
in terms of \gpus. The
edges between \gpbs are induced by the control flow of the procedure.
$\mtsym_f$
is constructed using the following steps:
        \begin{enumerate}
        \item \emph{creation}  of the initial \gpg, and \emph{inlining} optimized
              \gpgs of called procedures\footnote{
			This requires a bottom-up traversal of a spanning tree of the call graph 
			starting with its leaf nodes.}
		      within $\mtsym_f$,
        \item \emph{strength reduction} optimization to simplify the \gpus in $\mtsym_f$ by  
	       performing \emph{reaching \gpus analyses} and
               transforming \gpbs using \emph{\gpu reduction} based on the results of these analyses,
        \item \emph{redundancy elimination} optimizations to improve the compactness of 
		$\mtsym_f$.
        \end{enumerate}
%%Steps (2) and (3) are required to construct a compact \gpg for efficient analysis.
This section illustrates \gpg construction intuitively
using the motivating example in Figure~\ref{fig:mot_eg.cfgs}.
The formal details of these steps are provided in later sections.

\subsubsection{Creating a \gpg and Call Inlining}
\label{sec:gpg.creation}

In order to construct a \gpg from a \cfg, we first map the
\cfg naively into a \gpg by the following transformations:
\begin{itemize}
\item Non-pointer assignments and condition tests are removed (treating the latter as non-deterministic control flow).
	\gpg flow edges are induced from those of the \cfg.
\item Each pointer assignment labelled \flab is transliterated to its \gpu (denoted 
	$\edge_\flab$). Figure~\ref{fig:basic.gpg.edges} presented the \gpus for basic pointer assignments in C.
	\item A singleton \gpb is created for every pointer assignment in the \cfg.
\end{itemize}
Then procedure calls are replaced by the optimized \gpgs of the callees. 
The resulting \gpg may still contain unresolved calls in the case of recursion and function pointers
(Sections~\ref{sec:handling_recur} and~\ref{sec:handling_fp}).

\begin{example}{}
The initial \gpg for procedure $g$ of Figure~\ref{fig:mot_eg.cfgs} is given in Figure~\ref{fig:mot_eg.gpg.g}. 
Each assignment is replaced by its corresponding \gpu.
The initial \gpg for procedure $f$ is shown in Figure~\ref{fig:mot_eg.2} with the call to 
procedure $g$ on line 09 replaced by its optimized \gpg.
Examples~\ref{examp.rgp.analysis.1} to~\ref{exmp:coalescing} in the rest of this section explain the analyses and optimizations
over $\mtsym_f$ and $\mtsym_g$ at an intuitive level.
\end{example}

\begin{figure}[t]
%%\centering
\small
%%%%%%%%%%%%%%%%%%%%%%%%% Definitions of graphs %%%%%%%%%%%%%%%%%%%%%%%5
		\newcommand{\gpgA}{%
		\begin{pspicture}(1,0)(14,4)
		%\psframe(0,0)(16,6)
		\putnode{a1}{origin}{3}{2}{\pscirclebox[fillstyle=solid,fillcolor=white,framesep=1.22]{$r$}}
		\putnode{m1}{a1}{10}{0}{\pscirclebox[fillstyle=solid,fillcolor=white,framesep=1.22]{$a$}}
		\ncline[arrowsize=1.5]{->}{a1}{m1}
		\naput[labelsep=.25,npos=.5]{\scriptsize $1|0$}
		\nbput[labelsep=1,npos=.5]{\scriptsize 01}
		\end{pspicture}
		}%
		\newcommand{\gpgZ}{%
		\begin{pspicture}(1,0)(14,4)
		%\psframe(0,0)(16,6)
		\putnode{a1}{origin}{3}{2}{\pscirclebox[fillstyle=solid,fillcolor=white,framesep=1]{$q$}}
		\putnode{m1}{a1}{10}{0}{\pscirclebox[fillstyle=solid,fillcolor=white,framesep=.72]{$b$}}
		\ncline[arrowsize=1.5]{->}{a1}{m1}
		\naput[labelsep=.25,npos=.5]{\scriptsize $1|0$}
		\nbput[labelsep=1,npos=.5]{\scriptsize 03}
		\end{pspicture}
		}%

		\newcommand{\gpgB}{%
		\begin{pspicture}(1,0)(14,10)
		%\psframe(0,0)(16,6)
		\putnode{b1}{origin}{3}{8}{\pscirclebox[fillstyle=solid,fillcolor=white,framesep=.87]{$b$}}
		\putnode{m1}{b1}{10}{-3}{\pscirclebox[fillstyle=solid,fillcolor=white,framesep=.72]{$m$}}
		\putnode{z1}{b1}{0}{-6}{\pscirclebox[fillstyle=solid,fillcolor=white,framesep=1]{$q$}}
		%
		%\ncline[arrowsize=1.5]{->}{b1}{m1}
		\nccurve[angleA=15,angleB=135,nodesepB=-.8,arrowsize=1.5]{->}{b1}{m1}
		\naput[labelsep=0,npos=.5]{\scriptsize $1|0$}
		\nbput[labelsep=0,npos=.5]{\scriptsize 02}
		\nccurve[angleA=-15,angleB=225,nodesepB=-.8,arrowsize=1.5]{->}{z1}{m1}
		\naput[labelsep=0,npos=.5]{\scriptsize $2|0$}
		\nbput[labelsep=0,npos=.5]{\scriptsize 02}
		\end{pspicture}
		}
		\newcommand{\gpgC}{%
		\begin{pspicture}(1,0)(14,4)
		%\psframe(0,0)(16,6)
		\putnode{x1}{origin}{3}{2}{\pscirclebox[fillstyle=solid,fillcolor=white,framesep=1.1]{$e$}}
		\putnode{y1}{x1}{10}{0}{\pscirclebox[fillstyle=solid,fillcolor=white,framesep=.92]{$p$}}
		\ncline[arrowsize=1.5]{->}{x1}{y1}
		\naput[labelsep=.25,npos=.5]{\scriptsize $1|2$}
		\nbput[labelsep=1,npos=.5]{\scriptsize 04}
		\end{pspicture}
		}
		\newcommand{\gpgD}{%
		\begin{pspicture}(1,0)(14,4)
		%\psframe(0,0)(16,6)
		\putnode{z1}{origin}{3}{2}{\pscirclebox[fillstyle=solid,fillcolor=white,framesep=1]{$q$}}
		\putnode{x1}{z1}{10}{0}{\pscirclebox[fillstyle=solid,fillcolor=white,framesep=1.1]{$e$}}
		\ncline[arrowsize=1.5]{->}{z1}{x1}
		\naput[labelsep=.25,npos=.5]{\scriptsize $1|0$}
		\nbput[labelsep=1,npos=.5]{\scriptsize 05}
		\end{pspicture}
		}
	\newcommand{\gpfThree}{%
		\begin{pspicture}(1,0)(14,4)
		%\psframe(0,0)(20,8)
		\putnode{z1}{origin}{3}{2}{\pscirclebox[fillstyle=solid,fillcolor=white,framesep=1]{$q$}}
		\putnode{m1}{z1}{10}{0}{\pscirclebox[fillstyle=solid,fillcolor=white,framesep=.72]{$m$}}

		%
		%%%%%%%%%%%%%%%%%%%%%%%%%%%%%
		\nccurve[arrowsize=1.5,angleA=0,angleB=180,ncurv=1]{->}{z1}{m1}
		\naput[labelsep=.25,npos=.5]{\scriptsize $2|0$}
		\nbput[labelsep=.75,npos=.5]{\scriptsize 02}
		\end{pspicture}
	}
	\newcommand{\gpgNewE}{%
		\begin{pspicture}(0,0)(14,14)
		%\psframe(0,0)(16,6)
		\putnode{e}{origin}{3}{12}{\pscirclebox[fillstyle=solid,fillcolor=white,framesep=1.1]{$e$}}
		\putnode{p}{e}{10}{0}{\pscirclebox[fillstyle=solid,fillcolor=white,framesep=.92]{$p$}}
		\putnode{q}{e}{0}{-10}{\pscirclebox[fillstyle=solid,fillcolor=white,framesep=1]{$q$}}
		\ncline[arrowsize=1.5]{->}{e}{p}
		\naput[labelsep=.25,npos=.5]{\scriptsize $1|2$}
		\nbput[labelsep=.75,npos=.5]{\scriptsize 04}
		%
		%\nccurve[angleA=165, angleB=-65,arrowsize=1.5, nodesepA=-.2, nodesepB=-.2]{->}{q}{e}
		\ncline[arrowsize=1.5]{->}{q}{e}
		\naput[labelsep=.5,npos=.5]{\scriptsize $1|0$}
		\nbput[labelsep=.75,npos=.5]{\scriptsize 05}
		\end{pspicture}
		}
	\newcommand{\gpgNewB}{%
		\begin{pspicture}(1,0)(14,18)
		%\psframe(0,0)(16,6)
		\putnode{a1}{origin}{3}{16}{\pscirclebox[fillstyle=solid,fillcolor=white,framesep=1.22]{$r$}}
		\putnode{a2}{a1}{10}{0}{\pscirclebox[fillstyle=solid,fillcolor=white,framesep=1.22]{$a$}}
		\putnode{b1}{a1}{0}{-7}{\pscirclebox[fillstyle=solid,fillcolor=white,framesep=.95]{$b$}}
		\putnode{m1}{b1}{10}{0}{\pscirclebox[fillstyle=solid,fillcolor=white,framesep=.72]{$m$}}
		\putnode{z1}{b1}{0}{-7}{\pscirclebox[fillstyle=solid,fillcolor=white,framesep=1]{$q$}}
		%
		%\ncline[arrowsize=1.5]{->}{b1}{m1}
		\ncline{->}{a1}{a2}
		\naput[labelsep=0,npos=.5]{\scriptsize $1|0$}
		\nbput[labelsep=0,npos=.5]{\scriptsize 01}
		\ncline[arrowsize=1.5]{->}{b1}{m1}
		\naput[labelsep=.25,npos=.5]{\scriptsize $1|0$}
		\nbput[labelsep=.75,npos=.5]{\scriptsize 02}
		\nccurve[angleA=-15,angleB=255,arrowsize=1.5]{->}{z1}{m1}
		\naput[labelsep=0,npos=.5]{\scriptsize $2|0$}
		\nbput[labelsep=0,npos=.5]{\scriptsize 02}
		\end{pspicture}
		}
\begin{center}
\renewcommand{\tabcolsep}{4pt}
\begin{tabular}{cccc}
\hline
\rule{0em}{1em}%
\cfg
\rule[-1em]{0em}{2.5em}%
	& Initial \gpg $\mtsym_g$
	& \renewcommand{\arraystretch}{.7}%
		\begin{tabular}{@{}l@{}}
		$\mtsym_g$ after strength \\ reduction
		\end{tabular}
	& \renewcommand{\arraystretch}{.7}%
		\begin{tabular}{@{}l@{}}
	$\mtsym_g$ after redundancy \\ elimination
		\end{tabular}
	\\ \hline
\begin{tabular}{@{}c@{}}
\psset{unit=.25mm}
\begin{pspicture}(0,0)(130,194)
%\psframe(0,0)(120,194)
\small
\putnode{n0}{origin}{44}{180}{\psframebox[framesep=3]{$\Startscriptsize{g}$}}
\putnode{n1}{n0}{0}{-40}{\psframebox{$\renewcommand{\arraystretch}{.8}%
		\begin{array}{@{}l@{}}
			\tt r = \&a
				\\
			\tt *q = \&m 
		\end{array}$}}
	\putnode{w}{n1}{-36}{7}{01}
	\putnode{w}{n1}{-36}{-7}{02}
\putnode{n2}{n1}{30}{-40}{\psframebox{$\tt q = \&b$}}
	\putnode{w}{n2}{-35}{0}{03}
\putnode{n3}{n2}{-30}{-46}{\psframebox{$\renewcommand{\arraystretch}{.8}%
		\begin{array}{@{}l@{}}
			\tt e = *p \\
			\tt q = \&e
		\end{array}$}}
	\putnode{w}{n3}{-38}{7}{04}
	\putnode{w}{n3}{-38}{-7}{05}
\putnode{n4}{n3}{0}{-48}{\psframebox[framesep=3]{$\Endscriptsize{g}$}}
\ncline{->}{n0}{n1}
\ncline{->}{n1}{n2}
\nccurve[angleA=225,angleB=135,offset=5,nodesep=-3]{->}{n1}{n3}
%\ncline{->}{n2}{n3}
\ncloop[arm=8,offsetB=6,linearc=2,angleA=270,angleB=90,loopsize=-36]{->}{n2}{n1}
\ncline{->}{n3}{n4}
\end{pspicture}
\end{tabular}
&
\begin{tabular}{@{}c}
\begin{pspicture}(3,-10)(37,78)
%\psframe(0,0)(35,78)
\putnode{n0}{origin}{14}{74}{\psframebox[framesep=1]{$\Startscriptsize{g}$}}
\putnode{n1}{n0}{0}{-11}{\psframebox[fillstyle=solid,fillcolor=lightgray,framesep=2]{\gpgA}}
	\putnode{w}{n1}{-12}{0}{$\gpbsym_{01}$}
\putnode{n2}{n1}{0}{-13}{\psframebox[fillstyle=solid,fillcolor=lightgray,framesep=2]{\gpfThree}}
	\putnode{w}{n2}{-12}{0}{$\gpbsym_{02}$}
\putnode{nz}{n2}{8}{-13}{\psframebox[fillstyle=solid,fillcolor=lightgray,framesep=2]{\gpgZ}}
	\putnode{w}{nz}{-12}{0}{$\gpbsym_{03}$}
\putnode{n3}{nz}{-8}{-14}{\psframebox[fillstyle=solid,fillcolor=lightgray,framesep=2]{\gpgC}}
	\putnode{w}{n3}{-12}{0}{$\gpbsym_{04}$}
\putnode{n4}{n3}{0}{-13}{\psframebox[fillstyle=solid,fillcolor=lightgray,framesep=2]{\gpgD}}
	\putnode{w}{n4}{-12}{0}{$\gpbsym_{05}$}
\putnode{nn}{n4}{0}{-11}{\psframebox[framesep=1]{$\Endscriptsize{g}$}}
\psset{arrowsize=1.5,arrowinset=0}
\ncline{->}{n0}{n1}
\ncline{->}{n4}{nn}
\ncline{->}{n1}{n2}
\ncline{->}{n2}{nz}
\nccurve[angleA=225,angleB=125,ncurv=1]{->}{n2}{n3}
\ncloop[armB=3,armA=3,linearc=.5,angleA=270,angleB=90,loopsize=-9,offsetA=3.5,offsetB=2]{->}{nz}{n1}
\ncline{->}{n3}{n4}

	%\putnode{w}{n0}{-10}{-2}{$\mtsym_g$}
\end{pspicture}
\end{tabular}
&
\begin{tabular}{@{}c}
\begin{pspicture}(3,-10)(33,78)
%\psframe(0,0)(32,73)
\putnode{n0}{origin}{14}{74}{\psframebox[framesep=1]{$\Startscriptsize{g}$}}
\putnode{n1}{n0}{0}{-11}{\psframebox[fillstyle=solid,fillcolor=lightgray,framesep=2]{\gpgA}}
	\putnode{w}{n1}{-12}{0}{$\gpbsym_{01}$}
\putnode{n2}{n1}{0}{-16}{\psframebox[fillstyle=solid,fillcolor=lightgray,framesep=2]{\gpgB}}
	\putnode{w}{n2}{-12}{0}{$\gpbsym_{02}$}
\putnode{nz}{n2}{7}{-16}{\psframebox[fillstyle=solid,fillcolor=lightgray,framesep=2]{\gpgZ}}
	\putnode{w}{nz}{-12}{0}{$\gpbsym_{03}$}
\putnode{n3}{nz}{-8}{-13}{\psframebox[fillstyle=solid,fillcolor=lightgray,framesep=2]{\gpgC}}
	\putnode{w}{n3}{-12}{0}{$\gpbsym_{04}$}
\putnode{n4}{n3}{0}{-13}{\psframebox[fillstyle=solid,fillcolor=lightgray,framesep=2]{\gpgD}}
	\putnode{w}{n4}{-12}{0}{$\gpbsym_{05}$}
\putnode{nn}{n4}{0}{-11}{\psframebox[framesep=1]{$\Endscriptsize{g}$}}
%\ncline[doubleline=true,arrowsize=1.75,arrowinset=0]{->}{n0}{n1}
%\ncline[doubleline=true,arrowsize=1.75,arrowinset=0]{->}{n4}{nn}
\psset{arrowsize=1.5,arrowinset=0}
\ncline{->}{n0}{n1}
\ncline{->}{n4}{nn}
\ncline{->}{n1}{n2}
\ncline{->}{n2}{nz}
\nccurve[offset=-2,nodesepA=1,nodesepB=1,angleA=240,angleB=120,ncurv=1]{->}{n2}{n3}
%\ncline{->}{nz}{n3}
\ncline{->}{n3}{n4}
\ncloop[armA=3,armB=3,offsetA=3.5,offsetB=2,linearc=.5,angleA=270,angleB=90,loopsize=-9]{->}{nz}{n1}

	%\putnode{w}{n0}{-10}{-2}{$\mtsym_g$}
\end{pspicture}
\end{tabular}
&
\begin{tabular}{@{}c}
\begin{pspicture}(0,-10)(22,78)
%\psframe(0,0)(22,78)
\putnode{n0}{origin}{9}{74}{\psframebox[framesep=1]{$\Startscriptsize{g}$}}
\putnode{n1}{n0}{0}{-21}{\psframebox[fillstyle=solid,fillcolor=lightgray,framesep=2]{\gpgNewB}}
	\putnode{w}{n1}{-6}{14}{$\gpbsym_{11}$}
\putnode{n3}{n1}{0}{-28}{\psframebox[fillstyle=solid,fillcolor=lightgray,framesep=2]{\gpgNewE}}
	\putnode{w}{n3}{-6}{12}{$\gpbsym_{12}$}
\putnode{n2}{n1}{14}{-14}{\psframebox[fillstyle=solid,fillcolor=lightgray]{$\gpbsym_{16}$}}
\putnode{nn}{n3}{0}{-20}{\psframebox[framesep=1]{$\Endscriptsize{g}$}}
\putnode{w}{n3}{-4}{-30}{$\gpbsym_{16} = 
			\left\{
			\denew{r}{1|0}{a}{01},
			\denew{e}{1|2}{p}{04},
			\denew{q}{1|0}{e}{05}
			\right\}
			$}
%\ncline[doubleline=true,arrowsize=1.75,arrowinset=0]{->}{n0}{n1}
%\ncline[doubleline=true,arrowsize=1.75,arrowinset=0]{->}{n4}{nn}
\psset{arrowsize=1.5,arrowinset=0}
%\ncloop[doubleline=true,armA=3,armB=4,offset=2,linearc=.5,angleA=270,angleB=90,loopsize=-13]{->}{n0}{nn}
\ncangle[doubleline=true,armB=25,offsetA=2,linearc=.5,angleA=270,angleB=90,arrowsize=2]{->}{n0}{n2}
\ncangle[doubleline=true,armB=5,offsetB=2,linearc=.5,angleA=270,angleB=90,arrowsize=2]{->}{n2}{nn}
\ncline{->}{n0}{n1}
\ncline{->}{n3}{nn}
\ncline{->}{n1}{n3}
%\ncline{->}{n2}{n3}

	%\putnode{w}{n0}{-10}{-2}{$\mtsym_g$}
\end{pspicture}
\end{tabular}
\\ \hline
\end{tabular}
\end{center}
\caption{Constructing the \gpg for procedure $g$ (see Figure~\protect\ref{fig:mot_eg.cfgs}). 
The edges with double lines are not different from the control flow edges but have been shown
separately because they are introduced to represent definition-free paths for
		the sources of all \gpus that do not appear in 
		\gpb $\gpbsym_{16}$. Thus,
		it is a definition-free path for the
		sources \text{$(b,1)$} and \text{$(q,2)$} of \gpus 
		\protect\denew{b}{1|0}{m}{02} and \protect\denew{q}{2|0}{m}{02}.
		}
\label{fig:mot_eg.gpg.g}
\end{figure}
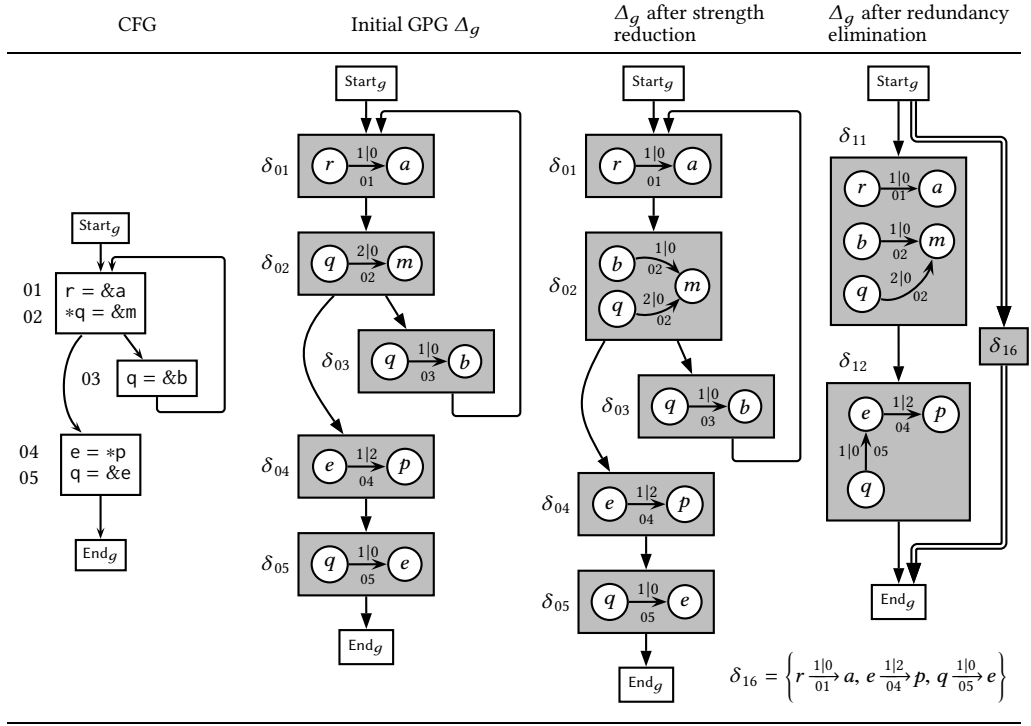

\begin{figure}[t]
%%\centering
\small
%%%%%%%%%%%%%%%%%%%%%%%%% Definitions of graphs %%%%%%%%%%%%%%%%%%%%%%%5
	\newcommand{\gpgInEndf}{%
		\begin{pspicture}(1,2)(30,39)
		%\psframe(0,0)(32,39)
		\putnode{a1}{origin}{16}{35}{\pscirclebox[fillstyle=solid,fillcolor=white,framesep=1.22]{$r$}}
		\putnode{a2}{a1}{10}{0}{\pscirclebox[fillstyle=solid,fillcolor=white,framesep=1.22]{$a$}}
		\putnode{b1}{a1}{0}{-7}{\pscirclebox[fillstyle=solid,fillcolor=white,framesep=1]{$b$}}
		\putnode{d1}{b1}{0}{-8}{\pscirclebox[fillstyle=solid,fillcolor=white,framesep=.8]{$d$}}
		\putnode{x1}{d1}{0}{-15}{\pscirclebox[fillstyle=solid,fillcolor=white,framesep=1.2]{$e$}}
		\putnode{c1}{x1}{0}{7}{\pscirclebox[fillstyle=solid,fillcolor=white,framesep=1.1]{$c$}}

		\putnode{m1}{b1}{12}{0}{\pscirclebox[fillstyle=solid,fillcolor=white,framesep=.72]{$m$}}
		\putnode{n1}{m1}{0}{-15}{\pscirclebox[fillstyle=solid,fillcolor=white,framesep=1.02]{$n$}}
		\putnode{o1}{x1}{12}{0}{\pscirclebox[fillstyle=solid,fillcolor=white,framesep=1.02]{$o$}}

		\putnode{z1}{x1}{-12}{0}{\pscirclebox[fillstyle=solid,fillcolor=white,framesep=1]{$q$}}
		\putnode{y1}{c1}{-12}{0}{\pscirclebox[fillstyle=solid,fillcolor=white,framesep=.9]{$p$}}
		%%%%%%%%%%%%%%%%%%%%%%%%%%%%%
		\ncline{->}{a1}{a2}
		\naput[labelsep=0,npos=.55]{\scriptsize $1|0$}
		\nbput[labelsep=0,npos=.5]{\scriptsize 01}
		\nccurve[arrowsize=1.5,angleA=0,angleB=180,nodesep=-.3]{->}{z1}{x1}
		\naput[labelsep=.25,npos=.5]{\scriptsize $1|0$}
		\nbput[labelsep=.75,npos=.5]{\scriptsize 05}
		\nccurve[arrowsize=1.5,angleA=0,angleB=180,nodesepA=-.4,nodesepB=-.5]{->}{b1}{m1}
		\naput[labelsep=.25,npos=.4]{\scriptsize $1|0$}
		\nbput[labelsep=.75,npos=.5]{\scriptsize 02}
		\nccurve[arrowsize=1.5,angleA=15,angleB=240,nodesepA=-.4,nodesepB=-.5]{->}{d1}{m1}
		\naput[labelsep=-.1,npos=.3]{\scriptsize $1|0$}
		\nbput[labelsep=0,npos=.7]{\scriptsize 02}
		\nccurve[arrowsize=1.5,angleA=-15,angleB=120,nodesepA=-.4,nodesepB=-.5]{->}{d1}{n1}
		\naput[labelsep=-.1,npos=.7]{\scriptsize $1|0$}
		\nbput[labelsep=0,npos=.5]{\scriptsize 08}
		\nccurve[angleA=0,angleB=180,arrowsize=1.5]{->}{y1}{c1}
		\naput[labelsep=.25,npos=.5]{\scriptsize $1|0$}
		\nbput[labelsep=.75,npos=.5]{\scriptsize 06}
		\nccurve[arrowsize=1.5,angleA=0,angleB=180,nodesepA=-.2,nodesepB=-.3]{->}{x1}{o1}
		\naput[labelsep=.25,npos=.5]{\scriptsize $1|0$}
		\nbput[labelsep=.75,npos=.5]{\scriptsize 10}
		\end{pspicture}
	}
		\newcommand{\gpgA}{%
		\begin{pspicture}(1,0)(14,4)
		%\psframe(0,0)(16,6)
		\putnode{a1}{origin}{3}{2}{\pscirclebox[fillstyle=solid,fillcolor=white,framesep=1.22]{$a$}}
		\putnode{m1}{a1}{10}{0}{\pscirclebox[fillstyle=solid,fillcolor=white,framesep=.72]{$m$}}
		\ncline[arrowsize=1.5]{->}{a1}{m1}
		\naput[labelsep=.25,npos=.5]{\scriptsize $1|0$}
		\nbput[labelsep=1,npos=.5]{\scriptsize 01}
		\end{pspicture}
		}%
		\newcommand{\gpgZ}{%
		\begin{pspicture}(1,0)(14,4)
		%\psframe(0,0)(16,6)
		\putnode{a1}{origin}{3}{2}{\pscirclebox[fillstyle=solid,fillcolor=white,framesep=1]{$q$}}
		\putnode{m1}{a1}{10}{0}{\pscirclebox[fillstyle=solid,fillcolor=white,framesep=.72]{$b$}}
		\ncline[arrowsize=1.5]{->}{a1}{m1}
		\naput[labelsep=.25,npos=.5]{\scriptsize $1|0$}
		\nbput[labelsep=1,npos=.5]{\scriptsize 03}
		\end{pspicture}
		}%

		\newcommand{\gpgB}{%
		\begin{pspicture}(1,0)(14,10)
		%\psframe(0,0)(16,6)
		\putnode{b1}{origin}{3}{8}{\pscirclebox[fillstyle=solid,fillcolor=white,framesep=.87]{$b$}}
		\putnode{m1}{b1}{10}{-3}{\pscirclebox[fillstyle=solid,fillcolor=white,framesep=.72]{$m$}}
		\putnode{z1}{b1}{0}{-6}{\pscirclebox[fillstyle=solid,fillcolor=white,framesep=1]{$q$}}
		%
		%\ncline[arrowsize=1.5]{->}{b1}{m1}
		\nccurve[angleA=15,angleB=135,nodesepB=-.8,arrowsize=1.5]{->}{b1}{m1}
		\naput[labelsep=0,npos=.5]{\scriptsize $1|0$}
		\nbput[labelsep=0,npos=.5]{\scriptsize 02}
		\nccurve[angleA=-15,angleB=225,nodesepB=-.8,arrowsize=1.5]{->}{z1}{m1}
		\naput[labelsep=0,npos=.5]{\scriptsize $2|0$}
		\nbput[labelsep=0,npos=.5]{\scriptsize 02}
		\end{pspicture}
		}
		\newcommand{\gpgC}{%
		\begin{pspicture}(1,0)(14,4)
		%\psframe(0,0)(16,6)
		\putnode{x1}{origin}{3}{2}{\pscirclebox[fillstyle=solid,fillcolor=white,framesep=1.1]{$e$}}
		\putnode{y1}{x1}{10}{0}{\pscirclebox[fillstyle=solid,fillcolor=white,framesep=.92]{$p$}}
		\ncline[arrowsize=1.5]{->}{x1}{y1}
		\naput[labelsep=.25,npos=.5]{\scriptsize $1|2$}
		\nbput[labelsep=1,npos=.5]{\scriptsize 04}
		\end{pspicture}
		}
		\newcommand{\gpgD}{%
		\begin{pspicture}(1,0)(14,4)
		%\psframe(0,0)(16,6)
		\putnode{z1}{origin}{3}{2}{\pscirclebox[fillstyle=solid,fillcolor=white,framesep=1]{$q$}}
		\putnode{x1}{z1}{10}{0}{\pscirclebox[fillstyle=solid,fillcolor=white,framesep=1.1]{$e$}}
		\ncline[arrowsize=1.5]{->}{z1}{x1}
		\naput[labelsep=.25,npos=.5]{\scriptsize $1|0$}
		\nbput[labelsep=1,npos=.5]{\scriptsize 05}
		\end{pspicture}
		}
		\newcommand{\gpgE}{%
		\begin{pspicture}(1,0)(14,12)
		%\psframe(0,0)(16,6)
		\putnode{e}{origin}{3}{10}{\pscirclebox[fillstyle=solid,fillcolor=white,framesep=1.1]{$e$}}
		\putnode{p}{e}{10}{0}{\pscirclebox[fillstyle=solid,fillcolor=white,framesep=.92]{$p$}}
		\putnode{q}{p}{0}{-8}{\pscirclebox[fillstyle=solid,fillcolor=white,framesep=1]{$q$}}
		\ncline[arrowsize=1.5]{->}{e}{p}
		\naput[labelsep=.25,npos=.5]{\scriptsize $1|2$}
		\nbput[labelsep=.25,npos=.5]{\scriptsize 04}
		\nccurve[angleA=165, angleB=-65,arrowsize=1.5, nodesepA=-.2, nodesepB=-.2]{->}{q}{e}
		\naput[labelsep=.25,npos=.5]{\scriptsize $1|0$}
		\nbput[labelsep=.25,npos=.5]{\scriptsize 05}
		\end{pspicture}
		}
		\newcommand{\gpgF}{%
		\begin{pspicture}(1,0)(14,12)
		%\psframe(0,0)(16,6)
		\putnode{e}{origin}{3}{10}{\pscirclebox[fillstyle=solid,fillcolor=white,framesep=1.1]{$e$}}
		\putnode{p}{e}{10}{0}{\pscirclebox[fillstyle=solid,fillcolor=white,framesep=1.32]{$c$}}
		\putnode{q}{p}{0}{-8}{\pscirclebox[fillstyle=solid,fillcolor=white,framesep=1]{$q$}}
		\ncline[arrowsize=1.5]{->}{e}{p}
		\naput[labelsep=.25,npos=.5]{\scriptsize $1|1$}
		\nbput[labelsep=.25,npos=.5]{\scriptsize 04}
		\nccurve[angleA=165, angleB=-65,arrowsize=1.5, nodesepA=-.2, nodesepB=-.2]{->}{q}{e}
		\naput[labelsep=.25,npos=.5]{\scriptsize $1|0$}
		\nbput[labelsep=.25,npos=.5]{\scriptsize 05}
		\end{pspicture}
		}

		\newcommand{\gpgfA}{%
		\begin{pspicture}(1,0)(14,4)
		%\psframe(0,0)(16,6)
		\putnode{a1}{origin}{3}{2}{\pscirclebox[fillstyle=solid,fillcolor=white,framesep=.92]{$p$}}
		\putnode{m1}{a1}{10}{0}{\pscirclebox[fillstyle=solid,fillcolor=white,framesep=1.32]{$c$}}
		\ncline[arrowsize=1.5]{->}{a1}{m1}
		\naput[labelsep=.25,npos=.5]{\scriptsize $1|0$}
		\nbput[labelsep=1,npos=.5]{\scriptsize 06}
		\end{pspicture}
		}
		\newcommand{\gpgfB}{%
		\begin{pspicture}(1,0)(14,4)
		%\psframe(0,0)(16,6)
		\putnode{a1}{origin}{3}{2}{\pscirclebox[fillstyle=solid,fillcolor=white,framesep=1]{$q$}}
		\putnode{m1}{a1}{10}{0}{\pscirclebox[fillstyle=solid,fillcolor=white,framesep=.92]{$d$}}
		\ncline[arrowsize=1.5]{->}{a1}{m1}
		\naput[labelsep=.25,npos=.5]{\scriptsize $1|0$}
		\nbput[labelsep=1,npos=.5]{\scriptsize 07}
		\end{pspicture}
		}
		\newcommand{\gpgfC}{%
		\begin{pspicture}(1,0)(14,4)
		%\psframe(0,0)(16,6)
		\putnode{a1}{origin}{3}{2}{\pscirclebox[fillstyle=solid,fillcolor=white,framesep=.92]{$d$}}
		\putnode{m1}{a1}{10}{0}{\pscirclebox[fillstyle=solid,fillcolor=white,framesep=1.12]{$n$}}
		\ncline[arrowsize=1.5]{->}{a1}{m1}
		\naput[labelsep=.25,npos=.5]{\scriptsize $1|0$}
		\nbput[labelsep=1,npos=.5]{\scriptsize 08}
		\end{pspicture}
		}
		\newcommand{\gpgfD}{%
		\begin{pspicture}(1,0)(14,4)
		%\psframe(0,0)(16,6)
		\putnode{x1}{origin}{3}{2}{\pscirclebox[fillstyle=solid,fillcolor=white,framesep=1.1]{$e$}}
		\putnode{y1}{x1}{10}{0}{\pscirclebox[fillstyle=solid,fillcolor=white,framesep=1.12]{$c$}}
		\ncline[arrowsize=1.5]{->}{x1}{y1}
		\naput[labelsep=.25,npos=.5]{\scriptsize $1|1$}
		\nbput[labelsep=1,npos=.5]{\scriptsize 04}
		\end{pspicture}
		}
		\newcommand{\gpgfE}{%
		\begin{pspicture}(1,0)(14,4)
		%\psframe(0,0)(16,6)
		\putnode{x1}{origin}{3}{2}{\pscirclebox[fillstyle=solid,fillcolor=white,framesep=1.1]{$e$}}
		\putnode{y1}{x1}{10}{0}{\pscirclebox[fillstyle=solid,fillcolor=white,framesep=1.12]{$o$}}
		\ncline[arrowsize=1.5]{->}{x1}{y1}
		\naput[labelsep=.25,npos=.5]{\scriptsize $1|0$}
		\nbput[labelsep=1,npos=.5]{\scriptsize 10}
		\end{pspicture}
		}
		\newcommand{\gpgfF}{%
		\begin{pspicture}(1,0)(14,4)
		%\psframe(0,0)(16,6)
		\putnode{x1}{origin}{3}{2}{\pscirclebox[fillstyle=solid,fillcolor=white,framesep=.9]{$q$}}
		\putnode{y1}{x1}{10}{0}{\pscirclebox[fillstyle=solid,fillcolor=white,framesep=1.12]{$o$}}
		\ncline[arrowsize=1.5]{->}{x1}{y1}
		\naput[labelsep=.25,npos=.5]{\scriptsize $2|0$}
		\nbput[labelsep=1,npos=.5]{\scriptsize 10}
		\end{pspicture}
		}
		\newcommand{\gpgfG}{%
		\begin{pspicture}(1,0)(14,10)
		%\psframe(0,0)(16,6)
		\putnode{b1}{origin}{3}{8}{\pscirclebox[fillstyle=solid,fillcolor=white,framesep=.87]{$b$}}
		\putnode{m1}{b1}{10}{-3}{\pscirclebox[fillstyle=solid,fillcolor=white,framesep=.72]{$m$}}
		\putnode{z1}{b1}{0}{-6}{\pscirclebox[fillstyle=solid,fillcolor=white,framesep=.72]{$d$}}
		%
		%\ncline[arrowsize=1.5]{->}{b1}{m1}
		\nccurve[angleA=15,angleB=135,nodesepB=-.8,arrowsize=1.5]{->}{b1}{m1}
		\naput[labelsep=0,npos=.5]{\scriptsize $1|0$}
		\nbput[labelsep=0,npos=.5]{\scriptsize 02}
		\nccurve[angleA=-15,angleB=225,nodesepB=-.8,arrowsize=1.5]{->}{z1}{m1}
		\naput[labelsep=0,npos=.5]{\scriptsize $1|0$}
		\nbput[labelsep=0,npos=.5]{\scriptsize 02}
		\end{pspicture}
		}
	\newcommand{\gpgNewE}{%
		\begin{pspicture}(0,0)(14,14)
		%\psframe(0,0)(16,6)
		\putnode{e}{origin}{3}{12}{\pscirclebox[fillstyle=solid,fillcolor=white,framesep=1.1]{$e$}}
		\putnode{p}{e}{10}{0}{\pscirclebox[fillstyle=solid,fillcolor=white,framesep=.92]{$p$}}
		\putnode{q}{e}{0}{-10}{\pscirclebox[fillstyle=solid,fillcolor=white,framesep=1]{$q$}}
		\ncline[arrowsize=1.5]{->}{e}{p}
		\naput[labelsep=.25,npos=.5]{\scriptsize $1|2$}
		\nbput[labelsep=.75,npos=.5]{\scriptsize 04}
		%
		%\nccurve[angleA=165, angleB=-65,arrowsize=1.5, nodesepA=-.2, nodesepB=-.2]{->}{q}{e}
		\ncline[arrowsize=1.5]{->}{q}{e}
		\naput[labelsep=.5,npos=.5]{\scriptsize $1|0$}
		\nbput[labelsep=.75,npos=.5]{\scriptsize 05}
		\end{pspicture}
		}
	\newcommand{\gpgNewNewE}{%
		\begin{pspicture}(0,0)(14,14)
		%\psframe(0,0)(16,6)
		\putnode{e}{origin}{3}{12}{\pscirclebox[fillstyle=solid,fillcolor=white,framesep=1.1]{$e$}}
		\putnode{p}{e}{10}{0}{\pscirclebox[fillstyle=solid,fillcolor=white,framesep=1.12]{$c$}}
		\putnode{q}{e}{0}{-10}{\pscirclebox[fillstyle=solid,fillcolor=white,framesep=1]{$q$}}
		\ncline[arrowsize=1.5]{->}{e}{p}
		\naput[labelsep=.25,npos=.5]{\scriptsize $1|1$}
		\nbput[labelsep=.75,npos=.5]{\scriptsize 04}
		%
		%\nccurve[angleA=165, angleB=-65,arrowsize=1.5, nodesepA=-.2, nodesepB=-.2]{->}{q}{e}
		\ncline[arrowsize=1.5]{->}{q}{e}
		\naput[labelsep=.5,npos=.5]{\scriptsize $1|0$}
		\nbput[labelsep=.75,npos=.5]{\scriptsize 05}
		\end{pspicture}
		}
	\newcommand{\gpgNewB}{%
		\begin{pspicture}(1,0)(14,18)
		%\psframe(0,0)(16,6)
		\putnode{a1}{origin}{3}{16}{\pscirclebox[fillstyle=solid,fillcolor=white,framesep=1.22]{$r$}}
		\putnode{a2}{a1}{10}{0}{\pscirclebox[fillstyle=solid,fillcolor=white,framesep=1.22]{$a$}}
		\putnode{b1}{a1}{0}{-7}{\pscirclebox[fillstyle=solid,fillcolor=white,framesep=.95]{$b$}}
		\putnode{m1}{b1}{10}{0}{\pscirclebox[fillstyle=solid,fillcolor=white,framesep=.72]{$m$}}
		\putnode{z1}{b1}{0}{-7}{\pscirclebox[fillstyle=solid,fillcolor=white,framesep=1]{$q$}}
		%
		%\ncline[arrowsize=1.5]{->}{b1}{m1}
		\ncline{->}{a1}{a2}
		\naput[labelsep=0,npos=.5]{\scriptsize $1|0$}
		\nbput[labelsep=0,npos=.5]{\scriptsize 01}
		\ncline[arrowsize=1.5]{->}{b1}{m1}
		\naput[labelsep=.25,npos=.5]{\scriptsize $1|0$}
		\nbput[labelsep=.75,npos=.5]{\scriptsize 02}
		\nccurve[angleA=-15,angleB=255,arrowsize=1.5]{->}{z1}{m1}
		\naput[labelsep=0,npos=.5]{\scriptsize $2|0$}
		\nbput[labelsep=0,npos=.5]{\scriptsize 02}
		\end{pspicture}
		}
	\newcommand{\gpgNewNewB}{%
		\begin{pspicture}(1,0)(14,18)
		%\psframe(0,0)(16,6)
		\putnode{a1}{origin}{3}{16}{\pscirclebox[fillstyle=solid,fillcolor=white,framesep=1.22]{$r$}}
		\putnode{a2}{a1}{10}{0}{\pscirclebox[fillstyle=solid,fillcolor=white,framesep=1.22]{$a$}}
		\putnode{b1}{a1}{0}{-7}{\pscirclebox[fillstyle=solid,fillcolor=white,framesep=.95]{$b$}}
		\putnode{m1}{b1}{10}{0}{\pscirclebox[fillstyle=solid,fillcolor=white,framesep=.72]{$m$}}
		\putnode{z1}{b1}{0}{-7}{\pscirclebox[fillstyle=solid,fillcolor=white,framesep=1]{$d$}}
		%
		%\ncline[arrowsize=1.5]{->}{b1}{m1}
		\ncline{->}{a1}{a2}
		\naput[labelsep=0,npos=.5]{\scriptsize $1|0$}
		\nbput[labelsep=0,npos=.5]{\scriptsize 01}
		\ncline[arrowsize=1.5]{->}{b1}{m1}
		\naput[labelsep=.25,npos=.5]{\scriptsize $1|0$}
		\nbput[labelsep=.75,npos=.5]{\scriptsize 02}
		\nccurve[angleA=-15,angleB=255,arrowsize=1.5]{->}{z1}{m1}
		\naput[labelsep=0,npos=.5]{\scriptsize $1|0$}
		\nbput[labelsep=0,npos=.5]{\scriptsize 02}
		\end{pspicture}
		}
\noindent
\begin{tabular}{@{}cccc@{}}
\hline
\rule{0em}{1em}%
\cfg
\rule[-1em]{0em}{2.5em}%
	& Initial \gpg $\mtsym_f$
	& \renewcommand{\arraystretch}{.7}%
		\begin{tabular}{@{}l@{}}
		$\mtsym_f$ after strength \\ reduction
		\end{tabular}
	& \renewcommand{\arraystretch}{.7}%
		\begin{tabular}{@{}l@{}}
		$\mtsym_f$ after redundancy \\ elimination
		\end{tabular}
	\\ \hline
\begin{tabular}{@{}c@{}}
\psset{unit=.25mm}
\begin{pspicture}(-1,0)(92,450)
%\psframe(0,0)(100,452)
\small
\putnode{n0}{origin}{50}{452}{\psframebox[framesep=3]{$\Startscriptsize{f}$}}
\putnode{n1}{n0}{0}{-48}{\psframebox{$\renewcommand{\arraystretch}{.8}%
		\begin{array}{@{}l@{}}
			\tt p = \&c \\
			\tt q = \&d \\
			\tt d = \&n \\
		\end{array}$}}
	\putnode{w}{n1}{-38}{13}{06}
	\putnode{w}{n1}{-38}{1}{07}
	\putnode{w}{n1}{-38}{-12}{08}
\putnode{n2}{n1}{0}{-52}{\psframebox{$\tt call\; g()$}}
	\putnode{w}{n2}{-40}{0}{09}
\putnode{q0}{n2}{28}{0}{}
\putnode{q1}{n2}{28}{9}{}
\putnode{q2}{n2}{28}{-9}{}
\putnode{n3}{n2}{0}{-38}{\psframebox{$ \tt *q = \&o $}}
	\putnode{w}{n3}{-40}{0}{10}
\putnode{n4}{n3}{0}{-34}{\psframebox[framesep=3]{$\Endscriptsize{f}$}}
\ncline{->}{n0}{n1}
\ncline{->}{n1}{n2}
\ncline{->}{n2}{n3}
\ncline{->}{n3}{n4}
\end{pspicture}
\end{tabular}
&
\begin{tabular}{@{}c}
\begin{pspicture}(3,-12)(28,111)
%\psframe(0,0)(24,111)
\putnode{n0}{origin}{14}{106}{\psframebox[framesep=1]{$\Startscriptsize{f}$}}
\putnode{n1}{n0}{0}{-11}{\psframebox[fillstyle=solid,fillcolor=lightgray,framesep=2]{\gpgfA}}
	\putnode{w}{n1}{-12}{0}{$\gpbsym_{06}$}
\putnode{p1}{n1}{0}{-13}{\psframebox[fillstyle=solid,fillcolor=lightgray,framesep=2]{\gpgfB}}
	\putnode{w}{p1}{-12}{0}{$\gpbsym_{07}$}
\putnode{p2}{p1}{0}{-13}{\psframebox[fillstyle=solid,fillcolor=lightgray,framesep=2]{\gpgfC}}
	\putnode{w}{p2}{-12}{0}{$\gpbsym_{08}$}
\putnode{p3}{p2}{0}{-21}{\psframebox[fillstyle=solid,fillcolor=lightgray,framesep=2]{\gpgNewB}}
	\putnode{w}{p3}{-12}{0}{$\gpbsym_{13}$}
\putnode{n2}{p3}{14}{-14}{\psframebox[fillstyle=solid,fillcolor=lightgray]{$\gpbsym_{16}$}}
%\putnode{n2}{p3}{0}{-16}{\psframebox[fillstyle=solid,fillcolor=lightgray,framesep=2]{\gpgB}}
	%\putnode{w}{n2}{-12}{0}{$\gpbsym_{13}$}
\putnode{n3}{p3}{0}{-26}{\psframebox[fillstyle=solid,fillcolor=lightgray,framesep=2]{\gpgNewE}}
	\putnode{w}{n3}{-12}{0}{$\gpbsym_{14}$}
\putnode{p4}{n3}{0}{-19}{\psframebox[fillstyle=solid,fillcolor=lightgray,framesep=2]{\gpgfF}}
	\putnode{w}{p4}{-12}{0}{$\gpbsym_{10}$}
\putnode{nn}{p4}{0}{-11}{\psframebox[framesep=1]{$\Endscriptsize{f}$}}
\putnode{q3}{n0}{2}{-70}{\psframebox[linestyle=dashed,dash=.5 .5,framearc=.2]{%
			\rule{32mm}{0mm}%
			\rule{0mm}{50mm}%			
			}}
\putnode{u}{q3}{-17}{9}{}
\putnode{w}{u}{-3}{-5}{$\mtsym_g$}
\putnode{q4}{q3}{-16}{24}{}
\putnode{q5}{q3}{-15}{-24}{}
%\ncline[linestyle=dashed,dash=.5 .5]{q1}{q4}
%\ncline[linestyle=dashed,dash=.5 .5]{q2}{q5}
%\ncline[doubleline=true,doublesep=1,offsetB=3.5,nodesepA=1,nodesepB=1.5]{->}{q0}{q3}
\psset{arrowsize=1.5,arrowinset=0}
\ncline{->}{n0}{n1}
\ncline{->}{p4}{nn}
\ncline{->}{n1}{p1}
\ncline{->}{p1}{p2}
\ncline{->}{p2}{p3}
\ncline{->}{p3}{n3}
%\ncline{->}{n2}{n3}
\ncline{->}{n3}{p4}
%\ncloop[armA=3,armB=4,offsetA=3.5,offsetB=3,linearc=.5,angleA=270,angleB=90,loopsize=-8]{->}{n2}{p3}

	%\putnode{w}{n0}{-10}{-2}{$\mtsym_g$}
%\ncloop[doubleline=true,armA=3,armB=4,offset=2,linearc=.5,angleA=270,angleB=90,loopsize=-11]{->}{p2}{p4}
\ncangle[doubleline=true,armB=24,offsetA=2,linearc=.5,angleA=270,angleB=90,arrowsize=2]{->}{p2}{n2}
\ncangle[doubleline=true,armB=4,offsetB=2,linearc=.5,angleA=270,angleB=90,arrowsize=2]{->}{n2}{p4}
\end{pspicture}
\end{tabular}
&
\begin{tabular}{@{}c}
\begin{pspicture}(-3,-12)(22,111)
%\psframe(0,0)(24,111)
\putnode{n0}{origin}{14}{106}{\psframebox[framesep=1]{$\Startscriptsize{f}$}}
\putnode{n1}{n0}{0}{-11}{\psframebox[fillstyle=solid,fillcolor=lightgray,framesep=2]{\gpgfA}}
	\putnode{w}{n1}{-13}{0}{$\gpbsym_{06}$}
\putnode{p1}{n1}{0}{-13}{\psframebox[fillstyle=solid,fillcolor=lightgray,framesep=2]{\gpgfB}}
	\putnode{w}{p1}{-13}{0}{$\gpbsym_{07}$}
\putnode{p2}{p1}{0}{-13}{\psframebox[fillstyle=solid,fillcolor=lightgray,framesep=2]{\gpgfC}}
	\putnode{w}{p2}{-13}{0}{$\gpbsym_{08}$}
\putnode{p3}{p2}{0}{-21}{\psframebox[fillstyle=solid,fillcolor=lightgray,framesep=2]{\gpgNewNewB}}
	\putnode{w}{p3}{-13}{0}{$\gpbsym_{13}$}
\putnode{n2}{p3}{14}{-14}{\psframebox[fillstyle=solid,fillcolor=lightgray]{$\gpbsym_{16}$}}
%\putnode{n2}{p3}{0}{-16}{\psframebox[fillstyle=solid,fillcolor=lightgray,framesep=2]{\gpgfG}}
	%\putnode{w}{n2}{-13}{0}{$\gpbsym_{13}$}
\putnode{n3}{p3}{0}{-26}{\psframebox[fillstyle=solid,fillcolor=lightgray,framesep=2]{\gpgNewNewE}}
	\putnode{w}{n3}{-13}{0}{$\gpbsym_{14}$}
\putnode{p4}{n3}{0}{-19}{\psframebox[fillstyle=solid,fillcolor=lightgray,framesep=2]{\gpgfE}}
	\putnode{w}{p4}{-13}{0}{$\gpbsym_{10}$}
\putnode{nn}{p4}{0}{-11}{\psframebox[framesep=1]{$\Endscriptsize{f}$}}
%\ncline[doubleline=true,arrowsize=1.75,arrowinset=0]{->}{n0}{n1}
%\ncline[doubleline=true,arrowsize=1.75,arrowinset=0]{->}{n4}{nn}
\psset{arrowsize=1.5,arrowinset=0}
\ncline{->}{n0}{n1}
\ncline{->}{p4}{nn}
\ncline{->}{n1}{p1}
\ncline{->}{p1}{p2}
\ncline{->}{p2}{p3}
\ncline{->}{p3}{n3}
%\ncline{->}{n2}{n3}
\ncline{->}{n3}{p4}
%\ncloop[armA=3,armB=4,offsetA=3.5,offsetB=3,linearc=.5,angleA=270,angleB=90,loopsize=-8]{->}{n2}{p3}

	%\putnode{w}{n0}{-10}{-2}{$\mtsym_g$}
%\ncloop[doubleline=true,armA=3,armB=4,offset=2,linearc=.5,angleA=270,angleB=90,loopsize=-11]{->}{p2}{p4}
\ncangle[doubleline=true,armB=24,offsetA=2,linearc=.5,angleA=270,angleB=90,arrowsize=2]{->}{p2}{n2}
\ncangle[doubleline=true,armB=4,offsetB=2,linearc=.5,angleA=270,angleB=90,arrowsize=2]{->}{n2}{p4}
\end{pspicture}
\end{tabular}
&
\begin{tabular}{@{}l@{}}
\begin{pspicture}(4,-12)(52,111)
%\psframe(4,-12)(52,111)
\putnode{p0}{origin}{26}{106}{\psframebox[framesep=1]{$\Startscriptsize{f}$}}
\putnode{p1}{p0}{0}{-30}{\psframebox[fillcolor=lightgray,fillstyle=solid]{\gpgInEndf}}
	\putnode{w}{p1}{-6}{23}{$\gpbsym_{15}$}
	%\putnode{w}{p0}{-16}{-2}{$\mtsym_f$}
\putnode{n2}{p1}{22}{0}{\psframebox[fillstyle=solid,fillcolor=lightgray]{$\gpbsym_{17}$}}
\putnode{pn}{p1}{0}{-31}{\psframebox[framesep=1]{$\Endscriptsize{f}$}}
%\ncline[doubleline=true,arrowsize=1.75,arrowinset=0]{->}{p0}{p1}
%\ncline[doubleline=true,arrowsize=1.75,arrowinset=0]{->}{p1}{pn}
\psset{arrowsize=1.5,arrowinset=0}
\ncline{->}{p0}{p1}
\ncline{->}{p1}{pn}
%\ncloop[doubleline=true,armA=3,armB=4,offset=2,linearc=.5,angleA=270,angleB=90,loopsize=-18]{->}{p0}{pn}
\ncangle[doubleline=true,armB=20,offsetA=2,linearc=.5,angleA=270,angleB=90,arrowsize=2]{->}{p0}{n2}
\ncangle[doubleline=true,armB=5,offsetB=2,linearc=.5,angleA=270,angleB=90,arrowsize=2]{->}{n2}{pn}
\putnode{w}{pn}{5}{-30}{\begin{minipage}{40mm}\raggedright
			$\gpbsym_{16}$ is as in Figure~\protect\ref{fig:mot_eg.gpg.g}.
			
			\bigskip
			\bigskip
			\renewcommand{\arraystretch}{1.8}	
			$\gpbsym_{17} =\big\{\begin{array}[t]{@{}l}\denew{r}{1|0}{a}{01}, \denew{d}{1|0}{m}{02}, \\
\denew{d}{1|0}{n}{08}, \denew{p}{1|0}{c}{06}, \\ \denew{q}{1|0}{e}{05}, \denew{e}{1|0}{o}{10}\big\}\end{array}$
%%			$\gpbsym_{17}$ contains all \gpus of $\mtsym_f$ except
%%			\denew{b}{1|0}{m}{02}. \gpu
%%			\denew{q}{2|0}{m}{02} which has a definition-free path in $\mtsym_g$,
%%			reduces to 
%%			\denew{d}{1|0}{m}{02}. Since $d$ is defined in $\gpbsym_{08}$ also,
%%                        it does not have a definition-free path in $\mtsym_f$.
			
			\end{minipage}}
\end{pspicture}
\end{tabular}
\\ \hline
\end{tabular}
\caption{Constructing the \gpg for procedure $f$ (see Figures~\protect\ref{fig:mot_eg.cfgs} 
and~\ref{fig:mot_eg.gpg.g}).
\gpbs $\gpbsym_{13}$ through $\gpbsym_{14}$ in the \gpg are the (renumbered) \gpbs representing
the inlined optimized \gpg of procedure $g$.
The statement labels in the \gpus of these \gpbs remain unchanged.
Redundancy elimination of $\mtsym_f$ coalesces all of its \gpbs creating a new \gpb $\gpbsym_{15}$. 
\gpb $\gpbsym_{17}$ is required for modelling definition-free paths.
The edges with double lines are control flow edges shown
separately because they are introduced to represent definition-free paths.
}
%\AMcomment{Somewhere we need to say that, in general we need to rename the block numbers (and indeed
%statement numbers) when a procedure is inlined multiple times, and note (quietly) that this isn't
%necessary in our example.}
\label{fig:mot_eg.2}
\end{figure}
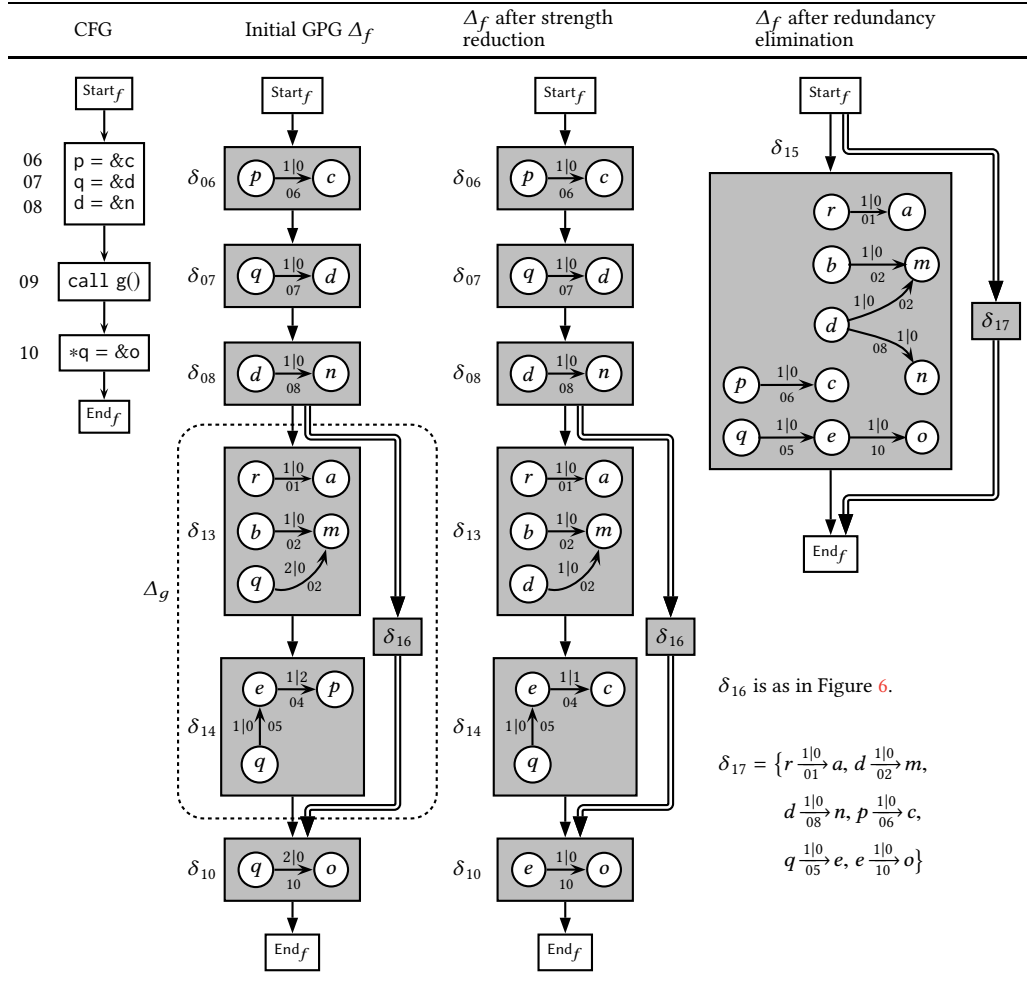

\subsubsection{Strength Reduction Optimization}

This step simplifies \gpb $\gpbsym_{\flab}$ for each statement \flab by 
\begin{itemize}
\item performing reaching \gpus analysis; this performs \gpu reduction \text{$\edge \rcomp \RIn{\flab}$} for each \text{$\edge \in \gpbsym_{\flab}$}
      which computes a set of \gpus that are equivalent to $\gpbsym_{\flab}$, and
\item replacing $\gpbsym_{\flab}$ by the resulting \gpus.
\end{itemize}

In some cases, the reaching \gpus analysis needs to \emph{block}
certain \gpus from participating in \gpu reduction
(as in Example~\ref{exmp:caller.dependence} in Section~\ref{sec:bottom.up.nomenclature})
to ensure the
soundness of strength reduction.
When this happens,
redundancy elimination optimizations need to know if the blocked \gpus in a \gpg are useful for potential composition
after the \gpg is inlined in the callers.
These two conflicting requirements 
(of ignoring some \gpus for strength reduction but remembering them for redundancy elimination)
are met by performing two variants of reaching \gpus
analysis: first with blocking, and then without blocking. There is no instance of blocking in our motivating example, hence we
provide an overview only of reaching \gpus analysis without blocking.

Effectively, strength reduction simplifies each \gpb as much as possible
given the absence of knowledge of aliasing in the caller
%%\AMcomment{Should we refer back to Section~\ref{sec:bottom.up.nomenclature}?}
(Example~\ref{exmp:caller.dependence} in Section~\ref{sec:bottom.up.nomenclature}).
In the process, data dependences are eliminated to the extent possible thereby paving way for redundancy
elimination (Section~\ref{sec:ds.opt}).

In order to reduce the \indlevs of the \gpus within a \gpb, we need to know the \gpus reaching the \gpb along all control flow paths
from the \Start{} \gpb of the procedure. We compute such \gpus through a data flow analysis in the spirit of the
classical reaching definitions analysis except that it is not a bit-vector framework because it computes sets of
\gpus by processing pointer assignments. 
This analysis annotates nodes $\gpbsym_{\flab}$ of the \gpg with $\RIn{\flab}, \ROut{\flab}, \RGen{\flab}$ and $\RKill{\flab}$.
It computes \RIn{\flab} as a union of \ROut{} of the predecessors of \flab. Then it computes \RGen{\flab} by
performing \gpu reduction \text{$\edge \, \rcomp  \RIn{\flab}$} for each \gpu \text{$\edge \in \gpbsym_{\flab}$}.
By construction, all resulting \gpus are equivalent to $\edge$ and have indirection levels that do not exceed that
of \edge. Because of the presence of \text{$\edge \in \gpbsym_{\flab}$}, some \gpus in \RIn{\flab} are killed and are not
included in \ROut{\flab}. This process may require a fixed-point computation in the presence of loops. 
Since this step follows inlining of
\gpgs of callee procedures, procedure calls have already been eliminated and hence this analysis
is effectively intraprocedural.

There is one last bit of detail which we allude to here and explain in Section~\ref{sec:may.must.xxp.edges} 
where the analysis is presented
formally: For the start \gpb of the \gpg, \RIn{} is initialized to \emph{boundary definitions}\footnote{The boundary definitions
represent \emph{boundary conditions}~\cite{Aho:2006:CPT:1177220}.} 
that help track definition-free
paths to identify variables that are upwards exposed (i.e. live on entry to the procedure
 and therefore may have additional pointees unknown to the current procedure). 
This is required for making
a distinction between strong and weak updates (Sections~\ref{sec:memory.updates} and~\ref{sec:may.must.xxp.edges}). 
For the purpose of this overview, we do not show boundary definitions in our example below.
They are explained in Example~\ref{exmp:motivating.exmp.rgpu.analysis}
in Section~\ref{sec:may.must.xxp.edges}.

%%\change{}{
%%Since this step follows inlining of
%%\gpgs of callee procedures, procedure calls have already been eliminated and hence this analysis
%%is effectively intraprocedural.
%%}

\begin{example}{examp.rgp.analysis.1}
We intuitively explain the reaching \gpus analysis for procedure $g$ 
over its initial \gpg (Figure~\ref{fig:mot_eg.gpg.g}). 
The final result is
shown later in Figure~\ref{fig:mot_eg.rpgs.g}.
Since we ignore boundary definitions for now, the analysis begins with
\text{$\RIn{01} = \emptyset$}. Further, since we compute the least fixed point, \ROut{} values are initialized to $\emptyset$ for all 
statements.
 The \gpu corresponding to the assignment in statement 01 \text{$\edge_1 \!:\!
\denew{r}{1|0}{a}{01}$}, forms \ROut{01} and \RIn{02}. 
For statement 02,
 \text{$\RIn{02} = \{\denew{r}{1|0}{a}{01}\}$} and \text{$\RGen{02} = \{\denew{q}{2|0}{m}{02}\}$}.
\text{$\RKill{02} = \emptyset$} and 
\ROut{02} is computed using \RIn{02} which also forms \RIn{03} which is 
\text{$\{\denew{r}{1|0}{a}{01}, \denew{q}{2|0}{m}{02}\}$}.
For  statement 03, \text{$\edge_3 \!:\! \denew{q}{1|0}{b}{03}$}
forms \RGen{03}. In the second iteration of the analysis over the loop, we have 
\text{$\RIn{01} = \ROut{03} = \{\denew{r}{1|0}{a}{01}, \denew{q}{2|0}{m}{02}, \denew{q}{1|0}{b}{03}\}$}.
\RIn{02} is also the same set.
Composing $\edge_2\!:\!\denew{q}{2|0}{m}{02}$ with \denew{q}{1|0}{b}{03} in \RIn{02} results in 
the \gpu \denew{b}{1|0}{m}{02}.
Also, the pointee information of $q$
is available only along one path (identified with the help of boundary definitions that are not shown here)
and hence the assignment causes a weak update and the \gpu \denew{q}{2|0}{m}{02} is also retained. Thus,
\RGen{02} is now updated and now contains two \gpus: \denew{b}{1|0}{m}{02} and \denew{q}{2|0}{m}{02}.
This process continues until the least fixed point is reached. 
Strength reduction optimization after reaching \gpus analysis gives the \gpg shown in the 
third column of Figure~\ref{fig:mot_eg.gpg.g}
(the fourth column represents the \gpg after redundancy elimination optimizations and is 
explained in Section~\ref{sec:ds.opt}).
\end{example}

\subsubsection{Redundancy Elimination Optimizations}
\label{sec:ds.opt}

This step performs  the following optimizations across \gpbs
to improve the compactness of a \gpg.

First, we perform dead \gpu elimination to remove
\emph{redundant} \gpus in $\gpbsym_{\flab}$, i.e.\ those that
are killed along every control flow path from \flab to 
the \End{} \gpb of the procedure.
If a \gpu \text{$\edge \notin \ROut{\Endscriptsize{}}$},
then \edge is removed from all \gpbs.
In the process, if a \gpb becomes empty, it is eliminated
by connecting its predecessors to its successors.

\begin{example}{exmp:mot.reaching.analysis}
In procedure $g$ of Figure~\ref{fig:mot_eg.gpg.g}, 
pointer $q$ is defined in statement 03 but is redefined in statement 05 and hence the \gpu \denew{q}{1|0}{b}{03} is eliminated. Hence 
the \gpb $\gpbsym_{03}$ becomes empty and is removed from the \gpg of procedure $g$ ($\mtsym_g$). 
Note that \gpu \denew{q}{2|0}{m}{02} does not define $q$ but its pointee and hence is not 
killed by statement 05. Thus it is not eliminated from $\mtsym_g$.

For procedure $f$ in Figure~\ref{fig:mot_eg.2}, the \gpu
\denew{q}{1|0}{d}{07} in $\gpbsym_{07}$ is killed by the \gpu \denew{q}{1|0}{e}{05} in $\gpbsym_{14}$. Hence the \gpu
\denew{q}{1|0}{d}{07} is eliminated from the
\gpb $\gpbsym_{07}$ which then becomes empty and is removed from the optimized \gpg.
Similarly, the \gpu \denew{e}{1|1}{c}{04} in \gpb $\gpbsym_{14}$ is removed because $e$ is redefined by the \gpu \denew{e}{1|0}{o}{10} in the \gpb $\gpbsym_{10}$
(after strength reduction in $\mtsym_f$). However, 
\gpu \denew{d}{1|0}{n}{08} in
\gpb $\gpbsym_{08}$ is not removed even though $\gpbsym_{13}$
contains a definition of $d$ expressed by \gpu \denew{d}{1|0}{m}{02}.
This is because $\gpbsym_{13}$ 
also contains \gpu \denew{b}{1|0}{m}{02}
 which defines $b$, indicating that
% $\gpbsym_{12}$ defines $d$ as well as $b$ implying that
$d$ is not defined along all paths.
%%\AMcomment{Somehow we don't really capture the importance that
%%\gpgs are both for statement 12 yet}
%%\UKcomment{The last sentence of Example~\ref{examp.rgp.analysis.1} says it for 02 which becomes 12 after inlining.}
Hence the previous definition of $d$ cannot be killed---giving
a weak update.
\end{example}

Finally, we eliminate the redundant control flow
in the \gpg by perform coalescing analysis 
(Section~\ref{sec:coalescing.analysis}).
It partitions the \gpbs of a \gpg (into \emph{parts}) such that
all \gpbs
in a part are coalesced (i.e., a new \gpb is formed by taking a union of the \gpus
of all \gpbs in the part)  and control flow is retained only across the new \gpbs representing the parts. 
Given a \gpb $\gpbsym_{\flab}$ in part $\group_i$, we can add its
adjacent \gpb $\gpbsym_{\slab}$ to $\group_i$
provided the \may property (Section~\ref{sec:gpg.def}) 
of $\group_i$ is preserved.
This is possible 
if the \gpus in $\group_i$ and  $\gpbsym_{\slab}$ do not have a data dependence between them.

The data dependences that can be identified using the information available 
within a procedure (or its callees) are eliminated
by strength reduction. However, 
when a \gpu involves an unresolved dereference
which requires information from calling contexts, its data dependences with other
\gpus is unknown.
Coalescing decisions involving such unknown data dependences are resolved using types.
The control flow is retained 
only when type matching indicates the possibility of \RaW or \WaW data dependence.
In all other cases the two \gpbs are coalesced. 

The new \gpb after coalescing is numbered with a new label
because \gpbs are distinguished using labels for maintaining control flow.
%%\gpb labels are used for maintaining control flow within a \gpg. 
A callee \gpg may be inlined at multiple call sites within a procedure.
Hence, we renumber the 
\gpb labels after call inlining and coalescing. 
Note that strength reduction does not create new \gpbs; it only creates new (equivalent) \gpus within the same \gpb.
The statement labels in \gpus remain unchanged because
they are unique across the program.

Coalescing two \gpbs that do not have control flow between them may eliminate a definition-free path for the \gpus
in it (see the Example~\ref{exmp:coalescing} below).
We handle this situation as follows:
We create an artificial \gpb by collecting
all \gpus that do not have a definition-free path in the \gpg.
We add a path from start to end via this \gpb.
This introduces a definition-free path for all \gpus that
do not appear in this \gpb.

%%we need to maintain the association between the \gpus
%%and the corresponding statements in the program (for computing points-to information).
%%Hence, the statement labels appearing within \gpus are never renumbered. 
%%The optimizations change the association of \gpbs with the program, hence they are
%%renumbered for convenience.

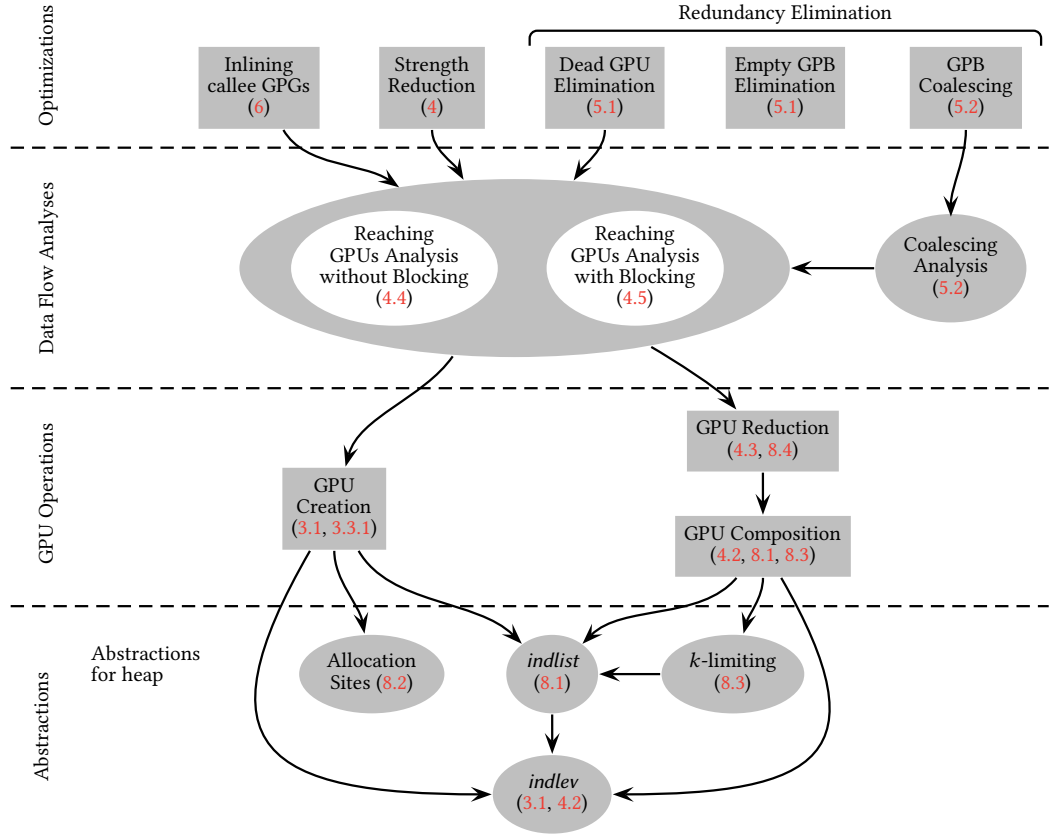
\begin{figure}[t]
\centering\small%\red
\begin{pspicture}(0,18)(138,131)
%\psframe(0,16)(138,131)
\psset{arrowsize=2.0,linewidth=.3}
%%%%%%%%%%%%%%%%%%%%%%%%%%%%%%%%%
\putnode{a1}{origin}{0}{15}{}
\putnode{w}{a1}{4}{18}{\rotatebox{90}{Abstractions}}
\putnode{w}{a1}{18}{26}{\renewcommand{\arraystretch}{.8}\begin{tabular}{l}Abstractions\\ for heap\end{tabular}}
%%%%% NOTE [l] removed
\putnode{b1}{a1}{138}{0}{}
%\ncline[linestyle=dashed]{a1}{b1}
%
\putnode{a2}{a1}{0}{34}{}
\putnode{w}{a2}{5}{15}{\rotatebox{90}{\gpu Operations}}
\putnode{b2}{a2}{138}{0}{}
\ncline[linestyle=dashed]{a2}{b2}
\putnode{a3}{a2}{0}{29}{}
\putnode{w}{a3}{5}{16}{\rotatebox{90}{Data Flow Analyses}}
\putnode{b3}{a3}{138}{0}{}
%\ncline[linestyle=dashed]{a3}{b3}
\ncline[linestyle=dashed]{a3}{b3}
\putnode{a4}{a3}{0}{32}{}
\putnode{w}{a4}{5}{11}{\rotatebox{90}{Optimizations}}
\putnode{b4}{a4}{138}{0}{}
\ncline[linestyle=dashed]{a4}{b4}
%%%%%%%%%%%%%%%%%%%%%%%%%%%%%%%%%%%%%
%\psset{linecolor=gray}
\putnode{n0}{origin}{67}{94}{\psovalbox[framesep=2.5,linestyle=none,fillstyle=solid,fillcolor=lightgray]{%
					\rule{48mm}{0mm}\rule{0mm}{13mm}}}
\putnode{n1}{origin}{51}{94}{\psovalbox[framesep=-.5,linestyle=none,fillstyle=solid,fillcolor=white]{%
			\renewcommand{\arraystretch}{.8}%
			\begin{tabular}{@{}c@{}}
			Reaching \\ \gpus 
			Analysis
			\\
			without Blocking
			\\
			(\ref{sec:reach.gpus.analysis})
			\end{tabular}}}
\putnode{n2}{n1}{32}{0}{\psovalbox[framesep=-.5,linestyle=none,fillstyle=solid,fillcolor=white]{%
			\renewcommand{\arraystretch}{.8}%
			\begin{tabular}{@{}c@{}}
			Reaching \\ \gpus
			Analysis
			\\
			with Blocking
			\\
			(\ref{sec:blocked.gpus.analysis})
			\end{tabular}}}
\putnode{n3}{n2}{42}{0}{\psovalbox[linestyle=none,fillstyle=solid,fillcolor=lightgray]{%
			\renewcommand{\arraystretch}{.8}%
			\begin{tabular}{@{}c@{}}
			Coalescing \\
			Analysis
			\\
			(\ref{sec:coalescing.analysis})
			\end{tabular}}}
%%
%%\nccurve[ncurv=.75,angleA=30,angleB=150,nodesepA=-1.5,nodesepB=-1.5]{->}{n1}{n2}
%%\nccurve[ncurv=.75,angleA=210,angleB=-25,nodesepA=-1.5,nodesepB=-1.1]{->}{n2}{n1}
%%\nccurve[ncurv=.75,angleA=150,angleB=30,nodesepB=-1.5,nodesepA=-2.1]{->}{n3}{n2}
\nccurve[ncurv=.4,angleA=180,angleB=0]{->}{n3}{n0}
\putnode{m1}{n0}{-34}{24}{\psframebox[linestyle=none,fillstyle=solid,fillcolor=lightgray]{%
			\renewcommand{\arraystretch}{.8}%
			\begin{tabular}{@{}c@{}}
			Inlining \\
			callee \gpgs
			\\
			(\ref{sec:interprocedural.extensions})
			\end{tabular}}}
\putnode{m2}{n0}{-11}{24}{\psframebox[linestyle=none,fillstyle=solid,fillcolor=lightgray]{%
			\renewcommand{\arraystretch}{.8}%
			\begin{tabular}{@{}c@{}}
			Strength \\ Reduction
			\\
			(\ref{sec:strength-reduction-optimization})
			\end{tabular}}}
\putnode{m3}{n0}{12}{24}{\psframebox[linestyle=none,fillstyle=solid,fillcolor=lightgray]{%
			\renewcommand{\arraystretch}{.8}%
			\begin{tabular}{@{}c@{}}
			Dead \gpu \\ Elimination
			\\
			(\ref{sec:dead.code.elimination})
			\end{tabular}}}
\putnode{m5}{n0}{60}{24}{\psframebox[linestyle=none,fillstyle=solid,fillcolor=lightgray]{%
			\renewcommand{\arraystretch}{.8}%
			\begin{tabular}{@{}c@{}}
			\gpb \\ Coalescing
			\\
			(\ref{sec:coalescing.analysis})
			\end{tabular}}}
\putnode{m0}{n0}{36}{24}{\psframebox[linestyle=none,fillstyle=solid,fillcolor=lightgray]{%
			\renewcommand{\arraystretch}{.8}%
			\begin{tabular}{@{}c@{}}
			Empty \gpb \\ Elimination
			\\
			(\ref{sec:dead.code.elimination})
			\end{tabular}}}
%%%
%%\ncbar[offsetA=-10,offsetB=-10,arm=2,angleA=90,angleB=90,linearc=1]{m1}{m1}
%%\nbput{Phase 0}
%%\ncbar[offsetA=-9,offsetB=-9,arm=2,angleA=90,angleB=90,linearc=1]{m2}{m2}
%%\nbput{Phase 1}
\ncbar[offsetA=10,offsetB=10,arm=2,angleA=90,angleB=90,linearc=1]{m3}{m5}
\naput[labelsep=2pt]{Redundancy Elimination}
%%%
\nccurve[angleA=300,angleB=145,nodesepB=-1.25]{->}{m1}{n0}
\nccurve[angleA=270,angleB=120,nodesepB=-.25]{->}{m2}{n0}
\nccurve[angleA=270,angleB=56,nodesepB=-.35]{->}{m3}{n0}
%%\nccurve[angleA=240,angleB=35,nodesepB=-1.25]{->}{m5}{n0}
\nccurve[angleA=270,angleB=90,nodesepB=-.5]{->}{m5}{n3}
\putnode{p0}{n1}{-8}{-32}{\psframebox[fillstyle=solid,fillcolor=lightgray,linestyle=none]{%
			\renewcommand{\arraystretch}{.8}%
			\begin{tabular}{@{}c@{}}
			\gpu \\ Creation
			\\
			(\ref{sec:gpg.def}, \ref{sec:gpg.creation})
			\end{tabular}}}
\putnode{p1}{n2}{17}{-23}{\psframebox[fillstyle=solid,fillcolor=lightgray,linestyle=none]{%
			\renewcommand{\arraystretch}{.8}%
			\begin{tabular}{@{}c@{}}
			\gpu Reduction
			\\
			(\ref{sec:edge.reduction}, \ref{sec:cycle-gpg})
			\end{tabular}}}
\putnode{p2}{p1}{0}{-14}{\psframebox[fillstyle=solid,fillcolor=lightgray,linestyle=none]{%
			\renewcommand{\arraystretch}{.8}%
			\begin{tabular}{@{}c@{}}
			\gpu Composition
			\\
			(\ref{sec:edge.composition}, \ref{sec:ind-list}, \ref{sec:k-limmiting})
			\end{tabular}}}
\nccurve[angleA=235,angleB=75,nodesepA=-.25]{->}{n0}{p0}
\ncline{->}{p1}{p2}
%%%%
%
\nccurve[angleA=330,angleB=130,nodesepA=-1.5,ncurv=.5]{->}{n0}{p1}
%%%%
%\putnode{r0}{p0}{-9}{-30}{\psovalbox[framesep=0,linestyle=none,fillstyle=solid,fillcolor=lightgray]{%
					%\rule{29mm}{0mm}\rule{0mm}{12mm}}}
\putnode{r1}{p2}{-28}{-33}{\psovalbox[fillstyle=solid,fillcolor=lightgray,linestyle=none]{%
			\renewcommand{\arraystretch}{.8}%
			\begin{tabular}{@{}c@{}}
				\indlev
			\\
			(\ref{sec:gpg.def}, \ref{sec:edge.composition})
				\end{tabular}}}
\putnode{r2}{p2}{-28}{-17}{\psovalbox[fillstyle=solid,fillcolor=lightgray,linestyle=none]{%
			\renewcommand{\arraystretch}{.8}%
			\begin{tabular}{@{}c@{}}
				\indlist
			\\
			(\ref{sec:ind-list})
				\end{tabular}}}
\putnode{r3}{r2}{24}{0}{\psovalbox[fillstyle=solid,fillcolor=lightgray,linestyle=none]{%
			\renewcommand{\arraystretch}{.8}%
			\begin{tabular}{@{}c@{}}
				$k$-limiting
			\\
			(\ref{sec:k-limmiting})
				\end{tabular}}}
\putnode{r4}{r2}{-24}{0}{\psovalbox[fillstyle=solid,fillcolor=lightgray,linestyle=none]{%
			\renewcommand{\arraystretch}{.8}%
			\begin{tabular}{@{}c@{}}
				Allocation  \\ Sites
			
			(\ref{sec:alloc-summ})
				\end{tabular}}}
\nccurve[angleA=270,angleB=75]{->}{p2}{r3}
\nccurve[angleA=230,angleB=45,nodesepB=-1.35]{->}{p2}{r2}
\nccurve[angleA=300,angleB=0,ncurv=1.5]{->}{p2}{r1}
\nccurve[angleA=240,angleB=180,ncurv=1.5]{->}{p0}{r1}
\nccurve[angleA=270,angleB=100]{->}{p0}{r4}
\nccurve[angleA=300,angleB=135,nodesepB=-1.35]{->}{p0}{r2}
\ncline{->}{r3}{r2}
\ncline{->}{r2}{r1}
\end{pspicture}
\caption{The big picture of \gpg construction
as a fleshed out version of
Figure~\ref{fig:overview.of.the.big.picture}. The
arrows show the dependence between 
specific instances of optimizations, analyses, operations, and abstractions.
The results of the two variants of reaching \gpus analysis are required together.
The optimization of empty \gpb removal does not depend 
on any data flow analysis.
The labels in parentheses refer to relevant sections.
}
\label{fig:analyses.optimizations}
\end{figure}

%%\AMcomment{Yes, but do the \gpus within get their statement number relabelled?  No!}
%%\AMcomment{I feel we should explicitly define this notion of `data dependence' between \gpus when we define \gpus.}
\begin{example}{exmp:coalescing}
For procedure $g$ in Figure~\ref{fig:mot_eg.gpg.g}, the \gpbs
$\gpbsym_1$ and $\gpbsym_2$ can be coalesced: there is no 
data dependence between their \gpus because 
\gpu \denew{r}{1|0}{a}{01} in $\gpbsym_1$ defines $r$ whose type is $\tt int\, *\!*$ whereas 
the \gpus in $\gpbsym_2$ read the address of $m$, pointer $b$, and pointee of $q$. The type of latter two is
$\tt int\, *$. Since types do not match, there is no data dependence. 

The \gpus in $\gpbsym_2$ and $\gpbsym_4$ contain a dereference whose data dependence
is unknown. We therefore use the
type information. 
Since both $q$ and $p$ have the same types,
there is a possibility of \RaW data dependence between the \gpus
\denew{q}{2|0}{m}{02} and \denew{e}{1|2}{p}{04} ($p$ and $q$ could be aliased in the caller).
Thus, we do not coalesce the 
\gpbs $\gpbsym_2$ and $\gpbsym_4$.
Also, there is no \RaW dependence between the \gpus in the \gpbs
$\gpbsym_4$ and $\gpbsym_5$ and we coalesce them; recall that potential \WaR dependence does not matter
because of the \may-property of \gpbs
(see Example~\ref{eg:war-dep}).

The \gpb resulting from coalescing \gpbs $\gpbsym_1$ and $\gpbsym_2$ is labelled $\gpbsym_{11}$. 
Similarly, the \gpb resulting from coalescing \gpbs $\gpbsym_4$ and $\gpbsym_5$ 
is labelled $\gpbsym_{12}$. 
The loop formed by the back edge \text{$\gpbsym_2 \rightarrow \gpbsym_1$} in the \gpg before coalescing now 
reduces to a self loop over $\gpbsym_{11}$. Since the \gpus in a \gpb do not have a dependence between them,
the self loop \text{$\gpbsym_{11} \rightarrow \gpbsym_{11}$} is
redundant and is removed.

For procedure $f$ in Figure~\ref{fig:mot_eg.2}, 
after performing dead \gpu elimination, the remaining
\gpbs in the \gpg of procedure $f$ are all coalesced into a single \gpb $\gpbsym_{15}$ because there is no data dependence within
the \gpus of its \gpbs.

As exemplified in Example~\ref{exmp:mot.reaching.analysis}, the sources of the \gpus
\denew{b}{1|0}{m}{02} and \denew{q}{2|0}{m}{02} in procedure $g$ are not defined along all paths from
\Start{g} to \End{g} leading to a weak update.
This is modelled by introducing a definition-free path 
(shown by edges with double lines in the fourth column of Figure~\ref{fig:mot_eg.gpg.g}).
Thus for procedure $g$, we have \gpb $\gpbsym_{16}$ that contains all \gpus of $\mtsym_g$ that
	are defined along all paths to create a definition-free path for those that are not.
Similarly, for procedure $f$, we have a definition-free path for the source of \gpu
\denew{b}{1|0}{m}{02} (as shown in the fourth column of Figure~\ref{fig:mot_eg.2}).
The \gpb $\gpbsym_{17}$ contains all \gpus of $\mtsym_f$ except
			\denew{b}{1|0}{m}{02}. \gpu
			\denew{q}{2|0}{m}{02} which has a definition-free path in $\mtsym_g$,
			reduces to 
			\denew{d}{1|0}{m}{02} in $\mtsym_f$. Since $d$ is also defined in $\gpbsym_{08}$,
                        it does not have a definition-free path in $\mtsym_f$.
\end{example}

\subsection{The Big Picture}

In this section, we have defined the concepts of \gpus, \gpbs, and \gpgs as memory transformers and described their semantics.
We have also provided an overview of \gpg construction in the context of our motivating example.

Figure~\ref{fig:analyses.optimizations} 
is a fleshed out version of Figure~\ref{fig:overview.of.the.big.picture}.
It provides the big picture of \gpg construction by listing specific
abstractions, operations, data flow analyses, and optimizations and
shows dependences between them.  The optimizations use the results of data flow analyses.
The two variants of reaching \gpus analysis are the key analyses; they
have been clubbed together because  their results are required together. They use the \gpu operations which are defined in
terms of key abstractions.
Empty \gpb removal does not require a data flow analysis. 

%Section~\ref{sec:strength-reduction-optimization} formalizes some of these operations and ties them together 
%for 
%reachability-based optimization of 
%a \gpg while Section~\ref{sec:structural-optimizations} does the same for dead \gpu elimination
%and structural optimizations of a 
%\gpg.

%%
%%\input{grand-fig-part-a}

\section{Strength Reduction Optimization}
\label{sec:strength-reduction-optimization}

In this section, we formalize the basic operations that compute the information required for performing 
strength reduction optimization of \gpbs in a \gpg.

\subsection{An Overview of Strength Reduction Optimization}
\label{sec:overview.local.optimizations}

Recall that the construction of a \gpg 
of a procedure begins
by transliterating each pointer assignment labelled \flab 
in the \cfg of the procedure into a \gpb $\gpbsym_{\flab}$ containing 
the singleton
%%\change{}{\footnote{\gpu reduction and \gpb coalescing 
%%may include multiple \gpus in a \gpb. Hence for generality, we treat a \gpb as a set of possibly multiple \gpus.}}
\gpu corresponding to the assignment.
Then the \gpus are simplified by composing them with other \gpus.
This simplification progressively converts a \gpu to a
classical points-to edge; as noted in Section~\ref{sec:bottom.up.nomenclature}.
Some simplifications can be done immediately
while others are blocked awaiting knowledge of aliasing in the callers and so
are postponed. They are reconsidered in the calling context after
the \gpg is inlined as a procedure summary in its callers.
The strength reduction optimization then replaces every \gpu \text{$\edge \in \gpbsym_{\flab}$} with its simplified version.

% The simplification of a \emph{consumer} \gpu \candedge is performed using a
% \emph{producer} \gpu \prevedge through an operation called \emph{\gpu composition} denoted
%%\AMcomment{I've italicized consumer/producer to clarify this introduces the terms.}
Based on the knowledge of a (\emph{producer}) \gpu \prevedge,
a \emph{consumer} \gpu \candedge is simplified
through an operation called \emph{\gpu composition} denoted
\text{$\candedge\, \ecompwt \prevedge$} (where \ecomptype is \sscomp or \tscomp). 
A consumer \gpu may require multiple \gpu compositions to reduce it 
to an equivalent \gpu with \indlev $1|0$ (a classical points-to edge). 
This is achieved by  
\emph{\gpu reduction} $\candedge \rcomp \flow$ which involves 
a series of \gpu compositions with appropriate producer \gpus in \flow in order to
simplify the consumer \gpu \candedge\ maximally.
The set \flow of \gpus used for simplification provides a context for \candedge and represents
generalized-points-to knowledge from previous statements. It is obtained
by performing a data flow analysis called the \emph{reaching \gpus analysis} 
which computes the sets
$\RIn{\flab}$, 
$\ROut{\flab}$,
$\RGen{\flab}$, and 
$\RKill{\flab}$. 
The set $\RGen{\flab}$ is semantically equivalent to 
$\gpbsym_{\flab}$ in the context of
\RIn{\flab} and may beneficially replace 
$\gpbsym_{\flab}$. We have two variants of reaching \gpus analysis for reasons
indicated earlier and described below.

\begin{figure}[t]
\centering\small
\begin{tabular}{@{}c|c}
\begin{tabular}{@{}c@{}}
\begin{pspicture}(0,0)(66,28)
%\psframe(0,0)(66,28)
\putnode{n1}{origin}{8}{20}{$\prevedge:\denew{\rnode{sp}{x}}{k|l}{\rnode{tp}{y}}{\flab}$}
\putnode{n2}{n1}{0}{-12}{$\candedge:\denew{\rnode{sc}{z}}{i|j}{\rnode{tc}{x}}{\slab}$}
\nccurve[linestyle=dashed,dash=.5 .5,angleA=275,angleB=100,nodesep=.75]{->}{sp}{tc}
\putnode{n3}{n2}{11}{0}{$\Rightarrow$}
\putnode{n4}{n3}{15}{0}{$\rededge:\denew{\rnode{sr}{z}}{i|(l+j-k)}{\rnode{tr}{y}}{\slab}$}
\nccurve[linestyle=dashed,dash=.5 .5,angleA=300,angleB=240,nodesep=.75]{->}{sc}{sr}
\nccurve[linestyle=dashed,dash=.5 .5,angleA=300,angleB=120,nodesepA=.75,nodesepB=.5]{->}{tp}{tr}
\putnode{w}{origin}{24}{26}{\footnotesize A generic illustration of \stscomp composition}
\putnode{w}{origin}{58}{26}{
\renewcommand{\arraystretch}{.8}%
\begin{tabular}{@{}l@{}}
\footnotesize An example%% specific \\ 
%%\footnotesize instance \\ 
%%\footnotesize of code
\end{tabular}}
\putnode{w}{origin}{59}{12}{\small
$\begin{array}{@{}l@{}}
\flab\!:\tt x  =  \&y \\
\slab\!:\tt z = x \\
	\multicolumn{1}{c}{\Downarrow} \\
\flab\!:\tt x  =  \&y \\
\slab\!:\tt z = \&y
\end{array}$
}
\psline[linewidth=.2mm](50,28)(50,2)
\end{pspicture}
\end{tabular}
&
\begin{tabular}{@{}c@{}}
\begin{pspicture}(2,0)(66,28)
%\psframe(0,0)(66,28)
\putnode{n1}{origin}{8}{20}{$\prevedge:\denew{\rnode{sp}{x}}{k|l}{\rnode{tp}{y}}{\flab}$}
\putnode{n2}{n1}{0}{-12}{$\candedge:\denew{\rnode{sc}{x}}{i|j}{\rnode{tc}{z}}{\slab}$}
\nccurve[linestyle=dashed,dash=.5 .5,angleA=270,angleB=90,nodesep=.75]{->}{sp}{sc}
\putnode{n3}{n2}{11}{0}{$\Rightarrow$}
\putnode{n4}{n3}{15}{0}{$\rededge:\denew{\rnode{sr}{y}}{(l+i-k)|j}{\rnode{tr}{z}}{\slab}$}
\nccurve[linestyle=dashed,dash=.5 .5,angleA=300,angleB=120,nodesepA=.75,nodesepB=.5]{->}{tp}{sr}
\nccurve[linestyle=dashed,dash=.5 .5,angleA=300,angleB=240,nodesepA=.75,nodesepB=.5,ncurv=.4]{->}{tc}{tr}
\putnode{w}{origin}{24}{26}{\footnotesize A generic illustration of \ssscomp composition}
\putnode{w}{origin}{58}{26}{
\renewcommand{\arraystretch}{.8}%
\begin{tabular}{@{}l@{}}
\footnotesize An example %% \\ 
%%\footnotesize instance \\ 
%%\footnotesize of code
\end{tabular}}
\putnode{w}{origin}{59}{12}{\small
$\begin{array}{@{}l@{}}
\flab\!:\tt x  =  \&y \\
\slab\!:\tt *x  =  z \\
	\multicolumn{1}{c}{\Downarrow} \\
\flab\!:\tt x  =  \&y \\
\slab\!:\tt y  =  z 
\end{array}$
}
\psline[linewidth=.2mm](50,28)(50,2)
\end{pspicture}
\end{tabular}
\\
\begin{minipage}{67mm}
\small %%footnotesize
\begin{itemize}
%%\item $\rededge = \candedge\, \ecomp^{\textrm{ts}} \prevedge$ 
%%	for consumer \candedge and producer \prevedge. 
%%	The resulting \gpu \rededge is semantically equivalent to \candedge.
\item The pivot $x$ is the target of \candedge and the source of \prevedge. 
\item There is a \RaW dependence if \text{$j \geq k$}.
%%\item A \stscomp composition can be performed between \candedge and \prevedge if \prevedge defines 
%%        a pointee of $x$ and \candedge reads it in the RHS of the assignment. This requires the \sindlevs of $x$ in 
%%	\candedge and \prevedge
%%	to satisfy the condition \text{$j\geq k$}.
\item \rededge is computed by adding $j-k$ to {\footnotesize\sindlev} of both source and target of \prevedge.
%%\item \rededge is computed by simplifying \candedge by replacing its pivot $x$ by the target of \prevedge.
%%	This requires balancing the \sindlev of the pivot in both the \gpus.  It is achieved by adding the difference
%%        $j-k$ to both numbers in the \sindlev of \prevedge.
%%	This allows us to connect the source of \candedge (i.e. $z$)  to the target of \prevedge (i.e. $y$) to form \rededge.
\end{itemize}
\end{minipage}
&
\begin{minipage}{68mm}
\small %%footnotesize
\raggedright
\begin{itemize}
%%\item $\rededge = \candedge\, \ecomp^{\textrm{ss}} \prevedge$ 
%%	for consumer \candedge and producer \prevedge.
%%	The resulting \gpu \rededge is semantically equivalent to \candedge.
\item The pivot $x$ is the source of both \candedge and \prevedge.
\item There is a \RaW dependence if \text{$i>k$}.
%%\item An \ssscomp composition can be performed between \candedge and \prevedge if \prevedge defines 
%%        a pointee of $x$ and \candedge reads it in the LHS of the assignment. This requires the \sindlevs of $x$ in 
%%	\candedge and \prevedge
%%	to satisfy the condition \text{$i>k$}.
\item \rededge is computed by adding $i-k$ to {\footnotesize\sindlev} of both source and target of \prevedge.

%%\item \rededge is computed by simplifying \candedge by replacing its pivot $x$ by the target of \prevedge.
%%	This requires balancing the \sindlev of the pivot in both the \gpus.  It is achieved by adding the difference
%%	$i-k$ to both numbers in the \sindlev of \prevedge.
%%	This allows us to connect the target of \prevedge (i.e. $y$)  to the target of \candedge (i.e. $z$) to form \rededge.
\end{itemize}
\end{minipage}
\end{tabular}
\caption{Composing a consumer \gpu \candedge with a  producer \gpu \prevedge
to compute a new \gpu \rededge which is equivalent to \candedge in the context of \prevedge.
Both \ssscomp and \stscomp compositions
exploit a \RaW dependence
of statement at \slab on the statement at \flab because the pointer defined in \prevedge is used to simplify a pointer used in
\candedge. 
%%The other two possible compositions \sttcomp and \sstcomp are less useful.
}
\label{fig:picture.edge.composition}
\end{figure}
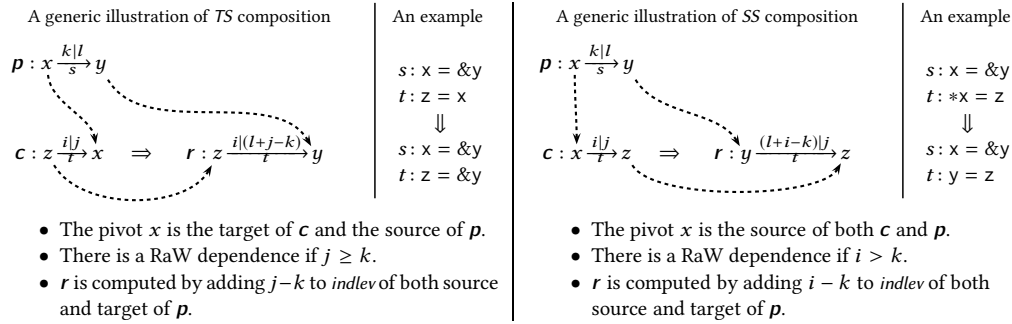

In some cases, the location read by \candedge could be different from the
 location defined by \prevedge due to the presence of 
a \gpu \baredge  (called a \emph{barrier}) corresponding to an intervening assignment.
The \gpu \prevedge may be updated by the \gpu
\baredge depending on the aliases in the calling context
(Section~\ref{sec:bottom.up.nomenclature}). 
This could happen because the \indlev of the source of \prevedge 
or \baredge is greater than 1 indicating that the pointer being defined by this \gpu is still not 
known. In such a situation (characterized formally in Section~\ref{sec:need.blocked.analysis}), 
replacing $\gpbsym_\flab$ by $\RGen{\flab}$ during strength reduction may be unsound.
To ensure soundness, we need to \emph{postpone} the
composition \text{$\candedge \ecompwt \prevedge$} explicitly by eliminating those \gpus from \flow which are blocked by 
a barrier.\footnote{Formally the term `barrier' applies to a \gpu, but we abuse this 
and refer to its associated statement as a barrier too.}
We do this by  performing a variant of reaching \gpus analysis called the 
\emph{reaching \gpus analysis with blocking} that identifies \gpus blocked by a barrier
%% to compute reaching \gpus that are not blocked 
(Section~\ref{sec:blocked.gpus.analysis}). 
We distinguish the two variants by using the phrase
\emph{reaching \gpus analysis without blocking}
for the earlier reaching \gpus analysis.
For  strength reduction, it is sufficient to perform reaching \gpus analysis with blocking. However, 
redundancy elimination optimizations need to know whether
the blocked \gpus in a \gpg are useful for potential composition
after the \gpg is inlined in the callers. These two conflicting requirements force us to
perform both the variants of reaching \gpus
analysis: first with blocking, and then without blocking.

%%Note: composition across an indirect assignment and composition with an indirect assignment.

%%These situations are rare as shown by our measurements in 
%%Section~\ref{sec:unknown}.

Section~\ref{sec:edge.composition} defines \gpu composition as a family of partial
operations.
Section~\ref{sec:edge.reduction} defines \gpu reduction.
Section~\ref{sec:reach.gpus.analysis} provides data flow equations for reaching \gpus analysis 
without blocking while 
Section~\ref{sec:blocked.gpus.analysis} provides data flow equations for reaching \gpus analysis
with blocking.

\newcommand{\gnode}[5]{
\putnode{#1}{#2}{#3}{#4}{\pscirclebox[fillstyle=solid,fillcolor=black,framesep=1]{}}
\putnode{w}{#1}{0}{-5}{#5}
\ifthenelse{\equal{#2}{origin}}{}{%
\ncline{->}{#2}{#1}
}
}

\newcommand{\figOneOne}{%
\psset{unit=.5mm,arrowsize=3}
\begin{tabular}{@{}c@{}}
\begin{pspicture}(0,0)(38,28)
%\psframe(0,0)(48,20)
\putnode{n1}{origin}{5}{18}{\pscirclebox[fillstyle=solid,fillcolor=black,framesep=1]{}}
	\putnode{w}{n1}{-1}{-5}{$x$}
\putnode{n2}{n1}{12}{0}{\pscirclebox[fillstyle=solid,fillcolor=black,framesep=1]{}}
\putnode{n3}{n2}{12}{0}{\pscirclebox[fillstyle=solid,fillcolor=black,framesep=1]{}}
	\putnode{w}{n3}{0}{5}{$y$}
	%\putnode{w}{n3}{6}{0}{$\locp$}
\putnode{n4}{n1}{8}{-10}{\pscirclebox[fillstyle=solid,fillcolor=black,framesep=1]{}}
	\putnode{w}{n4}{0}{-5}{$z$}
	%\putnode{w}{n4}{6}{0}{\locn}
%
\ncline[linestyle=dashed,dash=1 1]{->}{n1}{n2}
\ncline{->}{n2}{n3}
\ncline[nodesep=-.25]{->}{n1}{n4}
\end{pspicture}
\end{tabular}
}

\newcommand{\figOneTwo}{%
\psset{unit=.5mm,arrowsize=3}
\begin{tabular}{@{}c@{}}
\begin{pspicture}(0,0)(38,20)
%\psframe(0,0)(48,20)
\putnode{n1}{origin}{4}{8}{\pscirclebox[fillstyle=solid,fillcolor=black,framesep=1]{}}
	\putnode{w}{n1}{0}{-5}{$x$}
\putnode{n2}{n1}{14}{0}{\pscirclebox[fillstyle=solid,fillcolor=black,framesep=1]{}}
	\putnode{w}{n2}{0}{-5}{$y$}
	%\putnode{w}{n2}{0}{5}{\locp}
\putnode{n3}{n2}{14}{0}{\pscirclebox[fillstyle=solid,fillcolor=black,framesep=1]{}}
	\putnode{w}{n3}{0}{-5}{$z$}
	%\putnode{w}{n3}{0}{5}{\locn}
%
\ncline{->}{n1}{n2}
\ncline{->}{n2}{n3}
\end{pspicture}
\end{tabular}
}

\newcommand{\figOneThree}{%
\psset{unit=.5mm,arrowsize=3}
\begin{tabular}{@{}c@{}}
\begin{pspicture}(0,0)(38,20)
%\psframe(0,0)(48,20)
\putnode{n1}{origin}{4}{14}{\pscirclebox[fillstyle=solid,fillcolor=black,framesep=1]{}}
	\putnode{w}{n1}{0}{-5}{$x$}
\putnode{n2}{n1}{12}{0}{\pscirclebox[fillstyle=solid,fillcolor=black,framesep=1]{}}
\putnode{n3}{n2}{12}{0}{\pscirclebox[fillstyle=solid,fillcolor=black,framesep=1]{}}
	\putnode{w}{n3}{0}{5}{$y$}
	%\putnode{w}{n3}{6}{0}{\locp}
\putnode{n4}{n2}{8}{-10}{\pscirclebox[fillstyle=solid,fillcolor=black,framesep=1]{}}
	\putnode{w}{n4}{0}{-5}{$z$}
	%\putnode{w}{n4}{6}{0}{\locn}
%
\ncline{->}{n1}{n2}
\ncline[linestyle=dashed,dash=1 1]{->}{n2}{n3}
\ncline[nodesep=-.25]{->}{n2}{n4}
\end{pspicture}
\end{tabular}
}

\newcommand{\figThreeOne}{%
\psset{unit=.5mm,arrowsize=3}
\begin{tabular}{@{}c@{}}
\begin{pspicture}(0,0)(38,20)
%\psframe(0,0)(48,20)
\putnode{n1}{origin}{4}{9}{\pscirclebox[fillstyle=solid,fillcolor=black,framesep=1]{}}
	\putnode{w}{n1}{0}{5}{$x$}
\putnode{n2}{n1}{14}{0}{\pscirclebox[fillstyle=solid,fillcolor=black,framesep=1]{}}
	%\putnode{w}{n2}{0}{5}{\locn}
\putnode{n3}{n2}{14}{0}{\pscirclebox[fillstyle=solid,fillcolor=black,framesep=1]{}}
	\putnode{w}{n3}{0}{-5}{$y$}
	%\putnode{w}{n3}{0}{5}{\locp}
\putnode{n4}{n2}{-10}{-8}{\pscirclebox[fillstyle=solid,fillcolor=black,framesep=1]{}}
	\putnode{w}{n4}{-4}{0}{$z$}
\ncline{->}{n1}{n2}
\ncline{->}{n2}{n3}
\ncline[nodesep=-.25]{->}{n4}{n2}
\end{pspicture}
\end{tabular}
}

\newcommand{\figThreeTwo}{%
\psset{unit=.5mm,arrowsize=3}
\begin{tabular}{@{}c@{}}
\begin{pspicture}(0,0)(38,30)
%\psframe(0,0)(48,20)
\putnode{n1}{origin}{2}{18}{\pscirclebox[fillstyle=solid,fillcolor=black,framesep=1]{}}
	\putnode{w}{n1}{0}{-5}{$x$}
\putnode{n2}{n1}{12}{0}{\pscirclebox[fillstyle=solid,fillcolor=black,framesep=1]{}}
	%\putnode{w}{n2}{0}{5}{\locp}
	\putnode{w}{n2}{0}{-5}{$y$}
\putnode{n3}{n2}{12}{0}{\pscirclebox[fillstyle=solid,fillcolor=black,framesep=1]{}}
	%\putnode{w}{n3}{6}{0}{\locn}
\putnode{n4}{n3}{-8}{-10}{\pscirclebox[fillstyle=solid,fillcolor=black,framesep=1]{}}
	\putnode{w}{n4}{0}{-5}{$z$}
\ncline{->}{n1}{n2}
\ncline{->}{n2}{n3}
\ncline[nodesep=-.25]{->}{n4}{n3}
\end{pspicture}
\end{tabular}
}

\newcommand{\figThreeThree}{%
\psset{unit=.5mm,arrowsize=3}
\begin{tabular}{@{}c@{}}
\begin{pspicture}(0,0)(38,24)
%\psframe(0,0)(48,20)
\putnode{n1}{origin}{8}{16}{\pscirclebox[fillstyle=solid,fillcolor=black,framesep=1]{}}
	\putnode{w}{n1}{-4}{0}{$x$}
\putnode{n2}{n1}{14}{-6}{\pscirclebox[fillstyle=solid,fillcolor=black,framesep=1]{}}
	%\putnode{w}{n2}{6}{3}{\locp}
	%\putnode{w}{n2}{6}{-3}{\locn}
	\putnode{w}{n2}{0}{-5}{$y$}
\putnode{n3}{n1}{0}{-12}{\pscirclebox[fillstyle=solid,fillcolor=black,framesep=1]{}}
	\putnode{w}{n3}{-4}{0}{$z$}
\ncline{->}{n1}{n2}
\ncline{<-}{n2}{n3}
\end{pspicture}
\end{tabular}
}

\begin{figure}[t]
\setlength{\tabcolsep}{3.7pt}
\centering
\begin{tabular}{|c|c|c|
                |c|c|c|
		}
\hline
\multicolumn{3}{|c||}{\rule{0em}{1em}Possible \ssscomp Compositions}
	& \multicolumn{3}{c|}{Possible \stscomp Compositions}
	\\ \hline
\rule[-.9em]{0em}{2.4em}%
 \renewcommand{\arraystretch}{.8}\begin{tabular}{@{}c@{}} Statement \\ sequence\end{tabular}
	&  \renewcommand{\arraystretch}{.8}\begin{tabular}{@{}c@{}} Memory graph after \\ the stmt. sequence \end{tabular}
	&  \renewcommand{\arraystretch}{.8}\begin{tabular}{@{}c@{}}\gpus \end{tabular}
	&  \renewcommand{\arraystretch}{.8}\begin{tabular}{@{}c@{}} Statement \\ sequence\end{tabular}
	&  \renewcommand{\arraystretch}{.8}\begin{tabular}{@{}c@{}}Memory graph after \\ the stmt. sequence \end{tabular}
	&  \renewcommand{\arraystretch}{.8}\begin{tabular}{@{}c@{}}\gpus \end{tabular}
	\\ \hline\hline
\multicolumn{3}{|c||}{\rule[-.55em]{0em}{1.5em}$i < k$}
	& \multicolumn{3}{c|}{\rule[-.55em]{0em}{1.5em}$j < k$}
	\\ \hline
$
\begin{array}{@{}r@{\ }c@{\ }l@{}}
\multicolumn{3}{@{}l@{}}{\rule[-.67em]{0em}{1em}\psframebox{\text{Ex. \ssa}}}
	\\ 
%\rule{0em}{1em}
\tt*x &\tt = &\tt \& y \\
\tt x  &\tt = &\tt \&z
\end{array}
$
	& \figOneOne
	&
		\begin{tabular}{@{}r@{\ }l@{}}
		 \prevedge: & \de{x}{2|0}{y}
			\\
		 \candedge: & \de{x}{1|0}{z}
			\\
		   & (\invalid)
		\end{tabular}
&
$
\begin{array}{@{}r@{\ }c@{\ }l@{}}
		\multicolumn{3}{@{}l@{}}{\rule[-.67em]{0em}{1em}\psframebox{\text{Ex. \tsa}}}
		\\ 
%\rule{0em}{1em}
\tt*x &\tt = &\tt \& y \\
\tt z  &\tt = &\tt x
\end{array}
$
	& \figThreeOne
	&
		\begin{tabular}{@{}r@{\ }l@{}}
		 \prevedge: & \de{x}{2|0}{y}
			\\
		 \candedge: & \de{z}{1|1}{x}
			\\
	 	  & (\invalid)
		\end{tabular}
	
\\ \hline
\multicolumn{3}{|c||}{\rule[-.55em]{0em}{1.5em}$i > k$}
	& \multicolumn{3}{c|}{\rule[-.55em]{0em}{1.5em}$j > k$}
\\ \hline
		$
		\begin{array}{@{}r@{\ }c@{\ }l@{}}
		\multicolumn{3}{@{}l@{}}{\rule[-.67em]{0em}{1em}\psframebox{\text{Ex. \ssb}}}
		\\ 
%\rule{0em}{1em}
	\tt	x &\tt = &\tt \& y \\
	\tt	*x  &\tt = &\tt \&z
		\end{array}
		$
	& \figOneTwo
	&
		\begin{tabular}{@{}r@{\ }l@{}}
		 \prevedge: & \de{x}{1|0}{y}
			\\
		 \candedge: & \de{x}{2|0}{z}
			\\
		\rededge: & \de{y}{1|0}{z}
		\end{tabular}
	&
		$
		\begin{array}{@{}r@{\ }c@{\ }l@{}}
		\multicolumn{3}{@{}l@{}}{\rule[-.67em]{0em}{1em}\psframebox{\text{Ex. \tsb}}}
		\\ 
%\rule{0em}{1em}
	\tt	x &\tt = &\tt \& y \\
	\tt	z  &\tt = &\tt * x
		\end{array}
		$
	&  \figThreeTwo
	&
		\begin{tabular}{@{}r@{\ }l@{}}
		 \prevedge: & \de{x}{1|0}{y}
			\\
		 \candedge: & \de{z}{1|2}{x}
			\\
		 \rededge: & \de{z}{1|1}{y}
		\end{tabular}
\\ \hline
\multicolumn{3}{|c||}{\rule[-.55em]{0em}{1.5em}$i = k$}
	& \multicolumn{3}{c|}{\rule[-.55em]{0em}{1.5em}$j = k$}
\\ \hline
		$
		\begin{array}{@{}r@{\ }c@{\ }l@{}}
		\multicolumn{3}{@{}l@{}}{\rule[-.67em]{0em}{1em}\psframebox{\text{Ex. \ssc}}}
		\\ 
%\rule{0em}{1em}
	\tt	*x &\tt = &\tt \& y \\
	\tt	*x  &\tt = &\tt \&z
		\end{array}
		$
	& \figOneThree
	&
		\begin{tabular}{@{}r@{\ }l@{}}
		 \prevedge: & \de{x}{2|0}{y}
			\\
		 \candedge: & \de{x}{2|0}{z}
			\\
		  & (\invalid)
		\end{tabular}
	&
		$
		\begin{array}{@{}r@{\ }c@{\ }l@{}}
		\multicolumn{3}{@{}l@{}}{\rule[-.67em]{0em}{1em}\psframebox{\text{Ex. \tsc}}}
		\\ 
%\rule{0em}{1em}
	\tt	x &\tt = &\tt \& y \\
	\tt	z  &\tt = &\tt x
		\end{array}
		$
	& \figThreeThree
	&
		\begin{tabular}{@{}r@{\ }l@{}}
		 \prevedge: & \de{x}{1|0}{y}
			\\
		 \candedge: & \de{z}{1|1}{x}
			\\
		 \rededge: & \de{z}{1|0}{y}
		\end{tabular}
	\\ \hline
\end{tabular}
\caption{Illustrating the \svalidity of \ssscomp and \stscomp compositions based on the \sindlevs of
pivot ($x$ in these examples) in the consumer \gpu \candedge and producer \gpu \prevedge. 
%%\change{}{Memory graph edges that are likely to be killed are shown by dashed edges.}
}
\label{fig:edge.composition-new}
\end{figure}

\subsection{\gpu Composition}
\label{sec:edge.composition}

We define \gpu composition as a family of partial operations. These operations 
simplify a consumer \gpu $\candedge$ 
using a producer \gpu \prevedge and compute a semantically 
equivalent \gpu. 

\subsubsection{The Intuition Behind \gpu Composition}
\label{subsec:what_edge_comp}

The composition of a consumer \gpu \candedge and a producer \gpu \prevedge, denoted \text{$\candedge \,\ecompwt \prevedge$}, computes
a resulting \gpu \rededge by simplifying \candedge using \prevedge. This is possible when 
\candedge has a \RaW dependence on 
\prevedge through a common variable 
called the \emph{pivot} of composition. This requires the pivot to be the source of \prevedge
but it could be the source or the target of \candedge.

We name the compositions as \tscomp or \sscomp
where the first letter indicates the role of the pivot in \candedge and second letter
indicates its role in \prevedge.
If the pivot is the target of \candedge and the source of \prevedge, the composition is called a \tscomp composition.
If the pivot is the source of both \candedge and \prevedge, the composition is
called an \sscomp composition. 
We remark for completeness that there are two further
\gpu-composition operations which can be applied when
the pivot is the target of \prevedge.
These are called \stcomp and \ttcomp compositions 
which %%and correspond to output dependence. They 
are optional and we do not use them here.
However, \tscomp and \sscomp compositions are sufficient
to convert a \gpu to a classical points-to edge.
%For more details see Appendix~\ref{app:st.tt.comp}. 
%\AMcomment{I think the \emph{paper} reference should say ``See the first author's PhD thesis for details''!
%But by all means keep the Appendix, labelled as ``PhD thesis material only, not part of the paper'',
%until we are ready to submit the paper to Toplas.}

Figure~\ref{fig:picture.edge.composition} illustrates \tscomp and \sscomp compositions. For \tscomp composition,
consider \gpus \text{\candedge $\!:\!$ \denew{z}{i|j}{x}{\slab}} and \text{\prevedge $\!:\!$ \denew{x}{k|l}{y}{\flab}} with a
pivot $x$ which is the target of \candedge and the source of \prevedge.  The goal of \gpu composition is to join the source $z$ 
of \candedge and the target $y$ of \prevedge by using the pivot $x$ as a bridge.
This requires the \indlevs of $x$ to be made the same in the two \gpus.
For example, if \text{$j \geq k$} (other cases are explained later in the section),
this can be achieved by adding \text{$j-k$}
% (\text{$j \geq k$})
to the 
\indlevs of the source and target of \prevedge 
to view the base \gpu \prevedge{} in its derived form as \de{x}{j|(l+j-k)}{y}. This balances the 
\indlevs of $x$ in the two \gpus allowing us to create a simplified \gpu
$\rededge \!:\! \de{z}{i|(l+j-k)}{y}$. (Given a \gpu \denew{x}{i|j}{y}{\flab}, we can create a \gpu \denew{x}{(i+1)|(j+1)}{y}{\flab}
based on the type restrictions on the \indlevs of $x$ and $y$.)

%%{\revOne CHECK: SHOULD WE KEEP THIS AND THE APPENDIX?
%%Although this computes the transitive effect of \gpus, in general, it cannot be modelled using multiplication of matrices
%%representing graphs as explained in Appendix~\ref{sec:why.not.matrix.mult}.
%%}

\subsubsection{Defining \gpu Composition}
\label{sec:edge.comp.properties}

Before we define the \gpu composition formally, we need to establish the properties of \validity and \desirability
that allow us to characterize meaningful \gpu compositions. We say that 
a \gpu composition 
is \admissible if and only if it is \valid and \desirable.
\begin{enumerate}[(a)]
\item A composition \text{$\rededge = \candedge \,\ecompwt \prevedge$} is \valid only if \candedge reads a location defined by \prevedge
	and this read/write happens through the pivot of the composition.
	
\item A composition \text{$\rededge = \candedge \,\ecompwt \prevedge$} is \desirable only if the \indlev of \rededge does not exceed the \indlev of \candedge.
\end{enumerate}

\label{sec:relevant.edge.composition}

\Validity
requires the \indlev of the pivot in \candedge to be greater than 
the \indlev of pivot in \prevedge. For the generic \indlevs used in Figure~\ref{fig:picture.edge.composition}, this
requirement translates to the following constraints:
\begin{align}
j \geq k & & (\tscomp \text{ composition})
		\label{constraint:tscomp.relevance}
	\\
i > k & & (\sscomp \text{ composition})
		\label{constraint:sscomp.relevance}
\end{align}
Observe that \sscomp composition condition (\ref{constraint:sscomp.relevance}) prohibits equality unlike 
the condition for \tscomp composition (\ref{constraint:tscomp.relevance}). This is because of the fact that
\sscomp composition involves the source nodes of both the \gpus and when \text{$i=k$}, 
\candedge overwrites the location written by \prevedge; for a location written by \prevedge to be read by \candedge
in its source, $i$ must be strictly greater than $k$.

\begin{example}{}
	The following (attempted) compositions in Figure~\ref{fig:edge.composition-new}
	are \invalid because \candedge does not read a location defined by \prevedge.
	\begin{itemize}
	\item  In example \ssa (\sscomp composition), $k = 2$ and $i=1$
	        violating Constraint~(\ref{constraint:sscomp.relevance}). \gpu \candedge 
		redefines $x$ instead of reading a location defined by \prevedge.
	\item  In example \ssc (\sscomp composition), \text{$k = i = 2$}
	        violating Constraint~(\ref{constraint:sscomp.relevance}). \gpu \candedge 
		redefines $*x$ instead of reading a location defined by \prevedge.
	\item   In example \tsa (\tscomp composition), \text{$k = 2$} and \text{$j = 1$}
	        violating Constraint~(\ref{constraint:tscomp.relevance}). 
		\gpu
		\candedge reads $x$ instead of reading $*x$ defined by \prevedge.
                In other words, there is no data dependence between \candedge and \prevedge which is
  		evident from the fact that the order of the statements can be changed and yet the
		meaning of the program remains same.
	\end{itemize}
	The following compositions in Figure~\ref{fig:edge.composition-new}
	are \valid because \candedge reads a location defined by \prevedge.
	\begin{itemize}
	\item  In example \ssb (\sscomp composition), $k = 1$ and $i = 2$ satisfies 
			Constraint~(\ref{constraint:sscomp.relevance}).
	\item  In example \tsb (\tscomp composition), $k = 1$ and $j = 2$ satisfies 
			Constraint~(\ref{constraint:tscomp.relevance}).
	\item  In example \tsc (\tscomp composition), $k = 1$ and $j = 1$ satisfies 
			Constraint~(\ref{constraint:tscomp.relevance}).
	\end{itemize}
\end{example}

\begin{Definition}
\begin{center}
\psframebox[framesep=0pt,doubleline=true,doublesep=1.5pt,linewidth=.2mm]{%
\gpucompIndlevDef
}
\end{center}
\defcaption{\gpu Composition $\candedge \ecompwt \prevedge$}{def:gpu.composition.indlev}
\end{Definition}

\label{sec:relevant.useful.ec}
\label{sec:desirable.edge.composition}
The \desirability of \gpu composition characterizes progress in conversion of \gpus 
into classical points-to
edges
by ensuring that the \indlev of the new source  and the new target in \rededge
      does not exceed the corresponding \indlev in the consumer \gpu \candedge. 
This requires
the \indlev in the simplified \gpu \rededge and the consumer \gpu \candedge to satisfy the following constraints.
In each constraint, the first term in the conjunct compares the \indlevs of the sources of \candedge and \rededge
while the second term compares those of the targets
 (see
Figure~\ref{fig:picture.edge.composition}):
	\begin{align}
	(i \leq i)\; \wedge\; (l+j-k \leq j)
			& \quad \mbox{or equivalently} \quad l \leq k \;\;
			& (\tscomp \text{ composition})
		\label{eq:desirability.constraint.tscomp}
		\\
	(l+i-k \leq i)\; \wedge\; (j \leq j)
			& \quad \mbox{or equivalently} \quad l \leq k \;\;
			& (\sscomp \text{ composition})
		\label{eq:desirability.constraint.sscomp}
	\end{align}

\begin{example}{}
	Consider the statement sequence \text{$x=*y; z = x$}. 
	A \tscomp composition of the corresponding 
	\gpus \text{$\prevedge:\denew{x}{1|2}{y}{}$} and \text{$\candedge:\denew{z}{1|1}{x}{}$}
	is \valid because 
	\text{$j = k = 1$} satisfying Constraint~\ref{constraint:tscomp.relevance}. However,
	if we perform this composition, we get \text{$\rededge:\denew{z}{1|2}{y}{}$}. Intuitively, this \gpu is not useful for
        computing a points-to edge because the \indlev of \rededge is ``$1|2$'' which is greater than the \indlev of
        \candedge which is ``$1|1$''.  Formally, this composition is flagged \undesirable 
	because \text{$l=2$} which is greater than \text{$k=1$} violating 
	Constraint~\ref{eq:desirability.constraint.tscomp}.
\end{example}
\label{subsec:derivation.constraint}

We take a conjunction of the constraints of \validity (\ref{constraint:tscomp.relevance} and~\ref{constraint:sscomp.relevance})
and \desirability (\ref{eq:desirability.constraint.tscomp} and~\ref{eq:desirability.constraint.sscomp})
to characterize \admissible \gpu compositions.

\begin{align}
\label{cons:ts}
	l \leq k \leq j
	&& \text{(\tscomp composition)} 
	\\
\label{cons:ss}
	l \leq k < i
		&& \text{(\sscomp composition)} 
\end{align}
\label{sec:ecomp-partial-func}

Note that an \undesirable \gpu composition in a \gpg is \valid but \inadmissible. 
It will eventually become \desirable after the producer \gpu is simplified further through strength 
reduction optimization after the \gpg is inlined in a caller's \gpg.

Definition~\ref{def:gpu.composition.indlev} defines \gpu composition formally.
 It computes a simplified \gpu 
\text{$\rededge = \candedge \; \ecompwt \prevedge$} by balancing the \indlev of the pivot in both the \gpus
provided the composition (\tscomp or \sscomp) 
is \admissible.
Otherwise it fails---being a partial operation.
Note that \tscomp and \sscomp compositions are mutually exclusive for a given pair of
\candedge and \prevedge because a variable cannot occur both in the RHS and the LHS of 
a pointer assignment in the case of pointers to scalars.\footnote{
Since our language is modelled on C, 
\gpus for
statements such as 
\text{$*x=x$} or \text{$x=*x$}
are prohibited by typing rules; 
\gpus for statements such as \text{$*x=*x$} are ignored
as inconsequential. Further, we assume as allowed by C-standard \emph{undefined behaviour}
that the programmer has not abused type-casting to simulate such prohibited statements.
Section~\ref{sec:heap} considers the richer situation with structs and unions where
we can have an assignment \text{$x\rightarrow n = x$} which might have both
\stscomp and \ssscomp compositions with a \gpu \prevedge that defines $x$.
}

\begin{Definition}
\begin{center}
\psframebox[framesep=5pt,doubleline=true,doublesep=1.5pt,linewidth=.2mm]{%
\gpuReductionDef
}
\end{center}
\defcaption{\gpu Reduction $\candedge \rcomp \flow$}{def:edge.reduction.aa}
\end{Definition}

\subsection{\gpu Reduction}
\label{sec:edge.reduction}

\gpu reduction \text{$\candedge \rcomp \flow$} 
uses the \gpus in \flow to
compute a set of \gpus \GENEDGES whose \indlevs do not exceed that of
\candedge.
The result of \gpu reduction \text{$\candedge \rcomp \flow$} must
ensure the semantic equivalence of \GENEDGES with \candedge in the context of \flow.
The set \flow is computed using reaching \gpus analysis without blocking (Section~\ref{sec:reach.gpus.analysis}).
In some cases, we need to restrict \flow using the 
reaching \gpus analysis with blocking (Section~\ref{sec:blocked.gpus.analysis}) 
to ensure this semantic equivalence.

For \text{$\candedge \rcomp \flow$},
the \indlev of \candedge is reduced progressively using the \gpus from \flow through a series of \admissible \gpu compositions.
For example, a \gpu \de{x}{1|2}{y} requires two \tscomp compositions 
to transform it into a classical points-to edge: 
first one for identifying the pointees of $y$ and second one
for identifying the pointees of pointees of $y$. Similarly, for a \gpu \de{x}{2|1}{y}, 
an \sscomp composition is
required to identify the pointees of $x$ which are being defined and 
a \tscomp composition is required to identify
the pointees of $y$ whose addresses are being assigned.
Thus, the result of \gpu reduction is a fixed-point of cascaded \gpu compositions in the context of
\flow.

\subsubsection{Defining \gpu Reduction $\candedge \rcomp \flow$}
\label{sec:define.edge.reduction}

Definition~\ref{def:edge.reduction.aa} 
gives the algorithm for \gpu reduction. The worklist \WL is initialized to $\{\candedge\}$.
A reduced \gpu is added to \WL for further \gpu compositions. When a \gpu $w$ cannot be reduced any further, the flag
\WCompose remains \FALSE and $w$ is added to 
\GENEDGES (lines 17 and 18 of Definition~\ref{def:edge.reduction.aa}).
This algorithm assumes that the graph induced by the \gpus in \flow is acyclic. This holds for
scalar pointers.
However, in the presence of structures 
the graph may contain cycles via fields of structures;
Section~\ref{sec:cycle-gpg} extends the algorithm to handle cycles.

\begin{example}{}
 Consider the statements on the right.
For $\candedge\!:\!\denew{x}{1|2}{y}{23}$, 
\text{$\flow = \{\denew{y}{1|0}{a}{21}, \denew{a}{1|0}{b}{22}\}$}.
\setlength{\intextsep}{-.4mm}%
\setlength{\columnsep}{2mm}%
\begin{wrapfigure}{r}{21.5mm}
\setlength{\codeLineLength}{8mm}%
\renewcommand{\arraystretch}{.7}%
	\begin{tabular}{|rc}
	%\hline
	\codeLineOne{21}{0}{$\tt y =  \&a;$}{white}
	\codeLine{0}{$\tt a =  \&b;$   }{white}
	\codeLine{0}{$\tt x = *y; $   }{white}
	 %\hline
	\end{tabular}
\end{wrapfigure}%
The reduction \text{$\candedge \rcomp \flow$} involves two
consecutive \tscomp compositions. 
The first composition 
involves \denew{y}{1|0}{a}{21} as \prevedge, resulting in
$\rededge = \denew{x}{1|1}{a}{23}$ which is added to the worklist.
In the second iteration of the \textbf{while} loop on line 04 of Definition~\ref{def:edge.reduction.aa}, the reduced \gpu \denew{x}{1|1}{a}{23} 
in the previous iteration
now becomes the consumer \gpu. It is composed with \denew{a}{1|0}{b}{22} which results in 
a reduced \gpu \denew{x}{1|0}{b}{23}. This \gpu is added to the worklist. However, since it cannot be reduced 
further as it is already
in the classical points-to form, the loop terminates. 
The flag \WCompose remains \FALSE for the final \gpu \denew{x}{1|0}{b}{23} because no further composition
is possible and \text{$\GENEDGES = \{\denew{x}{1|0}{b}{23}\}$}.
\end{example}

The termination of \gpu reduction is guaranteed by the following reasons:
\begin{itemize}
	\item A \gpu $w$ extracted from the worklist will never be added to it again.
              If there is no reduction, then $w$ is added to \GENEDGES directly. This is 
		ensured by setting the flag \WCompose appropriately.
	\item  Reduction of \indlev of source and target of a \gpu $w$ is performed independently, hence there is no oscillation 
		across iterations of fixed-point computation.
	\item The process terminates only when the \gpus in \GENEDGES are either in their simplified 
		form or no more \gpus are available in \flow for further \gpu compositions.
\item The order in which a \gpu \edge is selected from \flow for composition with $w$ does not matter because of 
	the following properties of \flow{} that are established by the 
	reaching \gpus analysis with and without blocking (Sections~\ref{sec:reach.gpus.analysis}
	and~\ref{sec:blocked.gpus.analysis}).
	
	Consider two \gpus $\edge_1$ and $\edge_2$ in \flow. Then $\edge_1$ and $\edge_2$ cannot compose with each other:
	If the composition $\edge_2 \ecomp \edge_1$ were possible, it would have been performed during the
        reaching \gpus analysis (Section~\ref{sec:reach.gpus.analysis}) and $\edge_2$ would not exist in \flow because 
	it would be replaced by the result of the composition. 
	Similarly if the composition $\edge_1 \ecomp \edge_2$ were possible, $\edge_1$ would not exist in \flow. Hence
	we examine the possible reasons of existence of both $\edge_1$ and $\edge_2$ in \flow and explain why
        the order of performing the 
	compositions \text{$w \ecomp \edge_1$} and \text{$w \ecomp \edge_2$} does not matter.
	\begin{enumerate}[(a)]
	\item There is no data dependence between $\edge_1$ and $\edge_2$ because there is no pivot between them
		or one does not follow the other on any control flow path.
		Hence a composition 
               between them is ruled out. In this case,
		the order between \text{$w \ecomp \edge_1$} and \text{$w \ecomp \edge_2$} 
		is irrelevant because of the
		absence of data dependence between $\edge_1$ and $\edge_2$. 
	\item There is data dependence between $\edge_1$ and $\edge_2$ potentially enabling a composition. 
		Without any loss of generality, consider the composition $\edge_2 \ecomp
		\edge_1$. Then there are two possibilities that may have prohibited the composition:
		\begin{enumerate}[(i)]
		\item $\edge_2 \ecomp \edge_1$ is \inadmissible because it is \undesirable.  Then, $w \ecomp \edge_1$ also is
	              \undesirable because the \desirability constraint is based solely on the \indlev of $\edge_1$
(Constraints~\ref{eq:desirability.constraint.tscomp} and~\ref{eq:desirability.constraint.sscomp}).
			Thus $w$ may compose only with $\edge_2$ 
			and the issue of an order between \text{$w \ecomp \edge_1$} and \text{$w \ecomp \edge_2$} does not arise. 
		\item $\edge_2 \ecomp \edge_1$ is \admissible but has been postponed 
                      because of a 
                      barrier (introduced in Section~\ref{sec:bottom.up.nomenclature} and
			explained later in Section~\ref{sec:blocked.gpus.analysis}) between
		      $\edge_2$ and $\edge_1$. 
			In this case, the barrier also prohibits a composition of $w$ with $\edge_1$ and it can compose 
			only with $\edge_2$.
			Thus the issue of an order between \text{$w \ecomp \edge_1$} and \text{$w \ecomp \edge_2$} does not arise. 
		\end{enumerate}
	\end{enumerate}

\end{itemize}

\subsubsection{Modelling Caller-Defined Pointer Variables}
%%\subsection{Extending \flow to Support Strong Updates.}
\label{sec:may.must.xxp.edges}

In abstract memory, we may be uncertain as to which of several locations a variable points to.
Hence, for an indirect assignment ($*p=\&x$ say), \gpu reduction returns a set of \gpus which
define multiple pointers (or different pointees of the same pointer)
leading to a weak update. In this case we do not
overwrite any of its pointees, but merely add $\&x$ to the possible values they can contain.
Sometimes however, we may discover that $p$ has a single pointee within the procedure
and conclude that 
there is only one possible abstract location defined
by the assignment. In this case we may, in general, \emph{replace} the contents
of this location.  This is a strong update.
However, this is necessary but not sufficient for a strong update because the pointer may not be defined along 
all paths---there may be a path along which the pointer (or some pointee of the pointer) may not be defined
within the procedure but may be defined in a caller.
In the presence of such a definition-free path in a procedure, even if we find a single pointee of $p$ in the procedure, 
we cannot guarantee that a single abstract location is being defined. This 
makes it difficult to distinguish between strong and weak updates. %% as illustrated in Figure~\ref{fig:asummflow.s.update}.
%%\AMcomment{See my discussion of `?' earlier}
Also, the effect of definition-free paths has to be taken into account during strength reduction optimization:
if $\edge_1$ is simplified to $\edge_2$, $\edge_2$ can replace $\edge_1$ provided
there is no definition-free path reaching $\edge_1$; otherwise
$\edge_1$ should also be included with $\edge_2$ to allow the composition of $\edge_1$
with the \gpus in a caller.

\begin{example}{}
Figure~\ref{fig:mot_eg.gpg.g} shows the set of \gpus corresponding to statement 02 
($\gpbsym_{02}$ in the \gpg after strength reduction) of procedure $g$ of Figure~\ref{fig:mot_eg.cfgs}.
There is a definition-free path for $q$
meaning that $\gpbsym_{11}$ in the optimized $\mtsym_g$
must include \gpu
\denew{q}{2|0}{m}{02} along with its reduced \gpu \denew{b}{1|0}{m}{02}.
\end{example}

We identify definition-free paths by introducing \emph{boundary definitions} (explained below)
which also help us to preserve definition-free paths that may be eliminated by coalescing.

The boundary definitions 
are introduced for
global variables and formal parameters because they could be read in a procedure before being defined.
They
are symbolic in that they are not introduced in the \gpg of a procedure but are included in 
\RIn{} of the \Start{} \gpb during reaching \gpus analysis.
They 
are of the form
\denew{x}{\ell|\ell}{x'}{00} where $x'$ is a symbolic representation of the initial value of $x$ at the start of the procedure 
and $\ell$ ranges from 1 to the maximum depth of 
the indirection level which depends on the type of $x$, and
00 is the label of the \Start{} \gpb.
For type \text{(int $**$)}, $\ell$ ranges from 1 to 2.
Variable version $x'$ is called the \emph{upwards exposed}~\cite{dfa_book} version of $x$. 
This is similar to Hoare-logic style specifications in which postconditions use (immutable) \emph{auxiliary variables} $x'$
to be able to talk about the original value of variable $x$ (which may have since changed).
Our upwards-exposed versions serve a similar purpose, so that logically on entry to each procedure
the statement $x = x'$ provides a definition of $x$.

A reduced \gpu \denew{x}{i|j}{y}{\flab} along any path
kills the boundary definition $\denew{x}{i|i}{x'}{00}$ 
on that path
indicating that ${(i-1)}^{th}$ pointees of $x$ are redefined.
%%The inclusion of boundary definitions 
%%guarantees that an assignment defines a single pointer (or some pointee of a pointer)
%%only when the source is defined along every path thereby eliminating the dashed path in Figure~\ref{fig:asummflow.s.update}. 
Including boundary definitions at the start ensures that if a boundary definition 
$\denew{x}{i|i}{x'}{00}$ reaches a program point \flab, there is a definition-free path from \Start{} to \flab; 
its absence at \flab guarantees that the source of 
$\denew{x}{i|i}{x'}{00}$ has been defined along all paths reaching \flab.
This leads to a simple necessary and sufficient condition for strong updates: All \gpus 
corresponding to a statement \flab must define the same location.

The boundary definitions also participate in \gpu compositions thereby modelling the semantics of definition-free paths.
They enable strong updates thereby improving the precision of analysis.

\begin{example}{exmp:motivating.exmp.rgpu.analysis}
Consider reaching \gpus analysis for the \gpb corresponding to statement 02 
in the initial \gpg of procedure $g$ ($\gpbsym_{02}$ in Figure~\ref{fig:mot_eg.gpg.g}). 
We include the boundary definitions for each global variable and the parameters of a procedure 
as \RIn{} of the \Start{} \gpb of the \gpg of procedure $g$. 
Although Figure~\ref{fig:mot_eg.gpg.g} does not show
boundary definitions for simplicity, they are shown in Figure~\ref{fig:mot_eg.rpgs.g} for variable
$q$ (boundary definitions of other variables are not required for strong updates in this example). 
These boundary definitions capture the effect of definition-free paths
to distinguish between weak and strong updates.

The \gpu \text{$\edge_2 \!:\! \denew{q}{2|0}{m}{02}$} is composed with \gpus from \RIn{02} 
which contains a \gpu \denew{q}{1|0}{b}{03}  indicating that pointer $b$ is being defined by statement 02. 
However, this is not the case of strong update as $b$ is not the only pointer that is being defined by the assignment. 
There is a definition-free path along which pointee of $q$ is not available indicating that 
$q$ may have a definition the callers of procedure $g$ which is also required in statement 02 of $g$
but is currently unavailable. The presence of boundary definition 
\denew{q}{1|1}{q'}{00} in \RIn{02} 
indicates the presence of a definition-free path and the composition of this \gpu results in a reduced \gpu \denew{q'}{2|0}{m}{02} which is 
also a part of $\gpbsym_{02}$. 
The \gpu \denew{q'}{2|0}{m}{02} has been represented by the \gpu \denew{q}{2|0}{m}{02} in 
 Figure~\ref{fig:mot_eg.gpg.g} because it ignores boundary definitions.

At the call site in procedure $f$,
after the composition of \gpu \denew{q}{1|0}{d}{07} and \denew{q}{2|0}{m}{02} (the upwards-exposed version $q'$ is 
replaced by $q$ during call inlining; for more details see Section~\ref{sec:interprocedural.extensions}), 
the set of reduced \gpus corresponding to statement 02 in procedure $f$ (\gpb $\gpbsym_{13}$) 
contains two \gpus \denew{b}{1|0}{m}{02} and \denew{d}{1|0}{m}{02} (Figure~\ref{fig:mot_eg.2}).
Since, the assignment defines two pointers $d$ and $b$, no \gpu is removed 
and hence the \gpu \denew{d}{1|0}{n}{08} in \gpb $\gpbsym_{08}$ is retained owing to a weak update.
\end{example}

An important observation is that boundary definitions only appear in 
\RIn{} and \ROut{} of the reaching-\gpus analysis---they never appear
in the \gpbs or in \RGen{},
although the upwards-exposed versions of variables could be involved in the \gpus in \RGen{}.
Also, the algorithm for \gpu reduction does not change with the introduction of boundary definitions because a \gpu can be composed 
with boundary definitions just like with any other \gpus.

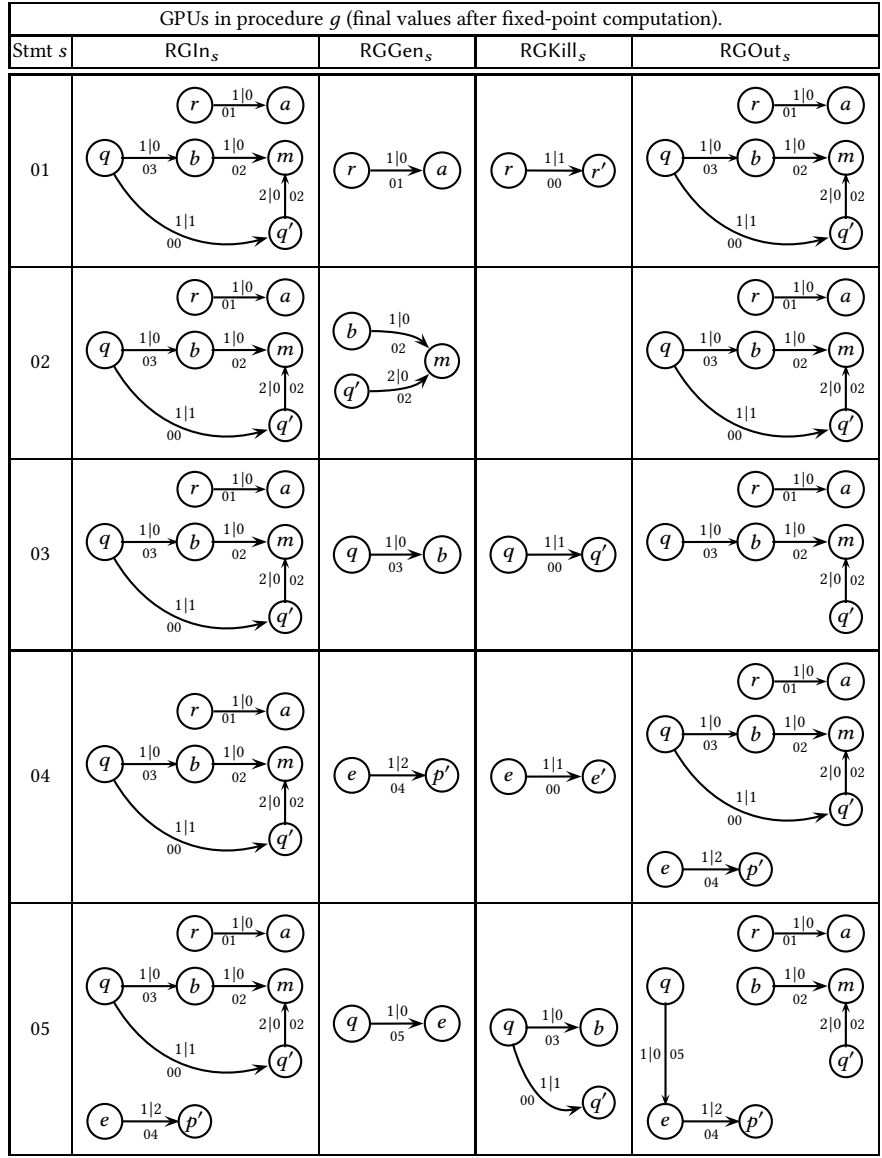
\begin{figure}[t]
%%\centering
\small
%%%%%%%%%%%%%%%%%%%%%%%%% Definitions of graphs %%%%%%%%%%%%%%%%%%%%%%%5
	\newcommand{\gpgInTwo}{%
		\begin{pspicture}(0,1)(32,8)
		%\psframe(0,0)(32,8)
		\putnode{a1}{origin}{16}{4}{\pscirclebox[fillstyle=solid,fillcolor=white,framesep=1.22]{$a$}}
		\putnode{b1}{a1}{0}{-7}{}

		\putnode{m1}{a1}{12}{0}{\pscirclebox[fillstyle=solid,fillcolor=white,framesep=.72]{$m$}}

		%%%%%%%%%%%%%%%%%%%%%%%%%%%%%
		\nccurve[arrowsize=1.5,angleA=0,angleB=180,ncurv=1]{->}{a1}{m1}
		\naput[labelsep=.25,npos=.5]{\scriptsize $1|0$}
		\nbput[labelsep=.75,npos=.5]{\scriptsize 01}
		\end{pspicture}
	}
	\newcommand{\gpgGenOne}{%
		\begin{pspicture}(0,1)(20,8)
		%\psframe(0,0)(20,8)
		\putnode{a1}{origin}{4}{4}{\pscirclebox[fillstyle=solid,fillcolor=white,framesep=1.22]{$r$}}
		\putnode{b1}{a1}{0}{-7}{}

		\putnode{m1}{a1}{12}{0}{\pscirclebox[fillstyle=solid,fillcolor=white,framesep=1.22]{$a$}}

		%%%%%%%%%%%%%%%%%%%%%%%%%%%%%
		\nccurve[arrowsize=1.5,angleA=0,angleB=180,ncurv=1]{->}{a1}{m1}
		\naput[labelsep=.25,npos=.5]{\scriptsize $1|0$}
		\nbput[labelsep=.75,npos=.5]{\scriptsize 01}
		\end{pspicture}
	}
	\newcommand{\gpgGenTwo}{%
		\begin{pspicture}(0,1)(20,8)
		%\psframe(0,0)(20,8)
		\putnode{z1}{origin}{4}{4}{\pscirclebox[fillstyle=solid,fillcolor=white,framesep=1]{$q$}}
		\putnode{b1}{z1}{12}{0}{\pscirclebox[fillstyle=solid,fillcolor=white,framesep=1]{$b$}}

		%
		%%%%%%%%%%%%%%%%%%%%%%%%%%%%%
		\nccurve[arrowsize=1.5,angleA=0,angleB=180,nodesepA=-.3]{->}{z1}{b1}
		\naput[labelsep=.25,npos=.5]{\scriptsize $1|0$}
		\nbput[labelsep=.75,npos=.5]{\scriptsize 03}
		\end{pspicture}
	}
	\newcommand{\gpgInThree}{%
		\begin{pspicture}(0,1)(32,15)
		%\psframe(0,0)(32,15)
		\putnode{a1}{origin}{16}{11}{\pscirclebox[fillstyle=solid,fillcolor=white,framesep=1.22]{$a$}}
		\putnode{b1}{a1}{0}{-7}{\pscirclebox[fillstyle=solid,fillcolor=white,framesep=1]{$b$}}

		\putnode{m1}{a1}{12}{0}{\pscirclebox[fillstyle=solid,fillcolor=white,framesep=.72]{$m$}}

		\putnode{z1}{b1}{-12}{0}{\pscirclebox[fillstyle=solid,fillcolor=white,framesep=1]{$q$}}
		%%%%%%%%%%%%%%%%%%%%%%%%%%%%%
		\nccurve[arrowsize=1.5,angleA=0,angleB=180,ncurv=1]{->}{a1}{m1}
		\naput[labelsep=.25,npos=.5]{\scriptsize $1|0$}
		\nbput[labelsep=.75,npos=.5]{\scriptsize 01}
		\nccurve[arrowsize=1.5,angleA=0,angleB=180,nodesepA=-.3]{->}{z1}{b1}
		\naput[labelsep=.25,npos=.5]{\scriptsize $1|0$}
		\nbput[labelsep=.75,npos=.5]{\scriptsize 03}
		\end{pspicture}
	}
	\newcommand{\gpfThree}{%
		\begin{pspicture}(0,1)(20,8)
		%\psframe(0,0)(20,8)
		\putnode{z1}{origin}{4}{4}{\pscirclebox[fillstyle=solid,fillcolor=white,framesep=1]{$q$}}
		\putnode{m1}{z1}{12}{0}{\pscirclebox[fillstyle=solid,fillcolor=white,framesep=.92]{$a$}}

		%
		%%%%%%%%%%%%%%%%%%%%%%%%%%%%%
		\nccurve[arrowsize=1.5,angleA=0,angleB=180,ncurv=1]{->}{z1}{m1}
		\naput[labelsep=.25,npos=.5]{\scriptsize $2|1$}
		\nbput[labelsep=.75,npos=.5]{\scriptsize 02}
		\end{pspicture}
	}
	\newcommand{\gpgGenThree}{%
		\begin{pspicture}(0,1)(20,14)
		%\psframe(0,0)(20,14)
		\putnode{b1}{origin}{4}{11}{\pscirclebox[fillstyle=solid,fillcolor=white,framesep=1]{$b$}}

		\putnode{m1}{b1}{12}{-4}{\pscirclebox[fillstyle=solid,fillcolor=white,framesep=.72]{$m$}}

		\putnode{z1}{b1}{0}{-8}{\pscirclebox[fillstyle=solid,fillcolor=white,framesep=.25]{$q'$}}
		%%%%%%%%%%%%%%%%%%%%%%%%%%%%%
		\nccurve[arrowsize=1.5,angleA=0,angleB=225,nodesepB=-.7]{->}{z1}{m1}
		\naput[labelsep=0,npos=.5]{\scriptsize $2|0$}
		\nbput[labelsep=.2,npos=.5]{\scriptsize 02}
		\nccurve[arrowsize=1.5,angleA=0,angleB=135,nodesepB=-.7]{->}{b1}{m1}
		\naput[labelsep=.25,npos=.4]{\scriptsize $1|0$}
		\nbput[labelsep=.75,npos=.5]{\scriptsize 02}
		\end{pspicture}
	}
	\newcommand{\gpgInFour}{%
		\begin{pspicture}(0,-3)(32,21)
		%\psframe(0,0)(32,21)
		\putnode{a1}{origin}{16}{17}{\pscirclebox[fillstyle=solid,fillcolor=white,framesep=1.22]{$r$}}
		\putnode{a2}{a1}{12}{0}{\pscirclebox[fillstyle=solid,fillcolor=white,framesep=1.22]{$a$}}
		\putnode{b1}{a1}{0}{-7}{\pscirclebox[fillstyle=solid,fillcolor=white,framesep=1]{$b$}}

		\putnode{m1}{b1}{12}{0}{\pscirclebox[fillstyle=solid,fillcolor=white,framesep=.72]{$m$}}

		\putnode{z1}{b1}{-12}{0}{\pscirclebox[fillstyle=solid,fillcolor=white,framesep=1]{$q$}}
		\putnode{sp}{m1}{0}{-10}{\pscirclebox[fillstyle=solid,fillcolor=white,framesep=.15]{$q'$}}
		%%%%%%%%%%%%%%%%%%%%%%%%%%%%%
		\ncline{->}{a1}{a2}
		\naput[labelsep=0,npos=.55]{\scriptsize $1|0$}
		\nbput[labelsep=0,npos=.3]{\scriptsize 01}
		%%
%%		\nccurve[arrowsize=1.5,angleA=0,angleB=180,nodesepA=-.3]{->}{sp}{b1}
		\ncline[nodesepA=-.4,nodesepB=-.4]{->}{z1}{b1}
		\naput[labelsep=0.2,npos=.5]{\scriptsize $1|0$}
		\nbput[labelsep=.5,npos=.5]{\scriptsize 03}
		\nccurve[arrowsize=1.5,angleA=-60,angleB=195,nodesepA=-.5,nodesepB=0,ncurv=.8]{->}{z1}{sp}
		\naput[labelsep=0,npos=.5]{\scriptsize $1|1$}
		\nbput[labelsep=.2,npos=.5]{\scriptsize 00}
		\ncline[nodesepA=-.4,nodesepB=-.4]{->}{sp}{m1}
		\naput[labelsep=0.3,npos=.5]{\scriptsize $2|0$}
		\nbput[labelsep=.3,npos=.5]{\scriptsize 02}
		\nccurve[arrowsize=1.5,angleA=0,angleB=180,nodesepA=-.4]{->}{b1}{m1}
		\naput[labelsep=.25,npos=.4]{\scriptsize $1|0$}
		\nbput[labelsep=.75,npos=.5]{\scriptsize 02}
		\end{pspicture}
	}
	\newcommand{\gpgInSix}{%
		\begin{pspicture}(0,-3)(32,21)
		%\psframe(0,0)(32,21)
		\putnode{a1}{origin}{16}{17}{\pscirclebox[fillstyle=solid,fillcolor=white,framesep=1.22]{$r$}}
		\putnode{a2}{a1}{12}{0}{\pscirclebox[fillstyle=solid,fillcolor=white,framesep=1.22]{$a$}}
		\putnode{b1}{a1}{0}{-7}{\pscirclebox[fillstyle=solid,fillcolor=white,framesep=1]{$b$}}

		\putnode{m1}{b1}{12}{0}{\pscirclebox[fillstyle=solid,fillcolor=white,framesep=.72]{$m$}}

		\putnode{z1}{b1}{-12}{0}{\pscirclebox[fillstyle=solid,fillcolor=white,framesep=1]{$q$}}
		\putnode{sp}{m1}{0}{-10}{\pscirclebox[fillstyle=solid,fillcolor=white,framesep=.15]{$q'$}}
		%%%%%%%%%%%%%%%%%%%%%%%%%%%%%
		\ncline{->}{a1}{a2}
		\naput[labelsep=0,npos=.55]{\scriptsize $1|0$}
		\nbput[labelsep=0,npos=.3]{\scriptsize 01}
		%%
%%		\nccurve[arrowsize=1.5,angleA=0,angleB=180,nodesepA=-.3]{->}{sp}{b1}
		\ncline[nodesepA=-.4,nodesepB=-.4]{->}{z1}{b1}
		\naput[labelsep=0.2,npos=.5]{\scriptsize $1|0$}
		\nbput[labelsep=.5,npos=.5]{\scriptsize 03}
		\ncline[nodesepA=-.4,nodesepB=-.4]{->}{sp}{m1}
		\naput[labelsep=0.3,npos=.5]{\scriptsize $2|0$}
		\nbput[labelsep=.3,npos=.5]{\scriptsize 02}
		\nccurve[arrowsize=1.5,angleA=0,angleB=180,nodesepA=-.4]{->}{b1}{m1}
		\naput[labelsep=.25,npos=.4]{\scriptsize $1|0$}
		\nbput[labelsep=.75,npos=.5]{\scriptsize 02}
		\end{pspicture}
	}
	\newcommand{\gpgGenFour}{%
		\begin{pspicture}(0,1)(20,8)
		%\psframe(0,0)(20,8)
		\putnode{x1}{origin}{4}{4}{\pscirclebox[fillstyle=solid,fillcolor=white,framesep=1.2]{$e$}}

		\putnode{y1}{x1}{12}{0}{\pscirclebox[fillstyle=solid,fillcolor=white,framesep=.2]{$p'$}}
		%%%%%%%%%%%%%%%%%%%%%%%%%%%%%
		\nccurve[arrowsize=1.5,angleA=0,angleB=180,nodesepA=-.2,nodesepB=-.3]{->}{x1}{y1}
		\naput[labelsep=.25,npos=.5]{\scriptsize $1|2$}
		\nbput[labelsep=.75,npos=.5]{\scriptsize 04}
		\end{pspicture}
	}
	\newcommand{\gpgKillOne}{%
		\begin{pspicture}(0,1)(20,8)
		%\psframe(0,0)(20,8)
		\putnode{x1}{origin}{4}{4}{\pscirclebox[fillstyle=solid,fillcolor=white,framesep=1.2]{$r$}}

		\putnode{y1}{x1}{12}{0}{\pscirclebox[fillstyle=solid,fillcolor=white,framesep=.45]{$r'$}}
		%%%%%%%%%%%%%%%%%%%%%%%%%%%%%
		\nccurve[arrowsize=1.5,angleA=0,angleB=180,nodesepA=-.2,nodesepB=-.3]{->}{x1}{y1}
		\naput[labelsep=.25,npos=.5]{\scriptsize $1|1$}
		\nbput[labelsep=.75,npos=.5]{\scriptsize 00}
		\end{pspicture}
	}
	\newcommand{\gpgKillThree}{%
		\begin{pspicture}(0,1)(20,8)
		%\psframe(0,0)(20,8)
		\putnode{x1}{origin}{4}{4}{\pscirclebox[fillstyle=solid,fillcolor=white,framesep=1]{$q$}}

		\putnode{y1}{x1}{12}{0}{\pscirclebox[fillstyle=solid,fillcolor=white,framesep=.25]{$q'$}}
		%%%%%%%%%%%%%%%%%%%%%%%%%%%%%
		\nccurve[arrowsize=1.5,angleA=0,angleB=180,nodesepA=-.2,nodesepB=-.3]{->}{x1}{y1}
		\naput[labelsep=.25,npos=.5]{\scriptsize $1|1$}
		\nbput[labelsep=.75,npos=.5]{\scriptsize 00}
		\end{pspicture}
	}

	\newcommand{\gpgInFive}{%
		\begin{pspicture}(0,-3)(32,29)
		%\psframe(0,0)(32,29)
		\putnode{a1}{origin}{16}{25}{\pscirclebox[fillstyle=solid,fillcolor=white,framesep=1.22]{$r$}}
		\putnode{a2}{a1}{12}{0}{\pscirclebox[fillstyle=solid,fillcolor=white,framesep=1.22]{$a$}}
		\putnode{b1}{a1}{0}{-7}{\pscirclebox[fillstyle=solid,fillcolor=white,framesep=1]{$b$}}
		\putnode{x1}{b1}{-12}{-18}{\pscirclebox[fillstyle=solid,fillcolor=white,framesep=1.2]{$e$}}

		\putnode{m1}{b1}{12}{0}{\pscirclebox[fillstyle=solid,fillcolor=white,framesep=.72]{$m$}}

		\putnode{z1}{b1}{-12}{0}{\pscirclebox[fillstyle=solid,fillcolor=white,framesep=1]{$q$}}
		\putnode{sp}{m1}{0}{-10}{\pscirclebox[fillstyle=solid,fillcolor=white,framesep=.15]{$q'$}}
		\putnode{y1}{x1}{12}{0}{\pscirclebox[fillstyle=solid,fillcolor=white,framesep=.2]{$p'$}}
		%%%%%%%%%%%%%%%%%%%%%%%%%%%%%
		\ncline{->}{a1}{a2}
		\naput[labelsep=0,npos=.55]{\scriptsize $1|0$}
		\nbput[labelsep=0,npos=.3]{\scriptsize 01}
		%%
%%		\nccurve[arrowsize=1.5,angleA=0,angleB=180,nodesepA=-.3]{->}{z1}{b1}
		\ncline[nodesepA=-.4,nodesepB=-.4]{->}{z1}{b1}
		\naput[labelsep=0.2,npos=.5]{\scriptsize $1|0$}
		\nbput[labelsep=.5,npos=.5]{\scriptsize 03}
		\ncline[nodesepA=-.4,nodesepB=-.4]{->}{sp}{m1}
		\naput[labelsep=0.3,npos=.5]{\scriptsize $2|0$}
		\nbput[labelsep=.3,npos=.5]{\scriptsize 02}
		\nccurve[arrowsize=1.5,angleA=0,angleB=180,nodesepA=-.4]{->}{b1}{m1}
		\naput[labelsep=.25,npos=.4]{\scriptsize $1|0$}
		\nbput[labelsep=.75,npos=.5]{\scriptsize 02}
		\nccurve[arrowsize=1.5,angleA=0,angleB=180,nodesepA=-.2,nodesepB=-.3]{->}{x1}{y1}
		\naput[labelsep=.25,npos=.5]{\scriptsize $1|2$}
		\nbput[labelsep=.75,npos=.5]{\scriptsize 04}
		\nccurve[arrowsize=1.5,angleA=-60,angleB=195,nodesepA=-.5,nodesepB=0,ncurv=.8]{->}{z1}{sp}
		\naput[labelsep=0,npos=.5]{\scriptsize $1|1$}
		\nbput[labelsep=.2,npos=.5]{\scriptsize 00}
		\end{pspicture}
	}
	\newcommand{\gpgGenFive}{%
		\begin{pspicture}(0,0)(20,8)
		%\psframe(0,0)(20,8)
		\putnode{z1}{origin}{4}{4}{\pscirclebox[fillstyle=solid,fillcolor=white,framesep=1]{$q$}}
		\putnode{x1}{z1}{12}{0}{\pscirclebox[fillstyle=solid,fillcolor=white,framesep=1.2]{$e$}}

		%
		%%%%%%%%%%%%%%%%%%%%%%%%%%%%%
		\nccurve[arrowsize=1.5,angleA=0,angleB=180]{->}{z1}{x1}
		\naput[labelsep=.25,npos=.5]{\scriptsize $1|0$}
		\nbput[labelsep=.75,npos=.5]{\scriptsize 05}
		\end{pspicture}
	}
	\newcommand{\gpgKillFour}{%
		\begin{pspicture}(0,1)(20,8)
		%\psframe(0,0)(20,8)
		\putnode{z1}{origin}{4}{4}{\pscirclebox[fillstyle=solid,fillcolor=white,framesep=1.2]{$e$}}
		\putnode{b1}{z1}{12}{0}{\pscirclebox[fillstyle=solid,fillcolor=white,framesep=.45]{$e'$}}

		%
		%%%%%%%%%%%%%%%%%%%%%%%%%%%%%
		\nccurve[arrowsize=1.5,angleA=0,angleB=180,nodesepA=-.3]{->}{z1}{b1}
		\naput[labelsep=.25,npos=.5]{\scriptsize $1|1$}
		\nbput[labelsep=.75,npos=.5]{\scriptsize 00}
		\end{pspicture}
	}
	\newcommand{\gpgKillFive}{%
		\begin{pspicture}(0,1)(20,8)
		%\psframe(0,0)(20,8)
		\putnode{z1}{origin}{4}{4}{\pscirclebox[fillstyle=solid,fillcolor=white,framesep=1]{$q$}}
		\putnode{b1}{z1}{12}{0}{\pscirclebox[fillstyle=solid,fillcolor=white,framesep=1]{$b$}}
		\putnode{q1}{b1}{0}{-10}{\pscirclebox[fillstyle=solid,fillcolor=white,framesep=.15]{$q'$}}

		%
		%%%%%%%%%%%%%%%%%%%%%%%%%%%%%
		\nccurve[arrowsize=1.5,angleA=0,angleB=180,nodesepA=-.3]{->}{z1}{b1}
		\naput[labelsep=.25,npos=.5]{\scriptsize $1|0$}
		\nbput[labelsep=.75,npos=.5]{\scriptsize 03}
	
		\nccurve[arrowsize=1.5,angleA=-75,angleB=195,nodesepA=-.4,nodesepB=-0.1,ncurv=.8]{->}{z1}{q1}
		\naput[labelsep=0,npos=.5]{\scriptsize $1|1$}
		\nbput[labelsep=.2,npos=.5]{\scriptsize 00}

%%		\ncline[nodesepA=-.6,nodesepB=-.7]{->}{z1}{q1}
%%		\naput[labelsep=0,npos=.65]{\scriptsize $1|1$}
%%		\nbput[labelsep=0,npos=.65]{\scriptsize 00}
		\end{pspicture}
	}
	\newcommand{\gpgInEndg}{%
		\begin{pspicture}(0,-3)(32,29)
		%\psframe(0,0)(32,29)
		\putnode{a1}{origin}{16}{25}{\pscirclebox[fillstyle=solid,fillcolor=white,framesep=1.22]{$r$}}
		\putnode{a2}{a1}{12}{0}{\pscirclebox[fillstyle=solid,fillcolor=white,framesep=1.22]{$a$}}
		\putnode{b1}{a1}{0}{-7}{\pscirclebox[fillstyle=solid,fillcolor=white,framesep=1]{$b$}}
		\putnode{x1}{b1}{-12}{-18}{\pscirclebox[fillstyle=solid,fillcolor=white,framesep=1.2]{$e$}}

		\putnode{m1}{b1}{12}{0}{\pscirclebox[fillstyle=solid,fillcolor=white,framesep=.72]{$m$}}

		\putnode{z1}{b1}{-12}{0}{\pscirclebox[fillstyle=solid,fillcolor=white,framesep=1]{$q$}}
		\putnode{sp}{m1}{0}{-10}{\pscirclebox[fillstyle=solid,fillcolor=white,framesep=.15]{$q'$}}
		\putnode{y1}{x1}{12}{0}{\pscirclebox[fillstyle=solid,fillcolor=white,framesep=.2]{$p'$}}
		%%%%%%%%%%%%%%%%%%%%%%%%%%%%%
		\ncline{->}{a1}{a2}
		\naput[labelsep=0,npos=.55]{\scriptsize $1|0$}
		\nbput[labelsep=0,npos=.3]{\scriptsize 01}
		%%
%%		\nccurve[arrowsize=1.5,angleA=0,angleB=180,nodesepA=-.3]{->}{z1}{b1}
		\ncline[nodesepA=-.4,nodesepB=-.4]{->}{z1}{x1}
		\nbput[labelsep=0.3,npos=.5]{\scriptsize $1|0$}
		\naput[labelsep=.3,npos=.5]{\scriptsize 05}
		\ncline[nodesepA=-.4,nodesepB=-.4]{->}{sp}{m1}
		\naput[labelsep=0.3,npos=.5]{\scriptsize $2|0$}
		\nbput[labelsep=.3,npos=.5]{\scriptsize 02}
		\nccurve[arrowsize=1.5,angleA=0,angleB=180,nodesepA=-.4]{->}{b1}{m1}
		\naput[labelsep=.25,npos=.4]{\scriptsize $1|0$}
		\nbput[labelsep=.75,npos=.5]{\scriptsize 02}
		\nccurve[arrowsize=1.5,angleA=0,angleB=180,nodesepA=-.2,nodesepB=-.3]{->}{x1}{y1}
		\naput[labelsep=.25,npos=.5]{\scriptsize $1|2$}
		\nbput[labelsep=.75,npos=.5]{\scriptsize 04}
		\end{pspicture}
	}
\begin{center}
\renewcommand{\tabcolsep}{1pt}
\begin{tabular}{|c|c|c|c|c|}
\hline
%\multirow{2}{*}{\cfgs of procedures}
	 \multicolumn{5}{|c|}{\gpus in procedure $g$ (final values after fixed-point computation). \rule[-.5em]{0em}{1.5em}}
	\\ \hline
Stmt \flab 
	& \RIn{\flab}\rule[-.5em]{0em}{1.5em}
	& \RGen{\flab}
	& \RKill{\flab}
	& \ROut{\flab}
	\\ \hline\hline
01 
	& 	\begin{tabular}{@{}c@{}}
		\rule[-.5em]{0em}{2.25em}%
		\gpgInFour
		\end{tabular}
	& 	\begin{tabular}{@{}c@{}}
		\gpgGenOne
		\end{tabular}
	& 	\begin{tabular}{@{}c@{}}
		\gpgKillOne
		\end{tabular}
	& 	\begin{tabular}{@{}c@{}}
		\rule[-.5em]{0em}{2.25em}%
		\gpgInFour
		\end{tabular}
	\\ \hline
02 
	& 	\begin{tabular}{@{}c@{}}
		\rule[-.5em]{0em}{2.25em}%
		\gpgInFour
		\end{tabular}
	& 	\begin{tabular}{@{}c@{}}
		\rule[-.5em]{0em}{2.25em}%
		\gpgGenThree
		\end{tabular}
	&
	& 	\begin{tabular}{@{}c@{}}
		\rule[-.5em]{0em}{2.25em}%
		\gpgInFour
		\end{tabular}
	\\ \hline
03 
	& 	\begin{tabular}{@{}c@{}}
		\rule[-.5em]{0em}{2.25em}%
		\gpgInFour
		\end{tabular}
	& 	\begin{tabular}{@{}c@{}}
		\gpgGenTwo
		\end{tabular}
	& 	\begin{tabular}{@{}c@{}}
		\gpgKillThree
		\end{tabular}
	& 	\begin{tabular}{@{}c@{}}
		\rule[-.5em]{0em}{2.25em}%
		\gpgInSix
		\end{tabular}
	\\ \hline
04 
	& 	\begin{tabular}{@{}c@{}}
		\rule[-.5em]{0em}{2.25em}%
		%%\gpgInSix
		\gpgInFour
		\end{tabular}
	& 	\begin{tabular}{@{}c@{}}
		\rule[-.5em]{0em}{2.25em}%
		\gpgGenFour
		\end{tabular}
	& 	\begin{tabular}{@{}c@{}}
		\gpgKillFour
		\end{tabular}
	& 	\begin{tabular}{@{}c@{}}
		\rule[-.5em]{0em}{2.25em}%
		\gpgInFive
		\end{tabular}
	\\ \hline
05 
	& 	\begin{tabular}{@{}c@{}}
		\rule[-.5em]{0em}{2.25em}%
		\gpgInFive
		\end{tabular}
	& 	\begin{tabular}{@{}c@{}}
		\rule[-.5em]{0em}{2.25em}%
		\gpgGenFive
		\end{tabular}
	& 	\begin{tabular}{@{}c@{}}
		\rule[-.5em]{0em}{2.25em}%
		\gpgKillFive
		\end{tabular}
	& 	\begin{tabular}{@{}c@{}}
		\rule[-.5em]{0em}{2.25em}%
		\gpgInEndg
		\end{tabular}
	\\ \hline
\end{tabular}

\caption{The data flow information computed by reaching \gpus analysis for procedure $g$ of 
the motivating example given in Figure~\ref{fig:mot_eg.cfgs}.
In \RIn{} and \ROut{}, we show only one boundary definition \protect\denew{q}{1|1}{q'}{00}
because other boundary definitions do not participate in \gpu reduction for this example.
However, the boundary definitions that are removed are shown in \RKill{}. 
}
\label{fig:mot_eg.rpgs.g}
\end{center}
\end{figure}

\subsection{Reaching \gpus Analysis without Blocking}
\label{sec:reach.gpus.analysis}

In this section, we 
% ignore the effect of barriers and 
present the data flow equations for computing \RIn{} and \ROut{} for every \gpb \gpbsym in the
\gpg of a procedure.
These equations ignore the effect of barriers;
Section~\ref{sec:blocked.gpus.analysis} incorporates the effects of
barriers and performs reaching \gpus analysis with blocking to compute \BRIn{} and \BROut{} for every \gpb \gpbsym. 
%The data flow information \BRGen{} computed by this analysis is then used to perform strength
%reduction optimization of a \gpg.

The reaching \gpus analysis is an intraprocedural forward
data flow analysis in the spirit of the classical reaching definitions analysis.
It computes the set \RIn{\flab} of \gpus  reaching a given \gpb $\gpbsym_{\flab}$
by processing the \gpbs that precede $\gpbsym_{\flab}$ on control flow paths reaching
$\gpbsym_{\flab}$. Then it
incorporates the effect of $\gpbsym_{\flab}$ 
on the \gpus in \RIn{\flab} 
through \gpu reduction to compute a set of \gpus
after \flab (\ROut{\flab}). The result of \gpu reduction, denoted 
\RGen{\flab}, is semantically equivalent to that of
$\gpbsym_{\flab}$.
The \gpus in \RGen{\flab} have \indlevs that do not exceed the \indlevs of the corresponding \gpus 
in $\gpbsym_{\flab}$. 
Thus, $\gpbsym_{\flab}$ can be replaced by \RGen{\flab} as a part of strength reduction optimization after the analysis 
reaches its fixed point.

\begin{Definition}
\begin{center}
\psframebox[framesep=0pt,doubleline=true,doublesep=1.5pt,linewidth=.2mm]{%
\reachingGPUdfaDef
}
\end{center}
\defcaption{Data flow equations for Reaching \gpus Analysis without Blocking}{def:dfaReachingGPUAnalysis}
\end{Definition}

\ROut{\flab} is computed using \RGen{\flab} and \RKill{\flab}.
\RGen{\flab} contains  all \gpus computed by \gpu reduction \text{$\edge \rcomp \RIn{\flab}$} 
(for all \text{$\edge \in \gpbsym_{\flab}$}).
\RKill{\flab} contains
the \gpus to be removed. They are under-approximated when a strong update cannot be 
performed. 
When a strong update is performed, we kill those 
\gpus of \RIn{\flab} whose source and \indlev match that of the
shared source of the reduced \gpus (identified by \text{$\match(\edge,\RIn{\flab})$}).
For a weak update, \text{$\conskill(\RGen{\flab},\RIn{\flab}) = \emptyset$}.

\gpu reduction allows us to model \conskill (i.e., \gpu removal from \RIn{})
in the case of strong update as follows: The reduced \gpus should
define the same pointer (or the same pointee of a given pointer) along every control flow path reaching the
statement represented by \edge. This is 
captured by the requirement \text{$|\Def(X, \edge)|=1$} in the definition of \text{$\conskill(X,\flow)$}
in Definition~\ref{def:dfaReachingGPUAnalysis}
where \text{$\Def(X, \edge)$} extracts the source nodes and their indirection levels of the \gpus
(i.e. pair $(x, i)$ for \gpu \denew{x}{i|j}{y}{\flab})
 in $X$ that are constructed 
for the same statement \flab.
The \gpus that are killed are determined by 
the \gpus in \RGen{\flab} and not those in $\gpbsym_{\flab}$. 
%%For defining \conskill, we need to identify the pointers defined by a statement. 
%%Since a \gpb may have \gpus corresponding to multiple pointer assignment statements on account of structural optimization of coalescing
%%of \gpbs, we partition
%%the set of \gpus in a \gpb  according to the assignments they correspond to.

\begin{example}{}
Figure~\ref{fig:mot_eg.rpgs.g} gives the final result of reaching \gpus analysis for procedure $g$
of our motivating example. We have shown the boundary \gpu \denew{q}{1|1}{q'}{00} for $q$.
Other boundary \gpus are not required for strong updates in this example and have been omitted.
This result has been used to construct \gpg $\mtsym_g$ shown in 
Figure~\ref{fig:mot_eg.gpg.g}. 
For procedure $f$, we do not show the complete result of the analysis but make some observations.
The \gpu \denew{q}{2|0}{o}{10} is composed with the \gpu
\denew{q}{1|0}{e}{05} to create a reduced \gpu \denew{e}{1|0}{o}{10}. 
Since, only a single pointer (in this case $e$) is being defined by the assignment, 
this is a case of strong update and hence kills \denew{e}{1|1}{c}{04}. 
The \gpu to be killed is identified 
by \text{$\match(\denew{e}{1|0}{o}{10}, \RIn{10})$} which matches the 
source and the \indlev of the \gpu to be killed to that of the 
reduced \gpu. Thus, kill is determined by the reduced \gpu (in this case \denew{e}{1|0}{o}{10}) and not the consumer \gpu (in
this case \denew{q}{2|0}{o}{10}).
\end{example}

\subsection{Reaching \gpus Analysis with Blocking}
\label{sec:blocked.gpus.analysis}

Given a \gpb $\gpbsym_{\flab}$, strength reduction
seeks to replace a consumer \gpu $\candedge \in \gpbsym_{\flab}$
with the \gpus obtained by reducing \candedge.
During \gpu reduction, it is possible that \candedge has an \admissible
composition with some producer \gpu \prevedge, but 
the location read by \candedge could be different from the
 location defined by \prevedge due to the presence of 
a barrier \gpu \baredge
(Sections~\ref{sec:bottom.up.nomenclature} and~\ref{sec:overview.local.optimizations}).
The barrier may change the pointer 
chain established by \prevedge thereby altering the data dependence
between \prevedge and \candedge.
In this case, \candedge should not be composed
with \prevedge and should be left unsimplified.
If \text{$\candedge \ecompwt \prevedge$} is performed, then \RGen{\flab} will not contain 
\candedge. Hence, when strength reduction optimization replaces $\gpbsym_{\flab}$ by \RGen{\flab},
\candedge will be replaced by the result of composition, possibly leading to unsoundness.

To ensure soundness, we perform a variant of reaching \gpus analysis that identifies barriers
and excludes blocked \gpus from the set of reaching \gpus. The unblocked \gpus are contained in 
the sets \BRIn{\flab} and \BROut{\flab} computed through a data flow analysis.
The data flow information \BRGen{\flab} computed by this analysis is then used to 
replace $\gpbsym_{\flab}$ thereby ensuring the soundness of strength
reduction optimization.

\subsubsection{The Need of Blocking}
\label{sec:need.blocked.analysis}

\begin{figure}[t]
\centering
\setlength{\codeLineLength}{45mm}%
\renewcommand{\arraystretch}{.7}%
\begin{tabular}{c|c}
	\begin{tabular}{rc}
	\codeLineNoNumber{1}{$\;\;\;\,\tt int\; a, b, *p, *q, *\!*\!x;$}{white} 
	\codeLineOne{1}{0}{$\tt void \; h()$}{white} 
	\codeLine{0}{\OB$\tt  p =  \&a;$  \rm\rule{4mm}{0mm} /* \gpu \prevedge */}{white}
	\codeLine{1}{$\tt *x =  \&b;  $   \rm\rule{2mm}{0mm} /* \gpu \baredge  */}{white}
	\codeLine{1}{$\tt q =  p;   $  \rm\rule{6.5mm}{0mm} /* \gpu \candedge */}{white}
	\codeLine{0}{\CB}{white}
	\end{tabular}
	&
	\begin{tabular}{rc}
	\codeLineNoNumber{1}{$\;\;\;\,\tt int\; a, b, *p, *q, *\!*\!x;$}{white} 
	\codeLineOne{1}{0}{$\tt void \; h()$}{white} 
	\codeLine{0}{\OB$\tt  *x = \&a;$  \rm\rule{2mm}{0mm} /* \gpu \prevedge */}{white}
	\codeLine{1}{$\tt p = \&b;    $   \rm\rule{3.5mm}{0mm} /* \gpu \baredge  */}{white}
	\codeLine{1}{$\tt q = *x;     $   \rm\rule{4.5mm}{0mm} /* \gpu \candedge */}{white}	
	\codeLine{0}{\CB}{white}
	\end{tabular}
\\
\begin{tabular}{@{}c@{}}
\begin{minipage}{63mm}
\raggedright\rule{0em}{1.5em}\footnotesize%
If $x$ points-to $p$ then $q$ points-to $b$ else $q$ points-to $a$.
\end{minipage}
\end{tabular}
	& \begin{tabular}{@{}c@{}}
	 \begin{minipage}{63mm}
	  \raggedright\rule{0em}{1.5em}\footnotesize%
	  If $x$ points-to $p$ then $q$ points-to $b$ else $q$ points-to $a$.
	\end{minipage}
	\end{tabular}
\\
\footnotesize\rule{0em}{1.5em}%
(a) Composition across an indirect \gpu \baredge
	&\footnotesize (b)\footnotesize\rule{0em}{1.5em}% 
		{Composition with an indirect \gpu across the \gpu \baredge}
\end{tabular}
\caption{
Risk of unsoundness in \gpu reduction caused by a barrier \gpu. 
}
\label{fig:replacing_n_by_r}
\end{figure}

The location read by a \gpu \candedge could be different from the 
location defined by \prevedge 
because of a combined effect of the \gpus in a calling context and the \gpus corresponding to the intervening
assignments on a control flow path from \prevedge to \candedge which may update the \gpu 
\prevedge. We characterize these situations by building on Section~\ref{sec:bottom.up.nomenclature}
and defining the notion of a \emph{barrier \gpu}
which \emph{blocks} certain \gpus so that \gpu compositions 
leading to potentially unsound strength reduction optimization
are \emph{postponed}.
After inlining the \gpg in a caller, more information may become available. Thus, 
it may resolve any uncertain data dependence between \candedge and \prevedge---so
a composition which was earlier postponed may now safely be performed.
% without causing unsoundness.
This is explained in the rest of the section.

We define a barrier as follows.
Let an \emph{indirect} \gpu refer to a \gpu whose \indlev of
the source is greater than 1 (i.e., the pointer being defined by the \gpu is not known).
Then, a \gpu \baredge corresponding to an assignment between \candedge and \prevedge on some control
flow path is a barrier if:
\begin{itemize}
\item \baredge is an indirect \gpu. This is
	a composition across an indirect \gpu \baredge (Figure~\ref{fig:replacing_n_by_r}(a)).  
\item \prevedge is an indirect \gpu (\baredge need not be an indirect \gpu).
	This is a composition with an indirect \gpu across the \gpu \baredge 
	(Figure~\ref{fig:replacing_n_by_r}(b)). 
\end{itemize}
We illustrate these situations in the following example.

\begin{example}{exmp:barrier}
Consider the procedure in Figure~\ref{fig:replacing_n_by_r}(a). 
The composition between the \gpus for statements 02 and 04
is \admissible. However, statement 03 
may cause a side-effect by indirectly defining $p$ (if $x$ points to $p$ in the calling context).
Thus, $q$ in statement 04 would point to $b$ if $x$ points to $p$; 
otherwise it would point to $a$.
If we replace the \gpu \denew{q}{1|1}{p}{04} by \denew{q}{1|0}{a}{04} (which is the result of composing 
\denew{q}{1|1}{p}{04} with \denew{p}{1|0}{a}{02}), then we would  miss the \gpu 
\denew{q}{1|0}{b}{04} if $x$ points to $p$ in the calling context---leading to unsoundness.
Since the calling context is not available during \gpg construction, we postpone this composition to eliminate
the possibility of unsoundness.
This is done by blocking the \gpu \denew{p}{1|0}{a}{02} by an indirect \gpu 
\denew{x}{2|0}{b}{03} which acts as a barrier. This corresponds to the first case described above.

For the second case, consider
statement 02 of  the procedure in Figure~\ref{fig:replacing_n_by_r}(b) which may indirectly 
define $p$ (if $x$ points to $p$). Statement 03
directly defines $p$.
Thus, $q$ in statement 04 would point to $b$ if $x$ points to $p$; 
otherwise it would point to $a$.
We postpone the composition 
\text{$\candedge \!:\! \denew{q}{1|2}{x}{04}$} with \text{$\prevedge \!:\! \denew{x}{2|0}{a}{02}$} 
by blocking the \gpu \prevedge where the \gpu \denew{p}{1|0}{b}{03}
acts as a barrier.
\end{example}

A barrier \gpu is likely to have a \WaW or \WaR dependence with some preceding \gpus which
cannot be ascertained without the alias information in the calling context.
In the absence of alias information from the calling context, we use the type information 
to identify some such \gpus as non-blocking.
The barrier blocks such \gpus, so that
the compositions of 
\candedge with them are
postponed (Section~\ref{sec:bottom.up.nomenclature}).
Consider a \gpu \prevedge originally 
blocked by a barrier \baredge where \prevedge or \baredge is an indirect \gpu.
After inlining the \gpg in its callers and performing reductions in the calling
contexts, the following situations could arise: 
\begin{enumerate}
\item The \indlev of the source of the indirect \gpu (\prevedge or \baredge) is reduced to 1 
	thereby identifying the pointer being defined by the \gpu. In this case, \baredge
        ceases being a barrier and so no longer blocks \prevedge
        leading to the following two situations:
	\begin{enumerate}[(a)]
	\item \baredge redefines the pointer defined by \prevedge, 
	killing \prevedge thereby obviating the composition \text{$\candedge\, \ecompwt \prevedge$}.
	\item \baredge does not redefine the pointer defined by \prevedge 
	thereby allowing the composition \text{$\candedge\, \ecompwt \prevedge$}.
	\end{enumerate}
\item The \indlev of the source of the indirect \gpu (\prevedge or \baredge) remains greater 
than 1. In this case, \baredge continues to block \prevedge awaiting further
inlining.
\end{enumerate}

In case 1(a), an eager reduction of \candedge without blocking \prevedge would cause \candedge 
to be replaced by the result of composition \text{$\candedge\, \ecompwt \prevedge$}, thereby causing
unsoundness. Reaching \gpus analysis with blocking helps to postpone the composition until all information becomes
available. 
Our measurements (Section~\ref{sec:emp-eval}) show that situation 1(a) rarely arises
in practice 
because it amounts to defining the same pointer multiple times 
through different aliases in the same context.

\begin{example}{}
Case 1(a) above could arise if $x$ points to $p$ in the calling context
of the procedure in Figure~\ref{fig:replacing_n_by_r}(a). 
As a result, 
\gpu \denew{p}{1|0}{a}{02} is killed by the barrier \gpu \denew{p}{1|0}{b}{03} (which is 
the simplified version of the barrier \gpu \denew{x}{2|0}{b}{03}) and hence
the composition is prohibited and $q$ points to $b$ for statement 04.
Case 1(b) could arise if $x$ points to any location other than $p$ in the calling context. 
In this case, the composition between \denew{q}{1|1}{p}{04}
and \denew{p}{1|0}{a}{02} is sound and $q$ points to $a$ for statement 04.
Case 2 could arise if pointee of $x$ is not available even in the calling context.
In this case, the barrier \gpu \denew{x}{2|0}{b}{03} continues to block
\denew{p}{1|0}{a}{02}.
\end{example}

\begin{example}{}
To see how reaching \gpus analysis with blocking helps,
consider the example in Figure~\ref{fig:replacing_n_by_r}(b).
The set of \gpus reaching the statement 04 is
\text{$\RIn{04} = \{\denew{x}{2|0}{a}{02}, \denew{p}{1|0}{b}{03}\}$}. 
The \gpu \denew{x}{2|0}{a}{02} is blocked by the barrier \gpu \denew{p}{1|0}{b}{03} 
and hence \text{$\BRIn{04} = \{\denew{p}{1|0}{b}{03}\}$}. 
Thus, \gpu reduction for \text{$w \!:\! \denew{q}{1|2}{x}{04}$} 
(in the context of \text{$\BRIn{04}$})
computes \GENEDGES as $\{w\}$ with the flag \WCompose set to \FALSE  
because $w$ cannot be reduced further within the \gpg of the procedure. However, $w$ 
is still not a points-to edge and can be simplified further after the \gpg
is inlined in its callers.
Hence we postpone the composition of $w$ with \text{$\prevedge \!:\! \denew{x}{2|0}{a}{02}$} 
until \prevedge is simplified.
\end{example}

\subsubsection{Data Flow Equations for Computing \BRIn{} and \BROut{}}
\label{sec:dfe.blocking}

\begin{Definition}
\begin{center}
\psframebox[framesep=0pt,doubleline=true,doublesep=1.5pt,linewidth=.2mm]{%
\reachingGPUWBdfaDef
}
\end{center}
\defcaption{Data flow equations for Reaching \gpus Analysis with Blocking.
%The main difference between \BROut{\flab} and \ROut{\flab} (Definition~\ref{def:dfaReachingGPUAnalysis})
%is that the former uses the function \blocked which in turn require
%\unblockall and \unblockindirect. These three functions are absent in
%Definition~\ref{def:dfaReachingGPUAnalysis}.
}{def:dfaReachingGPUWBAnalysis}
\end{Definition}

A barrier may not necessarily block all preceding \gpus. We use the type information
to identify absence of
data dependence between a barrier and the \gpus reaching it.
This allows us to minimize blocking by
identifying \gpus that need not be blocked.
A barrier \text{$\baredge \in \BRGen{\flab}$} may block a producer \gpu
\text{$\prevedge \in \BRIn{\flab}$} if it writes into a location read by or written by 
\prevedge.
Thus, they could share a \WaW or a \WaR data dependence.
Recall that a barrier \gpu \baredge is either an indirect 
\gpu or a \gpu that follows an indirect \gpu (Section~\ref{sec:need.blocked.analysis}). 
Thus the following \gpus should be blocked:
\begin{itemize}
\item If \BRGen{\flab} contains an indirect \gpu \baredge, 
	then all \gpus reaching $\gpbsym_{\flab}$ that share a data dependence with \baredge
	should be blocked regardless of the 
      nature of other \gpus  (if any) in \BRGen{\flab}.
\item If \BRGen{\flab} does not contain an indirect \gpu and is not $\emptyset$, 
	then all indirect \gpus reaching $\gpbsym_{\flab}$ that share a 
	data dependence with a \gpu in \BRGen{\flab} should be blocked.
\end{itemize}

We define a predicate $\bDdep(B, I)$ to check 
the presence of data dependence between the set of \gpus $B$ and $I$
(Definition~\ref{def:dfaReachingGPUWBAnalysis}).
When the types of \text{$\baredge \in B$} and \text{$\prevedge \in I$}
match\footnote{Although C11 standard allows type casting for pointers,
there is no guarantee of the expected behaviour if 
there is alignment mismatch. 
For example, the runtime behaviour of assigning 
`{$\tt int\, *$}' to `{$\tt float\, *$}' depends on the compiler and the architecture. 
However, assigning `{$\tt void\, *$}' to `{$\tt int\, *$}' does not result in misalignment.
In our implementation, we trust the types recorded in the GIMPLE IR used by gcc and 
assume that there is no undefined behaviour of the program.
}, we assume the possibility of data dependence and \baredge blocks \prevedge.
%\footnote{
%When the source of \baredge is a pointer to a void, if \baredge is an indirect 
%\gpu, it blocks all \gpus in \BRIn{\flab}, otherwise
%it blocks all indirect \gpus in \BRIn{\flab}.
%}
%%\begin{align*}
%%\bDdep(B, I) & \Leftrightarrow 
%%		\begin{array}[t]{@{}l}
%%		\LT(B) \cap 
%%		\left(\LT(I) \cup \RT(I)\right) \neq \emptyset
%%		\end{array}
%%\end{align*}
$\LT(B)$ is the set of types of locations being written by a barrier whereas
\text{$\left(\LT(I) \cup \RT(I)\right)$} represents the 
set of types of locations defined or read by the \gpus in $I$
thereby checking a \WaW and \WaR dependence.
The type of the $i^{th}$ pointee of $x$ is given by 
$\Type(x, i)$ defined as illustrated below.

\begin{example}{}
If the declaration of a pointer $x$ is `$\tt int\, *\!*\,x$', then \text{$\Type(x, 1)$}
is `$\tt int\, *\!*$' and \text{$\Type(x, 2)$} is `$\tt int\, *$'.	  
Note that \text{$\Type(x, 0)$} is not a pointer and \text{$\Type(x, 3)$} is undefined because
$x$ cannot be dereferenced thrice.
\end{example}

%%\begin{align*}
%%\LT(X) = & \left\{ \Type(x, i) \mid  \denew{x}{i|j}{y}{\flab} \, \in\, X \right\}
%%\\
%%\RT(X) = & \left\{ \Type(x, k) \mid  1 \leq k < i, \denew{x}{i|j}{y}{\flab} \in X \right\}
%%		\bigcup
%%		\\
%%	& \left\{ \Type(y, k) \mid  1 \leq k < j, \denew{x}{i|j}{y}{\flab} \in X \right\}
%%\end{align*}

The data flow equations in Definition~\ref{def:dfaReachingGPUWBAnalysis} 
identify the \gpus in \BRGen{\flab} that can act as a barrier.
The main difference between \BROut{\flab} (Definition~\ref{def:dfaReachingGPUWBAnalysis})
and \ROut{\flab} (Definition~\ref{def:dfaReachingGPUAnalysis})
is that the former uses function \blocked which 
computes blocked \gpus as follows:
\begin{itemize}
\item Case 1 in \blocked equation %in Definition~\ref{def:dfaBlockedGPUAnalysis}
      corresponds to not blocking any \gpu because \BRGen{\flab} is empty.
\item Case 2 in \blocked equation %in Definition~\ref{def:dfaBlockedGPUAnalysis})
      corresponds to blocking appropriate \gpus reaching \flab (i.e. \BRIn{\flab}) 
      because \BRGen{\flab} contains an indirect \gpu. 
\item Case 3 in \blocked equation %in Definition~\ref{def:dfaBlockedGPUAnalysis})
      corresponds to blocking appropriate indirect \gpus reaching \flab 
      because \BRGen{\flab} does not contains an indirect \gpu and is not $\emptyset$.
\end{itemize}

\begin{example}{}
For the procedure in Figure~\ref{fig:replacing_n_by_r}(b), \text{$\BRIn{02} = \emptyset$} and
\text{$\BRGen{02}$} is \text{$\{\denew{x}{2|0}{a}{02}\}$}.
Although \BRGen{02} contains an indirect \gpu, since no \gpus reach 02 (because it is the first statement), 
\BROut{02} is \text{$\{\denew{x}{2|0}{a}{02}\}$} indicating that no \gpus are blocked.

For statement 03, 
\text{$\BRIn{03} = \{\denew{x}{2|0}{a}{02}\}$} and \text{$\BRGen{03} = \{\denew{p}{1|0}{b}{03}\}$}. 
\BRGen{03} is non-empty and does not contain an indirect \gpu and thus
$\BROut{03} = \{\denew{p}{1|0}{b}{03}\}$ according to the third case in the \blocked equation in
Definition~\ref{def:dfaReachingGPUWBAnalysis} indicating that the
\gpu \denew{x}{2|0}{a}{02} is blocked and should not be used for composition by the later \gpus. The indirect \gpu in \BRIn{03}
is excluded from \BROut{03}. Note that the indirect \gpu \denew{x}{2|0}{a}{02} is blocked by the \gpu \denew{p}{1|0}{b}{03}
because $\Type(x, 2)$ matches 
with $\Type(p, 1)$ indicating a possibility of \WaW dependence.

For statement 04, \text{$\BRIn{04} = \{\denew{p}{1|0}{b}{03}\}$}
and \BRGen{04}  is \{\denew{q}{1|2}{x}{04}\}. For this statement, the composition \text{$(\denew{q}{1|2}{x}{04} \ecomp^{\textrm{ts}}
\denew{x}{2|0}{a}{02})$} is postponed
because the \gpu \denew{x}{2|0}{a}{02} is blocked. In this case, \BRGen{04} does not contain an indirect \gpu and 
\text{$\BROut{04} = \{\denew{p}{1|0}{b}{03}, \denew{q}{1|2}{x}{04}\}$}.

Similarly in Figure~\ref{fig:replacing_n_by_r}(a), the \gpu \denew{p}{1|0}{a}{02} is blocked by the 
barrier \gpu \denew{x}{2|0}{b}{03} because $\Type(p, 1)$ matches
with $\Type(x, 2)$. Hence, the
composition \text{$(\denew{q}{1|1}{p}{04} \ecomp^{\textrm{ts}} \denew{p}{1|0}{a}{02})$} is postponed. 
%%
%%Since the \gpu compositions are postponed indicating that the \gpus are not completely simplified, the \gpus
%%\denew{x}{1|2}{y}{09} and \denew{r}{1|1}{p}{12} are added to \queued.

In the \gpg of procedure $g$ (of our motivating example) shown in Figure~\ref{fig:mot_eg.gpg.g},
 the \gpus \denew{r}{1|0}{a}{01} and
\denew{q}{1|0}{b}{03} are not blocked by the \gpu
\denew{q}{2|0}{m}{02} because they have different types.
However, the \gpu \denew{e}{1|2}{p}{04} blocks the indirect \gpu \denew{q}{2|0}{m}{02} because there is a possible \WaW 
data dependence ($e$ and $q$ could be aliased in the callers of $g$). 
\end{example}

\section{Redundancy Elimination Optimizations}
\label{sec:structural-optimizations}

Recall that strength reduction simplifies \gpus and eliminates data dependences between them. This paves way for redundancy 
elimination optimizations which remove redundant \gpus and minimize control flow. As a consequence, they
improve the compactness of a \gpg
and reduce the repeated re-analysis of \gpbs caused by inlining at call sites.
They include:

\begin{itemize}
\item \emph{Dead \gpu and empty \gpb elimination.}
\item \emph{Coalescing of \gpbs.}
\end{itemize}

Recall that the strength reduction optimization may postpone the reduction of certain \gpus.
This requires us to postpone
optimizations such as dead \gpu elimination and coalescing in order to ensure soundness.
In this section, we describe each of the optimizations in detail and characterize when to
postpone them.

\subsection{Dead \gpu and Empty \gpb Elimination}
\label{sec:dead.code.elimination}

We perform dead \gpu elimination to remove a redundant \gpu \text{$\edge \in \gpbsym_{\flab}$} that 
is killed along every control flow path from \flab to the \End{} \gpb of the procedure.
However, the following two kinds of \gpus should not be removed even if they are killed in reaching
\gpus analyses:
\begin{inparaenum}[(a)]
\item \gpus that are blocked, or 
\item \gpus that are producer \gpus for \undesirable compositions 
that have been postponed (Section~\ref{sec:edge.comp.properties}).
\end{inparaenum}
For the former, we check that a \gpu considered for dead \gpu elimination 
does not belong to \ROut{\Endscriptsize{}} (the result of reaching \gpus analysis without 
blocking); for the latter we check that the 
\gpu is not a producer \gpu for a postponed composition.
We record such \gpus in the set \queued computed for every \gpg. It is computed during \gpu reduction.\footnote{The revised definition 
is available at 
\htmladdnormallink{https://www.cse.iitb.ac.in/~uday/soft-copies/gpg-pta-paper-appendix.pdf}{https://www.cse.iitb.ac.in/~uday/soft-copies/gpg-pta-paper-appendix.pdf}.}
%%{(Definition~\ref{def:edge.reduction.ab} in Appendix~\ref{sec:augmented.gpu.reduction})}. 
Thus, we perform dead \gpu elimination 
and remove a \gpu \text{$\edge \in \gpbsym_{\flab}$} if 
\text{$\edge \notin (\ROut{\Endscriptsize{}} \cup \queued)$}.

%%There are two reasons why \edge may not be in 
%%\BROut{\Endscriptsize{}}:
%%\begin{itemize}
%%\item \edge may be blocked by a barrier, or
%%\item \edge may be killed by some other \gpu.
%%\end{itemize}
%%In the former case, \edge should not be considered for dead \gpu elimination. In the latter case,
%%if \edge is a producer \gpu for an \undesirable composition that has been postponed, then
%%it should not be considered for dead \gpu elimination.

%%However, \BROut{\Endscriptsize{}} does not contain 
%%a producer \gpu \prevedge in \mtsym with which 
%%a consumer \gpu \candedge cannot be composed  in \mtsym
%%but may be composed after \mtsym is inlined in the body
%%of a caller:
%%\begin{enumerate}[(a)]
%%\item Composition \text{$\candedge\, \ecompwt \prevedge$} may be postponed because \prevedge may be blocked by 
%%	a barrier; in this case, it is possible that the barrier
%%   may be simplified after inlining in a caller and may not block \prevedge in 
%%	the body of the caller.
%%\item Composition \text{$\candedge\, \ecompwt \prevedge$} may be \undesirable; in this case,
%%	inlining in a caller may allow \prevedge to be simplified
%%    making the composition \desirable in the body of the caller.
%%\end{enumerate}
%%Such \gpus should not be
%%eliminated even if they do not reach the \End{} \gpb.

\begin{example}{}
In procedure $g$ of Figure~\ref{fig:mot_eg.gpg.g}, 
pointer $q$ is defined in statement 03 but is redefined in statement 05 and hence the \gpu \denew{q}{1|0}{b}{03} is killed
and does not reach the \End{} \gpb. 
%%Further, \denew{q}{1|0}{b}{03} is not blocked by 
%%\denew{q}{2|0}{m}{03} because they define different locations. Function \unblockall in the second case of \blocked equation
%%in Definition~\ref{def:dfaReachingGPUWBAnalysis} ensures this.
Since no composition with the \gpu \denew{q}{1|0}{b}{03} 
is postponed, it does not belong to set \queued either.  Hence 
the \gpu \denew{q}{1|0}{b}{03} is eliminated from the \gpb $\gpbsym_{03}$ as an instance of dead \gpu elimination.

Similarly, the \gpus \denew{q}{1|0}{d}{07} (in $\gpbsym_{07}$) and 
\denew{e}{1|1}{c}{04} (in $\gpbsym_{14}$) in 
the \gpg of procedure $f$ (Figure~\ref{fig:mot_eg.2}) are eliminated from their corresponding \gpbs.
%Observe that the \gpu \denew{d}{1|0}{n}{08} in \gpb $\gpbsym_{08}$ is not removed even though $\gpbsym_{12}$
%contains a definition of $d$ expressed by the \gpu \denew{d}{1|0}{m}{02}.
%This is because $\gpbsym_{12}$ 
%also contains \gpu \denew{b}{1|0}{m}{02}
% which defines $b$, indicating that
%$d$ is not defined along all paths.
%Hence the previous definition of $d$ cannot be killed giving
%a weak update. 
\end{example}

\begin{example}{}
For the procedure in Figure~\ref{fig:replacing_n_by_r}(a), the \gpu
\denew{p}{1|0}{a}{02} is not killed but is blocked by the barrier \denew{x}{2|0}{b}{03}; hence it is present
in $\ROut{05}$ but not in \BROut{05} (05 is the \End{} \gpb).
This \gpu may be required when the barrier \denew{x}{2|0}{b}{03} is reduced after call inlining (and
 ceases to block \denew{p}{1|0}{a}{02}).
Thus, it is not removed by dead \gpu elimination.
\end{example}

In the process of dead \gpu elimination, if a \gpb becomes empty, it is eliminated by connecting its predecessors to 
its successors.

\begin{example}{}
In the \gpg of procedure $g$ of Figure~\ref{fig:mot_eg.gpg.g}, 
the \gpb $\gpbsym_{03}$ becomes empty after dead \gpu elimination.  Hence, 
$\gpbsym_{03}$ can be removed by connecting its predecessors to successors. This 
transforms the back edge \text{$\gpbsym_{03} \rightarrow \gpbsym_{01}$} to 
\text{$\gpbsym_{02} \rightarrow \gpbsym_{01}$}.
Similarly, the \gpb $\gpbsym_{07}$ is deleted from the \gpg of procedure $f$
in Figure~\ref{fig:mot_eg.2}.
\end{example}

\subsection{Minimizing the Control Flow by Coalescing \gpbs}
\label{sec:coalescing.analysis}

Strength reduction eliminates data dependence between \gpus rendering the control flow redundant. 
Eliminating redundant control flow is important to make a \gpg as compact as possible---in the 
absence of control flow minimization, the size of the \gpg of a procedure tends to increase exponentially 
because of transitive inlinings of calls in the procedure. This effect is aggravated by the fact that 
many procedures are called multiple times in the same procedure.  Besides, recursion
causes multiple inlinings of the \gpgs of procedures in the cycle of recursion 
(Section~\ref{sec:handling_recur}).

\subsubsection{Coalescing \gpbs by Partitioning a \gpg}
We eliminate redundant control flow by coalescing adjacent \gpbs. This amounts to 
partitioning the set of \gpbs in a \gpg such that each part contains the \gpbs
whose \gpus do not have a data dependence between them and hence can be
seen essentially as executed non-deterministically in any order in accordance with
abstract semantics of a \gpb as a  \emph{may} property (Section~\ref{sec:gpg.def}).

Since partitioning is driven by preserving and exploiting the absence of data dependence, it 
is characterized by the following properties:
\begin{itemize}
\item A \gpg can be partitioned in multiple ways to minimize the control flow.
      The absence of data dependence is not a transitive relation: Consider \gpbs
	$\gpbsym_l$, $\gpbsym_m$, and $\gpbsym_n$ such that \text{$m \in succ(l)$}
	 and \text{$n \in succ(m)$}. Assume that 
      \text{$\edge_m \in \gpbsym_m$} does not have a data dependence with \text{$\edge_l \in \gpbsym_l$} and
      \text{$\edge_n \in \gpbsym_n$} does not have a data dependence with \text{$\edge_m \in \gpbsym_m$}.
      However, there may be a data dependence between \text{$\edge_l \in \gpbsym_l$} and 
      \text{$\edge_n \in \gpbsym_n$}. If the data dependence exists, then 
the following two partitions have minimal control flow:
      \text{$\partition_1 = \left\{\left\{\gpbsym_l, \gpbsym_m\right\}, \left\{\gpbsym_n\right\}\right\}$} and
      \text{$\partition_2 = \left\{\left\{\gpbsym_l\right\}, \left\{\gpbsym_m, \gpbsym_n\right\}\right\}$}. 
Our heuristics (described below) construct partition $\partition_1$.

\item The possibility of data dependence between \gpbs $\gpbsym_m$ and $\gpbsym_n$
      matters only if there is control flow between them. Otherwise,
      they are executed in different execution instances of the program and there is no 
      data dependence between them even if the variables or abstract locations 
	accessed by them are same.
      Hence the successors of a \gpb can be coalesced with each other in the
	same part
	provided there is no control flow between them.

\item As a design choice, a successor (predecessor) of a \gpb is included in the part containing the 
     \gpb 
      \emph{iff} \emph{all} successors (predecessors) of the
      \gpb are included in the part: Consider
      \gpbs $\gpbsym_l$, $\gpbsym_m$ and $\gpbsym_n$ such that \text{$succ(l) = \{m, n\}$} and
      neither $m$ is a successor of $n$ nor vice-versa. Let \text{$\gpbsym_l \in \group_i$}. 
      Since there is no control flow between 
       $\gpbsym_m$ and $\gpbsym_n$, including only one of them in $\group_i$ will create a spurious control
      flow between them. This ordering could introduce
      a spurious data dependence between their \gpus which may cause imprecision (through a \RaW dependence that 
      may create spurious \gpus).
\item Coalescing may eliminate a definition-free path for the source of a \gpu. This
	may convert the \gpu from \may-def (i.e., source is defined along some path) to \must-def
	(i.e., source is defined along all paths) in the \gpg.
	Consider \gpbs $\gpbsym_l$, $\gpbsym_m$, $\gpbsym_n$, and $\gpbsym_o$  such that 
	\text{$succ(l) = \{m, n\}$} and \text{$pred(o) = \{m, n\}$}.
      Let \text{$\group_i = \{\gpbsym_l, \gpbsym_m, \gpbsym_n\}$} and 
	\text{$\group_j = \{\gpbsym_o\}$}. 
	The source of some \gpu \text{$\edge_m \in \gpbsym_m$} may have a definition-free path
	\text{$\gpbsym_l \rightarrow \gpbsym_n \rightarrow \gpbsym_o$}. After coalescing, this definition-free path
	ceases to exist because of the control flow edge \text{$\group_i \rightarrow \group_j$}.
	This may lead to strong updates instead of weak updates thereby leading to unsoundness.
	Hence, we add a separate definition-free path for such \gpus.
\end{itemize}

Due to the possibility of multiple partitions satisfying the above criteria,
identifying the ``best'' partition would require defining a cost model.
Instead, we compute a unique partition by imposing additional restrictions described below. 
Our empirical measurements show significant compression by our
heuristic partitioning below and any attempt of finding the best partitioning may provide only marginal overall benefits  because the
process would become inefficient.
 Hence we use the following greedy heuristics: 
\begin{itemize}
\item \Start{} \gpb and \End{} \gpb form singleton parts and no other \gpb is included in these parts. 
	This is required for modelling
      definition-free paths from \Start{} to \End{} to distinguish between strong and weak updates by 
      a callee \gpg in a caller \gpg.
\item The process of identifying the partition begins with \Start{} \gpb. Thus \Start{} forms 
	{$\group_1 \in \partition$}. As a consequence,
      a part $\group_i \in \partition$ grows only in the ``forward'' direction including only successor \gpbs.
      It never grows in the ``backward'' direction by considering predecessors.
\item Consider $\gpbsym_n$ and $\gpbsym_s$, $s\in succ(n)$ such that 
      \text{$\gpbsym_n \rightarrow \gpbsym_s$} is a back edge. Then
      $\gpbsym_n$ and $\gpbsym_s$ belong to the same partition $\group_i$ \emph{iff} 
      all \gpbs in the loop formed by the back edge (i.e. all
     \gpbs that appear on all paths from $\gpbsym_s$ to $\gpbsym_n$) belong to $\group_i$. 
\end{itemize}

In principle, partitioning could be performed using a greedy process interleaved with coalescing
such that each part grows incrementally. 
However, this incremental expansion cannot be done by coalescing one successor at a time
because all successors and all predecessors of all these successors must be included in the
same partition, and this property needs to be applied transitively.
Hence, we %%use the usual dichotomy of analysis and transformations and
separate the process of discovering the partition (analysis) from the process of coalescing (transformation).
%%We perform a data flow analysis that computes the partitions by examining the data dependencies and enforcing
%%all requirements of partitioning. Actual coalescing is performed only after all partitions have been identified.
We define a data flow analysis that constructs a part $\group_i$ inductively by  
considering the possibility of including the successors of the \gpbs 
that are already in $\group_i$. 

\begin{figure}[t]
%%\centering
\small
%%%%%%%%%%%%%%%%%%%%%%%%% Definitions of graphs %%%%%%%%%%%%%%%%%%%%%%%5
	\newcommand{\gpgA}{%
		\begin{pspicture}(1,3)(15,14)
		%\psframe(1,2)(15,14)
		\putnode{x}{origin}{2}{12}{\pscirclebox[fillstyle=solid,fillcolor=white,framesep=1]{$x$}}
		\putnode{m}{x}{12}{0}{\pscirclebox[fillstyle=solid,fillcolor=white,framesep=.72]{$m$}}
		\putnode{y}{x}{0}{-7}{\pscirclebox[fillstyle=solid,fillcolor=white,framesep=1]{$y$}}
		\putnode{n}{y}{12}{0}{\pscirclebox[fillstyle=solid,fillcolor=white,framesep=1.02]{$n$}}
%%%%%%%%%
		\ncline[arrowsize=1.5]{->}{x}{m}
		\naput[labelsep=0.5,npos=.5]{\scriptsize $2|0$}
		\nbput[labelsep=0.5,npos=.5]{\scriptsize $12$}
		\ncline[arrowsize=1.5]{->}{y}{n}
		\naput[labelsep=.25,npos=.5]{\scriptsize $2|0$}
		\nbput[labelsep=.25,npos=.5]{\scriptsize $14$}
		\end{pspicture}
	}
	\newcommand{\gpgB}{%
		\begin{pspicture}(1,3)(15,14)
		%\psframe(0,0)(32,39)
		\putnode{z}{origin}{2}{12}{\pscirclebox[fillstyle=solid,fillcolor=white,framesep=1]{$z$}}
		\putnode{o}{z}{12}{0}{\pscirclebox[fillstyle=solid,fillcolor=white,framesep=1.02]{$o$}}
		\putnode{x}{z}{0}{-7}{\pscirclebox[fillstyle=solid,fillcolor=white,framesep=1]{$x$}}
		\putnode{m}{x}{12}{0}{\pscirclebox[fillstyle=solid,fillcolor=white,framesep=.72]{$m$}}
%%%%%%%%%
		\ncline[arrowsize=1.5]{->}{x}{m}
		\naput[labelsep=0.5,npos=.5]{\scriptsize $2|0$}
		\nbput[labelsep=0.5,npos=.5]{\scriptsize $12$}
		\ncline[arrowsize=1.5]{->}{z}{o}
		\naput[labelsep=0.5,npos=.5]{\scriptsize $2|0$}
		\nbput[labelsep=0.5,npos=.5]{\scriptsize $32$}
		\end{pspicture}
	}
	\newcommand{\gpgC}{%
		\begin{pspicture}(1,2)(15,6)
		%\psframe(0,0)(32,39)
		\putnode{x}{origin}{2}{4}{\pscirclebox[fillstyle=solid,fillcolor=white,framesep=1]{$x$}}
		\putnode{n}{x}{13}{0}{\pscirclebox[fillstyle=solid,fillcolor=white,framesep=1.02]{$n$}}
%%%%%%%%%
		\ncline[arrowsize=1.5]{->}{x}{n}
		\naput[labelsep=0.5,npos=.5]{\scriptsize $2|0$}
		\nbput[labelsep=0.5,npos=.5]{\scriptsize $12$}
		\end{pspicture}
	}
	\newcommand{\gpgD}{%
		\begin{pspicture}(1,2)(15,6)
		%\psframe(0,0)(32,39)
		\putnode{u}{origin}{2}{4}{\pscirclebox[fillstyle=solid,fillcolor=white,framesep=1.02]{$u$}}
		\putnode{v}{u}{13}{0}{\pscirclebox[fillstyle=solid,fillcolor=white,framesep=1.02]{$v$}}
%%%%%%%%%
		\ncline[arrowsize=1.5]{->}{u}{v}
		\naput[labelsep=0.5,npos=.5]{\scriptsize $2|0$}
		\nbput[labelsep=0.5,npos=.5]{\scriptsize $17$}
		\end{pspicture}
	}
	\newcommand{\gpgE}{%
		\begin{pspicture}(1,2)(15,6)
		%\psframe(0,0)(32,39)
		\putnode{p}{origin}{2}{4}{\pscirclebox[fillstyle=solid,fillcolor=white,framesep=1]{$p$}}
		\putnode{s}{p}{13}{0}{\pscirclebox[fillstyle=solid,fillcolor=white,framesep=1.02]{$s$}}
%%%%%%%%%
		\ncline[arrowsize=1.5]{->}{p}{s}
		\naput[labelsep=0.5,npos=.5]{\scriptsize $1|0$}
		\nbput[labelsep=0.5,npos=.5]{\scriptsize $36$}
		\end{pspicture}
	}
	\newcommand{\gpgF}{%
		\begin{pspicture}(1,2)(15,6)
		%\psframe(0,0)(32,39)
		\putnode{q}{origin}{2}{4}{\pscirclebox[fillstyle=solid,fillcolor=white,framesep=1]{$q$}}
		\putnode{t}{q}{13}{0}{\pscirclebox[fillstyle=solid,fillcolor=white,framesep=1.02]{$t$}}
%%%%%%%%%
		\ncline[arrowsize=1.5]{->}{q}{t}
		\naput[labelsep=0.5,npos=.5]{\scriptsize $1|0$}
		\nbput[labelsep=0.5,npos=.5]{\scriptsize $37$}
		\end{pspicture}
	}
	\newcommand{\gpgG}{%
		\begin{pspicture}(1,5)(17,30)
		%\psframe(1,5)(17,30)
		\putnode{x}{origin}{3}{28}{\pscirclebox[fillstyle=solid,fillcolor=white,framesep=1]{$x$}}
		\putnode{m}{x}{12}{0}{\pscirclebox[fillstyle=solid,fillcolor=white,framesep=.72]{$m$}}
		\putnode{y}{x}{0}{-7}{\pscirclebox[fillstyle=solid,fillcolor=white,framesep=1]{$y$}}
		\putnode{n}{y}{12}{0}{\pscirclebox[fillstyle=solid,fillcolor=white,framesep=1.02]{$n$}}
		\putnode{z}{y}{0}{-7}{\pscirclebox[fillstyle=solid,fillcolor=white,framesep=.72]{$z$}}
		\putnode{o}{z}{12}{0}{\pscirclebox[fillstyle=solid,fillcolor=white,framesep=1.02]{$o$}}	
		\putnode{u}{z}{0}{-7}{\pscirclebox[fillstyle=solid,fillcolor=white,framesep=1.02]{$u$}}
		\putnode{v}{u}{12}{0}{\pscirclebox[fillstyle=solid,fillcolor=white,framesep=1.02]{$v$}}			
%%%%%%%%%
		\ncline[arrowsize=1.5]{->}{x}{m}
		\naput[labelsep=0.5,npos=.5]{\scriptsize $2|0$}
		\nbput[labelsep=0.5,npos=.5]{\scriptsize $12$}
		\ncline[arrowsize=1.5]{->}{y}{n}
		\naput[labelsep=.5,npos=.5]{\scriptsize $2|0$}
		\nbput[labelsep=.5,npos=.5]{\scriptsize $14$}
		\ncline[arrowsize=1.5]{->}{z}{o}
		\naput[labelsep=.5,npos=.5]{\scriptsize $2|0$}		
		\nbput[labelsep=.5,npos=.5]{\scriptsize $32$}		
		\ncline[arrowsize=1.5]{->}{u}{v}
		\naput[labelsep=.5,npos=.5]{\scriptsize $2|0$}
		\nbput[labelsep=.5,npos=.5]{\scriptsize $17$}
		\end{pspicture}
	}
	\newcommand{\gpgI}{%
		\begin{pspicture}(1,6)(17,38)
		%\psframe(1,6)(17,38)
		\putnode{x}{origin}{3}{36}{\pscirclebox[fillstyle=solid,fillcolor=white,framesep=1]{$x$}}
		\putnode{m}{x}{12}{0}{\pscirclebox[fillstyle=solid,fillcolor=white,framesep=.72]{$m$}}
		\putnode{z}{x}{0}{-7}{\pscirclebox[fillstyle=solid,fillcolor=white,framesep=1]{$y$}}
		\putnode{o}{z}{12}{0}{\pscirclebox[fillstyle=solid,fillcolor=white,framesep=1.02]{$n$}}	
		\putnode{u}{z}{0}{-7}{\pscirclebox[fillstyle=solid,fillcolor=white,framesep=1.02]{$u$}}
		\putnode{v}{u}{12}{0}{\pscirclebox[fillstyle=solid,fillcolor=white,framesep=1.02]{$v$}}			
		\putnode{p}{u}{0}{-7}{\pscirclebox[fillstyle=solid,fillcolor=white,framesep=1]{$p$}}
		\putnode{s}{p}{12}{0}{\pscirclebox[fillstyle=solid,fillcolor=white,framesep=1.02]{$s$}}			
		\putnode{q}{p}{0}{-7}{\pscirclebox[fillstyle=solid,fillcolor=white,framesep=1]{$q$}}
		\putnode{t}{q}{12}{0}{\pscirclebox[fillstyle=solid,fillcolor=white,framesep=1.02]{$t$}}			
%%%%%%%%%
		\ncline[arrowsize=1.5]{->}{x}{m}
		\naput[labelsep=0.5,npos=.5]{\scriptsize $2|0$}
		\nbput[labelsep=0.5,npos=.5]{\scriptsize $12$}
		\ncline[arrowsize=1.5]{->}{z}{o}
		\naput[labelsep=.5,npos=.5]{\scriptsize $2|0$}		
		\nbput[labelsep=.5,npos=.5]{\scriptsize $14$}		
		\ncline[arrowsize=1.5]{->}{u}{v}
		\naput[labelsep=.5,npos=.5]{\scriptsize $2|0$}
		\nbput[labelsep=.5,npos=.5]{\scriptsize $17$}
		\ncline[arrowsize=1.5]{->}{p}{s}
		\naput[labelsep=0.5,npos=.5]{\scriptsize $1|0$}
		\nbput[labelsep=0.5,npos=.5]{\scriptsize $36$}
		\ncline[arrowsize=1.5]{->}{q}{t}
		\naput[labelsep=0.5,npos=.5]{\scriptsize $1|0$}
		\nbput[labelsep=0.5,npos=.5]{\scriptsize $37$}
		\end{pspicture}
	}
	\newcommand{\gpgH}{%
		\begin{pspicture}(1,2)(15,14)
		%\psframe(0,0)(32,39)
		\putnode{p}{origin}{2}{12}{\pscirclebox[fillstyle=solid,fillcolor=white,framesep=1]{$p$}}
		\putnode{s}{p}{13}{0}{\pscirclebox[fillstyle=solid,fillcolor=white,framesep=1.02]{$s$}}
		\putnode{q}{p}{0}{-8}{\pscirclebox[fillstyle=solid,fillcolor=white,framesep=1.02]{$q$}}
		\putnode{t}{q}{13}{0}{\pscirclebox[fillstyle=solid,fillcolor=white,framesep=1.02]{$t$}}
%%%%%%%%%
		\ncline[arrowsize=1.5]{->}{p}{s}
		\naput[labelsep=0.5,npos=.5]{\scriptsize $1|0$}
		\nbput[labelsep=0.5,npos=.5]{\scriptsize $36$}
		\ncline[arrowsize=1.5]{->}{q}{t}
		\naput[labelsep=0.5,npos=.5]{\scriptsize $1|0$}
		\nbput[labelsep=0.5,npos=.5]{\scriptsize $37$}
		\end{pspicture}
	}
\begin{center}
{
\renewcommand{\tabcolsep}{5pt}
\begin{tabular}{cc}
\begin{tabular}{l}
$\tt int\, *\!*\,x;$
	\\
$\tt float\, *\!*\,y;$
	\\
$\tt short\, *\!*\,z;$
\end{tabular}
&
\begin{tabular}{l}
$\tt int\, *\!*\!*u;$
	\\ 
$\tt int\, *\!p, *q, *v;$
	\\
$\tt int\; m, n, o, s, t;$
\end{tabular}
\end{tabular}
}
\end{center}

\smallskip

\begin{center}
\setlength{\tabcolsep}{4pt}
\begin{tabular}{@{}c|c|c@{}}
\hline
\rule[-1em]{0em}{2.5em}%
	\renewcommand{\arraystretch}{.7}%
		\begin{tabular}{@{}l@{}}
		$\mtsym_f$ before coalescing
		\end{tabular}
	& \renewcommand{\arraystretch}{.7}%
		\begin{tabular}{@{}l@{}}
		$\mtsym_f$ after coalescing 
		\end{tabular}
	& \renewcommand{\arraystretch}{.7}%
		\begin{tabular}{@{}l@{}}
	$\mtsym_f$ after modelling \\ definition-free paths
		\end{tabular}
	\\ \hline
\begin{tabular}{@{}c}
\begin{pspicture}(-10,0)(46,110)
%\psframe(-10,0)(46,110)
\putnode{n0}{origin}{18}{106}{\psframebox{\Start{}}}
\putnode{n1}{n0}{0}{-11}{\psframebox[fillstyle=solid,fillcolor=lightgray,framesep=1]{\makebox[10mm]{\rule{0mm}{4mm}}}}
	\putnode{w}{n1}{-12}{0}{$\gpbsym_{1}$}
\putnode{p1}{n1}{-14}{-16}{\psframebox[fillstyle=solid,fillcolor=lightgray,framesep=2]{\gpgA}}
	\putnode{w}{p1}{-12}{0}{$\gpbsym_{2}$}
\putnode{p2}{n1}{14}{-16}{\psframebox[fillstyle=solid,fillcolor=lightgray,framesep=2]{\gpgB}}
	\putnode{w}{p2}{-12}{0}{$\gpbsym_{3}$}
\putnode{p3}{n1}{0}{-37}{\psframebox[fillstyle=solid,fillcolor=lightgray,framesep=2]{\gpgA}}
	\putnode{w}{p3}{-12}{0}{$\gpbsym_{4}$}
\putnode{n2}{p3}{0}{-18}{\psframebox[fillstyle=solid,fillcolor=lightgray,framesep=2]{\gpgD}}
	\putnode{w}{n2}{-12}{0}{$\gpbsym_{5}$}
\putnode{n3}{n2}{0}{-13}{\psframebox[fillstyle=solid,fillcolor=lightgray,framesep=2]{\gpgE}}
	\putnode{w}{n3}{-12}{0}{$\gpbsym_{6}$}
\putnode{n4}{n3}{0}{-13}{\psframebox[fillstyle=solid,fillcolor=lightgray,framesep=2]{\gpgF}}
	\putnode{w}{n4}{-12}{0}{$\gpbsym_{7}$}
\putnode{nn}{n4}{0}{-11}{\psframebox{\End{}}}
\psset{arrowsize=1.5,arrowinset=0}
\ncline{->}{n0}{n1}
\ncline{->}{n4}{nn}
\ncline{->}{n1}{p1}
\ncline{->}{n1}{p2}
\ncline{->}{p1}{p3}
\ncline{->}{p2}{p3}
\ncline{->}{p3}{n2}
\ncline{->}{n2}{n3}
\ncline{->}{n3}{n4}
\nccurve[angleA=-65, angleB=45,offsetB=2,nodesepB=1.5]{->}{p2}{n2}
\ncloop[angleA=270,angleB=90,offsetA=2,offsetB=2,armB=3.5,armA=3,loopsize=-25,linearc=.5]{->}{n2}{n1}
\end{pspicture}
\end{tabular}
&
\begin{tabular}{@{}c}
\begin{pspicture}(0,0)(30,110)
%\psframe(0,0)(30,110)
\putnode{s1}{origin}{17}{106}{\psframebox{\Start{}}}
\putnode{p1}{s1}{0}{-23}{\psframebox[fillstyle=solid,fillcolor=lightgray,framesep=2]{\gpgG}}
	\putnode{w}{p1}{-13}{0}{$\gpbsym_{8}$}
\putnode{p2}{p1}{0}{-29}{\psframebox[fillstyle=solid,fillcolor=lightgray,framesep=2]{\gpgH}}
	\putnode{w}{p2}{-12}{0}{$\gpbsym_{9}$}
\putnode{pn}{p2}{0}{-17}{\psframebox{\End{}}}
\psset{arrowsize=1.5,arrowinset=0}
%%\ncline{->}{s1}{p1}
%%\ncline[arrowsize=2,angleA=-.2,linewidth=1]{->}{s1}{p3}
\ncline{->}{p1}{p2}
%%\ncline[arrowsize=2,angleA=-.2,linewidth=1]{->}{p3}{s2}
%%\ncline{->}{p2}{s2}
\ncline{->}{s1}{p1}
\ncline{->}{p2}{pn}
\end{pspicture}
\end{tabular}
&
\begin{tabular}{@{}c}
\begin{pspicture}(0,0)(40,110)
%\psframe(0,0)(40,110)
\putnode{s1}{origin}{17}{106}{\psframebox[framesep=1]{\Start{}}}
	%\putnode{w}{s1}{-9}{0}{$\Start{}$}
\putnode{p1}{s1}{0}{-23}{\psframebox[fillstyle=solid,fillcolor=lightgray,framesep=2]{\gpgG}}
	\putnode{w}{p1}{-13}{0}{$\gpbsym_{8}$}
\putnode{p2}{p1}{0}{-29}{\psframebox[fillstyle=solid,fillcolor=lightgray,framesep=2]{\gpgH}}
	\putnode{w}{p2}{-12}{0}{$\gpbsym_{9}$}
\putnode{p3}{p2}{18}{18}{\psframebox[fillstyle=solid,fillcolor=lightgray]{$\gpbsym_{10}$}}
\putnode{s2}{p2}{0}{-17}{\psframebox[framesep=1]{\End{}}}
\putnode{w}{s2}{3}{-20}{\begin{minipage}{34mm}\raggedright
		\renewcommand{\arraystretch}{1.8}	
		$\gpbsym_{10} =\big\{\begin{array}[t]{@{}l}\denew{x}{2|0}{m}{12}, \denew{y}{2|0}{n}{14}, \\
\denew{u}{2|0}{v}{17}, \denew{p}{1|0}{s}{36}, \\ \denew{q}{1|0}{t}{37}\big\}\end{array}$

%%		$\gpbsym_{10}$ contains all \gpus of $\mtsym_f$ other than \denew{z}{2|0}{o}{32}
			\end{minipage}}
\psset{arrowsize=1.5,arrowinset=0}
\ncline{->}{s1}{p1}
\ncangle[angleA=270,angleB=90,arrowsize=2,doubleline=true,armB=25,offsetA=2,linearc=.3]{->}{s1}{p3}
\ncline{->}{p1}{p2}
\ncangle[angle=270,angleB=90,arrowsize=2,doubleline=true,armB=4,offsetB=2,linearc=.3]{->}{p3}{s2}
\ncline{->}{p2}{s2}
\end{pspicture}
\end{tabular}
\\ \hline
\end{tabular}
\caption{An example demonstrating the effect of coalescing. 
The loop formed by the back edge $\gpbsym_5 \rightarrow \gpbsym_1$ reduces to a self loop over \gpb $\gpbsym_8$ after coalescing. 
Since self loops are redundant, they are eliminated. 
%The final $\mtsym_f$ after modelling definition-free paths is shown in 
%Figure~\ref{fig:eg.coalescing.def}.
%%A separately modelled definition-free path is shown by edges with double lines in the \gpg after coalescing.
Control flow edges with double lines represent definition-free paths.
}
\label{fig:eg.coalescing}
\end{center}
\end{figure}
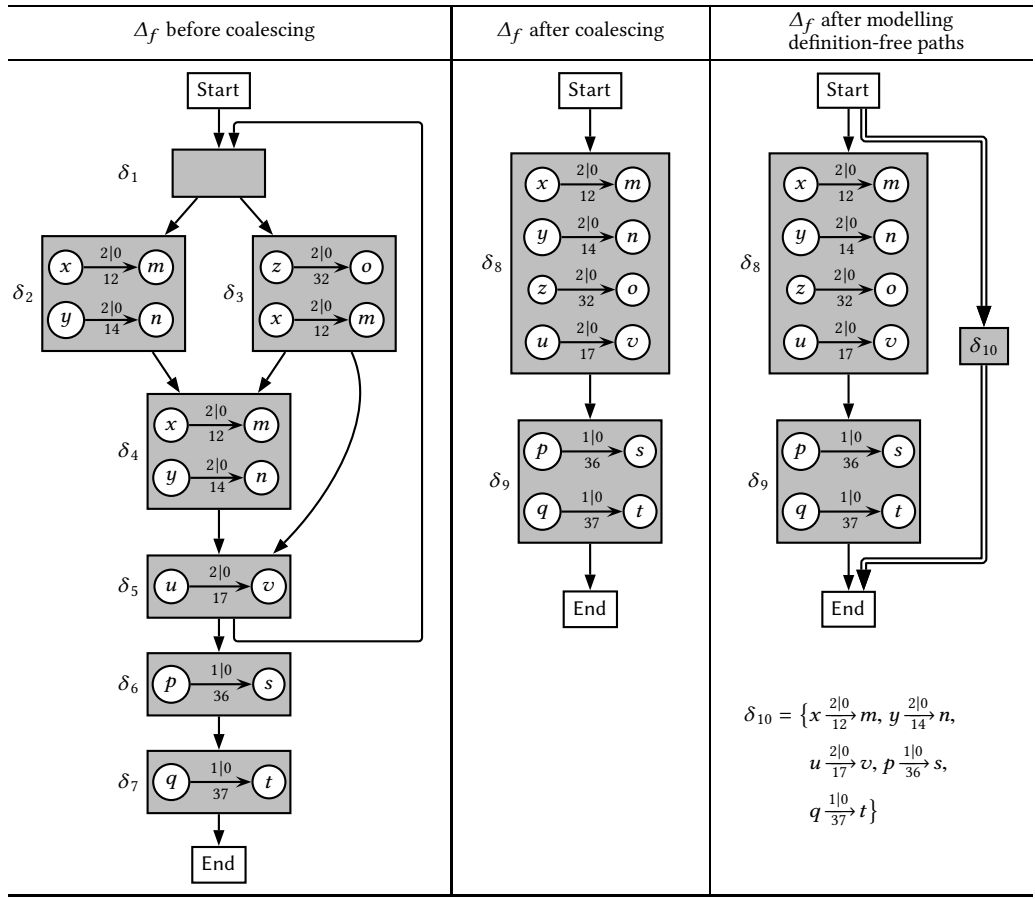

\subsubsection{The Role of Data Dependence in Blocking and Coalescing}
\label{sec:type.checking.for.coalescing}

The main differences between the use of data dependence for blocking
(Definition~\ref{def:dfaReachingGPUWBAnalysis} in Section~\ref{sec:blocked.gpus.analysis})
and for coalescing are:
\begin{itemize}
\item {\em The motivation behind using data dependence.}
      When analyzing for blocking, we identify the possibility of a barrier updating
      a location accessed by a previous \gpu. In coalescing we wish to establish that no
      control flow needs to be maintained between two \gpus.
\item {\em The way data dependence is used.}
      For blocking, we use the \emph{possible presence} of data dependence between
      a barrier and reaching \gpus to block some of the reaching \gpus. For coalescing, we use
      the \emph{guaranteed absence} of data dependence between the \gpus of a \gpb and those
      reaching it from within a part to coalesce the \gpb with the part.
\item {\em Relevant data dependences.}
	Coalescing removes control flow between two \gpus enabling their non-deterministic
      execution with respect to each other which is oblivious to any data dependence
	between the \gpus. Hence, a \RaW and \WaW
	dependences need to be preserved by prohibiting
	coalescing. However, a \WaR dependence is not affected by coalescing.
	On the other hand, blocking by a barrier does not involve \RaW 
	dependence (see the motivation above) and needs to handle only \WaW and
	\WaR dependences.  
 \item {\em The role of dereference in data dependence.}
      For blocking, only the write by a barrier is important and not a read.
      Hence, we check for a dereference only in the source of 
	a barrier \gpu. For coalescing analysis, we need to consider
      dereferences both in the source and the target.
\end{itemize}

These differences change the modelling of data dependence for coalescing in the following 
ways:
\begin{itemize}
\item We now include a check for a dereference within the predicate for data dependence check.
\item Consider a \gpb $\gpbsym_n$ for coalescing in a part $\group_i$. We now check for
      both reads and writes in the \gpus of $\gpbsym_n$ and only writes in the \gpus of
      $\group_i$.
\end{itemize}
Compare the predicates \bDdep (Definition~\ref{def:dfaReachingGPUWBAnalysis}) 
for blocking and \Ddep (Definition~\ref{def:dfacoalescingAnalysis})
for coalescing to see the above differences.
For establishing the absence of dependence, we match the types of
\text{$\edge_1 \in X$} with the types of \text{$\edge_2 \in Y$}.
This is meaningful only when \text{$\edge_1 \neq \edge_2$}.
The term $X-Y$ in the definition of predicate \Ddep ensures this.

\begin{Definition}
\begin{center}
\psframebox[framesep=0pt,doubleline=true,doublesep=1.5pt,linewidth=.2mm]{%
\coalescingdfaDef
}
\end{center}
\defcaption{Data flow equations for Coalescing Analysis.
}{def:dfacoalescingAnalysis}
\end{Definition}

\subsubsection{Partitioning Analysis}
\label{sec:part.analysis}

We define two interdependent data flow analyses that inductively  
\begin{itemize}
\item construct part $\group_i$ using data flow variables \InB{n}/\OutB{n}, and
\item compute the \gpus accumulated in $\Ggroup$ in data flow variables \InT{n}/\OutT{n}.
\end{itemize}
The latter is required to
identify the \RaW or \WaW data dependence between the \gpus in part $\group_i$.

Unlike the usual data flow variables that typically compute a set of facts, 
\InB{n}/\OutB{n} are predicates.
If \InB{n} is \true, it indicates that $\gpbsym_n$ belongs to the same part as that of \emph{all} of its predecessors.
If \OutB{n} is \true, it indicates that $\gpbsym_n$ belongs to the same part as that of \emph{all} of its successors.
%%It is assumed that the definition-free paths are modelled externally.
Thus our analysis does not enumerate the parts as sets of \gpbs explicitly; instead, parts are computed implicitly by
setting predicates \InB{}/\OutB{} of adjacent \gpbs.

The data flow equations to compute \InB{n}/\OutB{n} are given in 
Definition~\ref{def:dfacoalescingAnalysis}.
The initialization is \true for all \gpbs.
Predicate $\coalescePred(p, n)$ uses $\TypeFlow(p, n)$ to 
check if \gpus \Ggroupp are allowed to 
flow from $p$ to $n$---if yes,
then $p$ and $n$ belong to the same part. If $\OutT{p}$ 
is $\emptyset$, they belong to the same part regardless of $\TypeFlow(p, n)$.
The presence of \OutB{p} in the equation of \coalescePred (Definition~\ref{def:dfacoalescingAnalysis})
ensures that
\gpb $\gpbsym_p$ is considered for coalescing with $\gpbsym_n$ only if
$\gpbsym_p$ has not been found to be a ``boundary'' in coalescing because it cannot coalesce 
with some successor.

Another striking difference between the equations for \InB{}/\OutB{}
in Definition~\ref{def:dfacoalescingAnalysis}
and the usual data flow equations is that the data flow variables \InB{n} and \OutB{n}  for 
\gpb $n$ are 
independent of each other---\InB{n} depends only on the \OutB{} of its predecessors and \OutB{n} depends 
only on the \InB{} of its successors. 
Intuitively, this form of data flow equations attempts to \emph{melt} the boundaries 
of \gpb $n$ to explore fusing it with its successors and predecessors. 
\begin{itemize}
\item When \InB{n} is true, it melts 
the boundary at the top of the \gpb and glues it with all its predecessors that are already in
the part. Thus, a part grows in a forward direction.
\item When \OutB{n} 
is true, it melts the boundary at the bottom of the \gpb and includes all its successors in the
part thereby growing a part in the forward direction.
\end{itemize}

The incremental expansion of a part in a forward direction influences the flow of \gpus
accumulated in a part leading to a forward data flow analysis for computing \Ggroup using
data flow variables \InT{n}/\OutT{n}.
The data flow equations to compute them are given in Definition~\ref{def:dfacoalescingAnalysis}.
Function $\TypeFlow(p, n)$ in the equation for \InT{} computes the set of \gpus \Ggroupp
that flow from $p$ to $n$.
It establishes the absence of data dependences using 
predicate $\Ddep$ defined in Section~(\ref{sec:type.checking.for.coalescing}).
If no data dependence exists, the \gpus accumulated in $\OutT{p}$ are propagated to $n$.
The presence of $\neg\InB{p}$ in equation for \TypeFlow ensures that
\gpus in \OutT{p} are propagated to $\gpbsym_n$ only if
$\gpbsym_n$ has not been found to be a ``boundary'' in coalescing because it cannot coalesce 
with some predecessor.

\begin{figure}
 \newcommand{\IP}{\text{int$*$}\xspace}
 \newcommand{\FP}{\text{float$*$}\xspace}
 \newcommand{\SP}{\text{short$*$}\xspace}
 \newcommand{\IDP}{\text{int$**$}\xspace}

\begin{center}
{
\renewcommand{\tabcolsep}{5pt}
\begin{tabular}{c}
%%\begin{tabular}{ll}
%%\hline
%%\multicolumn{2}{c}{\rule{0em}{.9em}%
%%	Name for types}
%%	\\ \hline
%%\ITP & $\tt int\, *\!*\,*$
%%	\\
%%\IDP & $\tt int\, *\!*$
%%	\\
%%\IP & $\tt int\; *$
%%	\\
%%\FDP & $\tt float\, *\!*$
%%	\\
%%\SDP & $\tt short\, *\!*$
%%\end{tabular}
%%&
%%&
\begin{tabular}{ll|ll|ll|ll|ll|ll}
\hline
\multicolumn{12}{c}{\rule{0em}{.9em}%
Name for \gpus. Statement ids do not matter}
	\\ \hline
\rule{0em}{1.5em}%
$\edge_1$ & \denew{x}{2|0}{m}{12}
	&$\edge_2$& \denew{y}{2|0}{n}{14}
	& $\edge_3$& \denew{z}{2|0}{o}{32}
	& $\edge_4$ & \denew{u}{2|0}{v}{17}
	& $\edge_5$ & \denew{p}{1|0}{s}{36}
	& $\edge_6$ & \denew{q}{1|0}{t}{37}
\end{tabular}
\end{tabular}
}

\bigskip

\begin{tabular}{|c|c|c|c|c|c|c|}
\hline
%%\multicolumn{7}{|c|}{\rule{0em}{1.3em} \text{$\edge_1: \protect\denew{x}{[*,a]|[\,]}{m}{}$}, 
%%\text{$\edge_2: \protect\denew{x}{[*,b]|[\,]}{n}{}$},
%%\text{$\edge_3: \protect\denew{z}{[*,c]|[\,]}{o}{}$}, \text{$\edge_4: \protect\denew{y}{[*,d]|[\,]}{u}{}$}, \text{$\edge_5:
%%\protect\denew{p}{[*]|[\,]}{s}{}$},
%%\text{$\edge_6: \protect\denew{q}{[*]|[\,]}{t}{}$}}
%%\\ \hline
\gpb $n$ 
	& \LT(n)\rule[-.5em]{0em}{1.5em}
	& \RT(n)
	& \InT{n}
	& \OutT{n}
	& \InB{n}
	& \OutB{n}
	\\ \hline\hline
\rule{0em}{1em}
$\gpbsym_1$ 
	& $\emptyset$
	& $\emptyset$
	& $\{\edge_1, \edge_2, \edge_3, \edge_4\}$
	& $\{\edge_1, \edge_2, \edge_3, \edge_4\}$
	& F 
	& T
	\\ \hline
\rule{0em}{1em}
$\gpbsym_2$ 
	& $\{\IP, \FP\}$
	& $\emptyset$
	& $\{\edge_1, \edge_2, \edge_3, \edge_4\}$
	& $\{\edge_1, \edge_2, \edge_3, \edge_4\}$
	& T
	& T
	\\ \hline
\rule{0em}{1em}
$\gpbsym_3$ 
	& $\{\SP, \IP\}$
	& $\emptyset$
	& $\{\edge_1, \edge_2, \edge_3, \edge_4\}$
	& $\{\edge_1, \edge_2, \edge_3, \edge_4\}$
	& T
	& T
	\\ \hline
\rule{0em}{1em}
$\gpbsym_4$ 
	& $\{\IP, \FP\}$
	& $\emptyset$
	& $\{\edge_1, \edge_2, \edge_3, \edge_4\}$
	& $\{\edge_1, \edge_2, \edge_3, \edge_4\}$
	& T
	& T
	\\ \hline
\rule{0em}{1em}
$\gpbsym_5$ 
	& $\{\IDP\}$
	& $\emptyset$
	& $\{\edge_1, \edge_2, \edge_3, \edge_4\}$
	& $\{\edge_1, \edge_2, \edge_3, \edge_4\}$
	& T
	& F
	\\ \hline
\rule{0em}{1em}
$\gpbsym_6$ 
	& $\{\IP\}$
	& $\emptyset$
	& $\emptyset$
	& $\{\edge_5\}$
%%	& \begin{tabular}{c} $\{\IP\}$ \end{tabular}
	& F
	& T
	\\ \hline
\rule{0em}{1em}
$\gpbsym_7$ 
	& $\{\IP\}$
	& $\emptyset$
	& $\{\edge_5\}$
	& $\{\edge_5, \edge_6\}$
	& T
	& F
	\\ \hline
\end{tabular}
\end{center}
\caption{The data flow information computed by coalescing analysis for example in Figure~\protect\ref{fig:eg.coalescing}.
% Type names are abbreviated by their first letter.
The \InB{} and \OutB{} values indicate that 
\gpbs $\gpbsym_1$, $\gpbsym_2$, $\gpbsym_3$, $\gpbsym_4$, $\gpbsym_5$ can be coalesced. Similarly, \gpbs $\gpbsym_6$ and $\gpbsym_7$ can be coalesced.
\gpbs $\gpbsym_5$ and $\gpbsym_6$ must remain in different coalesced groups.
}
\label{fig:eg.coalescing.dfv}
\end{figure}

\begin{example}{}
Figure~\ref{fig:eg.coalescing.dfv} gives 
the data flow information for the example of Figure~\ref{fig:eg.coalescing}. 
\gpbs $\gpbsym_1$ and $\gpbsym_2$ can be coalesced because \OutB{1} 
is \true
and \OutT{1} is $\emptyset$. Thus, $\Ddep(1, 2)$ returns \false indicating
that 
types do not match and hence there is no possibility of a data dependence
between the \gpus of $\gpbsym_1$ and $\gpbsym_2$.
Similarly, \gpbs $\gpbsym_1$ and $\gpbsym_3$ can be coalesced. 
Thus \OutB{1}, \InB{2}, and \InB{3} are \true.
We check the data dependence between the \gpus 
of \gpbs $\gpbsym_2$ and $\gpbsym_4$ using the type information.
However, $\Ddep(2, 4)$ returns \false because the term
\text{$(\OutT{2} - \gpbsym_4)$} is $\emptyset$.
Thus, \gpbs $\gpbsym_2$ and $\gpbsym_4$ belong to the same part and can be coalesced.
For \gpbs $\gpbsym_3$ and $\gpbsym_4$, the possibility of data dependence is resolved 
based on the type
information. The term \text{$(\OutT{3} - \gpbsym_4)$} returns \text{$\denew{z}{2|0}{o}{32}$} 
whose $\Type(z, 1)$ does not match that of the pointers being read 
in the \gpus in $\gpbsym_4$.
Thus, 
\gpbs $\gpbsym_3$ and $\gpbsym_4$ can be coalesced. 
\gpbs $\gpbsym_4$ and $\gpbsym_5$ both contain a \gpu with a dereference, however \text{$\Ddep(\gpbsym_4, \gpbsym_5)$} returns \false
indicating that there is no type matching and hence no possibility of data dependence, thereby allowing the coalescing of the two \gpbs.
The \text{$\Ddep(\gpbsym_5, \gpbsym_6)$} returns \true (type of source of the \gpu $\denew{x}{2|0}{m}{12} \in \OutT{5}$ matches the
source of the \gpu $\denew{p}{1|0}{s}{36} \in \gpbsym_6$) indicating a possibility of data dependence in the caller 
through aliasing and hence the two \gpbs cannot be coalesced. Thus, the first part in the partition contains only \gpbs $\gpbsym_1$, 
$\gpbsym_2$, $\gpbsym_3$, $\gpbsym_4$, and $\gpbsym_5$. \gpb $\gpbsym_6$ now marks the first \gpb of the new part. 
\gpbs $\gpbsym_6$ and $\gpbsym_7$ can be coalesced as there is no data dependence between their \gpus. 
The loop $\gpbsym_5 \rightarrow \gpbsym_1$
before coalescing now reduces to self loop over \gpb $\gpbsym_8$ after coalescing. The self loop is redundant and hence eliminated.
\gpbs $\gpbsym_5$ and $\gpbsym_1$ can be coalesced because all the \gpbs of the loop belong to the same part.
\end{example}

Observe that some \gpus appear in multiple \gpbs of a \gpg (before coalescing). This is because we could have multiple calls to the
same procedure. Thus, even though the \gpbs are renumbered, the statement labels in the \gpus remain unchanged resulting in repetitive occurrence of a
\gpu.
This is a design choice because it helps us to accumulate the points-to information
of a particular statement in all contexts.

\begin{example}{}
In the example of Figure~\ref{fig:mot_eg.gpg.g},
\gpbs $\gpbsym_1$ and $\gpbsym_2$ can be coalesced because 
\text{$\Ddep(\gpbsym_1, \gpbsym_2)$} returns 
\false indicating that there is no type matching
and hence no possible data dependence between their \gpus.
Thus, \OutB{1} and \InB{2} are set to \true.
The loop formed by the back edge $\gpbsym_2 \rightarrow \gpbsym_1$
reduces to a self loop over \gpb $\gpbsym_{11}$ after coalescing. 
The self loop is redundant and hence it is eliminated.
For \gpbs $\gpbsym_2$ and $\gpbsym_4$,
\text{$\Ddep(\gpbsym_2, \gpbsym_4)$} returns \true because 
$\Type(q, 2)$ (for the \gpu \denew{q}{2|0}{m}{02} in $\gpbsym_{02}$) matches 
$\Type(p, 2)$ (for the \gpu \denew{e}{1|2}{p}{04} in $\gpbsym_{04}$) which is
$\tt int\, *$. This indicates the possibility of a data dependence between 
the \gpus of \gpbs $\gpbsym_2$ and $\gpbsym_4$ ($q$ and $p$ could be aliased in the caller) and hence
these \gpbs cannot be coalesced. Thus, \OutB{2} and \InB{4} are set to \false.
For \gpbs $\gpbsym_4$ and $\gpbsym_5$, 
\text{$\Ddep(\gpbsym_4, \gpbsym_5)$} returns \false because there is no possible 
data dependence. Hence \OutB{4} and \InB{5} are set to \true and the two
\gpbs can be coalesced.
\end{example}

Recall that our coalescing heuristics requires us to prohibit 
\begin{itemize}
\item coalescing with \Start{} and \End{} \gpbs so that definition-free paths can be modelled, and 
\item coalescing of the source and target \gpbs of a back edge unless all \gpbs in the 
loop formed by the back edge are included in the same part.
\end{itemize}
The data flow equations for Coalescing (\InB{}/\OutB{} in Definition~\ref{def:dfacoalescingAnalysis}) do not have any provision of these requirements;
they are enforced separately during the actual transformation. 

\subsubsection{Preserving Definition-Free Paths}

Consider a \gpu \edge that reaches the exit of a \gpg along some path but not all. It means that there is some path in the \gpg
along which the source of \edge is not defined (i.e., the source of \edge is \may-defined
in the \gpg). According to our heuristics of coalescing, a \gpb is coalesced 
either with all its successors or with none. Hence,
after coalescing with all successors, a definition-free path may get subsumed and
\edge may reach the exit of a \gpg along all paths 
indicating that the source of \edge is now \must-defined. This would lead to a strong update instead of a weak update thereby introducing unsoundness. 
Hence, we need to 
add an explicit definition-free path for such \gpus.
The \gpus with definition-free paths are identified by the corresponding boundary definitions.
A definition-free path for the source of \text{$\gpu: \denew{x}{i|j}{y}{\flab}$} 
exists in a \gpg 
only if the boundary definition \text{$\denew{x}{i|i}{x'}{00}$} reaches the exit of the \gpg.

\begin{example}{}
In the example of Figure~\ref{fig:eg.coalescing}, 
the definition-free path is shown by edges with 
double lines in the \gpg obtained after coalescing.
The \gpu \denew{z}{2|0}{o}{32} does not reach the exit along the path \text{$\gpbsym_1 \rightarrow \gpbsym_2
\rightarrow \gpbsym_4 \rightarrow \gpbsym_5 \rightarrow \gpbsym_6 \rightarrow \gpbsym_7$} 
which forms the definition-free path. We add a definition-free path between \Start{} and \End{} \gpbs of a \gpg with a \gpb that contains 
all \gpus that do not have any definition-free path. Thus, we have a \gpb $\gpbsym_{10}$
which contains all \gpus except \denew{z}{2|0}{o}{32}.
\end{example}

\begin{example}{}
In Figures ~\ref{fig:mot_eg.gpg.g} and ~\ref{fig:mot_eg.2}, definition-free paths are shown by 
edges with double lines in the \gpgs of procedures $f$ and $g$ obtained after coalescing.
For procedure $g$,
the \gpus \denew{b}{1|0}{m}{02} and \denew{q}{2|0}{m}{02} undergo a weak update and hence
do not kill their corresponding boundary definitions. This indicates that
the source of these \gpus are \may-defined and hence a definition-free path is required for these
\gpus. Thus, we add a definition-free path
between \Start{} and \End{} \gpbs of $\mtsym_g$ with \gpb $\gpbsym_{16}$ which contains the set of \gpus
\text{$\{\denew{r}{1|0}{a}{01}, \denew{e}{1|2}{p}{04}, \denew{q}{1|0}{e}{05}\}$}.

For procedure $f$,
the boundary definition \denew{b}{1|1}{b'}{00} 
reaches the exit of $\mtsym_f$ indicating
that $b$ is 
\may-defined. Hence a definition-free path is added with \gpb
$\gpbsym_{17}$ containing all \gpus of $\mtsym_f$ except
			\denew{b}{1|0}{m}{02}. \gpu
			\denew{q}{2|0}{m}{02}, which has a definition-free path in $\mtsym_g$,
			reduces to 
			\denew{d}{1|0}{m}{02} in $\mtsym_f$. However, 
			$d$ is defined in $\gpbsym_{08}$ also, hence 
                        it does not have a definition-free path in $\mtsym_f$.
\end{example}

\section{Call Inlining}
\label{sec:interprocedural.extensions}

%%We have discussed the construction of \gpgs in the absence of calls. 
In order to construct the \gpg of a procedure, the optimized \gpgs of its
callees are inlined at the call sites and the resulting \gpg of the procedure is then optimized.
After a \gpg is inlined at a call site, its \gpbs undergo another round of optimization in the calling
context.  This repeated optimization in the context of each transitive caller of a \gpg,
gives us our efficiency.

The \gpg of a procedure can be constructed completely only
when \begin{inparaenum}[(a)] \item all callees are known, and \item their \gpgs have been constructed completely. \end{inparaenum}
The first condition is violated by a call through function pointer
and the second condition is violated by a recursive call.
We classify procedure calls into the following three categories and
explain the handling of the first two in this section.
The third category is handled in Section~\ref{sec:handling_fp}
because it requires the concepts introduced in Section~\ref{sec:dfv_compute}.
\begin{itemize}
\item Callee is known and the call is non-recursive. %%Direct non-recursive calls. 
\item Callee is known and the call is recursive. %%Direct recursive calls. 
\item Callee is not known. %%Indirect calls. 
\end{itemize}

\begin{figure}[t]
%%\centering
\setlength{\codeLineLength}{20mm}
\renewcommand{\arraystretch}{.9}
\noindent
\begin{tabular}{@{}c@{}c@{}|c@{}}
\begin{tabular}{c}
\begin{pspicture}(1,0)(39,44)
%\psframe(0,0)(40,44)
\putnode{m}{origin}{22}{41}{\psframebox{\Start{p}}}  %%{{\psframebox{{\Start{f}}}}}
\putnode{a}{m}{-10}{-18}{\psframebox{$y = \&a$}} %%{\psframebox{$y = \&a;$}}
	\putnode{w}{a}{-9}{0}{$01$}
\putnode{b}{m}{10}{-18}{\psframebox{$\;\,q();\;\,$}}
	\putnode{w}{b}{-8}{0}{$02$}
\putnode{d}{m}{0}{-36}{\psframebox{{\End{p}}}}
%%%
\psset{arrowsize=1.5,arrowinset=0}
\ncline{->}{m}{a}
\ncline{->}{m}{b}
\ncline{->}{b}{d}
\ncline{->}{a}{d}
\end{pspicture}
\end{tabular}
&
\begin{tabular}{c}
\begin{pspicture}(0,0)(19,44)
%\psframe(0,0)(22,40)
\putnode{m}{origin}{10}{41}{\psframebox{\Start{q}}}  %%{{\psframebox{{\Start{f}}}}}
\putnode{a}{m}{0}{-12}{\psframebox{$y = \&b$}} %%{\psframebox{$y = \&a;$}}
	\putnode{w}{a}{-9}{0}{$11$}
\putnode{c}{a}{0}{-12}{\psframebox{$\;\;\,p();\;\;\,$}}
	\putnode{w}{c}{-9}{0}{$12$}
\putnode{d}{c}{0}{-12}{\psframebox{\End{q}}}
%%%
\ncline{->}{m}{a}
\ncline{->}{a}{c}
\ncline{->}{c}{d}
\end{pspicture}
\end{tabular}
&
%%\begin{tabular}{c}
%%\begin{pspicture}(-2,0)(40,44)
%%%%\psframe(0,0)(40,40)
%%\putnode{m}{origin}{22}{41}{\pscirclebox[framesep=.2]{\begin{tabular}{@{}c@{}} \text{main} \end{tabular}}}
%%\putnode{a}{m}{0}{-14}{\pscirclebox[framesep=2]{$p$}}
%%\putnode{b}{a}{0}{-13}{\pscirclebox[framesep=2]{$q$}}
%%\ncline{->}{m}{a}
%%\ncline{->}{a}{b}
%%\nccurve[angleA=-45, angleB=45,nodesepA=-1.2, nodesepB=-1.2]{->}{b}{a}
%%\end{pspicture}
%%\end{tabular}
\begin{tabular}{c}
\begin{pspicture}(0,0)(66,42)
%\psframe(0,0)(66,42)
\putnode{m}{origin}{8}{38}{\pscirclebox[framesep=.2]{\begin{tabular}{@{}c@{}} \text{main} \end{tabular}}}
\putnode{a1}{m}{0}{-14}{\pscirclebox[framesep=2]{$p$}}
\putnode{b1}{a1}{0}{-15}{\pscirclebox[framesep=2]{$q$}}
{
\psset{arrowsize=1.5,arrowinset=0}
\ncline{->}{m}{a1}
\ncline{->}{a1}{b1}
\nccurve[ncurv=1,angleA=-45, angleB=45,nodesepA=-1.2, nodesepB=-1.2]{->}{b1}{a1}
}
%%%%%%
\putnode{a2}{a1}{15}{8}{$\mtsym^1_p$}
\putnode{a3}{a2}{16}{0}{$\mtsym^2_p$}
\putnode{a4}{a3}{16}{0}{$\mtsym^3_p$}
\putnode{w}{a4}{7}{0}{$\equiv\mtsym_p$}
%\putnode{a5}{a4}{16}{0}{$\mtsym^4_p$}
%
\putnode{b2}{b1}{15}{4}{$\mtsym^1_q$}
\putnode{b3}{b2}{16}{0}{$\mtsym^2_q$}
\putnode{b4}{b3}{16}{0}{$\mtsym^3_q$}
%%\putnode{b5}{b4}{16}{0}{$\mtsym^4_q$}
\putnode{w}{b4}{7}{0}{$\equiv\mtsym_q$}
\putnode{t}{b2}{0}{10}{$\mtsym_{\top}$}
{
\psset{doubleline=true,nodesepA=1,nodesepB=.25}
\ncline[nodesepA=0,nodesepB=-.5]{->}{t}{b3}
\ncline{->}{b3}{a3}
\ncline[nodesepA=.5,nodesepB=-1]{->}{a3}{b4}
\ncline{->}{b4}{a4}
\ncline[nodesepA=.5,nodesepB=-.5]{->}{a4}{b5}
%\ncline{->}{b5}{a5}
}
%%%%%%%
\psset{linestyle=dashed,dash=.6 .6,nodesepA=1,nodesepB=.25,arrowsize=1.5}
\ncline{->}{a2}{a3}
\ncline{->}{b2}{b3}
\nccurve[angleA=-20,angleB=200]{->}{b2}{b4}
\nccurve[angleA=-30,angleB=210]{->}{b2}{b5}
\nccurve[angleA=20,angleB=160,nodesepB=0]{->}{a2}{a4}
%\nccurve[angleA=30,angleB=150,nodesepB=0]{->}{a2}{a5}
\end{pspicture}
\end{tabular}
\\
\multicolumn{2}{c|}{}
	&
\\
\multicolumn{2}{c|}{\small (a) Mutually recursive procedures}
	&\small  (b) Call graph and the order of constructing \gpgs
\end{tabular}
\caption{Constructing \gpgs for recursive procedures by successive refinements.
}
\label{fig:recur_eg}
\end{figure}
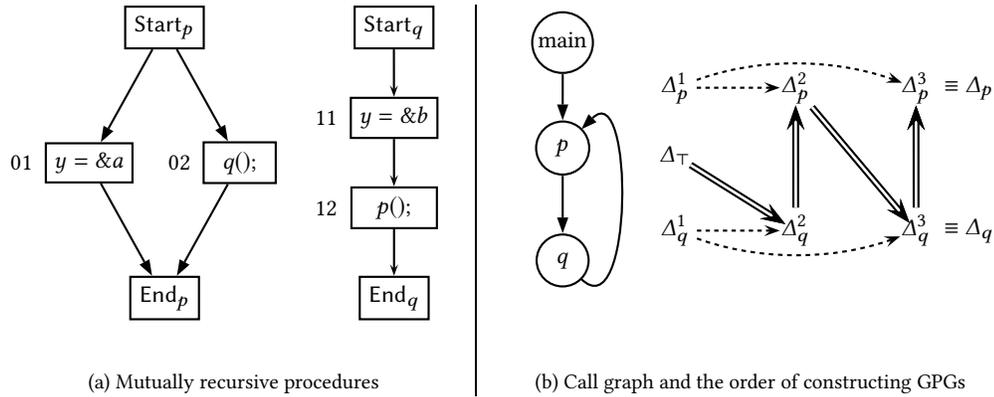

\subsection{Callee is Known and the Call is Non-Recursive}
\label{sec:func_calls}

In this case, the \gpg of the callee can be constructed completely before the \gpg of its callers
if we traverse the call graph bottom up. 
%%For example, in Figure~\ref{fig:mot_eg.2}, the \gpg $\mtsym_g$ of the callee procedure $g$ is 
%%inlined in the initial \gpg of procedure $f$ to construct $\mtsym_f$.  

We inline the optimized \gpgs of the callees at the call sites in the caller procedures.
\gpb labels are used for maintaining control flow within a \gpg. Hence, we renumber the 
\gpb labels after call inlining and coalescing. 
Note that if a \gpg is inlined multiple times then each inlining uses a fresh numbering.
Since the statement labels are unique across procedures, their
occurrences in \gpus do not change by inlining even if a \gpg is inlined at two
different call sites within the same procedure.
As noted earlier, 
this is a design choice because it helps us to accumulate the points-to information
of a particular statement in all contexts.

When inlining a callee's (optimized) \gpg, we add two new \gpbs, a predecessor to 
its \Start{} \gpb and a successor to
its \End{} \gpb. These new \gpbs contain respectively:
\begin{itemize}
\item \gpus that correspond to the actual-to-formal-parameter mapping.
\item A \gpu that maps the return variable of the callee to the receiver variable 
	of the call in the caller (or zero \gpus for a \texttt{void} function).
\end{itemize}
Some \gpus in the \gpg of the callee may have upwards-exposed versions of variables.
For example, if the callee reads a global variable $x$ defined in the caller, it
would have a \gpu referring to the initial value $x'$  (see Section~\ref{sec:may.must.xxp.edges}).
Hence when a \gpg is inlined in a caller procedure,
we substitute the callee's upwards-exposed variable $x'$ occurring in a callee's \gpu 
by the original variable $x$ when the \gpu is included in the caller's \gpg.
Note that $x$ may be a global variable or a formal parameter.

Inlining of procedure calls with the callee's optimized \gpg allows 
reaching \gpus analyses to remain intraprocedural analyses.
However, recursive and indirect calls need to be handled specially.
These cases are discussed in Section~\ref{sec:handling_recur} immediately below
and Section~\ref{sec:handling_fp}.

\begin{Definition} 
\begin{center}
\psframebox[framesep=5pt,doubleline=true,doublesep=1.5pt,linewidth=.2mm]{%
\handlingRecursionDefNew
}
\end{center}
\defcaption{Computing \gpgs for Recursive Procedures by Successive Refinement}{def:recursive.proc.gpg.new}
\end{Definition}

\begin{figure}[t]
%%\centering
\centering
\setlength{\codeLineLength}{20mm}
\renewcommand{\arraystretch}{.9}
\setlength{\tabcolsep}{1.8pt}
\begin{tabular}{|c|cc|cc|}
\hline
\rule[-.6em]{0em}{1.75em}%
\multirow{2}{*}{$\mtsym^{1}_{q}$}
&
\multicolumn{2}{c|}{$\mtsym^{2}_{q}$}
&
\multicolumn{2}{c|}{$\mtsym^{3}_{q}$}
%%&
%%\multicolumn{2}{c|}{$\mtsym^{4}_{q}$}
\\ \cline{2-5}
\rule[-.5em]{0em}{1.5em}
	& Unoptimized & Optimized 
	& Unoptimized & Optimized 
%%	& Unoptimized & Optimized 
\\ \hline\hline
\begin{tabular}{@{}c@{}}
\begin{pspicture}(0,-2)(28,42)
%\psframe(0,-2)(18,42)
\putnode{m}{origin}{15}{37}{\psframebox[framesep=2.5]{\;\;\;\;\;\;}}  %%{{\psframebox{{\Start{f}}}}}
	\putnode{w}{m}{-9}{0}{$\gpbsym_1$}
\putnode{a}{m}{0}{-12}{\psframebox{$\denew{y}{1|0}{b}{11}$}} %%{\psframebox{$y = \&a;$}}
	\putnode{w}{a}{-9}{0}{$\gpbsym_2$}
\putnode{b}{a}{0}{-12}{\psframebox[framesep=1]{\makebox[10mm]{$p()$}}} %%{\psframebox{$y = \&a;$}}
	\putnode{w}{b}{-9}{0}{$\gpbsym_3$}
\putnode{d}{b}{0}{-11}{\psframebox[framesep=2.5]{\;\;\;\;\;\;}}
	\putnode{w}{d}{-9}{0}{$\gpbsym_4$}
%%%
\psset{arrowsize=1.5,arrowinset=0}
\ncline{->}{m}{a}
\ncline{->}{a}{b}
\ncline{->}{b}{d}
\end{pspicture}
\end{tabular}
&
\begin{tabular}{@{}c@{}}
\begin{pspicture}(0,-2)(28,42)
%\psframe(0,-2)(18,42)
\putnode{m}{origin}{15}{37}{\psframebox[framesep=2.5]{\;\;\;\;\;\;}}  %%{{\psframebox{{\Start{f}}}}}
	\putnode{w}{m}{-9}{0}{$\gpbsym_1$}
\putnode{a}{m}{0}{-12}{\psframebox{$\denew{y}{1|0}{b}{11}$}} %%{\psframebox{$y = \&a;$}}
	\putnode{w}{a}{-9}{0}{$\gpbsym_2$}
\putnode{b}{a}{0}{-12}{\psframebox[framesep=1]{\makebox[10mm]{$\mtsym_{\top}$}}} %%{\psframebox{$y = \&a;$}}
	\putnode{w}{b}{-9}{0}{$\gpbsym_3$}
\putnode{d}{b}{0}{-11}{\psframebox[framesep=2.5]{\;\;\;\;\;\;}}
	\putnode{w}{d}{-9}{0}{$\gpbsym_4$}
%%%
\psset{arrowsize=1.5,arrowinset=0}
\ncline{->}{m}{a}
\ncline{->}{a}{b}
\ncline{->}{b}{d}
\end{pspicture}
\end{tabular}
&
\begin{tabular}{@{}c@{}}
\begin{pspicture}(0,-2)(18,42)
%\psframe(0,-2)(18,42)
\putnode{m}{origin}{9}{20}{\psframebox[framesep=1]{\makebox[10mm]{$\mtsym_{\top}$}}} %%{\psframebox{$y = \&a;$}}
\end{pspicture}
\end{tabular}
&
\begin{tabular}{@{}c@{}}
\begin{pspicture}(0,-2)(28,42)
%\psframe(0,-2)(18,42)
\putnode{m}{origin}{15}{37}{\psframebox[framesep=2.5]{\;\;\;\;\;\;}}  %%{{\psframebox{{\Start{f}}}}}
	\putnode{w}{m}{-9}{0}{$\gpbsym_1$}
\putnode{a}{m}{0}{-12}{\psframebox{$\denew{y}{1|0}{b}{11}$}} %%{\psframebox{$y = \&a;$}}
	\putnode{w}{a}{-9}{0}{$\gpbsym_2$}
\putnode{b}{a}{0}{-12}{\psframebox[framesep=1]{\denew{y}{1|0}{a}{01}}} %%{\psframebox{$y = \&a;$}}
	\putnode{w}{b}{-9}{0}{$\gpbsym_3$}
\putnode{d}{b}{0}{-11}{\psframebox[framesep=2.5]{\;\;\;\;\;\;}}
	\putnode{w}{d}{-9}{0}{$\gpbsym_4$}
%%%
\psset{arrowsize=1.5,arrowinset=0}
\ncline{->}{m}{a}
\ncline{->}{a}{b}
\ncline{->}{b}{d}
\end{pspicture}
\end{tabular}
&
\begin{tabular}{@{}c@{}}
\begin{pspicture}(0,-2)(28,42)
%\psframe(0,-2)(18,42)
\putnode{m}{origin}{15}{37}{\psframebox[framesep=2.5]{\;\;\;\;\;\;}}  %%{{\psframebox{{\Start{f}}}}}
	\putnode{w}{m}{-9}{0}{$\gpbsym_1$}
\putnode{a}{m}{0}{-12}{\psframebox{$\denew{y}{1|0}{a}{01}$}} %%{\psframebox{$y = \&a;$}}
	\putnode{w}{a}{-9}{0}{$\gpbsym_3$}
\putnode{d}{a}{0}{-11}{\psframebox[framesep=2.5]{\;\;\;\;\;\;}}
	\putnode{w}{d}{-9}{0}{$\gpbsym_4$}
%%%
\psset{arrowsize=1.5,arrowinset=0}
\ncline{->}{m}{a}
\ncline{->}{a}{d}
\end{pspicture}
\end{tabular}
%%&
%%\begin{tabular}{@{}c@{}}
%%\begin{pspicture}(0,-15)(18,42)
%%%\psframe(0,-15)(18,42)
%%\putnode{m}{origin}{9}{33}{\psframebox[framesep=2.5]{\;\;\;\;\;\;}}  %%{{\psframebox{{\Start{f}}}}}
%%	\putnode{w}{m}{-4}{5}{$\gpbsym_1$}
%%\putnode{a}{m}{0}{-14}{\psframebox{$\denew{y}{1|0}{b}{11}$}} %%{\psframebox{$y = \&a;$}}
%%	\putnode{w}{a}{-4}{7}{$\gpbsym_2$}
%%\putnode{b}{a}{0}{-16}{\psframebox[framesep=1]{\denew{y}{1|0}{a}{01}}} %%{\psframebox{$y = \&a;$}}
%%	\putnode{w}{b}{-4}{7}{$\gpbsym_3$}
%%\putnode{d}{b}{0}{-14}{\psframebox[framesep=2.5]{\;\;\;\;\;\;}}
%%	\putnode{w}{d}{-4}{5}{$\gpbsym_4$}
%%%%%
%%\psset{arrowsize=1.5,arrowinset=0}
%%\ncline{->}{m}{a}
%%\ncline{->}{a}{b}
%%\ncline{->}{b}{d}
%%\end{pspicture}
%%\end{tabular}
%%&
%%\begin{tabular}{@{}c@{}}
%%\begin{pspicture}(0,-15)(18,42)
%%%\psframe(0,-15)(18,42)
%%\putnode{m}{origin}{9}{33}{\psframebox[framesep=2.5]{\;\;\;\;\;\;}}  %%{{\psframebox{{\Start{f}}}}}
%%	\putnode{w}{m}{-4}{5}{$\gpbsym_1$}
%%\putnode{a}{m}{0}{-14}{\psframebox{$\denew{y}{1|0}{a}{01}$}} %%{\psframebox{$y = \&a;$}}
%%	\putnode{w}{a}{-4}{7}{$\gpbsym_3$}
%%\putnode{d}{a}{0}{-14}{\psframebox[framesep=2.5]{\;\;\;\;\;\;}}
%%	\putnode{w}{d}{-4}{5}{$\gpbsym_4$}
%%%%%
%%\psset{arrowsize=1.5,arrowinset=0}
%%\ncline{->}{m}{a}
%%\ncline{->}{a}{d}
%%\end{pspicture}
%%\end{tabular}
\\ \hline
\end{tabular}
%%\caption{Series of \gpgs of procedure $q$ 
%%of Figure~\protect\ref{fig:recur_eg}.
%%$\mtsym^1_q$ is the initial \gpg containing a recursive call to $p$.
%%$\mtsym^2_q$ is constructed by inlining call to $p$ with $\mtsym_{\top}$ (second column)
%%and $\mtsym^3_q$ is constructed by inlining call to $p$ with $\mtsym^2_{p}$ (third column).
%%Th final optimized $\mtsym_q$ is shown in the fourth column.
%%%% in Figure~\protect\ref{fig:recur_proc_q}.
%%}
%%\label{fig:recur_proc_q}
%%\end{center}
%%\end{figure}
%%
%%\begin{figure}[t]
%%\centering
%%\setlength{\codeLineLength}{20mm}
%%\renewcommand{\arraystretch}{.9}

\bigskip
\setlength{\tabcolsep}{3pt}
\begin{tabular}{|c|cc|cc|}
\hline
\rule[-.6em]{0em}{1.75em}%
\multirow{2}{*}{$\mtsym^{1}_{p}$}
&
\multicolumn{2}{c|}{$\mtsym^{2}_{p}$}
&
\multicolumn{2}{c|}{$\mtsym^{3}_{p}$}
\\ \cline{2-5}
\rule[-.5em]{0em}{1.5em}
	& Unoptimized & Optimized 
	& Unoptimized & Optimized 
\\ \hline\hline
\begin{tabular}{@{}c@{}}
\begin{pspicture}(1,3)(29,40)
%\psframe(1,0)(29,40)
\putnode{m}{origin}{16}{35}{\psframebox[framesep=2.5]{\;\;\;\;\;\;}}  %%{{\psframebox{{\Start{f}}}}}
	\putnode{w}{m}{-8}{0}{$\gpbsym_5$}
\putnode{a}{m}{-7}{-14}{\psframebox{$\denew{y}{1|0}{a}{01}$}} %%{\psframebox{$y = \&a;$}}
	\putnode{w}{a}{-2}{7}{$\gpbsym_6$}
\putnode{b}{m}{7}{-14}{\psframebox[framesep=2]{$q()$}} %%{\psframebox{$y = \&a;$}}
	\putnode{w}{b}{2}{7}{$\gpbsym_7$}
\putnode{d}{m}{0}{-28}{\psframebox[framesep=2.5]{\;\;\;\;\;\;}}
	\putnode{w}{d}{-9}{0}{$\gpbsym_8$}
%%%
\psset{arrowsize=1.5,arrowinset=0}
\ncline{->}{m}{a}
\ncline{->}{m}{b}
\ncline{->}{a}{d}
\ncline{->}{b}{d}
\end{pspicture}
\end{tabular}
&
\begin{tabular}{@{}c@{}}
\begin{pspicture}(2,3)(30,40)
%\psframe(2,0)(30,40)
\putnode{m}{origin}{17}{35}{\psframebox[framesep=2.5]{\;\;\;\;\;\;}}  %%{{\psframebox{{\Start{f}}}}}
	\putnode{w}{m}{-8}{0}{$\gpbsym_5$}
\putnode{a}{m}{-7}{-14}{\psframebox{$\denew{y}{1|0}{a}{01}$}} %%{\psframebox{$y = \&a;$}}
	\putnode{w}{a}{-2}{7}{$\gpbsym_6$}
\putnode{b}{m}{7}{-14}{\psframebox[framesep=2]{$\mtsym_{\top}$}} %%{\psframebox{$y = \&a;$}}
	\putnode{w}{b}{2}{7}{$\gpbsym_7$}
\putnode{d}{m}{0}{-28}{\psframebox[framesep=2.5]{\;\;\;\;\;\;}}
	\putnode{w}{d}{-8}{0}{$\gpbsym_8$}
%%%
\psset{arrowsize=1.5,arrowinset=0}
\ncline{->}{m}{a}
\ncline{->}{m}{b}
\ncline{->}{a}{d}
\ncline{->}{b}{d}
\end{pspicture}
\end{tabular}
&
\begin{tabular}{@{}c@{}}
\begin{pspicture}(0,3)(19,40)
%\psframe(2,0)(18,40)
\putnode{m}{origin}{11}{35}{\psframebox[framesep=2.5]{\;\;\;\;\;\;}}  %%{{\psframebox{{\Start{f}}}}}
	\putnode{w}{m}{-8}{0}{$\gpbsym_5$}
\putnode{a}{m}{0}{-14}{\psframebox{$\denew{y}{1|0}{a}{01}$}} %%{\psframebox{$y = \&a;$}}
	\putnode{w}{a}{-4}{7}{$\gpbsym_9$}
\putnode{d}{a}{0}{-14}{\psframebox[framesep=2.5]{\;\;\;\;\;\;}}
	\putnode{w}{d}{-8}{0}{$\gpbsym_8$}
%%%
\psset{arrowsize=1.5,arrowinset=0}
\ncline{->}{m}{a}
\ncline{->}{a}{d}
\end{pspicture}
\end{tabular}
&
\begin{tabular}{@{}c@{}}
\begin{pspicture}(1,3)(33,40)
%\psframe(1,0)(33,40)
\putnode{m}{origin}{17}{35}{\psframebox[framesep=2.5]{\;\;\;\;\;\;}}  %%{{\psframebox{{\Start{f}}}}}
	\putnode{w}{m}{-8}{0}{$\gpbsym_5$}
\putnode{a}{m}{-8}{-14}{\psframebox{$\denew{y}{1|0}{a}{01}$}} %%{\psframebox{$y = \&a;$}}
	\putnode{w}{a}{-2}{7}{$\gpbsym_6$}
\putnode{b}{m}{8}{-14}{\psframebox{$\denew{y}{1|0}{a}{01}$}} %%{\psframebox{$y = \&a;$}}
	\putnode{w}{b}{2}{7}{$\gpbsym_7$}
\putnode{d}{m}{0}{-28}{\psframebox[framesep=2.5]{\;\;\;\;\;\;}}
	\putnode{w}{d}{-8}{0}{$\gpbsym_8$}
%%%
\psset{arrowsize=1.5,arrowinset=0}
\ncline{->}{m}{a}
\ncline{->}{m}{b}
\ncline{->}{a}{d}
\ncline{->}{b}{d}
\end{pspicture}
\end{tabular}
&
\begin{tabular}{@{}c@{}}
\begin{pspicture}(0,3)(19,40)
%\psframe(2,0)(18,40)
\putnode{m}{origin}{11}{35}{\psframebox[framesep=2.5]{\;\;\;\;\;\;}}  %%{{\psframebox{{\Start{f}}}}}
	\putnode{w}{m}{-8}{0}{$\gpbsym_5$}
\putnode{a}{m}{0}{-14}{\psframebox{$\denew{y}{1|0}{a}{01}$}} %%{\psframebox{$y = \&a;$}}
	\putnode{w}{a}{-4}{7}{$\gpbsym_9$}
\putnode{d}{a}{0}{-14}{\psframebox[framesep=2.5]{\;\;\;\;\;\;}}
	\putnode{w}{d}{-8}{0}{$\gpbsym_8$}
%%%
\psset{arrowsize=1.5,arrowinset=0}
\ncline{->}{m}{a}
\ncline{->}{a}{d}
\end{pspicture}
\end{tabular}
\\ \hline
\end{tabular}
\caption{Series of \gpgs of procedures $p$  and $q$
of Figure~\protect\ref{fig:recur_eg}. 
They are computed in the order shown in 
Figure~\protect\ref{fig:recur_eg}(b).
See Example~\protect\ref{eg:fixed-pt-recur} for explanation.
%%The data flow values reaching \End{p} are identical for 
%%$\mtsym^2_p$ and $\mtsym^3_p$ hence the optimized version of $\mtsym^3_p$ is considered the final $\mtsym_p$. 
%%Although the data flow values reaching \End{} \gpb of $\mtsym^3_q$ are different from those
%%in $\mtsym^2_q$, $\mtsym^3_q$ uses $\mtsym^2_p$ whose effect is same as that of $\mtsym^3_p$.
%%Hence, $\mtsym^4_q$ will have the same effect as $\mtsym^3_q$ and hence is not constructed.
}
\label{fig:recur_proc_p_and_q}
\end{figure}
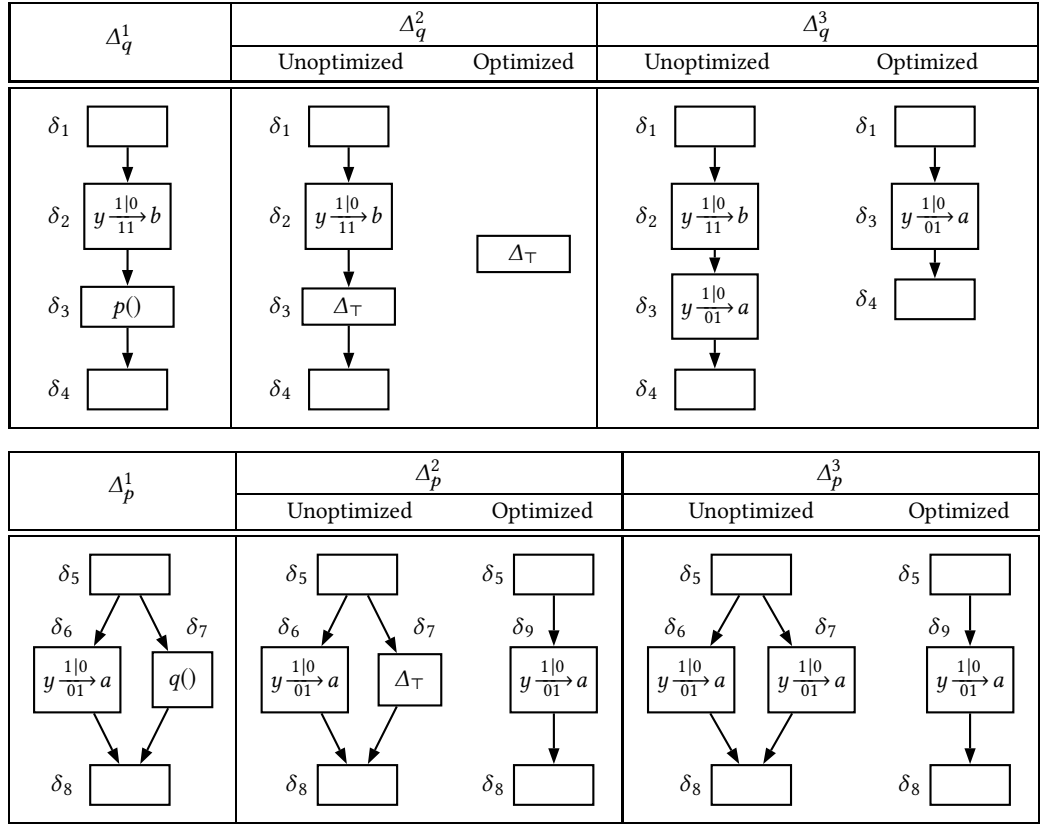

\subsection{Callee is Known and the Call is Recursive}
\label{sec:handling_recur}

Consider Figure~\ref{fig:recur_eg} in which
  procedure $p$ calls procedure $q$ and $q$ calls $p$.
The \gpg of $q$ depends on that of $p$ and vice-versa leading to
\emph{incomplete} \gpgs:  the \gpgs of the callees 
of some calls either have not been constructed or are incomplete.
We handle this mutual dependency by successive refinement of incomplete \gpgs of $p$ and $q$ through a fixed-point computation.

A set of recursive procedures is represented by a strongly connected component in a call graph 
    which is formed by a collection of back edges that represent recursive calls.
Since we traverse a call graph bottom up, the construction of \gpgs for a set of recursive
procedures begins with the
procedures that are the sources of back edges.
The \gpgs of some callees of these procedures (i.e. the callees that are targets of back edges in the call graph)
have not been constructed yet.
We handle such situations by using a special \gpg
 $\mtsym_{\top}$ that represents the effect of a call when the callee's \gpg is not available.
The \gpg $\mtsym_{\top}$ is the $\top$ element of the lattice of all possible procedure summaries.
It kills all \gpus and generates none (thereby, when applied,
computes the $\top$ value--- $\emptyset$---of the lattice
for \may points-to analysis)~\cite{dfa_book}.
Semantically, 
$\mtsym_{\top}$ corresponds to 
the call to a procedure that
never returns (e.g.\ loops forever).
It consists of a special \gpb called the 
\emph{call} \gpb whose flow functions are constant functions
computing the empty set of \gpus for both variants of reaching \gpus analysis.

We perform the reaching \gpus analyses over incomplete \gpgs containing recursive calls
by repeated inlining of callees starting with $\mtsym_{\top}$ as their initial \gpgs, 
until no further inlining is required. This is achieved as follows:
Since data flow analysis over incomplete \gpgs under-approximates the effect of
some calls through $\mtsym_{\top}$, the data flow values so computed 
need to be refined further. This is achieved by
inlining the calls by including incomplete \gpgs of the callees to compute a 
new \gpg over which the data flow analysis is repeated. 
Let $\mtsym^1_p$ denote the \gpg of procedure $p$ in which all the calls to the procedures that are not 
part of the strongly connected component 
are inlined by their respective optimized \gpgs. 
Note that the \gpgs of these procedures have already been constructed because
of the bottom up traversal over the call graph. The calls to procedures that are part of the strongly connected component are retained
in $\mtsym^1_p$. 
In each step of refinement, the recursive calls in $\mtsym^1_p$ are inlined either
\begin{itemize}
\item by $\mtsym_{\top}$ when no \gpg of the callee has been constructed, or
\item by an incomplete \gpg of a callee in which some calls are under-approximated using 
$\mtsym_{\top}$. 
\end{itemize}

Thus we compute a series of \gpgs $\mtsym^i_p$, $i>1$ for every procedure $p$ in a
strongly connected component until the termination of fixed-point computation.
For this purpose, we initialize a worklist with all procedures in a strongly connected
component. This worklist is ordered by the postorder relation between the procedures in the
call graph. A procedure is added to the worklist based on the following criterion;  
the process terminates when the worklist becomes empty.
Once $\mtsym^i_p$ is constructed, we decide to construct $\mtsym^j_q$
	for a caller $q$ of $p$ if the data flow values of the \End{} \gpb of 
	$\mtsym^{i}_p$ differ from
	those of the \End{} \gpb of $\mtsym^{i-1}_p$. This is because the overall 
	effect of a procedure on its callers is reflected by the values reaching its \End{} \gpb
	(because of forward flow of information in points-to analysis). 
If the data values of the \End{} \gpbs of $\mtsym^{i-1}_p$ and $\mtsym^i_p$ 
are same, then they would have identical effect	on their callers. Thus, 
the \gpgs are semantically identical as procedure summaries 
even if they differ structurally. This step is described in
Definition~\ref{def:recursive.proc.gpg.new}. 

The convergence of this fixed-point computation differs subtly from the usual fixed-point 
computation in the following manner:
in each step of computation, the \gpgs continue to change. {And yet,} we stop 
the fixed-point computation when the \emph{data flow} values of the \End{} \gpb
converge
across the changing \gpgs, not when the resultant GPGs converge. 

\begin{example}{eg:fixed-pt-recur}
In the example of Figure~\ref{fig:recur_eg}, the sole
strongly connected component contains procedures $p$ and $q$. Since
procedure $q$ is the source of the back edge in the call graph, the \gpg of procedure $q$ is constructed first.
There are no calls in procedure $q$ to procedures outside the strongly connected component.
Thus, $\mtsym^1_q$ contains a single call to procedure $p$ whose \gpg is not constructed yet and hence
the construction of $\mtsym^2_q$ requires inlining of $\mtsym_{\top}$. 
Since $\mtsym_{\top}$ represents a procedure call which never returns,
the \gpb \End{g} becomes unreachable from the rest of the \gpbs in $\mtsym^2_q$.
The optimized $\mtsym^2_q$ is $\mtsym_{\top}$ because all \gpbs that
no longer appear on a control flow path from the \Start{} \gpb to the
\End{} \gpb are removed from the \gpg, thereby garbage-collecting unreachable \gpbs.
$\mtsym^1_p$ contains a single call to procedure $q$ whose incomplete \gpg 
$\mtsym^2_q$, which is $\mtsym_{\top}$,
is inlined during construction of $\mtsym^2_p$.
The optimized version of $\mtsym^2_p$ is shown in Figure~\ref{fig:recur_proc_p_and_q}.
Then,
$\mtsym^2_p$ is used to construct $\mtsym^3_q$.
Reaching \gpus analyses with and without blocking 
are performed on $\mtsym^2_q$ and $\mtsym^3_q$. 
The data flow values for $\mtsym^2_q$ are
\text{\Rprev $ = \RprevBar = \emptyset$} 
whereas the data flow values for $\mtsym^3_q$ are
\text{$\Rcur = \RcurBar = \{\denew{y}{1|0}{a}{01}\}$}.
Since the data flow values have
changed, caller of $q$ i.e., $p$ is pushed on the worklist and $\mtsym^3_p$ is constructed by inlining $\mtsym^3_q$. 
%%{\revTwo Also, an additional round of inlining is required in $\mtsym^1_q$ after constructing $\mtsym^3_p$.}
The data flow values computed for $\mtsym^2_p$ and $\mtsym^3_p$ are identical
\text{$\Rprev = \RprevBar = \Rcur = \RcurBar = \{\denew{y}{1|0}{a}{01}\}$} and hence caller of $p$ i.e., procedure $q$ is not added to the worklist.
The worklist becomes empty and hence the process terminates.
Note that the data flow values of $\mtsym^2_q$ and $\mtsym^3_q$ differ and yet we do not construct the \gpg $\mtsym^4_q$. This is because
$\mtsym^4_q$ constructed by inlining $\mtsym^3_p$ will have the same effect as that of $\mtsym^3_q$ constructed by inlining $\mtsym^2_p$ since 
the impact of $\mtsym^2_p$ and $\mtsym^3_p$ is identical. 
\end{example}

The process of fixed-point computation
is guaranteed to terminate because of the finiteness of 
the set of \gpus \Rprev, \RprevBar, \Rcur, \RcurBar:
For two variables $x$ and $y$, the number of \gpus \denew{x}{i|j}{y}{\flab} depends on the 
number of possible \indlevs ($i|j$) and the number of statements.
Since the number of statements is finite,  we need to examine the number of \indlevs.
For pointers to scalars, the number of \indlevs between any two variables is bounded
because of
type restrictions. For pointers to structures 
(Section~\ref{sec:heap}), 
\indlevs are replaced by indirection lists (\indlists). 
Sections~\ref{sec:alloc-summ} and~\ref{sec:k-limmiting} 
summarize \indlists restricting them to a finite number.
Hence the number of \gpus is also finite.

\section{Computing Points-to Information using \gpgs}
\label{sec:dfv_compute}

The second phase of a bottom-up approach which uses procedure summaries created in the
first phase, is redundant in our method. This is
because our first phase computes the points-to information
as a side-effect of the construction of \gpgs.

Since we also need points-to information for statements that read pointers but do not define them,
we model them as \emph{use} statements.
Consider a use of a pointer variable in a non-pointer assignment or an expression. We
represent such a use with a \gpu whose source is a fictitious node
\usenode with \indlev 1 and the target is the pointee which is being read. 
Thus a condition `$\tt if \; (x == *y)$' where both $x$ and $y$ are pointers,
is modelled as a \gpb \text{$\left\{\denew{\usenode}{1|1}{x}{\flab}, \denew{\usenode}{1|2}{y}{\flab}\right\}$} whereas
an integer assignment `$\tt *x = 5;$' is modelled as a \gpb 
\text{$\left\{\denew{\usenode}{1|2}{x}{\flab}\right\}$}. 

\begin{example}{}
Consider the code snippet on the right.
There is a non-pointer assignment in
\setlength{\intextsep}{0mm}%
\setlength{\columnsep}{2mm}%
\begin{wrapfigure}{r}{24mm}
\setlength{\codeLineLength}{15mm}%
\renewcommand{\arraystretch}{.7}%
	\begin{tabular}{|rc}
	%\hline
   	\codeLineOne{1}{0}{$\tt x = \&a;$}{white}
	\codeLine{0}{$\tt *x = 5;$}{white} 
	 %\hline
	\end{tabular}
\end{wrapfigure}
which the
pointee of $x$ (which is the location $a$) is being defined. 
A client analysis would like to know the pointees of $x$ for statement 02. 
We model this use of pointee of $x$ as a \gpu \denew{\usenode}{1|2}{x}{02}.
This \gpu can be composed with \denew{x}{1|0}{a}{01} to get a reduced
\gpu \denew{\usenode}{1|1}{a}{02} indicating that pointee of $x$ in statement 2 is $a$. 
\end{example}

When a use involves multiple pointers such as `$\tt if \; (x == *y)$', the corresponding \gpb
contains multiple \gpus. If the exact pointer-pointee relationship
is required, rather than just the reduced form of the use (devoid of pointers), we need
additional minor bookkeeping to record \gpus and the corresponding pointers.

With the provision of a \gpu for a use statement, the process of computing 
points-to information can be seen as a two step process:
\begin{itemize}
\item creating def-use or use-def chains for pointers to view producer \gpus as definitions of pointers and consumer \gpus
      as the use of pointers, and 
\item performing strength reduction of the consumer \gpus using the information from the producer \gpus to reduce the
      \indlevs of the consumer \gpus.
\end{itemize}
Since our
first phase does this for constructing procedure summaries, it is sufficient to
compute points-to information in the first phase.

This process is easy to visualize if the definitions and uses are in the same procedure. 
Consider a producer \gpu \prevedge and
a consumer \gpu \candedge that are not in the same procedure. 
We can facilitate strength reduction involving them by
\begin{enumerate}[(a)]
\item propagating \prevedge to the procedure containing \candedge,
\item propagating \candedge to the procedure containing \prevedge, or
\item propagating both \prevedge and \candedge to a common procedure.
\end{enumerate}
The propagation of information in cases (a) and (b) is similar to that in a 
top-down analysis; case (a) corresponds to a forward analysis and case (b) 
corresponds to a backward analysis. However, 
case (c) is only possible in 
bottom-up analysis.

A typical second phase of a bottom-up approach involves propagation of 
information similar to cases (a) and (b). 
This is illustrated in Example~\ref{eg.phase2.pta}. 
We use propagation similar to case (c)
which is subsumed in the first phase of a bottom-up approach rendering
the second phase redundant.  It is illustrated in Example~\ref{eg.phase1.pta}. 

\begin{example}{eg.phase2.pta}
Consider procedures $f$, $g$, $h$ and $s$ defined in Figure~\ref{fig:pta}. 
We can facilitate strength reduction in the following ways for cases (a) and (b):
\begin{itemize}
\item \emph{Propagating \prevedge to the procedure containing \candedge}.
      A top-down forward analysis would propagate the \gpu
	\denew{x}{1|0}{a}{1} from procedure $f$ to  procedure $g$.
\item \emph{Propagating \candedge to the procedure containing \prevedge}.
      A top-down backward analysis in the spirit of liveness could propagate the \gpu
	\denew{y}{1|1}{x}{4} from procedure $g$ to  procedure $f$.
\end{itemize}
\end{example}

We handle case (c) by interleaved call inlining and strength reduction.
Call inlining enhances the opportunities for strength reduction by providing more information from the callers.
The interleaving of strength reduction and call inlining 
gradually converts a \gpu
\denew{x}{i|j}{y}{\flab} to a set of points-to edges 
\text{$\{\denew{a}{1|0}{b}{\flab} \mid a$ is $i^{th}$ pointee of $x$, $b$ is $j^{th}$ pointee of $y$\}}.
This is achieved by propagating the \emph{use}
of a pointer\footnote{This use could be in a pointer assignment or a use statement.} and its \emph{definitions} to a common context.
This may require propagating:
\begin{enumerate}
\item a consumer \gpu (i.e. a use of a pointer variable) to a caller,
\item a producer \gpu (i.e. a definition of a pointer variable) to a caller, 
\item both consumer and producer \gpus involving a pointer variable to a caller, and
\item neither (if they are same in the procedure).
\end{enumerate}

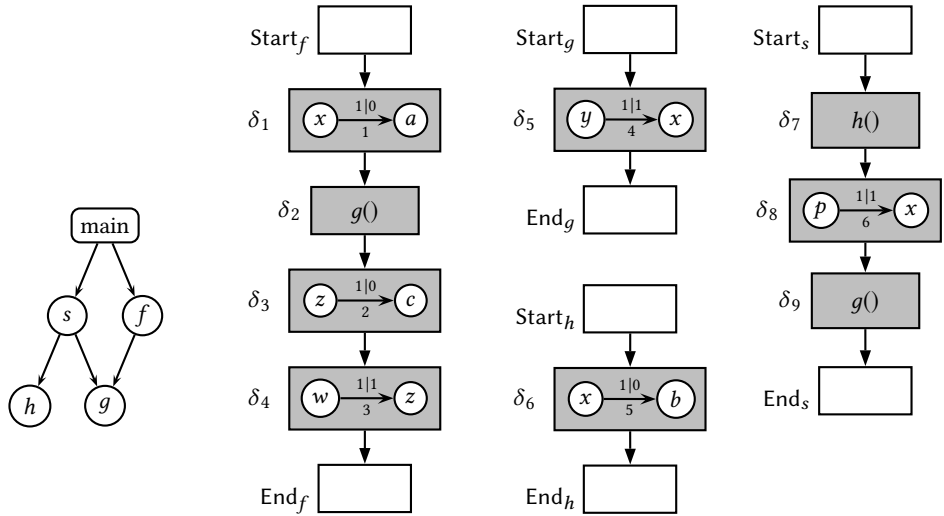
\begin{figure}
	\newcommand{\gpgA}{%
		\begin{pspicture}(0,1)(17,7)
		%\psframe(0,0)(32,39)
		\putnode{x}{origin}{3}{4}{\pscirclebox[fillstyle=solid,fillcolor=white,framesep=1]{$x$}}
		\putnode{a}{x}{12}{0}{\pscirclebox[fillstyle=solid,fillcolor=white,framesep=1]{$a$}}
%%%%%%%%%
		\ncline[arrowsize=1.5]{->}{x}{a}
		\naput[labelsep=0.5,npos=.5]{\scriptsize $1|0$}
		\nbput[labelsep=0.5,npos=.5]{\scriptsize $1$}
		\end{pspicture}
	}
	\newcommand{\gpgB}{%
		\begin{pspicture}(0,1)(17,7)
		%\psframe(0,0)(32,39)
		\putnode{x}{origin}{3}{4}{\pscirclebox[fillstyle=solid,fillcolor=white,framesep=1]{$z$}}
		\putnode{a}{x}{12}{0}{\pscirclebox[fillstyle=solid,fillcolor=white,framesep=1]{$c$}}
%%%%%%%%%
		\ncline[arrowsize=1.5]{->}{x}{a}
		\naput[labelsep=0.5,npos=.5]{\scriptsize $1|0$}
		\nbput[labelsep=0.5,npos=.5]{\scriptsize $2$}
		\end{pspicture}
	}
	\newcommand{\gpgC}{%
		\begin{pspicture}(0,1)(17,7)
		%\psframe(0,0)(32,39)
		\putnode{x}{origin}{3}{4}{\pscirclebox[fillstyle=solid,fillcolor=white,framesep=1]{$w$}}
		\putnode{a}{x}{12}{0}{\pscirclebox[fillstyle=solid,fillcolor=white,framesep=1]{$z$}}
%%%%%%%%%
		\ncline[arrowsize=1.5]{->}{x}{a}
		\naput[labelsep=0.5,npos=.5]{\scriptsize $1|1$}
		\nbput[labelsep=0.5,npos=.5]{\scriptsize $3$}
		\end{pspicture}
	}
	\newcommand{\gpgD}{%
		\begin{pspicture}(0,1)(17,7)
		%\psframe(0,0)(32,39)
		\putnode{x}{origin}{3}{4}{\pscirclebox[fillstyle=solid,fillcolor=white,framesep=1]{$y$}}
		\putnode{a}{x}{12}{0}{\pscirclebox[fillstyle=solid,fillcolor=white,framesep=1]{$x$}}
%%%%%%%%%
		\ncline[arrowsize=1.5]{->}{x}{a}
		\naput[labelsep=0.5,npos=.5]{\scriptsize $1|1$}
		\nbput[labelsep=0.5,npos=.5]{\scriptsize $4$}
		\end{pspicture}
	}
	\newcommand{\gpgE}{%
		\begin{pspicture}(0,1)(17,7)
		%\psframe(0,0)(32,39)
		\putnode{x}{origin}{3}{4}{\pscirclebox[fillstyle=solid,fillcolor=white,framesep=1]{$x$}}
		\putnode{a}{x}{12}{0}{\pscirclebox[fillstyle=solid,fillcolor=white,framesep=1]{$b$}}
%%%%%%%%%
		\ncline[arrowsize=1.5]{->}{x}{a}
		\naput[labelsep=0.5,npos=.5]{\scriptsize $1|0$}
		\nbput[labelsep=0.5,npos=.5]{\scriptsize $5$}
		\end{pspicture}
	}
	\newcommand{\gpgF}{%
		\begin{pspicture}(0,1)(17,7)
		%\psframe(0,0)(32,39)
		\putnode{x}{origin}{3}{4}{\pscirclebox[fillstyle=solid,fillcolor=white,framesep=.8]{$p$}}
		\putnode{a}{x}{12}{0}{\pscirclebox[fillstyle=solid,fillcolor=white,framesep=1]{$x$}}
%%%%%%%%%
		\ncline[arrowsize=1.5]{->}{x}{a}
		\naput[labelsep=0.5,npos=.5]{\scriptsize $1|1$}
		\nbput[labelsep=0.5,npos=.5]{\scriptsize $6$}
		\end{pspicture}
	}
\begin{tabular}{cccc}
\begin{tabular}{@{}c}
\begin{pspicture}(0,21)(30,90)
\putnode{n1}{origin}{15}{60}{\psframebox[framearc=.5]{main}}
\putnode{n2}{n1}{-5}{-12}{\pscirclebox[framesep=1.5]{$s$}}
\putnode{n3}{n1}{5}{-12}{\pscirclebox[framesep=	.7]{$f$}}
\putnode{n4}{n2}{-5}{-12}{\pscirclebox[framesep=1.2]{$h$}}
\putnode{n5}{n2}{5}{-12}{\pscirclebox[framesep=1.2]{$g$}}
\ncline[nodesepB=-.25]{->}{n1}{n2}
\ncline[nodesepB=-.25]{->}{n1}{n3}
\ncline[nodesep=-.25]{->}{n2}{n4}
\ncline[nodesep=-.25]{->}{n2}{n5}
\ncline[nodesep=-.25]{->}{n3}{n5}
\end{pspicture}
\end{tabular}
\begin{tabular}{@{}c}
\begin{pspicture}(0,21)(30,90)
%\psframe(0,21)(30,90)
\putnode{n0}{origin}{17}{95}{}
\putnode{f1}{n0}{0}{-9}{{\Start{f}$\;$\psframebox[framesep=1]{\rule{10mm}{0mm}\rule{0mm}{4mm}}\white$\;$\Start{f}}}
\putnode{n1}{f1}{0}{-12}{\psframebox[fillstyle=solid,fillcolor=lightgray,framesep=1]{\gpgA}}
	\putnode{w}{n1}{-14}{0}{$\gpbsym_{1}$}
\putnode{p1}{n1}{0}{-12}{\psframebox[fillstyle=solid,fillcolor=lightgray,framesep=1]{\makebox[12mm]{\rule[-1mm]{0mm}{4mm}$g()$}}}
	\putnode{w}{p1}{-10}{0}{$\gpbsym_{2}$}
\putnode{p3}{p1}{0}{-12}{\psframebox[fillstyle=solid,fillcolor=lightgray,framesep=1]{\gpgB}}
	\putnode{w}{p3}{-14}{0}{$\gpbsym_{3}$}
\putnode{n2}{p3}{0}{-13}{\psframebox[fillstyle=solid,fillcolor=lightgray,framesep=1]{\gpgC}}
	\putnode{w}{n2}{-14}{0}{$\gpbsym_{4}$}
\putnode{f2}{n2}{0}{-12}{\End{f}$\;$\psframebox[framesep=1]{\rule{10mm}{0mm}\rule{0mm}{4mm}}\white$\;$\End{f}}
\psset{arrowsize=1.5,arrowinset=0}
\ncline{->}{f1}{n1}
\ncline{->}{n1}{p1}
\ncline{->}{p1}{p3}
\ncline{->}{p3}{n2}
\ncline{->}{n2}{f2}
\end{pspicture}
\end{tabular}
&
\begin{tabular}{@{}c}
\begin{pspicture}(0,57)(30,90)
%\psframe(0,57)(30,90)
\putnode{n0}{origin}{17}{96}{}
\putnode{f1}{n0}{0}{-9}{{\Start{g}$\;$\psframebox[framesep=1]{\rule{10mm}{0mm}\rule{0mm}{4mm}}\white$\;$\Start{g}}}
\putnode{n1}{f1}{0}{-12}{\psframebox[fillstyle=solid,fillcolor=lightgray,framesep=1]{\gpgD}}
	\putnode{w}{n1}{-14}{0}{$\gpbsym_{5}$}
\putnode{f2}{n1}{0}{-12}{\End{g}$\;$\psframebox[framesep=1]{\rule{10mm}{0mm}\rule{0mm}{4mm}}\white$\;$\End{g}}
\psset{arrowsize=1.5,arrowinset=0}
\ncline{->}{f1}{n1}
\ncline{->}{n1}{f2}
\end{pspicture}
\\
\begin{pspicture}(0,57)(30,90)
%%\psframe(0,0)(30,95)
\putnode{n0}{origin}{17}{93}{}
\putnode{f1}{n0}{0}{-9}{{\Start{h}$\;$\psframebox[framesep=1]{\rule{10mm}{0mm}\rule{0mm}{4mm}}\white$\;$\Start{h}}}
\putnode{n1}{f1}{0}{-12}{\psframebox[fillstyle=solid,fillcolor=lightgray,framesep=1]{\gpgE}}
	\putnode{w}{n1}{-14}{0}{$\gpbsym_{6}$}
\putnode{f2}{n1}{0}{-12}{\End{h}$\;$\psframebox[framesep=1]{\rule{10mm}{0mm}\rule{0mm}{4mm}}\white$\;$\End{h}}
\psset{arrowsize=1.5,arrowinset=0}
\ncline{->}{f1}{n1}
\ncline{->}{n1}{f2}
\end{pspicture}
\end{tabular}
&
\begin{tabular}{@{}c}
\begin{pspicture}(4,21)(30,90)
%%\psframe(0,0)(30,95)
\putnode{n0}{origin}{17}{95}{}
\putnode{f1}{n0}{0}{-9}{{\Start{s}$\;$\psframebox[framesep=1]{\rule{10mm}{0mm}\rule{0mm}{4mm}}\white$\;$\Start{s}}}
\putnode{n1}{f1}{0}{-12}{\psframebox[fillstyle=solid,fillcolor=lightgray,framesep=1]{\makebox[12mm]{\rule[-1.5mm]{0mm}{5mm}$h()$}}}
	\putnode{w}{n1}{-10}{0}{$\gpbsym_{7}$}
\putnode{p1}{n1}{0}{-12}{\psframebox[fillstyle=solid,fillcolor=lightgray,framesep=1]{\gpgF}}
	\putnode{w}{p1}{-13}{0}{$\gpbsym_{8}$}
\putnode{p2}{p1}{0}{-12}{\psframebox[fillstyle=solid,fillcolor=lightgray,framesep=1]{\makebox[12mm]{\rule[-1.5mm]{0mm}{5mm}$g()$}}}
	\putnode{w}{p2}{-10}{0}{$\gpbsym_{9}$}
\putnode{f2}{p2}{0}{-12}{\End{s}$\;$\psframebox[framesep=1]{\rule{10mm}{0mm}\rule{0mm}{4mm}}\white$\;$\End{s}}
\psset{arrowsize=1.5,arrowinset=0}
\ncline{->}{f1}{n1}
\ncline{->}{n1}{p1}
\ncline{->}{p1}{p2}
\ncline{->}{p2}{f2}
\end{pspicture}
\end{tabular}
\end{tabular}
\caption{Computing points-to information using \gpgs. The first column gives the call graph while the other columns give \gpgs
before call inlining. The \gpg of procedure \emph{main} has been omitted.}
\label{fig:pta}
\end{figure}

Since statement numbers are unique across all procedures
and 
are not renamed on inlining,
the points-to edges computed across different contexts for a given statement 
represent the flow- and context-sensitive points-to information for the statement.

\begin{example}{eg.phase1.pta}
The four variants of hoisting
\prevedge and \candedge to a common procedure in the first phase of a bottom-up method are illustrated below.
Effectively, they make the second phase redundant.
%%Effectively, they obviate the second phase.
	\begin{enumerate}[(c.1)]
         \item When $\mtsym_g$ is inlined in $f$, 
	       \candedge\!:\,\denew{y}{1|1}{x}{4} from procedure $g$ is hoisted to procedure 
		$f$ that contains 
	       \gpu \prevedge\!:\,\denew{x}{1|0}{a}{1} thereby propagating the use of pointer $x$ in
		procedure $g$ to caller $f$.
		Strength reduction reduces \candedge to \denew{y}{1|0}{a}{4}.
	\item 
		When $\mtsym_h$ is inlined in $s$, 
	      \prevedge\!:\,\denew{x}{1|0}{b}{5} from procedure $h$ is 
		hoisted to procedure $s$ that contains 
	       \candedge\!:\,\denew{p}{1|1}{x}{6} thereby propagating the definition
		of $x$ in procedure $h$ to the caller $s$. 
		Strength reduction reduces \candedge to \denew{p}{1|0}{b}{6}.
	\item When $\mtsym_g$ and $\mtsym_h$ are inlined in $s$, 
	      \candedge\!:\,\denew{y}{1|1}{x}{4} in procedure $g$ 
	      and \prevedge\!:\,\denew{x}{1|0}{b}{5} in procedure $h$ are both 
              hoisted to procedure $s$ thereby propagating both the use and definition of $x$
		in procedure $s$.
		Strength reduction reduces \candedge to \denew{y}{1|0}{b}{4}.
	\item Both the definition and use of pointer $z$ are available in procedure $f$ with
	      \candedge\!:\,\denew{w}{1|1}{z}{3} and \text{\prevedge\!:\,\denew{z}{1|0}{c}{2}}.
		Strength reduction reduces \candedge to \denew{w}{1|0}{c}{3}.
	\end{enumerate}
Thus, $y$ points-to $a$ along the call from procedure $f$ and it points-to $b$ along the call from procedure $s$. Thus, the points-to information 
\text{$\{\denew{y}{1|0}{a}{}, \denew{y}{1|0}{b}{}\}$} represents flow- and context-sensitive information for statement 4.
\end{example}

\section{Handling Heap for Points-to Analysis using \gpg{}s}
\label{sec:heap}

So far we have created the concept of \gpgs for pointers to scalars
allocated on the stack or in the static area. 
This section extends the concepts to 
data \emph{structures} containing named fields created using C style \textbf{struct} or \textbf{union} and possibly 
allocated on the heap (as well as on the stack or in static memory).
For clarity, in this section, we show only the set of \gpus reaching a given statement and do not show the 
complete \gpg of a procedure. 

Extending \gpgs to handle structures and heap-allocated data
requires the following changes:
\begin{itemize}
\item The concept of \indlevs is generalized to indirection lists (\indlists) 
	to handle structures and heap accesses field sensitively.
\item Heap locations are abstracted using allocation sites. In this abstraction, all locations allocated at a 
	particular  allocation site are treated alike. 
	This approximation allows us to 
	handle the unbounded nature of heap as if it were bounded~\cite{Kanvar:2016:HAS:2966278.2931098}. Hence only weak updates can
	be performed on heap locations.\footnote{\label{footnote:address.escaped}We also perform weak updates for address-escaped variables
	(Section ~\ref{sec:implementation.meas}) because they share many similarities with heap locations. 
Like heap locations, address-escaped variables could outlive the lifetime
of the procedures that create them. They potentially represent multiple concrete locations 
because of multiple calls to the procedure. Further, this number could be
unbounded in case of recursive calls.
}  
\item When the \gpg of a procedure is being constructed, 
	the allocation sites may appear in a caller procedure and hence may not be known.
	We deal with this by an additional summarization based on $k$-limiting to 
	bound the accesses in a loop. Both these summarization techniques are required to
	create a decidable version of our method of constructing procedure summaries in the 
	form of \gpgs. The resulting points-to analysis is a precise 
	flow-sensitive, field-sensitive, and context-sensitive analysis
        (relative to these two summarization techniques).\footnote{
	In a top-down analysis, $k$-limiting is not required because allocation sites 
	are propagated from callers to callees. 	
	While the use of $k$-limiting in a bottom-up approach
	seems like an additional restriction, 
	unless the locations involved in a pointer chain are allocated by $m > k$ distinct
	allocation sites, there is no loss of precision compared to a top-down
	approach.
	}
\item Introduction of \indlists and $k$-limiting summarization requires extending the
	concept of \gpu composition to handle them. 
\item The allocation-site-based abstraction and $k$-limiting summarization 
	may create cycles in \gpus; a simple extension to \gpu reduction handles them naturally.
\end{itemize}
The optimizations performed on \gpgs and the required analyses remain the same.
Hence, the discussion in these sections is driven mainly by examples that illustrate how
the theory developed earlier is adapted to handle structures
(typically, but not necessarily, heap-allocated).

\subsection{Extending \gpu Composition to Indirection Lists}
\label{sec:ind-list}

\begin{figure}[t]
\centering
\setlength{\tabcolsep}{1.25mm}
\psset{arrowsize=1.5}
\begin{tabular}{|l|c|l|}
%%\begin{tabular}{|c|c|c|c|c|c|}
\hline
\begin{tabular}{c} Pointer assignment \end{tabular} & \begin{tabular}{c} \gpu \end{tabular} & \multicolumn{1}{c|}{Remark}
\\ \hline \hline
\rule{0em}{1.4em} 
$x = \text{\em malloc}(\ldots)$ & \de{x}{[*]|[\;]}{h_i} & The allocation site name is $i$
	\\ \hline
$x = \text{\small NULL}$ & \de{x}{[*]|[\;]}{\text{\small NULL}} & {\small NULL} is distinguished location
\\ \hline 
\rule{0em}{1.4em} 
$x = y.n$ &  \de{x}{[*]|[n]}{y} &
\\ \hline
\rule{0em}{1.4em} 
$x.n = y$ & \de{x}{[n]|[*]}{y} &
\\ \hline
\rule{0em}{1.4em} 
$x = y \rightarrow n$ & \de{x}{[*]|[*,n]}{y} & 
\\ \hline
\rule{0em}{1.4em} 
$x \rightarrow n = y$ & \de{x}{[*,n]|[*]}{y} &
\\ \hline
\end{tabular}
\caption{\gpus with indirection lists (\sindlist) for basic pointer assignments in C for structures.}
%  and heap.}
\label{fig:basic.gpg.edges.heap}
\end{figure}

The \indlev{} \ ``$i|j$'' of a \gpu \denew{x}{i|j}{y}{\flab} represents $i$ dereferences of $x$ and $j$ 
dereferences of $y$ using the dereference operator $*$. We can also view the \indlev ``$i|j$'' as lists (also referred to as indirection list or 
\indlist) containing 
$i$ and $j$ occurrences of $*$.
This representation naturally allows field-sensitive
handling of structures by using indirection lists containing field dereferences. 
Consider the statements \text{$x = *y$} and \text{$x = y\!\rightarrow\! n$} involving pointer dereferences.
Since \text{$x = y\!\rightarrow\! n$} is equivalent to
\text{$x = (*y).n$},
we can represent the two statements by \gpus as shown below:

\begin{center}
\begin{tabular}{c|c|c|c}
\hline
Statement \rule[-3.5mm]{0mm}{9mm} 
	& \begin{tabular}{@{}l@{}}
		Field-sensitive 
		\\
		representation 
		\end{tabular}
	& \begin{tabular}{@{}l@{}}
		Field-insensitive 
		\\
		representation 
		\end{tabular}
	& Our choice
\\ \hline
$x = *y$ 
	& {\de{x}{[*]|[*,*]}{y}} 
	& {\de{x}{1|2}{y}} 
	& {\de{x}{1|2}{y}} 
\\ \hline
$x = y\!\rightarrow\! n$ 
	& {\de{x}{[*]|[*,n]}{y}}
	& {\de{x}{1|2}{y}}
	& {\de{x}{[*]|[*,n]}{y}}
\\ \hline
\end{tabular}
\end{center}
We achieve field sensitivity by enumerating field names.
Having a
field-insensitive representation which does not distinguish between different fields, makes no
difference for a statement $x = *y$, but loses precision for a statement $x = y\!\rightarrow\! n$. 
Figure~\ref{fig:basic.gpg.edges.heap} illustrates the \gpus corresponding to the basic pointer 
assignments involving structures.
% and heap.

The dereference 
in the pointer expression
$y\! \rightarrow\! n$ is represented by an \indlist{} written as $[*,n]$
associated with pointer variable $y$. 
It means that, first the address in $y$ is read and then the address in field $n$
is read.
On the other hand,
the access $y.n$ as shown in the third row of Figure~\ref{fig:basic.gpg.edges.heap}
can be mapped to location by adding the offset of field $n$ to the virtual address of $y$ at compile time. Hence, it
can be treated as a separate variable which is represented
by a node $y.n$ with an \indlist $[*]$. 
We can also represent $y.n$ with a node $y$ and an \indlist $[n]$. 
For our implementation, we chose the former representation. 
However, the latter representation is more convenient for explaining the \gpu compositions 
and hence we use it in the rest of the paper.
For structures,%  and heap,
we ensure field sensitivity by maintaining \indlist in terms of field names.
We choose to handle unions field-insensitively to capture aliasing between its fields.

Recall that a \gpu composition \text{$\candedge\, \ecompwt \prevedge$} 
involves balancing the \indlev of the pivot in \candedge and \prevedge 
(Section~\ref{sec:edge.composition}). 
With \indlist  replacing \indlev, the operations remain similar in spirit, although now they become operations on lists rather than
operations on numbers. To motivate the operations on \indlists, let us recall the operations on \indlevs:
\gpu composition \text{$\candedge\, \ecompwt \prevedge$} requires balancing \indlevs of the pivot which involves 
computing the difference between the \indlev of the pivot in \candedge and \prevedge. This difference is then added to the \indlev of 
the non-pivot node in \prevedge.  Recall that a
\gpu composition is \valid (Section~\ref{sec:relevant.useful.ec})
only when the \indlev of the pivot in \candedge is greater than or equal to the \indlev of the pivot in \prevedge. 
For convenience, we illustrate it again in the following example.

\begin{example}{}
Consider $\prevedge \!:\! \de{y}{1|0}{x}$ and $\candedge \!:\! \de{w}{1|2}{y}$ where
$y$ is the pivot. Then a \tscomp composition \text{$\candedge \ecomp^{\textrm{ts}} \prevedge$} is \valid because 
\indlev of $y$ in \candedge (which is 2) is greater than \indlev of $y$ in \prevedge (which is 1). The difference ($2-1$) is added
to the \indlev of $x$ (which then becomes 1)
resulting in a reduced \gpu $\rededge \!:\! \de{w}{1|(2-1+0)}{x}$, i.e. $\rededge \!:\! \de{w}{1|1}{x}$.
\end{example}

We define similar operations for \indlists. A \gpu composition is \valid if 
the \indlist of the pivot in \gpu \prevedge is a prefix of the \indlist of the pivot in \gpu \candedge. 
For example, the \indlist \; ``\text{$[*]$}'' is a prefix of
the \indlist \; ``\text{$[*,n]$}''. 
The addition (\text{$+$}) of the difference (\text{$-$}) in the \indlevs of the pivot
to the \indlev of one of the other two nodes is replaced by the list-append operation denoted \text{@}. 

Similarly computing the difference (\text{$-$}) in the \indlev of the pivot is replaced by the `list-difference'
or `list-remainder' operation, $\rem: \indlist \times \indlist \to \indlist$\hspace*{1.5pt};
this takes two \indlists as its arguments
where the first is a prefix of the 
second and returns the suffix 
of the second \indlist{}
that remains after removing the first \indlist from it.
Given \text{$il_2 = il_1 \;\text{@}\; il_3$},
\text{$\rem(il_1, il_2) = il_3$}. 
When $il_1 = il_2$, the remainder
$il_3$ is an empty \indlist (denoted $[\;]$). 
A \gpu composition is \valid only when $il_1$ is a prefix of $il_2$; 
$\rem(il_1,il_2)$ is computed only
for \valid \gpu compositions.
This is again a natural generalization of the integer \indlev formulation earlier.

\begin{example}{}
Consider the statement sequence \text{$y=x; w=y\rightarrow n;$}.
In order to compose the corresponding \gpus $\prevedge \!:\! \de{y}{[*]|[*]}{x}$ and $\candedge \!:\! \de{w}{[*]|[*,n]}{y}$ 
we find the list remainder of the \indlists of $y$ in the two \gpus. This operation (\text{$\rem([*], [*,n])$}
returns $[n]$ which 
is appended to the \indlist of node $x$ (which is $[*]$) resulting in a new \indlist $[*]\;\text{@}\; [n] = [*,n]$  and 
thus, we get a reduced \gpu \de{w}{[*]|[*,n]}{x} representing \text{$w=x\rightarrow n$}.
\end{example}

The formal definition of \gpu composition using \indlists is similar to that using \indlevs 
(Definition~\ref{def:gpu.composition.indlev})
and is given in Definition~\ref{def:gpu.composition.indlist}.
Note that for \tscomp and \sscomp compositions in the equations, the pivot is $x$. 
Besides, for \sscomp composition, the condition \text{$il_6 \neq [\,]$}
(generalizing the strict inequality `$<$' in Definition~\ref{def:gpu.composition.indlev})
ensures that the consumer \gpu does not
redefine the location defined by the producer \gpu.
Unlike the case of pointers to scalars, \tscomp and \sscomp compositions are not mutually exclusive
for pointers to structures. For example,
an assignment \text{$x\rightarrow n = x$} could have both
\stscomp and \ssscomp compositions with a \gpu \prevedge defining $x$.
The two compositions are independent because \ssscomp composition resolves
the source of a \gpu whereas \stscomp composition resolves the target of the \gpu.
Hence, they can be performed in any order.

\begin{Definition}
\begin{center}
\psframebox[framesep=0pt,doubleline=true,doublesep=1.5pt,linewidth=.2mm]{%
\gpucompIndlistDef
}
\end{center}
\defcaption{\gpu Composition $\candedge \ecompwt \prevedge$ using \indlists}{def:gpu.composition.indlist}
\end{Definition}

A \gpu composition is \desirable if the \indlev of \rededge does not exceed that of \candedge.
Similarly, in the case of \indlists, a \gpu composition is \desirable if \indlists
%(\indlist of both the source and the target) 
of \rededge (say $il_1|il_2$) does not exceed that of \candedge (say $il_1'|il_2'$), i.e.\ 
\text{$|il_1| \leq |il_1'|\; \wedge\; |il_2| \leq |il_2'|$} where \text{$|il|$} denotes
the length of \indlist $il$. 
Note that, for \desirability,  we only need a smaller length and not a
prefix relation between \indlists.
In fact, the \indlist in \rededge is always a suffix of the
\indlist in \candedge as illustrated by the following example.

\begin{example}{}
Consider the code snippet on right.  The effect of 
statement 22
in the context of
\setlength{\intextsep}{-.4mm}%
\setlength{\columnsep}{2mm}%
\begin{wrapfigure}{r}{29.5mm}
$
\setlength{\arraycolsep}{3pt}
\begin{array}{|lrcl|}
\hline
\rule{0em}{0.85em}
{\footnotesize \darkgray\sf 21:}& x &= & \&y;
	\\
{\footnotesize \darkgray\sf 22:}& z &= & x \rightarrow n;
	\\ \hline
\end{array}
$
\end{wrapfigure}
statement 21
can be seen as an 
assignment 
\text{$z = y.n$}.
The composition of \gpus \text{$\candedge\!:\!\denew{z}{[*]|[*,n]}{x}{22}$} 
and \text{$\prevedge\!:\!\denew{x}{[*]|[\;]}{y}{21}$} results in the \gpu
\text{$\rededge\!:\!\denew{z}{[*]|[n]}{y}{22}$}. 
The \indlist of the target ($y$) of \rededge is not a prefix of that of target ($x$)
of \candedge but is a suffix. 
\end{example}

\begin{figure}[t]
\psset{arrowsize=1.5mm}%
\setlength{\codeLineLength}{45mm}%
\renewcommand{\arraystretch}{.9}%
\noindent%
\begin{tabular}{@{}c}
	\begin{tabular}{rc}
	\codeLineNoNumber{0}{$\tt struct \; node \; *x;$}{white}
	\codeLineNoNumber{0}{}{white}
	\codeLineOne{1}{0}{$\tt struct \; node \; \OB$}{white} 
%	\codeLine{0}{\OB }{white}
	\codeLine{1}{$\tt struct \; node \; *n;$}{white}
	\codeLine{1}{$\tt int\; d;$}{white}
	\codeLine{0}{\CB;}{white}
	\codeLineNoNumber{0}{}{white}
	\codeLine{0}{$\tt void\; g()\; \OB$}{white}
%	\codeLine{0}{\OB}{white}
	\codeLine{1}{$\tt struct \; node \; *y;$}{white}
	\codeLine{1}{$\tt while\;(...)\; \OB$}{white}
	\codeLine{2}{$\tt print \; x \rightarrow d;$}{white}
	\codeLine{2}{$\tt x = x \rightarrow n;$}{white}
	\codeLine{1}{\CB}{white}
	\codeLine{0}{\CB}{white}
	\end{tabular}

	\begin{tabular}{rc}
	\codeLine{0}{$\tt void \; f() \; \OB$}{white} 
% 	\codeLine{0}{\OB }{white}
	\codeLine{1}{$\tt struct \; node \; *y;$}{white}
	\codeLine{1}{$\tt y = malloc(\ldots);$}{white}
	\codeLine{1}{$\tt x = y;$}{white}
	\codeLine{1}{$\tt while \;(...) \;\OB$}{white}
	\codeLine{2}{$\tt y \rightarrow n = malloc(\ldots);$}{white}
	\codeLine{2}{$\tt y = y \rightarrow n;$}{white}
	\codeLine{1}{\CB}{white}
	\codeLine{1}{$\tt g();$}{white}
	\codeLine{0}{\CB}{white}
	\end{tabular}
\\
\rule[-.5em]{0em}{2.5em}\small
\begin{tabular}{l@{}l}
(a) \; & A program for creating a linked list and traversing it. We have omitted the null 
	assignment  for the last node \\ &of the list and the associated \gpus \end{tabular}
\\ \hline
\begin{tabular}{@{}c|c|c}
\begin{pspicture}(-36,0)(2,37)
\small
%\psframe(-36,0)(2,37)
\putnode{x}{origin}{-28}{24}{\pscirclebox[framesep=1.52]{$x$}}
\putnode{ix}{x}{24}{0}{\pscirclebox[framesep=.7]{$x'$}}
\putnode{l}{x}{11}{-21}{\small \begin{tabular}{l@{}l} (b) \; & $\BROut{11}$ (\gpus reaching \\ & the \End{} of $g$ for $k = 3$) \end{tabular}}
\nccurve[angleA=15,angleB=165]{->}{x}{ix}
\naput[labelsep=.5]{$[*]|[*,\!n,\!n]$}
\nbput[labelsep=.5]{09}
\nbput[labelsep=0,npos=.2]{\pscirclebox[linestyle=none,fillstyle=solid,fillcolor=lightgray,framesep=.2]{$g_3$}}
\nccurve[ncurv=1,angleA=65,angleB=115,nodesepA=-.3,nodesepB=-.4]{->}{x}{ix}
\naput[labelsep=.5]{$[*]|[*]$}
\nbput[labelsep=.5]{00}
\naput[labelsep=0,npos=.2]{\pscirclebox[linestyle=none,fillstyle=solid,fillcolor=lightgray,framesep=.2]{$g_1$}}
\nccurve[ncurv=1,angleA=-65,angleB=235,nodesepA=-.3,nodesepB=-.8]{->}{x}{ix}
\naput[labelsep=1]{$[*]|[*,\!n]$}
\nbput[labelsep=.5]{09}
\nbput[labelsep=0,npos=.2]{\pscirclebox[linestyle=none,fillstyle=solid,fillcolor=lightgray,framesep=.2]{$g_2$}}
\end{pspicture}
&
\begin{pspicture}(-34,0)(13,31)
%%\psframe(-34,0)(13,31)
\putnode{m1}{origin}{-30}{24}{\psframebox[framesep=1.1]{$d\;$}}
\putnode{n1}{m1}{5}{0}{\,\psframebox[framesep=1.5]{$n$}}
\putnode{x}{n1}{-3}{12}{$x$}
\putnode{y}{n1}{-3}{-12}{$y$}
\putnode{m2}{n1}{9}{0}{\psframebox[framesep=1.1]{$d\;$}}
\putnode{n2}{m2}{5}{0}{\,\psframebox[framesep=1.5]{$n$}}
\putnode{m3}{n2}{9}{0}{\psframebox[framesep=1.1]{$d\;$}}
\putnode{n3}{m3}{5}{0}{\,\psframebox[framesep=1.5]{$n$}}
\putnode{l}{n3}{9}{0}{$\ldots$}
\putnode{h1}{m1}{-1}{5}{$h_{14}$}
\putnode{h2}{m2}{-1}{5}{$h_{17}$}
\putnode{h2}{m3}{-1}{5}{$h_{17}$}
\putnode{h5}{y}{15}{-9}{\small\begin{tabular}{l@{}l} (c) \; & Linked list created by \\ & procedure $f$ \end{tabular}}
\ncline[nodesepA=1.2,nodesepB=.5,offsetB=-1.8]{->}{x}{n1}
\ncline[nodesepA=1.2,nodesepB=.5,offsetB=1.8]{->}{y}{n1}
\ncline[nodesepA=-1.3]{->}{n1}{m2}
\ncline[nodesepA=-1.3]{->}{n2}{m3}
\ncline[nodesepA=-1.3]{->}{n3}{l}
\end{pspicture}
&
\begin{pspicture}(-37,0)(8,42)
%\psframe(-38,0)(8,38)
\small
\putnode{y}{origin}{-34}{27}{\pscirclebox[framesep=1.52]{$y$}}
\putnode{x}{y}{5}{-15}{\pscirclebox[framesep=1.52]{$x$}}
\putnode{h1}{y}{17}{0}{\pscirclebox[framesep=.22]{$h_{14}$}}
\putnode{h2}{h1}{17}{0}{\pscirclebox[framesep=.22]{$h_{17}$}}
\putnode{l}{h1}{1}{-24}{\small\begin{tabular}{l@{}l} (d)\;& $\BRIn{20}$ (\gpus reaching \\
				& the call to $g$ on line 20)
				 \end{tabular}}
\ncline{->}{y}{h1}
\naput[labelsep=.5]{$[*],[\;]$}
\nbput[labelsep=.5,npos=.65]{14}
\nbput[labelsep=-.1,npos=.3]{\pscirclebox[linestyle=none,fillstyle=solid,fillcolor=lightgray,framesep=0]{$f_1$}}
\ncline[nodesepA=-.9,nodesepB=-1.]{->}{x}{h1}
\naput[labelsep=.2,npos=.3]{$[*]|[\;]$}
\nbput[labelsep=.2,npos=.2]{15}
\nbput[labelsep=-.7,npos=.6]{\pscirclebox[linestyle=none,fillstyle=solid,fillcolor=lightgray,framesep=0]{$f_2$}}
\ncline{->}{h1}{h2}
\naput[labelsep=.5]{$[n]|[\;]$}
\nbput[labelsep=.5,npos=.65]{17}
\nbput[labelsep=-.1,npos=.3]{\pscirclebox[linestyle=none,fillstyle=solid,fillcolor=lightgray,framesep=0]{$f_3$}}
\nccurve[ncurv=.5,angleA=65,angleB=120,nodesepA=-.4,nodesepB=-.5]{->}{y}{h2}
\naput[labelsep=.5]{$[*]|[\;]$}
\nbput[labelsep=.5]{18}
\naput[labelsep=-.1,npos=.3]{\pscirclebox[linestyle=none,fillstyle=solid,fillcolor=lightgray,framesep=0]{$f_4$}}
%%\nccurve[angleA=225,angleB=-35,nodesepA=-1.2,nodesepB=-.8]{->}{h2}{h1}
%%\naput[labelsep=.5]{$[*],[\;]$}
\nccurve[angleA=60,angleB=300,ncurv=6.3,nodesepA=-.8,nodesepB=-.7]{->}{h2}{h2}
\naput[labelsep=.1,npos=.15]{$[n]|[\;]$}
\nbput[labelsep=0,npos=.4]{17}
\naput[labelsep=.1,npos=.85]{\pscirclebox[linestyle=none,fillstyle=solid,fillcolor=lightgray,framesep=0]{$f_5$}}
\end{pspicture}
\end{tabular}
\end{tabular}
\caption{An example demonstrating the need of $k$-limiting summarization technique in addition to allocation-site-based 
abstraction for the heap. $h_{14}$ and $h_{17}$ are the heap nodes allocated on lines 14 and 17 respectively.}
\label{fig:heap_eg}
\end{figure}
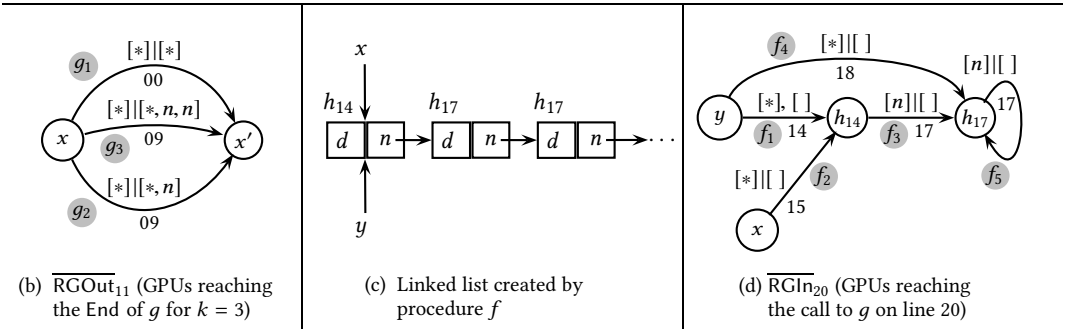

\subsection{Summarization Using Allocation Sites}
\label{sec:alloc-summ}

Under the allocation-site-based abstraction for the heap, the objects 
created by an allocation statement are collectively named by the allocation site and undergo
	weak update.
Thus, a statement \text{$x = \text{\em malloc}(\ldots)$} 
is represented by a \gpu
\denew{x}{[*]|[\;]}{\text{h$_{i}$}}{i} 
where \text{h$_{i}$} is the heap location created at the allocation 
site $i$.
The example below illustrates how this bounds an unbounded heap in a \gpg.
For convenience, we identify \gpus using procedure names.

\begin{example}{eg:allocation.sites}
For procedure $f$ shown in Figure~\ref{fig:heap_eg} we create 
heap objects $h_{14}$ and $h_{17}$ allocated at line
numbers 14 and 17. 
The \gpu set \BRIn{20}
in procedure $f$ represents a linked list with $x$ as its head
pointer (Figure~\ref{fig:heap_eg}(d)) and $h_{14}$ as its first node. 
The remaining nodes in the list are represented by the heap location $h_{17}$ and are summarized 
by a self-loop over
the node. This set of \gpus is 
computed as follows:
The \gpu \text{$f_1 \!:\! \denew{y}{[*]|[\;]}{h_{14}}{14}$} is created
for allocation-site 14.
The \gpu \denew{x}{[*]|[*]}{y}{15} composes with $f_1$ (under 
       \tscomp composition) to create a new \gpu 
       \text{$f_2 \!:\! \denew{x}{[*]|[\;]}{h_{14}}{15}$}.
When statement 17 is processed for the first time, \gpu 
\denew{y}{[*,n]|[\;]}{h_{17}}{17}
composes with $f_1$ (under \sscomp composition) to create a 
\gpu \text{$f_3 \!:\! \denew{h_{14}}{[n]|[\;]}{h_{17}}{17}$}. 
When statement 18 is processed for the first time, 
the \gpu \denew{y}{[*]|[*,n]}{y}{18}
composes with $f_1$ (under \tscomp composition) to create a \gpu 
\denew{y}{[*]|[n]}{h_{14}}{18} which is further composed with $f_3$ (under
\tscomp composition) to create a \gpu \text{$f_4 \!:\! \denew{y}{[*]|[\;]}{h_{17}}{18}$}. 
\gpu $f_4$ kills \gpu $f_1$ because $y$ is redefined by statement 18.
This completes the first iteration of the loop and the set of \gpus \BROut{19} is 
       \text{$\{f_2,f_3,f_4\}$} representing the following information:
	\begin{itemize}
	\bitem $f_2$ indicates that $x$ points to the head of the linked list.
	\bitem $f_3$ indicates that the field $n$ of heap location $h_{14}$ points to 
	      heap location $h_{17}$.
	\bitem $f_4$ indicates that $y$ points to heap location $h_{17}$.
	\end{itemize} 

In the second iteration of the reaching \gpus analysis over the loop, 
\BROut{15} and \BROut{19}
are merged to compute \BRIn{16} as \text{$\{f_1,f_2,f_3,f_4\}$}.
When statement 17 is processed for the second time, the \gpu \denew{y}{[*,n]|[\;]}{h_{17}}{17} 
composes with 
\begin{itemize}
       \bitem  $f_1$ (under \sscomp composition) to create $f_3$, and with
       \bitem  $f_4$ (under \sscomp composition) to create 
              \text{$f_5 \!:\! \denew{h_{17}}{[n]|[\;]}{h_{17}}{17}$}. 
 \end{itemize}	
 When statement 18 is processed for the second time, $f_4$ is recreated killing $f_1$. 
 This completes the second iteration of the loop and the set of \gpus \BRIn{20} is
 \text{$\{f_1,f_2,f_3,f_4,f_5\}$}. The new \gpu $f_5$ implies that the field 
 $n$ of heap location $h_{17}$ holds the address of heap location $h_{17}$.
 The self loop represents an unbounded list 
 \text{$\big(h_{17} \!\xrightarrow{n}\! h_{17} \!\xrightarrow{n}\! h_{17} \!\xrightarrow{n}\! 
 h_{17} \ldots\big)$} under the allocation-site-based abstraction.	
 The third iteration of reaching \gpus analysis over the loop does not add any new information 
 and reaching \gpus analysis reaches a fixed point.
\end{example}

\noindent The following example discusses the absence of blocking in the procedures in
Figure~\ref{fig:heap_eg}. 

\begin{example}{}
The \gpus in \BRIn{14} reach statement 17 unblocked because there is no barrier.
Since the pointee of $y$ is available, the set \BRGen{14} does not contain any indirect \gpus and hence do not contribute to
the blocking of any \gpus.
If the allocation site at statement 14 was not available, then the \gpu for statement 17
would not have been reduced and hence the set \BRGen{17}
would contain an indirect \gpu \denew{y}{[*, n]|[\,]}{h_{17}}{17}. 
This \gpu would block all \gpus in \BRIn{18} and in turn would
be blocked by the \gpus in  \BRGen{18} so that it cannot be used for reduction of any successive \gpus.
%%However, the \gpus in \BRGen{20} that reach statement 21 are used for reducing the \gpus
%%for statement 21 as the two statements being consecutive do not have any intervening assignment to act as a barrier.
%%If there was an intervening assignment between statements 20 and 21, then 
%%its corresponding \gpu would act as barrier, thereby postponing the composition between
%%\gpus corresponding to statements 20 and 21.
\end{example}

\subsection{Summarization Using $k$-Limiting}
\label{sec:k-limmiting}

This section shows why allocation-site-based abstraction is not sufficient for a bottom-up
points-to analysis although it serves the purpose well in a top-down analysis.

\subsubsection{The Need for $k$-Limiting}
In some cases, the allocation site may not be available during the
construction of the \gpg of a procedure. For our example in Figure~\ref{fig:heap_eg}, when the \gpg is constructed for procedure $g$, we do
not know the allocation site because the accesses to heap in procedure $g$ refer to the data-structure created in procedure $f$. Thus
allocation-site-based abstraction is not applicable for procedure $g$ and the indirection lists grow without bound.

In a top-down analysis, $k$-limiting is not required because allocation sites 
	are propagated from callers to callees. 	
	
\begin{example}{eg:k.limiting}
When the \gpg for procedure $g$ in Figure~\ref{fig:heap_eg}
is constructed, we have a boundary definition \text{$g_1 \!:\! \denew{x}{[*]|[*]}{x'}{00}$} 
at the start of the procedure. In the first iteration of the analysis over the loop, the \gpu 
\denew{x}{[*]|[*,n]}{x}{09}
composes with $g_1$ (under \tscomp composition) creating a reduced \gpu 
\text{$g_2 \!:\! \denew{x}{[*]|[*,n]}{x'}{09}$}. The \gpu $g_2$ kills \gpu $g_1$ because $x$ 
is redefined by statement at 09. 
However, the merge at the top of the loop reintroduces it.
In the second iteration, the \gpu \denew{x}{[*]|[*,n]}{x}{09} composes with
$g_1$ to recreate $g_2$, and with
$g_2$ to create \text{$g_3 \!:\! \denew{x}{[*]|[*,n,n]}{x'}{09}$}.
In the third iteration, we get an additional \gpu 
\text{$g_4 \!:\! \denew{x}{[*]|[*,n,n,n]}{x'}{09}$} 
apart from $g_2$ and $g_3$.
This continues and the indirection lists of the \gpus between $x$ and $x'$ grow without bound leading to non-termination. 
\end{example}

There are two ways of handling traversals of data structures created in some other procedure.
\begin{itemize}
\item As the above example illustrates, we perform compositions involving upwards
	exposed variables inspite of these compositions being \valid but \undesirable.
\item Alternatively, we can postpone these compositions (as suggested before) until call inlining
	enables their reduction.
\end{itemize}
We use the first approach and bound the length of indirection lists using $k$-limiting.
This limits the participation of the \gpus in the fixed-point computation
for the procedures containing them.
The second approach requires the \gpus to participate in the fixed-point computations
for the callers as well.
This could cause inefficiency.

While the use of $k$-limiting in a bottom-up approach
	seems like an additional restriction, 
	unless the locations involved in a pointer chain are allocated by $m > k$ distinct
	allocation sites, there is no loss of precision compared to a top-down
	approach.

%%Observe that the compositions in the above example are \undesirable because the \indlists of the resulting \gpus are longer
%%than the \indlists of the consumer \gpus. 
%%Recall that such compositions are postponed and are performed after the
%%\gpg is inlined in the callers when they do not remain \undesirable any more. 
%%%%\deleted{
%%%%This requires retaining the loops and so back edges 
%%%%cannot be removed (Section~\ref{sec:back.edge.removal}). 
%%%%The presence of back edges requires
%%%%fixed-point computation when the \gpg is inlined in a caller but it guarantees same precision as that of the 
%%%%top-down interprocedural approach.
%%%%}
%%However, in case of pointers to structures, the analysis would be less efficient because it could 
%%require $k$ iterations for reaching the $k$-limiting for achieving fixed 
%%point.
%%Hence, we allow \undesirable \gpu compositions that are \valid,
%%thereby allowing the \indlist of the reduced \gpu to grow. 
%%However, this requires the \indlists to be bounded for termination of analysis.
%%
%%We use the $k$-limiting technique for bounding the length of indirection lists.
%%With this provision, the \gpg of a recursive procedure 
%%(that has been converted to a self-recursive procedure, see Section~\ref{sec:handling_recur})
%%may be inlined a maximum of $k$ times within itself.

\subsubsection{Incorporating $k$-Limiting}
We limit the length of \indlists to $k$ such that
the \indlist is exact up to $k-1$ dereferences and approximate for $k$ or more dereferences in terms of an unbounded number of
dereferences. Besides, the dereferences are
field-insensitive beyond $k$. This summarization is implemented by redefining the list concatenation operator
@ such that for \text{$il_1\,@\,il_2$}, the result is a $k$-limited prefix of the concatenation of $il_1$ and $il_2$.

\begin{example}{}
The set of \gpus \BROut{11} reaching the \End{} of procedure $g$ of Figure~\ref{fig:heap_eg}, 
for $k=3$ is given in the Figure~\ref{fig:heap_eg}(b). 
A \gpu between $x$ and $x'$ has an \indlist \text{$[*,n]$} of length 2 and all
\indlists of length $\ge 3$ are approximated by \text{$[*,n,n]$}.

\gpu \text{$g_1 \!:\! \denew{x}{[*][*]}{x'}{00}$} in the \gpg for procedure $g$ represents the effect of \textbf{while} loop 
not executed even once. \gpu \text{$g_2\!:\! \denew{x}{[*]|[*,n]}{x'}{09}$} 
represents the effect of 
the first iteration of the \textbf{while} loop. The \gpu 
\text{$g_3 \!:\! \denew{x}{[*]|[*,n,n]}{x'}{09}$} 
represents the combined effect of the second and all subsequent iterations of the \textbf{while} loop.
The \gpg of procedure $g$ ($\mtsym_g$) contains a single \gpb which in turn contains a set of \gpus
\text{$\{g_2, g_3\}$}.
\end{example}

Note that an explicit summarization is required only for heap locations and 
address-escaped stack locations in recursive procedures because the \indlists can
grow without bound only in these cases (see Footnote~\ref{footnote:address.escaped}).

%%The cycle of the second kind may be $k$-limited indicating $k$ or an unbounded number of dereferences. 
The \gpu  composition defined in Section~\ref{sec:ind-list} 
(Definition~\ref{def:gpu.composition.indlist})
is extended to handle $k$-limited \indlists in the following manner:
The removal of a 
prefix from a $k$-limited \indlist in the \rem operation is over-approximated by
suffixing special field-insensitive dereferences denoted by ``\fisd'' 
where \fisd represents any field.
For an operation \text{$\rem(il_1,il_2)$}, $il_1$ must be a prefix of $il_2$ as
explained in Section~\ref{sec:ind-list}.
Let \text{$il_2 = il_1 \,@\, il_3$} for \text{$\rem(il_1,il_2)$}.  
We define a summarized list-remainder operation
\text{$\srem: \indlist \times \indlist \to 2^{\indlist}$} 
which takes two \indlists as its arguments and computes a set of \indlists as shown
below:

\[
\srem(il_1,il_2) = \begin{cases} 
		\{ il_3 \mid 
		il_2 = il_1 \,@\, il_3
		\}
		& |il_2| < k
		\\
		\{ il_3 \,@ \,\sigma \mid 
		il_2 = il_1 \,@\, il_3,
		\sigma \text{ is a sequence of \fisd}, 0 \leq |\sigma| \leq |il_1| 
		\}
		& \text{otherwise}
		\end{cases}
\]

Observe that \srem is a generalization of \rem defined in 
Section~\ref{sec:ind-list} because it computes a set of \indlists when its second argument is
a $k$-limited \indlist{\,}; for non $k$-limited \indlist, \srem returns a singleton set.
The longest \indlist in the set  computed by \srem represents a summary whereas the other \indlists
are exact in length but approximate in terms of fields because of field insensitivity introduced by \fisd.\footnote{This is somewhat similar to materialization~\cite{Sagiv:1998:SSP:271510.271517}
which extracts copies out of summary representation
of an object to create some exact objects.} This is illustrated in the example below.

\begin{example}{}
For $k = 3$, some examples of the sets of \indlists computed by the \srem operation are shown below:
\begin{align*}
& \srem([*],[*,n,n])  = \{ [n,n], [n,n,\fisd]\}
      	\\
& \srem([*,n],[*,n,n])  = \{ [n], [n, \fisd], [n,\fisd,\fisd]\}
       	\\
& \srem([*,n,n],[*,n,n])  = \{ [\;], [\fisd], [\fisd,\fisd], [\fisd,\fisd,\fisd]\}
\end{align*}
For the last case, the \srem operation can be viewed as an operation that creates an intermediate set
\text{$S = \{ 
[*,n,n],
[*,n,n,\fisd],
[*,n,n,\fisd,\fisd],
[*,n,n,\fisd,\fisd,\fisd]
\}$} 
obtained by adding upto 3 occurrences of \fisd (because $k=3$).
The \srem operation can then be viewed as a collection of \text{$\rem([*,n,n], \sigma)$} for each
$\sigma$ in this set:
\begin{align*}
\srem([*,n,n],[*,n,n])  = \{ \rem([*,n,n],\sigma) \mid \sigma \in S \}
\end{align*}
The first two cases in this example can also be explained in a similar manner.
\end{example}

\begin{Definition}
\begin{center}
\psframebox[framesep=5pt,doubleline=true,doublesep=1.5pt,linewidth=.2mm]{%
\edgeReductionDefHeap
}
\end{center}
\defcaption{\gpu Reduction $\candedge \rcomp \flow$ for Handling Heap}{def:edge.reduction.heap}
\end{Definition}

\gpu composition using \indlevs (Section~\ref{sec:edge.comp.properties}) or using
\indlists (Section~\ref{sec:ind-list}) is a partial operation
defined to compute a single \gpu as its result when it succeeds. Since we do not have a representation for an ``invalid'' \gpu, we model failure by defining \gpu composition
as a partial function for \gpus containing \indlevs or non-$k$-limited \indlists.
However, when \indlists are summarized using 
$k$-limiting, \srem naturally computes a set of \indlists (unlike \rem which computes a single \indlist).
This allows us to define \gpu composition as a total function, since we can express the previous partiality
simply by returning an empty set.

% Thus composition inherently computes multiple \gpus as its result.
% Since With a set of \gpus as the result of composition, when no \gpu can be computed,
% the failure of composition can be modelled in terms of an empty set. Thus \gpu composition is now a total function.

\subsection{Extending \gpu Reduction to Handle Cycles in \gpus}
\label{sec:cycle-gpg}

\begin{figure}[t]
\centering
\begin{tabular}{@{}c@{}c@{}}
\begin{tabular}{@{}c@{}}
\begin{pspicture}(0,-2)(70,15)
\psset{arrowsize=1.5}
%\psframe(0,-2)(68,15)
\putnode{n0}{origin}{3}{5}{}
\putnode{w}{n0}{36}{-3}{\psframebox[linestyle=none,fillstyle=solid,fillcolor=lightgray,framearc=.8]{\rule{54mm}{0mm}\rule{0mm}{8mm}}}
\putnode{n1}{n0}{0}{0}{\pscirclebox[framesep=2]{}}
\putnode{n2}{n1}{12}{0}{\pscirclebox[framesep=2]{}}
\putnode{n3}{n2}{12}{0}{\pscirclebox[framesep=2]{}}
\putnode{n4}{n3}{12}{0}{\pscirclebox[framesep=2]{}}
\putnode{n5}{n4}{12}{0}{\ldots}
\putnode{nn}{n5}{12}{0}{\pscirclebox[framesep=2]{}}
\ncline{->}{n1}{n2}
\nbput[labelsep=.5]{\candedge}
\ncline{->}{n2}{n3}
\nbput[labelsep=.5]{$\prevedge_1$}
\ncline{->}{n3}{n4}
\nbput[labelsep=.5]{$\prevedge_2$}
\ncline{->}{n4}{n5}
\nbput[labelsep=.5]{$\prevedge_3$}
\ncline{->}{n5}{nn}
\nbput[labelsep=.5]{$\prevedge_{n\!-\!1}$}
\nccurve[angleA=30,angleB=150,nodesepA=-.4,nodesepB=-.5]{->}{n1}{n3}
\naput[labelsep=0,npos=.9]{$\rededge_1$}
\nccurve[angleA=40,angleB=140,nodesepA=-.5,nodesepB=-.5,ncurv=.6]{->}{n1}{n4}
\naput[labelsep=0,npos=.9]{$\rededge_2$}
%\nccurve[angleA=50,angleB=130,nodesepA=-.5,nodesepB=-.5,ncurv=.5]{->}{n1}{n5}
%\naput[labelsep=0,npos=.9]{$\rededge_3$}
\nccurve[angleA=60,angleB=120,nodesepA=-.4,nodesepB=-.4,ncurv=.4]{->}{n1}{nn}
\naput[labelsep=0,npos=.92]{$\rededge_{n-1}$}
\nccurve[angleA=240,angleB=-60,nodesepA=-.4,nodesepB=-.4,ncurv=.35]{->}{nn}{n2}
\nbput[labelsep=.2,npos=.5]{$\prevedge_{n}$}
\nccurve[angleA=-45,angleB=225,nodesepA=-.7,nodesepB=-.7,ncurv=1]{->}{n1}{n2}
\nbput[labelsep=.2,npos=.5]{$\rededge_{n}$}
\end{pspicture}
\end{tabular}
&
\begin{tabular}{@{}c@{}}
\begin{minipage}{69mm}
\small\raggedright

\begin{itemize}
\item The shaded part shows the \gpus in \BRIn{}.

\item Let $\rededge_0 = \candedge$. Then \text{$\rededge_i = \rededge_{i-1}\ecompwt\prevedge_i$}, \text{$i>0$}.
\item For simplicity, the directions chosen in the \gpus illustrate only \tscomp compositions.
\end{itemize}
\end{minipage}
\end{tabular}
\end{tabular}
\caption{Series of compositions and its consequence when
the graph induced by the \gpus in \BRIn{} (shown by the shaded part) has a cycle. The compositions may
happen more than the required number of times, resulting in a points-to edge.}
\label{fig:series.of.compositions}
\end{figure}

In the presence of a heap, the graph induced by the set of \gpus reaching a \gpb can contain cycles of the following two kinds:
\begin{itemize}
\bitem Cycles arising out of creation of a recursive data structure in a procedure under allocation-site-based 
	abstraction. This manifests itself in the form of a cycle involving
       heap nodes $h_i$ as illustrated in Example~\ref{eg:allocation.sites} in Section~\ref{sec:alloc-summ}.
	These cycles are closed form representations of acyclic unbounded paths in the memory.
%%\bitem Cycles arising out of repeated access of a data structure defined in the caller. This 
%%       manifests itself in the form of a \gpu \de{x}{[*]|[*,\sigma]}{x'} where $\sigma$ is a 
%%       non-empty sequence of field dereferences.
%%       A \gpu \de{x}{[*]|[*]}{x'} in a \gpg for a statement \flab indicates that 
%%       $x$ has the same pointees at \flab as those at the start of the procedure. However, a \gpu
%%       \de{x}{[*]|[*,\sigma]}{x'} in a \gpg for a statement \flab indicates that, at \flab, $x$
%%       points to some transitive pointee of $x$ at the start of the procedure. 
%%       This is illustrated in Example~\ref{eg:k.limiting} in Section~\ref{sec:k-limmiting}.
\bitem Cycles arising out of cyclic data structures.
	These cycles represent cycles in the memory.
\end{itemize}

Both these cases of cycles are handled by \gpu composition using 
\srem operation over indirection lists. 
Definition~\ref{def:edge.reduction.heap}  extends the algorithm for \gpu reduction
to use the new definition of \gpu composition 
which computes a set of \gpus instead of a single \gpu.

For \gpu reduction \text{$\candedge \rcomp \flow$}, an \admissible composition 
\text{$\rededge_1 = \candedge\, \ecompwt \prevedge_1$} (where \text{$\prevedge_1 \in \BRIn{}$})
may lead to another composition 
\text{$\rededge_2 = \rededge_1\, \ecompwt \prevedge_2$} (where \text{$\prevedge_2 \in \BRIn{}$}). 
This in turn may lead to another composition thereby creating a chain of compositions.
If the graph induced by the reaching \gpus (i.e. \gpus in \BRIn{}) has a cycle 
(as illustrated in Example~\ref{eg:allocation.sites} in Section~\ref{sec:alloc-summ}),
some $\prevedge_m$ must be adjacent to
$\prevedge_1$ with the length of the cycle being $m+1$ 
as illustrated in Figure~\ref{fig:series.of.compositions}.
The lengths of \indlists in $\rededge_i$ would be smaller than (or equal to) those in 
$\rededge_{i-1}$ because of \admissibility.
If the length of an \indlist in \candedge exceeds $m$, 
the series of compositions would resume with $\prevedge_1$ after the composition with $\prevedge_m$. In other words,
after computing $\rededge_{m-1}$ using the composition \text{$\rededge_{m-2} \ecomp \prevedge_m$}, 
the next \gpu $\rededge_m$ would be computed using the composition \text{$\rededge_{m-1} \ecomp \prevedge_1$} and the process will
continue until some $\rededge_j$, $j\geq m$ is a points-to edge.\footnote{%
Note that this happens for reducing a single \gpu \candedge in the context of \BRIn{} and
does not require a cycle in the \gpg.}
Thus, we will have more compositions than required 
and the result of \gpu reduction may not represent the updates of locations that are updated by the 
original \gpu \candedge.
In order to prohibit this, we allow 
a \gpu \prevedge to be used only once in a chain of compositions.

Hence, the new definition of \gpu reduction (Definition~\ref{def:edge.reduction.heap}) uses
an additional argument, \used, which maintains a set of \gpus that have been used in a chain of \gpu compositions.
For the top level non-recursive call to \edgeReduction, \text{$\used = \emptyset$}.
In the case of pointers to scalars, a graph induced by a set of \gpus cannot have a cycle, hence 
a \gpu \prevedge
cannot be used multiple times in a series of \gpu compositions.
Therefore, we did not need set \used for defining \gpu reduction 
in the case of pointers to scalars (Definition~\ref{def:edge.reduction.aa}).

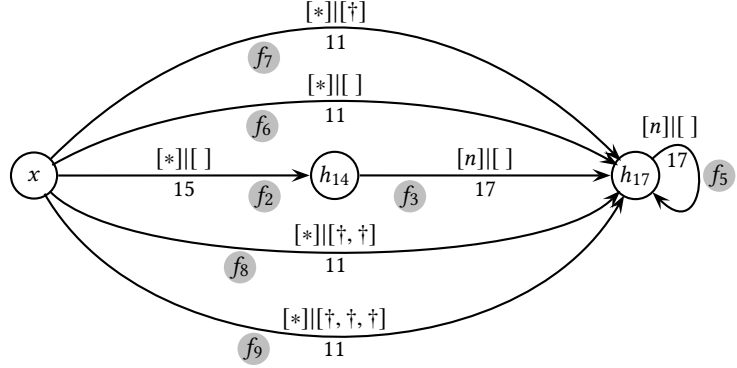
\begin{figure}[t]
\centering
\psset{arrowsize=1.75mm}
\begin{pspicture}(-43,0)(68,56)
%%\psframe(-43,0)(68,56)
\putnode{y}{origin}{-34}{28}{}
\putnode{x}{y}{0}{0}{\pscirclebox[framesep=1.82]{$x$}}
\putnode{h1}{y}{40}{0}{\pscirclebox[framesep=.52]{$h_{14}$}}
\putnode{h2}{h1}{40}{0}{\pscirclebox[framesep=.52]{$h_{17}$}}
\ncline{->}{x}{h1}
\naput[labelsep=.5]{$[*]|[\;]$}
\nbput[labelsep=.5]{15}
\nbput[labelsep=.2,npos=.82]{\pscirclebox[linestyle=none,fillstyle=solid,fillcolor=lightgray,framesep=0]{$f_2$}}
\ncline{->}{h1}{h2}
\naput[labelsep=.5]{$[n]|[\;]$}
\nbput[labelsep=.5]{$17$}
\nbput[labelsep=.2,npos=.2]{\pscirclebox[linestyle=none,fillstyle=solid,fillcolor=lightgray,framesep=0]{$f_3$}}
%%\naput[labelsep=.5]{$[*],[\;]$}
\nccurve[angleA=45,angleB=-45,ncurv=5.3,nodesepA=-1.2,nodesepB=-1.1]{->}{h2}{h2}
\naput[labelsep=.2,npos=.15]{$[n]|[\;]$}
\nbput[labelsep=.5,npos=.2]{17}
\naput[labelsep=.2,npos=.5]{\pscirclebox[linestyle=none,fillstyle=solid,fillcolor=lightgray,framesep=0]{$f_5$}}
%%%
\nccurve[ncurv=.6,angleA=30,angleB=150,nodesepA=-.8,nodesepB=-1]{->}{x}{h2}
\naput[labelsep=.5]{$[*]|[\;]$}
\nbput[labelsep=.5]{$11$}
\nbput[labelsep=0,npos=.38]{\pscirclebox[linestyle=none,fillstyle=solid,fillcolor=lightgray,framesep=0]{$f_6$}}
\nccurve[ncurv=.87,angleA=45,angleB=135,nodesepA=-1,nodesepB=-1.4]{->}{x}{h2}
\naput[labelsep=.5]{$[*]|[\fisd]$}
\nbput[labelsep=.5]{$11$}
\nbput[labelsep=-.2,npos=.37]{\pscirclebox[linestyle=none,fillstyle=solid,fillcolor=lightgray,framesep=0]{$f_7$}}
%
%%\nccurve[ncurv=1.1,angleA=60,angleB=120,nodesepA=-.8,nodesepB=-.6]{->}{x}{h2}
%%\naput[labelsep=.5]{$[*]|[\fisd,\fisd]$}
%%\nbput[labelsep=.5]{$11$}
%%\nbput[labelsep=-.2,npos=.37]{\pscirclebox[linestyle=none,fillstyle=solid,fillcolor=lightgray,framesep=0]{$f_8$}}
%%%
%%\nccurve[ncurv=1.35,angleA=75,angleB=105,nodesepA=-.4,nodesepB=-.5]{->}{x}{h2}
%%\naput[labelsep=.5]{$[*]|[\fisd,\fisd,\fisd]$}
%%\nbput[labelsep=.5]{$11$}
%%\nbput[labelsep=-.2,npos=.39]{\pscirclebox[linestyle=none,fillstyle=solid,fillcolor=lightgray,framesep=0]{$f_9$}}
%%%
\nccurve[ncurv=.4,angleA=-45,angleB=225,nodesepA=-1,nodesepB=-1.3]{->}{x}{h2}
\naput[labelsep=.5]{$[*]|[\fisd,\fisd]$}
\nbput[labelsep=.5]{$11$}
\nbput[labelsep=-.2,npos=.37]{\pscirclebox[linestyle=none,fillstyle=solid,fillcolor=lightgray,framesep=0]{$f_8$}}
%%\naput[labelsep=.5]{$[*]|[n]$}
%%\nbput[labelsep=.5]{$11$}
%%\nbput[labelsep=-.2,npos=.39]{\pscirclebox[linestyle=none,fillstyle=solid,fillcolor=lightgray,framesep=0]{$f_{\tiny 10}$}}
%
\nccurve[ncurv=.75,angleA=-60,angleB=240,nodesepA=-.8,nodesepB=-.7]{->}{x}{h2}
\naput[labelsep=.5]{$[*]|[\fisd,\fisd,\fisd]$}
\nbput[labelsep=.5]{$11$}
\nbput[labelsep=-.2,npos=.39]{\pscirclebox[linestyle=none,fillstyle=solid,fillcolor=lightgray,framesep=0]{$f_9$}}
%%\naput[labelsep=.5]{$[*]|[n,\fisd]$}
%%\nbput[labelsep=.5]{$11$}
%%\nbput[labelsep=-.2,npos=.39]{\pscirclebox[linestyle=none,fillstyle=solid,fillcolor=lightgray,framesep=0]{$f_{\tiny 11}$}}
%
%%\nccurve[ncurv=1.05,angleA=-75,angleB=255,nodesepA=-.4,nodesepB=-.5]{->}{x}{h2}
%%\naput[labelsep=.5]{$[*]|[n,\fisd,\fisd]$}
%%\nbput[labelsep=.5]{$11$}
%%\nbput[labelsep=-.2,npos=.39]{\pscirclebox[linestyle=none,fillstyle=solid,fillcolor=lightgray,framesep=0]{$f_{\tiny 12}$}}
%
\end{pspicture}
\caption{The set of \gpus \BROut{20} after the call to procedure $g$ in procedure $f$ of Figure~\ref{fig:heap_eg}. Local variable $y$ has been eliminated.}
\label{fig:heap_final.gpg.f}
\end{figure}

\begin{example}{}
This example illustrates \gpu reduction with 3-limited \indlists using
\gpu $g_3$ of $\mtsym_g$ shown in Figure~\ref{fig:heap_eg}(b).
At the call site 20 in 
procedure $f$ of Figure~\ref{fig:heap_eg}(a), the upwards-exposed variable 
$x'$ in $\mtsym_g$ is substituted by $x$ in
$\mtsym_{f}$ (see Section~\ref{sec:interprocedural.extensions}). All \gpu compositions 
for this examples are \tscomp compositions.
The \gpus in \BRIn{20} (Figure~\ref{fig:heap_eg}(d)) are used for composition. The set
\BROut{20} is same as \BROut{21} shown in Figure~\ref{fig:heap_final.gpg.f} except that \BROut{20} also contains the
\gpus involving $y$ which is a local variable of $f$ and is not in the scope of the
caller procedures.

The \gpu composition \text{$g_2 \ecomp f_2$} for 
\text{$f_2 \!:\! \denew{x}{[*]|[\;]}{h_{14}}{15}$} and \text{$g_2 \!:\! \denew{x}{[*]|[*,n]}{x}{11}$} (with $x$
substituting for $x'$)
creates a reduced \gpu \denew{x}{[*]|[n]}{h_{14}}{11} which is further composed with 
\text{$f_3 \!:\! \denew{h_{14}}{[n]|[\;]}{h_{17}}{17}$} to create a 
reduced \gpu \text{$f_6 \!:\! \denew{x}{[*]|[\;]}{h_{17}}{11}$} (Figure~\ref{fig:heap_final.gpg.f}). 

Now \gpu $g_3$ must be composed with $f_2$, $f_3$ and $f_5$.
The composition \text{$g_3 \ecomp f_2$} for \text{$g_3 \!:\! \denew{x}{[*]|[*,n,n]}{x}{11}$} creates
two \gpus \denew{x}{[*]|[n,n]}{h_{14}}{11} and \denew{x}{[*]|[n,n,\fisd]}{h_{14}}{11}.
The newly created \gpu  \denew{x}{[*]|[n,n]}{h_{14}}{11} is further composed with $f_3$ to create \gpu \denew{x}{[*]|[n]}{h_{17}}{11}
which is further composed with $f_5$ to recreate \gpu \text{$f_6 \!:\! \denew{x}{[*]|[\;]}{h_{17}}{11}$}. 
The \gpu  composition between the other newly created \gpu \denew{x}{[*]|[n,n,\fisd]}{h_{14}}{11} and $f_3$ 
creates \gpus \denew{x}{[*]|[n,\fisd]}{h_{17}}{11} and \denew{x}{[*]|[n,\fisd,\fisd]}{h_{17}}{11}. 
The \gpu \denew{x}{[*]|[n,\fisd]}{h_{17}}{11} further composes with $f_5$ creating a \gpu \text{$f_7 \!:\! \denew{x}{[*]|[\fisd]}{h_{17}}{11}$}
while the composition between \gpus  \de{x}{[*]|[n,\fisd,\fisd]}{h_{17}}  and $f_5$ creates two reduced \gpus
\text{$f_8 \!:\! \denew{x}{[*]|[\fisd,\fisd]}{h_{17}}{11}$} and \text{$f_9 \!:\! \denew{x}{[*]|[\fisd,\fisd,\fisd]}{h_{17}}{11}$}.

%%\gpu $f_6$ is already in its reduced form and hence no further \gpu compositions are required. 
%%The composition \text{$g_3 \ecomp f_6$} eventually creates three \gpus 
%%\text{$f_{10} \!:\! \denew{x}{[*]|[n]}{h_{17}}{11}$}, \text{$f_{11} \!:\! \denew{x}{[*]|[n,\fisd]}{h_{17}}{11}$} and 
%%\text{$f_{12} \!:\! \denew{x}{[*]|[n,\fisd,\fisd]}{h_{17}}{11}$} through a series of compositions.
Note that \gpu $f_5$ is used only once in a series of compositions (Example~\ref{eg:use.gpu.once} explains this).

The final reduced \gpus $f_6$, $f_7$, $f_8$, and, $f_9$ are 
members of the set \BROut{21} containing the \gpus reaching the \End{} of procedure $f$ (as shown in Figure~\ref{fig:heap_final.gpg.f}). 
%%Note that no kill is performed on account of the presence of \gpu.
These reduced \gpus represent the following information:
\begin{itemize}
\item $f_6$ implies that $x$ now points-to heap location $h_{17}$.
\item $f_7$ imply that $x$ points-to heap locations that are one dereference away from $h_{17}$. 
\item $f_8$ imply that $x$ points-to heap locations that are two dereferences away from $h_{17}$. 
\item $f_9$ imply that $x$ points-to heap locations that are beyond two dereferences from $h_{17}$. 
\end{itemize}
Thus, $x$ points to every node in the linked list.
\end{example}

\begin{example}{eg:use.gpu.once}
To see why \gpu reduction in Definition~\ref{def:edge.reduction.heap} excludes a \gpu used for composition once,
observe that \gpus $f_7$, $f_8$ and $f_9$ can be further composed with \gpu $f_5$.
The composition of $f_7$ with $f_5$ creates \gpu $f_6$. Similarly, repetitive compositions of $f_8$
with $f_5$ also creates \gpu $f_6$. This indicates that $x$ points to only $h_{17}$
and misses out the fact that $x$ points to every location in the linked list which is represented by $h_{17}$
and is represented by \gpus $f_7$, $f_8$ and $f_9$.
\end{example}

A cycle in a graph induced by a set of \gpus could also occur because of a cyclic data structure. 

\begin{example}{}
Let an assignment $y \rightarrow n = x$ be inserted in procedure $f$ after line 19
in Figure~\ref{fig:heap_eg}. This creates a 
circular linked list instead of a simple linked list. 
This will cause inclusion of the \gpu \denew{h_{17}}{[n]|[\;]}{h_{14}}{}
in Figure~\ref{fig:heap_eg}(d), thereby creating a cycle
between the nodes $h_{14}$ and $h_{17}$. 
\end{example}

\section{Handling Calls through Function Pointers}
\label{sec:handling_fp}
\label{sec:level_3}

\begin{figure}[t]
%%\centering
\begin{center}
\setlength{\codeLineLength}{20mm}
\renewcommand{\arraystretch}{.9}
\begin{tabular}{ccc}
\begin{tabular}{c}
\begin{pspicture}(0,0)(30,76)
%\psframe(0,0)(30,76)
\putnode{m}{origin}{17}{73}{\psframebox{\Start{f}}}
\putnode{e}{m}{0}{-10}{\psframebox{$\tt fp = p;$}}
	\putnode{w}{e}{-10}{0}{01}
\putnode{g}{e}{0}{-10}{\psframebox{$\tt x = \&a;$}}
	\putnode{w}{g}{-10}{0}{02}
\putnode{a}{g}{0}{-10}{\psframebox[framesep=2]{$\tt g(fp);$}}
	\putnode{w}{a}{-10}{0}{03}
\putnode{c}{a}{0}{-10}{\psframebox{$\tt fp = q;$}}
	\putnode{w}{c}{-10}{0}{04}
\putnode{h}{c}{0}{-10}{\psframebox{$\tt z = \&b;$}} 
	\putnode{w}{h}{-10}{0}{05}
\putnode{d}{h}{0}{-10}{\psframebox{$\tt g(fp);$}}
	\putnode{w}{d}{-10}{0}{06}
\putnode{f}{d}{0}{-10}{\psframebox{\End{f}}}
%%%
\ncline{->}{m}{e}
\ncline{->}{e}{g}
\ncline{->}{g}{a}
\ncline{->}{a}{c}
\ncline{->}{c}{h}
\ncline{->}{h}{d}
\ncline{->}{d}{f}
\end{pspicture}
\end{tabular}
&
\begin{tabular}{c}
\begin{pspicture}(0,0)(15,44)
%%\psframe(0,0)(22,40)
\putnode{m}{origin}{10}{35}{\psframebox{\Start{g}}}
\putnode{a}{m}{0}{-12}{\psframebox{$\;\;\tt fp();\;\;$}}
	\putnode{w}{a}{-10}{0}{07}
\putnode{d}{a}{0}{-12}{\psframebox{\End{g}}}
%%%
\ncline{->}{m}{a}
\ncline{->}{a}{d}
\end{pspicture}
\end{tabular}
&
\begin{tabular}{c}
\begin{tabular}{c}
\begin{pspicture}(0,0)(30,34)
%%\psframe(0,0)(22,40)
\putnode{m}{origin}{13}{30}{\psframebox{\Start{p}}}
\putnode{a}{m}{0}{-12}{\psframebox{$\tt y = x;$}}
	\putnode{w}{a}{-10}{0}{08}
\putnode{d}{a}{0}{-12}{\psframebox{\End{p}}}
%%%
\ncline{->}{m}{a}
\ncline{->}{a}{d}
\end{pspicture}
\end{tabular}
\\
\begin{tabular}{c}
\begin{pspicture}(0,0)(30,34)
%%\psframe(0,0)(22,40)
\putnode{m}{origin}{13}{28}{\psframebox{\Start{q}}}
\putnode{a}{m}{0}{-12}{\psframebox{$\tt y = z;$}}
	\putnode{w}{a}{-10}{0}{09}
\putnode{d}{a}{0}{-12}{\psframebox{\End{q}}}
%%%
\ncline{->}{m}{a}	
\ncline{->}{a}{d}
\end{pspicture}
\end{tabular}
\end{tabular}
\end{tabular}
\caption{An example demonstrating the handling of function pointers.}
\label{fig:fp_eg}
\end{center}
\end{figure}
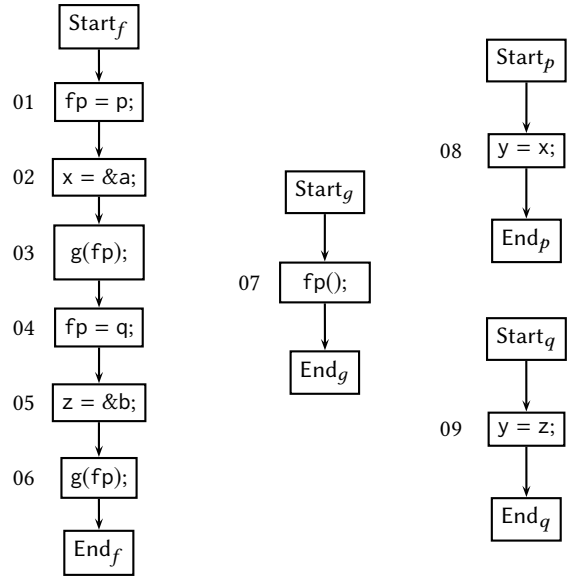

Recall that in the case of recursion, we may have incomplete \gpgs 
because the \gpgs of the callees are incomplete.
Similarly, in the presence of a call through a function pointer,  
we have incomplete \gpgs{} for a different reason---the callee procedure of such a call is not 
known.  
We model a call through function pointer (say {\em fp}) at call site \flab as a use  statement
with a \gpu \denew{\usenode}{1|1}{\text{\em fp}}{\flab} (Section~\ref{sec:dfv_compute}). 

%%For the purpose of reaching \gpus analyses,
%%these calls are represented by $\mtsym_{\top}$.
%%However, for constructing \gpg, the calls are represented
%%by a \emph{call} \gpb in the case of recursion and \emph{an indirect call} \gpb 
%%in the case of call through a function pointer.  
%%An indirect call \gpb differs from a call \gpb 
%%in that the former may represent multiple calls---one for each possible pointee of the function pointer.

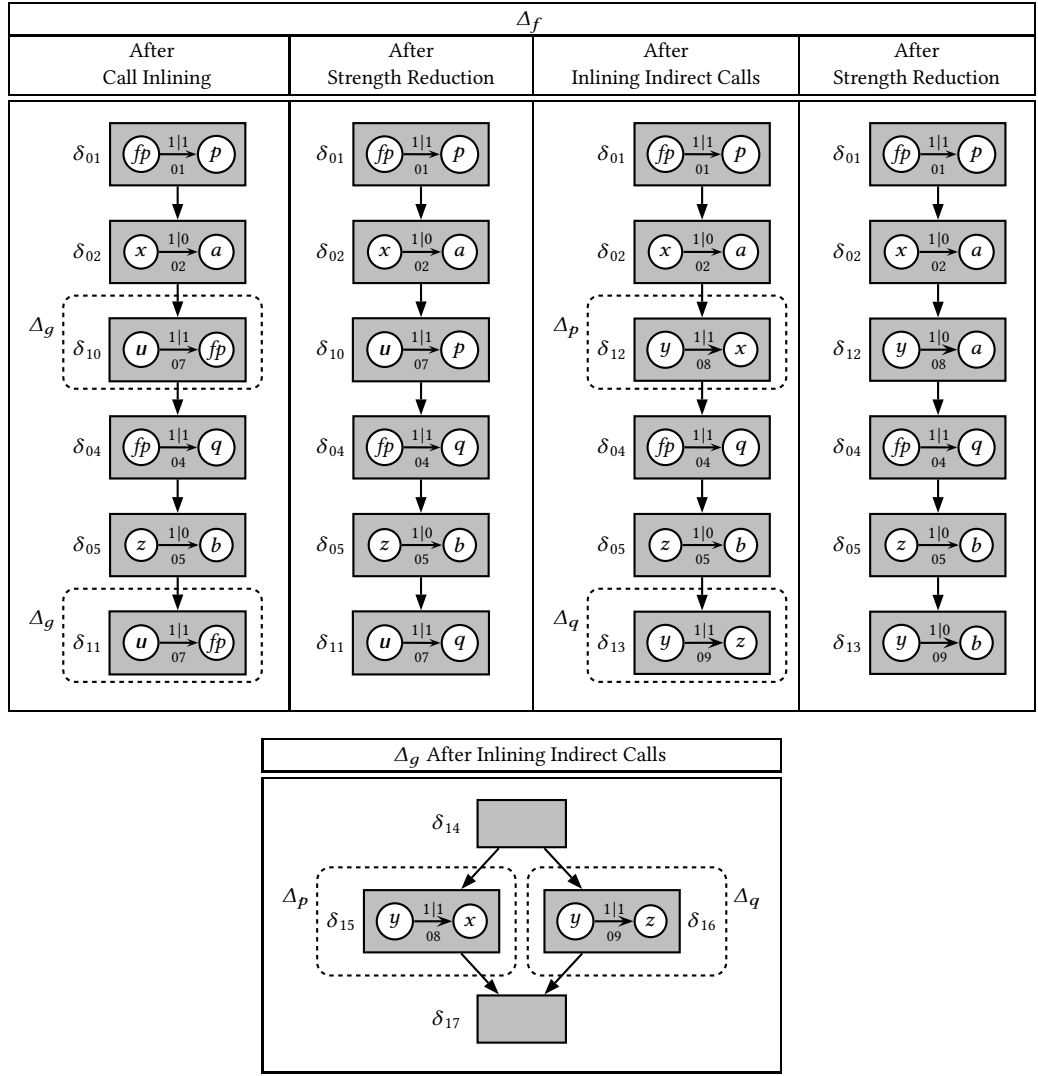
\begin{figure}[t]
%%\centering
\small
%%%%%%%%%%%%%%%%%%%%%%%%% Definitions of graphs %%%%%%%%%%%%%%%%%%%%%%%5
		\newcommand{\gpgA}{%
		\begin{pspicture}(1,0)(14,4)
		%\psframe(0,0)(16,6)
		\putnode{a1}{origin}{3}{2}{\pscirclebox[fillstyle=solid,fillcolor=white,framesep=1]{$y$}}
		\putnode{m1}{a1}{10}{0}{\pscirclebox[fillstyle=solid,fillcolor=white,framesep=1.22]{$a$}}
		\ncline[arrowsize=1.5]{->}{a1}{m1}
		\naput[labelsep=.25,npos=.5]{\scriptsize $1|0$}
		\nbput[labelsep=1,npos=.5]{\scriptsize 08}
		\end{pspicture}
		}%
		\newcommand{\gpgB}{%
		\begin{pspicture}(1,0)(14,4)
		%\psframe(0,0)(16,6)
		\putnode{a1}{origin}{3}{2}{\pscirclebox[fillstyle=solid,fillcolor=white,framesep=1]{$y$}}
		\putnode{m1}{a1}{10}{0}{\pscirclebox[fillstyle=solid,fillcolor=white,framesep=.72]{$b$}}
		\ncline[arrowsize=1.5]{->}{a1}{m1}
		\naput[labelsep=.25,npos=.5]{\scriptsize $1|0$}
		\nbput[labelsep=1,npos=.5]{\scriptsize 09}
		\end{pspicture}
		}%
		\newcommand{\gpgC}{%
		\begin{pspicture}(1,0)(14,4)
		%\psframe(0,0)(16,6)
		\putnode{b1}{origin}{3}{2}{\pscirclebox[fillstyle=solid,fillcolor=white,framesep=.5]{$\text{\em fp}$}}
		\putnode{m1}{b1}{10}{0}{\pscirclebox[fillstyle=solid,fillcolor=white,framesep=1]{$p$}}
		\ncline{->}{b1}{m1}
		\naput[labelsep=.25,npos=.5]{\scriptsize $1|1$}
		\nbput[labelsep=1,npos=.5]{\scriptsize 01}
		\end{pspicture}
		}
		\newcommand{\gpgD}{%
		\begin{pspicture}(1,0)(14,4)
		%\psframe(0,0)(16,6)
		\putnode{x}{origin}{3}{2}{\pscirclebox[fillstyle=solid,fillcolor=white,framesep=1]{$x$}}
		\putnode{z1}{x}{10}{0}{\pscirclebox[fillstyle=solid,fillcolor=white,framesep=1.1]{$a$}}
		\ncline{->}{x}{z1}
		\naput[labelsep=.25,npos=.5]{\scriptsize $1|0$}
		\nbput[labelsep=1,npos=.5]{\scriptsize 02}
		\end{pspicture}
		}
		\newcommand{\gpgE}{%
		\begin{pspicture}(1,0)(14,4)
		%\psframe(0,0)(16,6)
		\putnode{b1}{origin}{3}{2}{\pscirclebox[fillstyle=solid,fillcolor=white,framesep=.5]{$\text{\em fp}$}}
		\putnode{m1}{b1}{10}{0}{\pscirclebox[fillstyle=solid,fillcolor=white,framesep=1]{$q$}}
		\ncline{->}{b1}{m1}
		\naput[labelsep=.25,npos=.5]{\scriptsize $1|1$}
		\nbput[labelsep=1,npos=.5]{\scriptsize 04}
		\end{pspicture}
		}
		\newcommand{\gpgF}{%
		\begin{pspicture}(1,0)(14,4)
		%\psframe(0,0)(16,6)
		\putnode{x}{origin}{3}{2}{\pscirclebox[fillstyle=solid,fillcolor=white,framesep=1]{$z$}}
		\putnode{z1}{x}{10}{0}{\pscirclebox[fillstyle=solid,fillcolor=white,framesep=.72]{$b$}}
		\ncline{->}{x}{z1}
		\naput[labelsep=.25,npos=.5]{\scriptsize $1|0$}
		\nbput[labelsep=1,npos=.5]{\scriptsize 05}
		\end{pspicture}
		}
		\newcommand{\gpgG}{%
		\begin{pspicture}(1,0)(14,4)
		%\psframe(0,0)(16,6)
		\putnode{x1}{origin}{3}{2}{\pscirclebox[fillstyle=solid,fillcolor=white,framesep=1]{$y$}}
		\putnode{y1}{x1}{10}{0}{\pscirclebox[fillstyle=solid,fillcolor=white,framesep=.3]{$x'$}}
		\ncline[arrowsize=1.5]{->}{x1}{y1}
		\naput[labelsep=.25,npos=.5]{\scriptsize $1|1$}
		\nbput[labelsep=1,npos=.5]{\scriptsize 08}
		\end{pspicture}
		}
		\newcommand{\gpgH}{%
		\begin{pspicture}(1,0)(14,4)
		%\psframe(0,0)(16,6)
		\putnode{z1}{origin}{3}{2}{\pscirclebox[fillstyle=solid,fillcolor=white,framesep=1]{$y$}}
		\putnode{x1}{z1}{10}{0}{\pscirclebox[fillstyle=solid,fillcolor=white,framesep=.35]{$z'$}}
		\ncline[arrowsize=1.5]{->}{z1}{x1}
		\naput[labelsep=.25,npos=.5]{\scriptsize $1|1$}
		\nbput[labelsep=1,npos=.5]{\scriptsize 09}
		\end{pspicture}
		}
		\newcommand{\gpgNewA}{%
		\begin{pspicture}(0,0)(15,6)
		%\psframe(0,0)(16,6)
		\putnode{b1}{origin}{3}{3}{\pscirclebox[fillstyle=solid,fillcolor=white,framesep=1.1]{\usenode}}
		\putnode{m1}{b1}{10}{0}{\pscirclebox[fillstyle=solid,fillcolor=white,framesep=.4]{$\text{\em fp}$}}
		\ncline{->}{b1}{m1}
		\naput[labelsep=.25,npos=.5]{\scriptsize $1|1$}
		\nbput[labelsep=1,npos=.5]{\scriptsize 07}
		\end{pspicture}
		}
		\newcommand{\gpgNewB}{%
		\begin{pspicture}(0,0)(15,6)
		%\psframe(0,0)(16,6)
		\putnode{b1}{origin}{3}{3}{\pscirclebox[fillstyle=solid,fillcolor=white,framesep=1.1]{\usenode}}
		\putnode{m1}{b1}{10}{0}{\pscirclebox[fillstyle=solid,fillcolor=white,framesep=1]{$p$}}
		\ncline{->}{b1}{m1}
		\naput[labelsep=.25,npos=.5]{\scriptsize $1|1$}
		\nbput[labelsep=1,npos=.5]{\scriptsize 07}
		\end{pspicture}
		}
		\newcommand{\gpgNewC}{%
		\begin{pspicture}(0,0)(15,6)
		%\psframe(0,0)(16,6)
		\putnode{b1}{origin}{3}{3}{\pscirclebox[fillstyle=solid,fillcolor=white,framesep=1.1]{\usenode}}
		\putnode{m1}{b1}{10}{0}{\pscirclebox[fillstyle=solid,fillcolor=white,framesep=1]{$q$}}
		\ncline{->}{b1}{m1}
		\naput[labelsep=.25,npos=.5]{\scriptsize $1|1$}
		\nbput[labelsep=1,npos=.5]{\scriptsize 07}
		\end{pspicture}
		}
		\newcommand{\gpgNewD}{%
		\begin{pspicture}(1,0)(14,4)
		%\psframe(0,0)(16,6)
		\putnode{x1}{origin}{3}{2}{\pscirclebox[fillstyle=solid,fillcolor=white,framesep=1]{$y$}}
		\putnode{y1}{x1}{10}{0}{\pscirclebox[fillstyle=solid,fillcolor=white,framesep=1]{$x$}}
		\ncline[arrowsize=1.5]{->}{x1}{y1}
		\naput[labelsep=.25,npos=.5]{\scriptsize $1|1$}
		\nbput[labelsep=1,npos=.5]{\scriptsize 08}
		\end{pspicture}
		}
		\newcommand{\gpgNewE}{%
		\begin{pspicture}(1,0)(14,4)
		%\psframe(0,0)(16,6)
		\putnode{z1}{origin}{3}{2}{\pscirclebox[fillstyle=solid,fillcolor=white,framesep=1]{$y$}}
		\putnode{x1}{z1}{10}{0}{\pscirclebox[fillstyle=solid,fillcolor=white,framesep=1]{$z$}}
		\ncline[arrowsize=1.5]{->}{z1}{x1}
		\naput[labelsep=.25,npos=.5]{\scriptsize $1|1$}
		\nbput[labelsep=1,npos=.5]{\scriptsize 09}
		\end{pspicture}
		}
\begin{center}
\renewcommand{\tabcolsep}{4pt}
\begin{tabular}{c}
\begin{tabular}{|c|c|c|c|}
\hline
\multicolumn{4}{|c|}{\rule[.5em]{0em}{0.5em} \rule[-.5em]{0.5em}{0em}$\mtsym_f$} 
\\ \hline
\rule[.5em]{0em}{0.5em} After 
	& After 
	& After 
	& After 
	\\
\rule[-.5em]{0.5em}{0em}
Call Inlining 
	& Strength Reduction
	& Inlining Indirect Calls 
	& Strength Reduction
\\ \hline\hline
\begin{tabular}{@{}c}
\begin{pspicture}(-2,0)(31,80)
%\psframe(0,0)(28,80)
\putnode{n0}{origin}{19}{78}{}
\putnode{n1}{n0}{0}{-5}{\psframebox[fillstyle=solid,fillcolor=lightgray,framesep=2]{\gpgC}}
	\putnode{w}{n1}{-12}{0}{$\gpbsym_{01}$}
\putnode{n2}{n1}{0}{-13}{\psframebox[fillstyle=solid,fillcolor=lightgray,framesep=2]{\gpgD}}
	\putnode{w}{n2}{-12}{0}{$\gpbsym_{02}$}
\putnode{n3}{n2}{0}{-13}{\psframebox[fillstyle=solid,fillcolor=lightgray]{\gpgNewA}}
	\putnode{w}{n3}{-12}{0}{$\gpbsym_{10}$}
	\putnode{pa}{n3}{-2}{1}{\psframebox[linestyle=dashed, dash=.6 .6,framearc=.25]{\rule{24mm}{0mm}\rule{0mm}{10mm}}}
	\putnode{w}{n3}{-18}{3}{$\mtsym_{g}$}
\putnode{n4}{n3}{0}{-13}{\psframebox[fillstyle=solid,fillcolor=lightgray,framesep=2]{\gpgE}}
	\putnode{w}{n4}{-12}{0}{$\gpbsym_{04}$}
\putnode{n5}{n4}{0}{-13}{\psframebox[fillstyle=solid,fillcolor=lightgray,framesep=2]{\gpgF}}
	\putnode{w}{n5}{-12}{0}{$\gpbsym_{05}$}
\putnode{n6}{n5}{0}{-13}{\psframebox[fillstyle=solid,fillcolor=lightgray]{\gpgNewA}}
	\putnode{w}{n6}{-12}{0}{$\gpbsym_{11}$}
	\putnode{pa}{n6}{-2}{1}{\psframebox[linestyle=dashed, dash=.6 .6,framearc=.25]{\rule{24mm}{0mm}\rule{0mm}{10mm}}}
	\putnode{w}{n6}{-18}{3}{$\mtsym_{g}$}
\psset{arrowsize=1.5,arrowinset=0}
\ncline{->}{n1}{n2}
\ncline{->}{n2}{n3}
\ncline{->}{n3}{n4}
\ncline{->}{n4}{n5}
\ncline{->}{n5}{n6}
\end{pspicture}
\end{tabular}
&
\begin{tabular}{@{}c}
\begin{pspicture}(0,0)(28,80)
%\psframe(0,0)(28,80)
\putnode{n0}{origin}{16}{78}{}
\putnode{n1}{n0}{0}{-5}{\psframebox[fillstyle=solid,fillcolor=lightgray,framesep=2]{\gpgC}}
	\putnode{w}{n1}{-12}{0}{$\gpbsym_{01}$}
\putnode{n2}{n1}{0}{-13}{\psframebox[fillstyle=solid,fillcolor=lightgray,framesep=2]{\gpgD}}
	\putnode{w}{n2}{-12}{0}{$\gpbsym_{02}$}
\putnode{n3}{n2}{0}{-13}{\psframebox[fillstyle=solid,fillcolor=lightgray]{\gpgNewB}}
	\putnode{w}{n3}{-12}{0}{$\gpbsym_{10}$}
	%\putnode{pa}{n3}{-2}{1}{\psframebox[linestyle=dashed, dash=.6 .6,framearc=.25]{\rule{24mm}{0mm}\rule{0mm}{10mm}}}
	%\putnode{w}{n3}{-18}{3}{$\mtsym_{g}$}
\putnode{n4}{n3}{0}{-13}{\psframebox[fillstyle=solid,fillcolor=lightgray,framesep=2]{\gpgE}}
	\putnode{w}{n4}{-12}{0}{$\gpbsym_{04}$}
\putnode{n5}{n4}{0}{-13}{\psframebox[fillstyle=solid,fillcolor=lightgray,framesep=2]{\gpgF}}
	\putnode{w}{n5}{-12}{0}{$\gpbsym_{05}$}
\putnode{n6}{n5}{0}{-13}{\psframebox[fillstyle=solid,fillcolor=lightgray]{\gpgNewC}}
	\putnode{w}{n6}{-12}{0}{$\gpbsym_{11}$}
	%\putnode{pa}{n6}{-2}{1}{\psframebox[linestyle=dashed, dash=.6 .6,framearc=.25]{\rule{24mm}{0mm}\rule{0mm}{10mm}}}
	%\putnode{w}{n6}{-18}{3}{$\mtsym_{g}$}
\psset{arrowsize=1.5,arrowinset=0}
\ncline{->}{n1}{n2}
\ncline{->}{n2}{n3}
\ncline{->}{n3}{n4}
\ncline{->}{n4}{n5}
\ncline{->}{n5}{n6}
\end{pspicture}
\end{tabular}
&
\begin{tabular}{@{}c}
\begin{pspicture}(-2,0)(29,80)
%\psframe(0,0)(28,80)
\putnode{n0}{origin}{19}{78}{}
\putnode{n1}{n0}{0}{-5}{\psframebox[fillstyle=solid,fillcolor=lightgray,framesep=2]{\gpgC}}
	\putnode{w}{n1}{-12}{0}{$\gpbsym_{01}$}
\putnode{n2}{n1}{0}{-13}{\psframebox[fillstyle=solid,fillcolor=lightgray,framesep=2]{\gpgD}}
	\putnode{w}{n2}{-12}{0}{$\gpbsym_{02}$}
\putnode{n3}{n2}{0}{-13}{\psframebox[fillstyle=solid,fillcolor=lightgray,framesep=2]{\gpgNewD}}
	\putnode{w}{n3}{-12}{0}{$\gpbsym_{12}$}
	\putnode{w}{n3}{-18}{3}{$\mtsym_{p}$}
	\putnode{pa}{n3}{-2}{1}{\psframebox[linestyle=dashed, dash=.6 .6,framearc=.25]{\rule{24mm}{0mm}\rule{0mm}{10mm}}}
\putnode{n4}{n3}{0}{-13}{\psframebox[fillstyle=solid,fillcolor=lightgray,framesep=2]{\gpgE}}
	\putnode{w}{n4}{-12}{0}{$\gpbsym_{04}$}
\putnode{n5}{n4}{0}{-13}{\psframebox[fillstyle=solid,fillcolor=lightgray,framesep=2]{\gpgF}}
	\putnode{w}{n5}{-12}{0}{$\gpbsym_{05}$}
\putnode{n6}{n5}{0}{-13}{\psframebox[fillstyle=solid,fillcolor=lightgray,framesep=2]{\gpgNewE}}
	\putnode{w}{n6}{-12}{0}{$\gpbsym_{13}$}
	\putnode{w}{n6}{-18}{3}{$\mtsym_{q}$}
	\putnode{pb}{n6}{-2}{1}{\psframebox[linestyle=dashed, dash=.6 .6,framearc=.25]{\rule{24mm}{0mm}\rule{0mm}{10mm}}}
\psset{arrowsize=1.5,arrowinset=0}
\ncline{->}{n1}{n2}
\ncline{->}{n2}{n3}
\ncline{->}{n3}{n4}
\ncline{->}{n4}{n5}
\ncline{->}{n5}{n6}
\end{pspicture}
\end{tabular}
&
\begin{tabular}{@{}c}
\begin{pspicture}(2,0)(29,80)
%\psframe(0,0)(28,80)
\putnode{n0}{origin}{19}{78}{}
\putnode{n1}{n0}{0}{-5}{\psframebox[fillstyle=solid,fillcolor=lightgray,framesep=2]{\gpgC}}
	\putnode{w}{n1}{-12}{0}{$\gpbsym_{01}$}
\putnode{n2}{n1}{0}{-13}{\psframebox[fillstyle=solid,fillcolor=lightgray,framesep=2]{\gpgD}}
	\putnode{w}{n2}{-12}{0}{$\gpbsym_{02}$}
\putnode{n3}{n2}{0}{-13}{\psframebox[fillstyle=solid,fillcolor=lightgray,framesep=2]{\gpgA}}
	\putnode{w}{n3}{-12}{0}{$\gpbsym_{12}$}
	%\putnode{w}{n3}{-18}{3}{$\mtsym_{p}$}
	%\putnode{pa}{n3}{-2}{1}{\psframebox[linestyle=dashed, dash=.6 .6,framearc=.25]{\rule{24mm}{0mm}\rule{0mm}{10mm}}}
\putnode{n4}{n3}{0}{-13}{\psframebox[fillstyle=solid,fillcolor=lightgray,framesep=2]{\gpgE}}
	\putnode{w}{n4}{-12}{0}{$\gpbsym_{04}$}
\putnode{n5}{n4}{0}{-13}{\psframebox[fillstyle=solid,fillcolor=lightgray,framesep=2]{\gpgF}}
	\putnode{w}{n5}{-12}{0}{$\gpbsym_{05}$}
\putnode{n6}{n5}{0}{-13}{\psframebox[fillstyle=solid,fillcolor=lightgray,framesep=2]{\gpgB}}
	\putnode{w}{n6}{-12}{0}{$\gpbsym_{13}$}
	%\putnode{w}{n6}{-18}{3}{$\mtsym_{q}$}
	%\putnode{pb}{n6}{-2}{1}{\psframebox[linestyle=dashed, dash=.6 .6,framearc=.25]{\rule{24mm}{0mm}\rule{0mm}{10mm}}}
\psset{arrowsize=1.5,arrowinset=0}
\ncline{->}{n1}{n2}
\ncline{->}{n2}{n3}
\ncline{->}{n3}{n4}
\ncline{->}{n4}{n5}
\ncline{->}{n5}{n6}
\end{pspicture}
\end{tabular}
\\ \hline
\end{tabular}
\\
\\
\begin{tabular}{|c|}
\hline
\rule[.5em]{0em}{0.5em} \rule[-.5em]{0.5em}{0em}$\mtsym_g$
After Inlining Indirect Calls
\\ \hline\hline
\begin{pspicture}(0,0)(66,38)
%\psframe(0,0)(50,38)
\putnode{n0}{origin}{33}{40}{}
\putnode{n1}{n0}{0}{-8}{\psframebox[fillstyle=solid,fillcolor=lightgray,framesep=3]{\;\;\;\;\;\;\;}}
	\putnode{w}{n1}{-10}{0}{$\gpbsym_{14}$}
\putnode{n2}{n1}{-12}{-13}{\psframebox[fillstyle=solid,fillcolor=lightgray,framesep=2]{\gpgNewD}}
	\putnode{w}{n2}{-12}{0}{$\gpbsym_{15}$}
	\putnode{pc}{n2}{-2}{0}{\psframebox[linestyle=dashed, dash=.6 .6,framearc=.25]{\rule{24mm}{0mm}\rule{0mm}{12mm}}}
	\putnode{w}{n2}{-18}{3}{$\mtsym_{p}$}
\putnode{nz}{n1}{12}{-13}{\psframebox[fillstyle=solid,fillcolor=lightgray,framesep=2]{\gpgNewE}}
	\putnode{w}{nz}{12}{0}{$\gpbsym_{16}$}
	\putnode{pd}{nz}{2}{0}{\psframebox[linestyle=dashed, dash=.6 .6,framearc=.25]{\rule{24mm}{0mm}\rule{0mm}{12mm}}}
	\putnode{w}{nz}{18}{3}{$\mtsym_{q}$}
\putnode{n3}{n1}{0}{-26}{\psframebox[fillstyle=solid,fillcolor=lightgray,framesep=3]{\;\;\;\;\;\;\;}}
	\putnode{w}{n3}{-10}{0}{$\gpbsym_{17}$}
{
\psset{arrowsize=1.5,arrowinset=0}
\ncline{->}{n1}{n2}
\ncline{->}{n1}{nz}
\ncline{->}{n2}{n3}
\ncline{->}{nz}{n3}
}
%%\nccurve[nodesepA=1.5,linecolor=gray,offsetA=-4,angleA=90,angleB=240,doubleline=true]{->}{qa}{pc}
\end{pspicture}
\\ \hline
%%$\mtsym_q$\rule[-.75em]{0em}{2em}
%%\\ \hline
%%\begin{pspicture}(0,0)(22,28)
%%%\psframe(0,0)(24,58)
%%\putnode{n1}{origin}{12}{15}{\psframebox[fillstyle=solid,fillcolor=lightgray,framesep=2]{\gpgH}}
%%	\putnode{w}{n1}{-12}{0}{$\gpbsym_{09}$}
%%%%%%%
%%\nccurve[nodesepA=1.5,linecolor=gray,angleA=60,angleB=270,doubleline=true]{->}{n1}{pd}
%%\nccurve[nodesepA=1.5,linecolor=gray,angleA=240,angleB=-5,doubleline=true]{->}{n1}{pb}
%%\end{pspicture}
\end{tabular}
\end{tabular}
\caption{
Handling function pointers for the example in Figure~\ref{fig:fp_eg}. First, the direct calls are inlined leading to the discovery of 
pointees of the function pointer {\em fp} causing further inlining and strength reduction.
See Example~\protect\ref{eg:func-ptr} for explanation.
%%In the first column, $\mtsym_g$ is inlined in $\mtsym_f$ at call sites 03 and 06.
%%	Strength reduction of this version of $\mtsym_f$ identifies $p$ and $q$ as 
%%	the pointees of {\em fp} at call sites 03 and 06 respectively (second column).
%%	Thus, $\mtsym_p$ and $\mtsym_q$ are inlined in $\mtsym_f$ (third column).
%%	The fourth column gives $\mtsym_f$ after further strength reduction. Redundancy
%%	elimination optimizations over $\mtsym_f$ would create a single \gpb with all the
%%	\gpus (not shown in the figure).
%%	Similarly, $\mtsym_g$ is shown after converting the indirect call at call site 07 into
%%	two direct calls and inlining the corresponding \gpgs ($\mtsym_p$ and $\mtsym_q$).
}
\label{fig:fp_eg_gpgs}
\end{center}
\end{figure}

Our goal is to convert a call through a function pointer into a direct call
for every pointee of the function pointer. Interleaving of strength reduction and call inlining
reduces the \gpu \denew{\usenode}{1|1}{\text{\em fp}}{\flab} and provides the pointees of {\em fp}.
This is identical to computing points-to information (Section~\ref{sec:dfv_compute}).
Until the pointees become available, the \gpu \denew{\usenode}{1|1}{\text{\em fp}}{\flab}
acts as a barrier.
Once the pointees become available, the indirect call converts to a set of direct calls
and are handled as explained in Section~\ref{sec:interprocedural.extensions}.

\begin{example}{eg:func-ptr}
Figure~\ref{fig:fp_eg} provides an example of procedures containing calls through function
pointers.
Figure~\ref{fig:fp_eg_gpgs} provides the \gpgs of the procedures before and after 
resolving all calls through function pointers.
Procedure $g$ has an indirect call through function pointer \emph{fp} in statement 07 and is 
modelled by a \gpb containing a single \gpu \denew{\usenode}{1|1}{\text{\em fp}}{07} where
\usenode models a use (Section~\ref{sec:dfv_compute}).
This \gpg is inlined in procedure $f$ in statement 03 as $\gpbsym_{10}$
and in statement 06 as $\gpbsym_{11}$. 

Since we have
\text{$\denew{\text{\em fp}}{1|1}{p}{01} \in \BRIn{10}$}, the \gpu in $\gpbsym_{10}$
reduces to \text{$\denew{\usenode}{1|1}{p}{07}$}
indicating that the callee of this indirect call is $p$. 
Similarly, the callee for the indirect call in $\gpbsym_{11}$ is $q$.
Hence we inline $\mtsym_p$ in $\gpbsym_{10}$ which then becomes $\gpbsym_{12}$. 
Similarly, $\mtsym_q$ is inlined in $\gpbsym_{11}$ which then becomes $\gpbsym_{13}$.
This information is reflected in $g$ by recording $p$ and $q$ as the
pointees of {\em fp} in statement 07. The indirect call in $g$ is converted to two direct
calls leading to the inlining of $\mtsym_p$ and $\mtsym_q$ in $\mtsym_g$.

In $\gpbsym_{03}$ in procedure $f$,
only procedure $p$ is called because \emph{fp} points to $p$ in statement 03 whereas in $\gpbsym_{06}$, only $q$ is called because 
\emph{fp} points to $q$
in statement $06$. However, in procedure $g$, either $p$ is called in the context of call at $03$ 
(represented by the \gpb $\gpbsym_{15}$ in the final \gpg) or $q$ is called 
in the context of call at $06$ (represented by the \gpb $\gpbsym_{16}$ in the final \gpg).
\end{example}

\section{Empirical Evaluation}
\label{sec:emp-eval}

The main motivation of our implementation was to 
evaluate the effectiveness of our optimizations in handling the following challenge for practical programs:
\begin{quote}
A procedure summary for flow- and context-sensitive points-to analysis needs to model the accesses of pointees defined in 
the callers and needs to maintain control flow between memory updates when the data dependence between them is not known.
Thus, the size of a summary can be potentially large. This effect is exacerbated by the transitive inlining of the
summaries of the callee procedures which can increase the size of a summary exponentially thereby hampering the
scalability of analysis.
\end{quote}

Section~\ref{sec:implementation.meas} describes our implementation, 
Section~\ref{sec:measurements.data} describes the metrics that we have used for our measurements,
Section~\ref{sec:observations.meas}
describes our empirical observations, and Section~\ref{sec:discussion.meas} analyzes our observations and
describes the lessons learnt.

\begin{table}[t]
\centering
\renewcommand{\tabcolsep}{2pt}
\begin{tabular}{|l|c|c|c|c|c|c|c|c|c|c|c|c|c|c|}
\hline 
\multirow{2}{*}{
		\begin{tabular}{@{}c@{}} \\ Program \\ \end{tabular}
		}
	& 
\multirow{2}{*}{
		\begin{tabular}{@{}c@{}} \\ kLoC \\ \end{tabular}
		}

	%%%%%%%%%%%%%%%%% Statistics
	&  \multirow{3}{*}{
		\renewcommand{\arraystretch}{.8}%
		\begin{tabular}{@{}c@{}} \# of \\ pointer \\ stmts \end{tabular}
		}
	&  \multirow{3}{*}{
		\renewcommand{\arraystretch}{.8}%
		\begin{tabular}{@{}c@{}} \# of \\ call \\ sites \end{tabular}
		}
	&  \multirow{4}{*}{
		\renewcommand{\arraystretch}{.8}%
	  	\begin{tabular}{@{}c@{}} \# of \\ procs.\end{tabular}
		}
	& \multicolumn{4}{c|}{
	 	%\multirow{1}{*}
		{
		\renewcommand{\arraystretch}{.8}%
		\begin{tabular}{@{}c@{}} 
		%	\# of procs. with \\different  no. of \\ uses of \asummflow 
		Proc. count for \\ different buckets of \\ \# of calls %\\ reuse of \gpgs)
			\end{tabular}}
		}
	%%%%%%%%%%%%%%%%% Aliasing Patterns
 	& \multicolumn{6}{c|}{
		\renewcommand{\arraystretch}{.8}%
		\begin{tabular}{@{}c@{}} 
		\rule{0em}{.9em}
			\# of procs. requiring different \\
			no. of PTFs  based on the  \\ no. of aliasing patterns
		\end{tabular}}

\\ \cline{6-15}
\rule{0em}{1em}
	& & & & & 2-5 & 5-10 & 10-20 & 20+
 	& 2-5 & 6-10 & 11-15 & 15+ & 2-5 & 15+
\\ \hline
\rule[-.1em]{0em}{1em} &
\multicolumn{1}{c}{$A$} & \multicolumn{1}{|c}{$B$} & \multicolumn{1}{c}{$C$} & \multicolumn{1}{|c}{$D$} & \multicolumn{4}{|c}{$E$} & \multicolumn{4}{|c}{$F$} & \multicolumn{2}{|c|}{$G$}
\\ \hline\hline
lbm & 0.9 & 370 & 30 & 19 & 5 & 0 & 0 & 0 & 8 & 0 & 0 & 0 & 13 & 0
\\ \hline
mcf & 1.6 & 480 & 29 & 23 & 11 & 0 & 0 & 0 & 0 & 0 & 0 & 0 & 4 & 0
\\ \hline
libquantum & 2.6 & 340 & 277 & 80 & 24 & 11 & 4 & 3 & 7 & 3 & 1 & 0 & 14 & 4
\\ \hline
bzip2 & 5.7 & 1650 & 288 & 89 & 35 & 7 & 2 & 1 & 22 & 0 & 0 & 0 & 28 & 2
\\ \hline
milc & 9.5 & 2540 & 782 & 190 & 60 & 15 & 9 & 1 & 37 & 8 & 0 & 1 & 35 & 25
\\ \hline
sjeng & 10.5 & 700 & 726 & 133 & 46 & 20 & 5 & 6 & 14 & 3 & 1 & 3 & 10 & 14
\\ \hline
hmmer & 20.6 & 6790 & 1328 & 275 & 93 & 33 & 22 & 11 & 62 & 5 & 3 & 4 & 88 & 32
\\ \hline
h264ref & 36.1 & 17770 & 2393 & 566 & 171 & 60 & 22 & 16 & 85 & 17 & 5 & 3 & 102 & 46
\\ \hline
gobmk & 158.0 & 212830 & 9379 & 2699 & 317 & 110 & 99 & 134 & 206 & 30 & 9 & 10 & 210 & 121
\\ \hline

\end{tabular}
\caption{Benchmark characteristics relevant to our analysis.}
\label{fig:stats}
\end{table}

\subsection{Implementation and Experiments}
\label{sec:implementation.meas}

We have implemented \gpg-based points-to analysis in GCC 4.7.2 using the 
LTO framework and have carried out measurements on 
SPEC CPU2006 benchmarks on a machine with 16 GB RAM with eight 64-bit Intel i7-4770 CPUs running at 3.40GHz.

Our method eliminates non-address-taken local variables
using the def-use chains explicated by the SSA-form.
Although we construct \gpus involving such variables, they are used for computing the 
points-to information within the procedure and do not appear in the \gpg of the procedure.
If a \gpu defining a global variable or a parameter reads a non-address-taken local variable, 
we identify the corresponding producer \gpus by traversing the def-use chains transitively.
This eliminates the need for filtering out the local variables from the \gpgs for inlining them 
in the callers.
As a consequence,
 a \gpg of a procedure consists of \gpus that involve global variables\footnote{
From now on we regard static, heap-summary nodes, and address-taken local variables as `global  variables'.},
parameters of the procedure, and the return variable which
is visible in the scope of its callers.
Since non-address-taken local variables have SSA versions,
storing the \gpus that define them flow-insensitively results in no loss of precision.

All address-taken local variables in a procedure
 are treated as global variables because they can escape the scope of the procedure. However, these variables are not strongly updated because
they could represent multiple locations.

\begin{figure}[t]
\begin{tabular}{@{}c|c@{}}
\includegraphics[width=5.8cm,height=6cm]{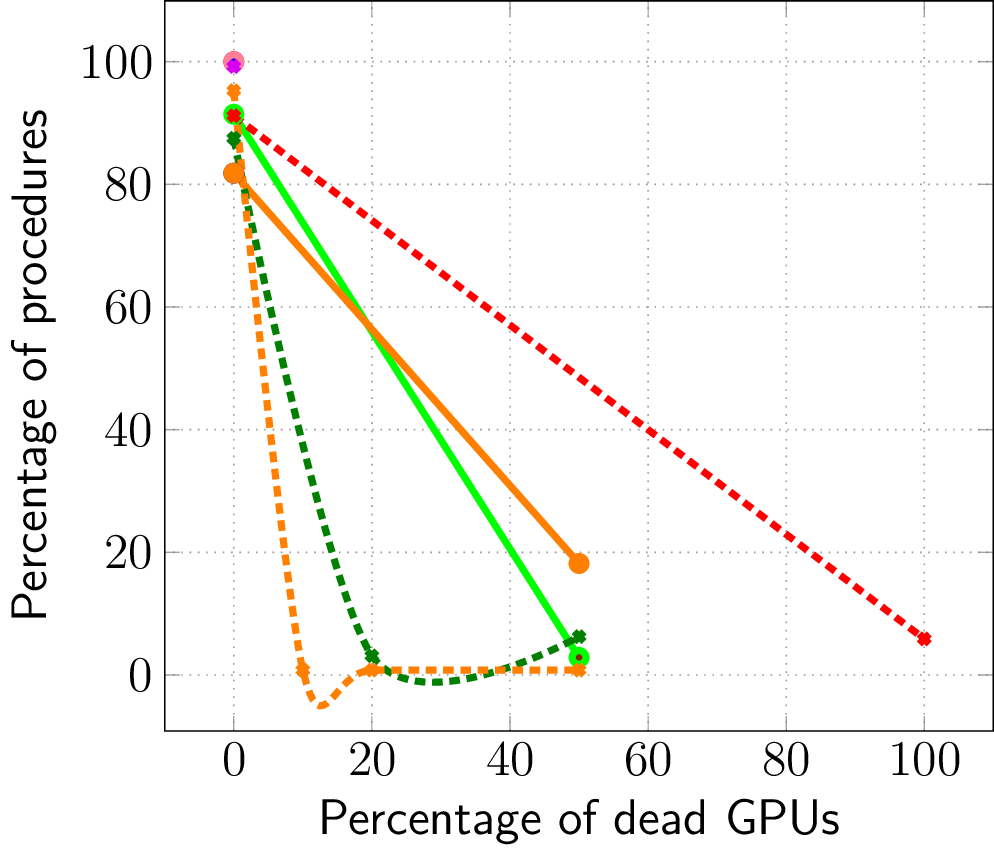}
\;
&
\includegraphics[width=7.4cm,height=6cm]{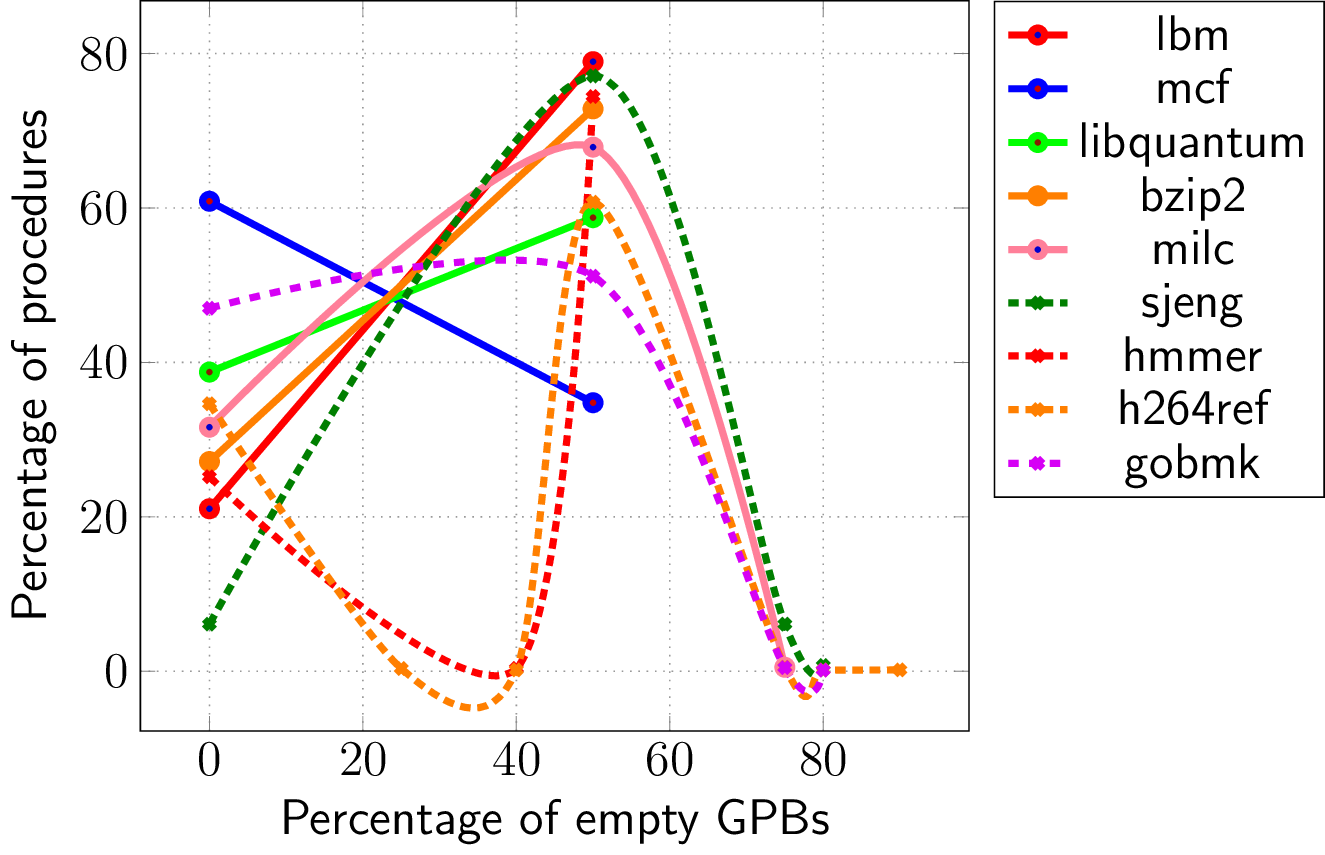}
\\ \hline 
\rule{0em}{17.5em}%
\includegraphics[width=5.8cm,height=6cm]{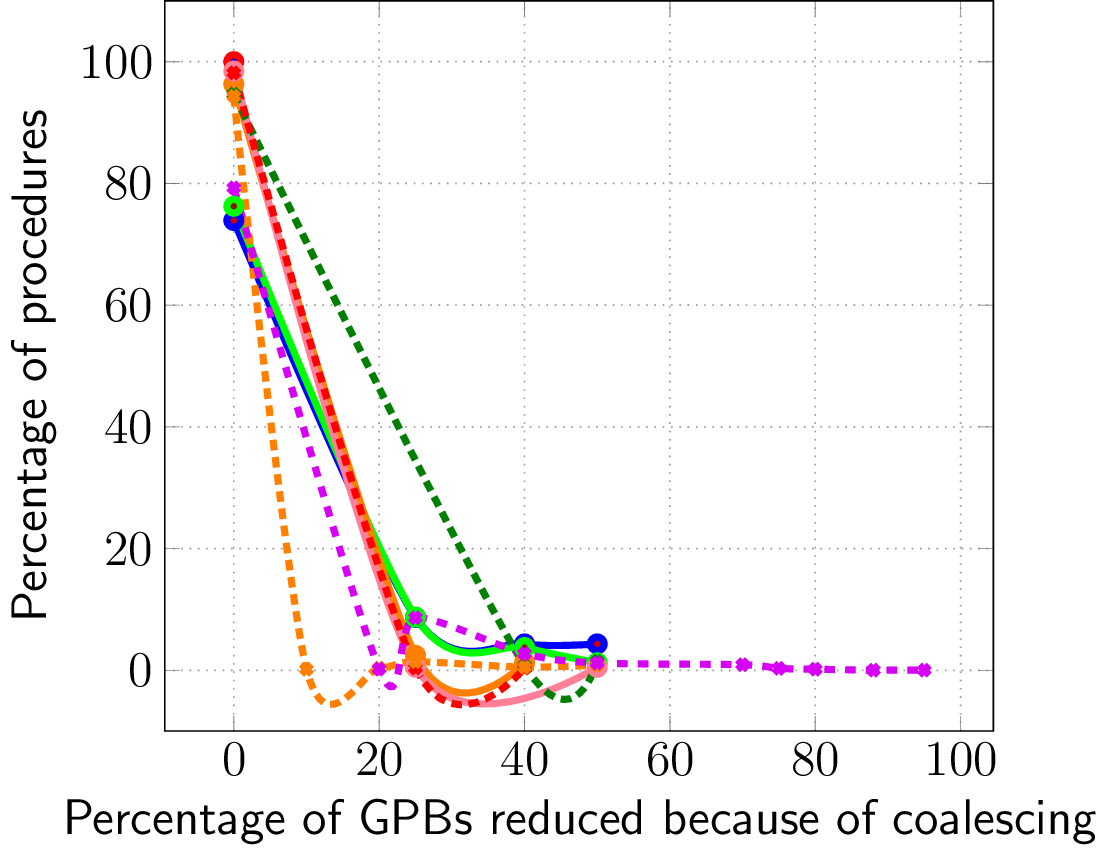}
\;
&
\includegraphics[width=7.4cm,height=6cm]{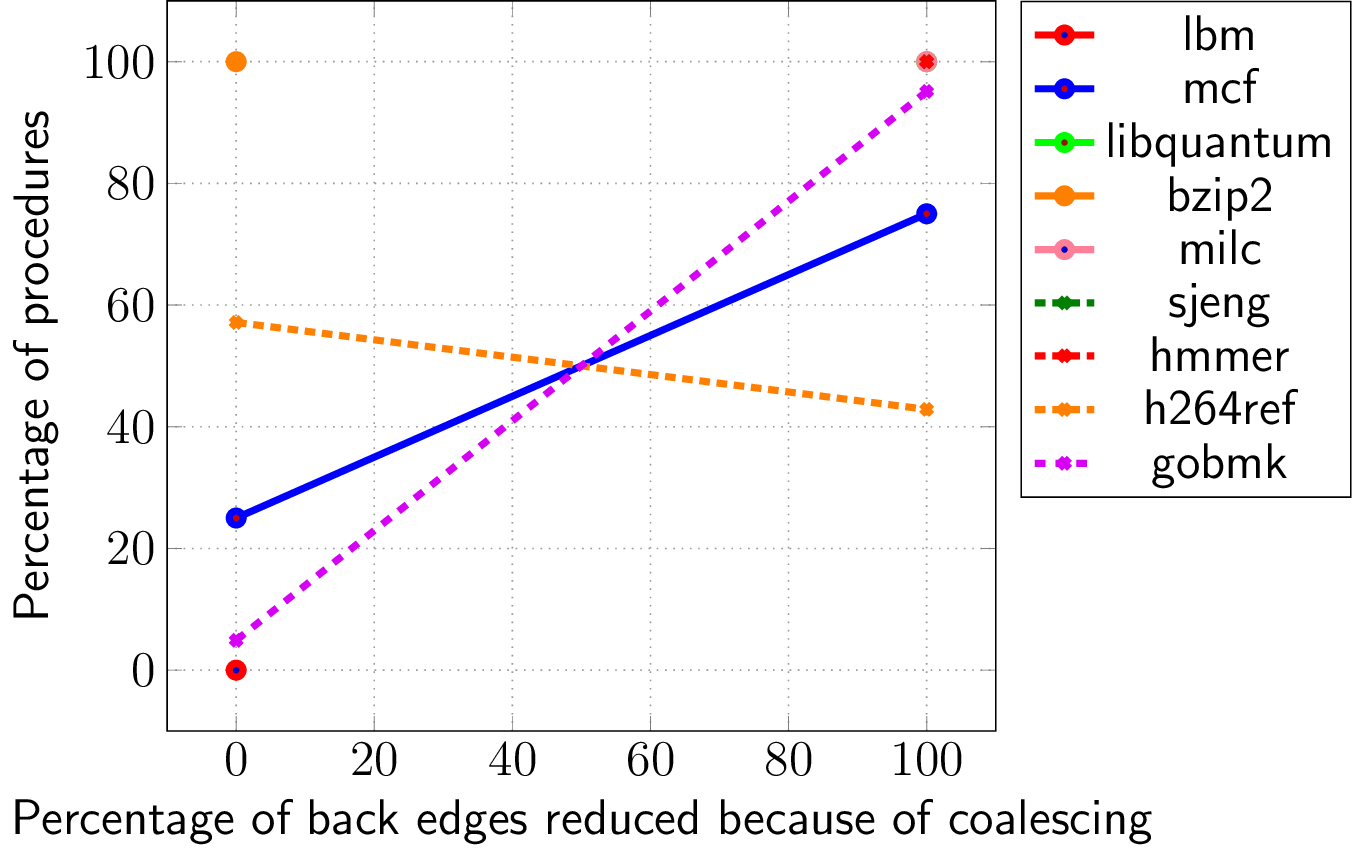}
\\ \hline 
\end{tabular}
\caption{Effectiveness of redundancy elimination optimizations. Benchmarks libquantum, milc, sjeng, and hmmer have all procedures whose all back
edges are eliminated because of coalescing shown by the same point (100, 100) in the fourth
plot. Hence they are not visible separately.}
\label{plot:dead-coal}
\end{figure}

We approximate the heap memory by maintaining $k$-limited indirection lists of field dereferences
for $k=3$ (see Section~\ref{sec:heap}).
An array is treated as a single variable in the following sense: accessing
a particular element is seen as accessing every possible element and updates are
treated as weak updates.
This applies to both when arrays of pointers are manipulated, as well as when arrays are
accessed through pointers.
Since there is no kill owing to weak update, arrays are maintained flow-insensitively by our analysis.

For pointer arithmetic involving a pointer to an array, 
we approximate the pointer being defined to point to every
element of the array.
For pointer arithmetic involving other pointers, 
we approximate the pointer being defined to point to every
possible location.
Our current implementation handles only locally defined
function pointers (Section~\ref{sec:handling_fp})
but can be easily extended to
handle function pointers defined in the calling contexts too.

\begin{figure}[t]
\begin{tabular}{@{}c|c@{}}
\includegraphics[width=5.8cm,height=6cm]{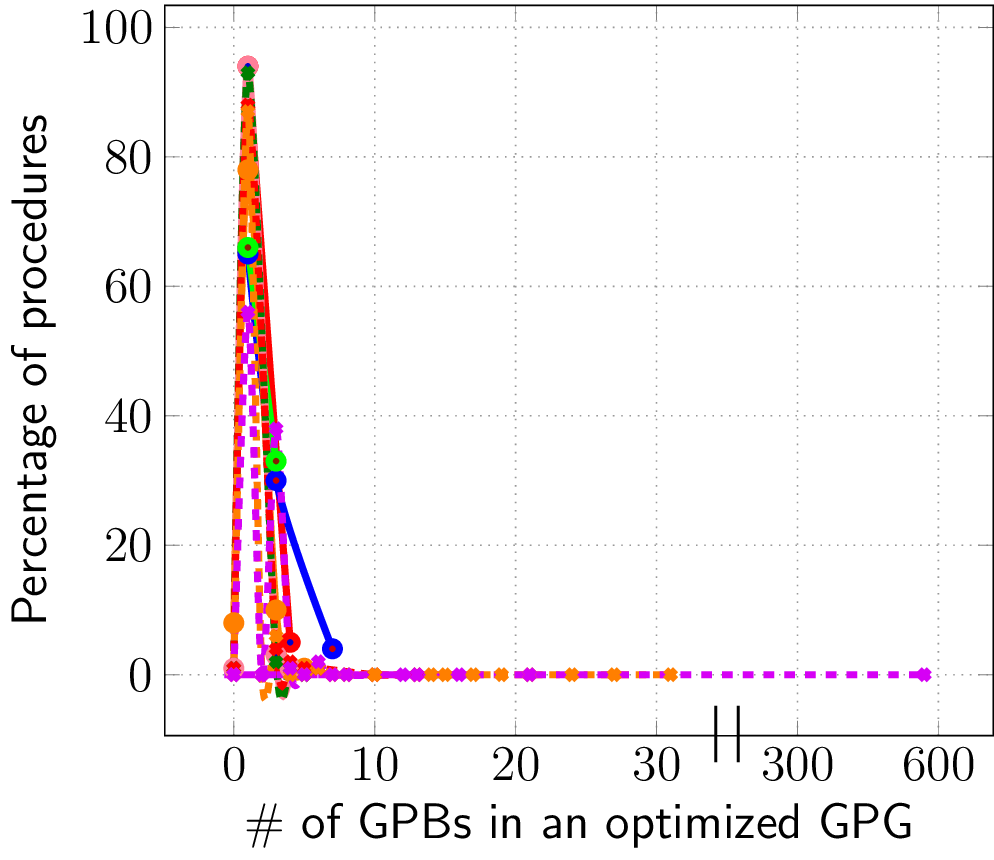}
\;
&
\includegraphics[width=7.4cm,height=6cm]{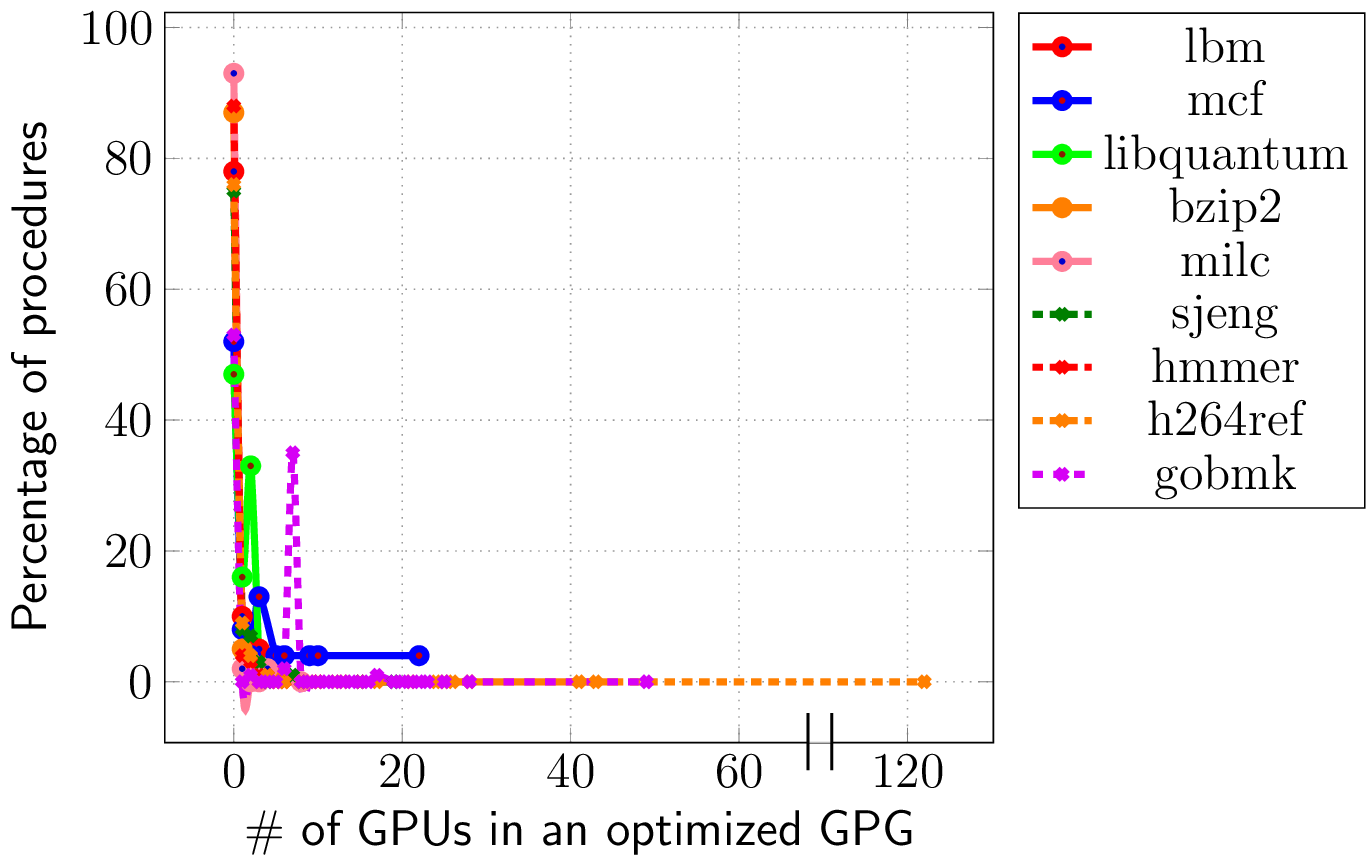}
\\ \hline
\multicolumn{2}{c}{\rule{0em}{17.5em}\includegraphics[width=7cm,height=5.8cm]{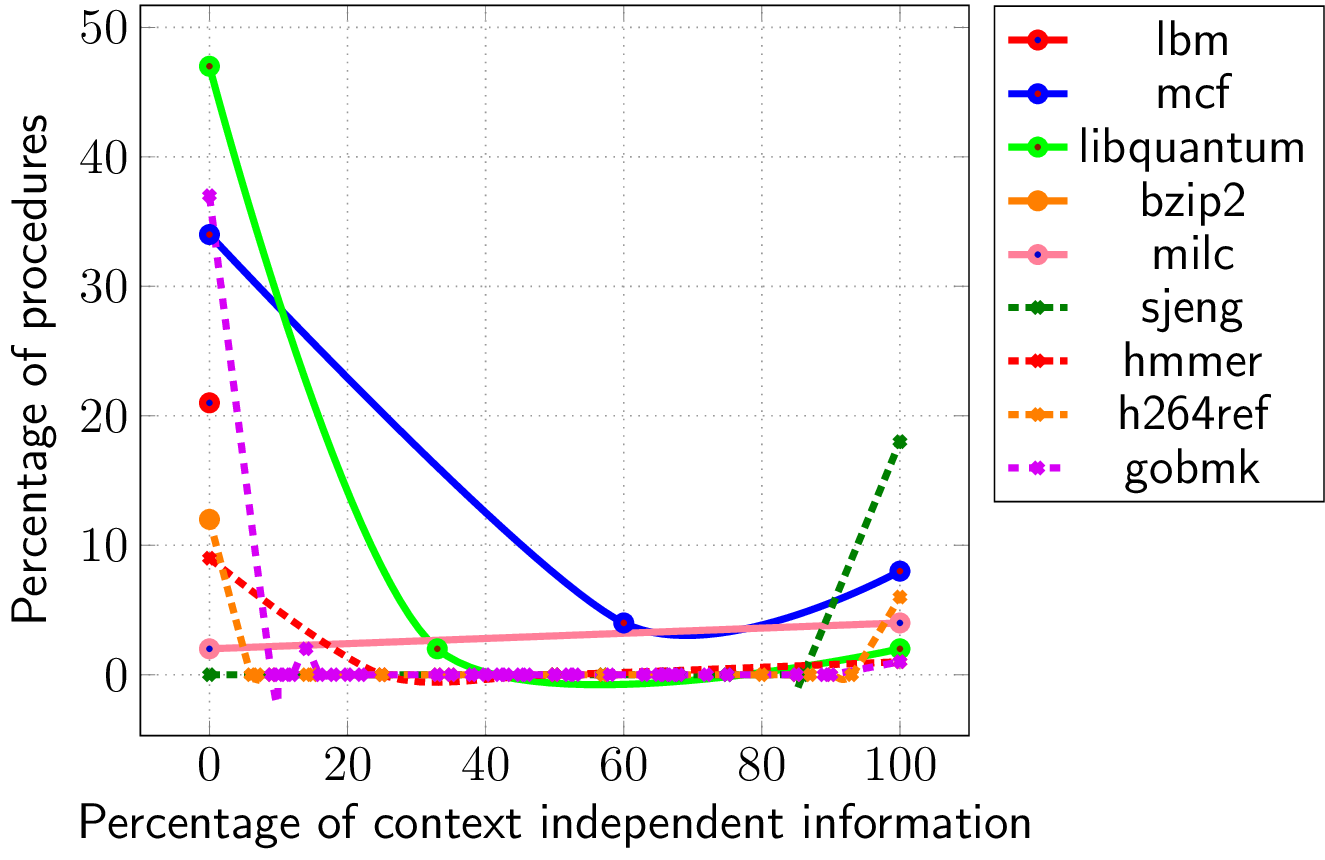}}
\\ \hline
\end{tabular}
\caption{Goodness measure of procedure summaries. A break in X-axis shown by two parallel 
lines is a 
discontinuity necessitated by wide variation in the number of \gpus and \gpbs across benchmarks.}
\label{plot:stats1}
\end{figure}

We have also implemented flow-insensitive points-to analysis 
by collecting the \gpus
in a \emph{\gpg store} which differs from a \gpb in that \gpus within a store can compose with each other whereas those in \gpb cannot.  
This allowed us to implement the following variants:
\begin{itemize}
	\item Flow- and context-insensitive (FICI) points-to analysis. For each benchmark program, we collected all \gpus
	across all procedures in a common store and performed all possible reductions. The resulting \gpus were 
	classical points-to edges representing the flow- and context-insensitive points-to information.
	\item Flow-insensitive and context-sensitive (FICS) points-to analysis. For each procedure of a benchmark program, all \gpus
	within the procedure were collected in a store for the procedure and all possible reductions were performed.
	The resulting store was used as a summary in the callers of the procedure giving context-sensitivity. 
	In the process the \gpus are reduced to classical points-to edges using the information from the calling context. This
	represents the flow-insensitive and context-sensitive points-to information for the procedure.
\end{itemize}

The third variant i.e., flow-sensitive and context-insensitive (FSCI) points-to analysis can be modelled by constructing
a supergraph by joining the control flow graphs of all procedures such that calls and returns are replaced by gotos.
This amounts to a top-down approach (or a bottom-up approach with a single summary for the entire program instead of
separate summaries for each procedure).
For practical programs,  this initial \gpg is too large for our analysis to scale. Our analysis achieves scalability
by keeping the \gpgs as small as possible at each stage. Therefore, we did not implement this variant of points-to analysis.
Note that the FICI variant  is also not a bottom-up approach because a separate summary is not constructed
for every procedure.  However, it was easy to implement because of a single store.

\subsection{Measurements}
\label{sec:measurements.data}

%%\change{}{We have evaluated our implementation on SPEC CPU 2006 benchmarks.}
We have measured the following for each benchmark program. 
The number of procedures varies significantly across the benchmark
programs. Besides, the number of \gpus and \gpbs varies across \gpgs. Hence we have plotted such data in terms of 
percentages.\footnote{The actual procedure counts are available at
\htmladdnormallink{https://www.cse.iitb.ac.in/~uday/soft-copies/gpg-pta-paper-appendix.pdf}{https://www.cse.iitb.ac.in/~uday/soft-copies/gpg-pta-paper-appendix.pdf}.}
%%The actual procedure counts are given in Appendix~\ref{app:add-data}. 

\begin{enumerate}[1)]
\item Characteristics of benchmark programs (Table~\ref{fig:stats}).
\item Effectiveness of redundancy elimination optimizations (Figure~\ref{plot:dead-coal}): 
      \begin{enumerate}[a)]
	\item The number of dead \gpus for each procedure.
	\item The number of empty \gpbs for each procedure created by strength reduction, call inlining and dead GPU elimination.
	\item A reduction in the number of \gpbs due to coalescing.
	\item A reduction in the number of back edges due to coalescing.
	\end{enumerate}
\item The goodness metric of the optimized procedure summaries (Figure~\ref{plot:stats1}):
      \begin{enumerate}[a)]
	\item Number of \gpbs in the optimized \gpgs.
	\item Number of \gpus in the optimized \gpgs.
	\item Number of \gpus that are dependent on locally defined pointers alone.
	\end{enumerate}
\item The number of \gpbs in a \gpg (Figure~\ref{plot:cmp-gpbs}): 
      \begin{enumerate}[a)]
	\item After call inlining, relative to the number of basic blocks in the \cfg.
	\item After all optimizations, relative to the number of basic blocks in the \cfg.
	\item After all optimizations,
              relative to the number of \gpbs after call inlining.
	\end{enumerate}
\item The number of \gpus in a \gpg (Figure~\ref{plot:cmp-gpus}):
      \begin{enumerate}[a)]
	\item After call inlining, relative to the number of pointer assignments in the \cfg.
	\item After all optimizations, relative to the number of pointer assignments in the \cfg.
	\item After all optimizations,
              relative to the number of \gpus in the \gpg after call inlining.
	\end{enumerate}
\item The number of control flow edges in a \gpg (Figure~\ref{plot:cmp-cf-edges}):
      \begin{enumerate}[a)]
	\item After call inlining, relative to the number of edges in the \cfg.
	\item After all optimizations, relative to the number of edges in the \cfg.
	\item After all optimizations, relative to the number of edges in the \gpg after call inlining.
	\end{enumerate}
\item Miscellaneous data about \gpgs (Table~\ref{tab:stats2}).
\item Time measurements (Figure~\ref{plot:time-meas-ptinfo}):
      \begin{enumerate}[a)]
	\item FSCS (with and without blocking), FICI, and FICS variants of points-to analyses (second plot). 
	\item Time for different optimizations without blocking (third plot).
	\item Time for different optimizations with blocking (fourth plot).
      \end{enumerate}
\item Average points-to pairs per procedure in FSCS, FICI, and FICS variants of points-to analyses.
      This data is plotted in the first plot of Figure~\ref{plot:time-meas-ptinfo}. 
\end{enumerate}

\begin{figure}[t]
\begin{tabular}{c|c}
\includegraphics[width=5.8cm,height=6cm]{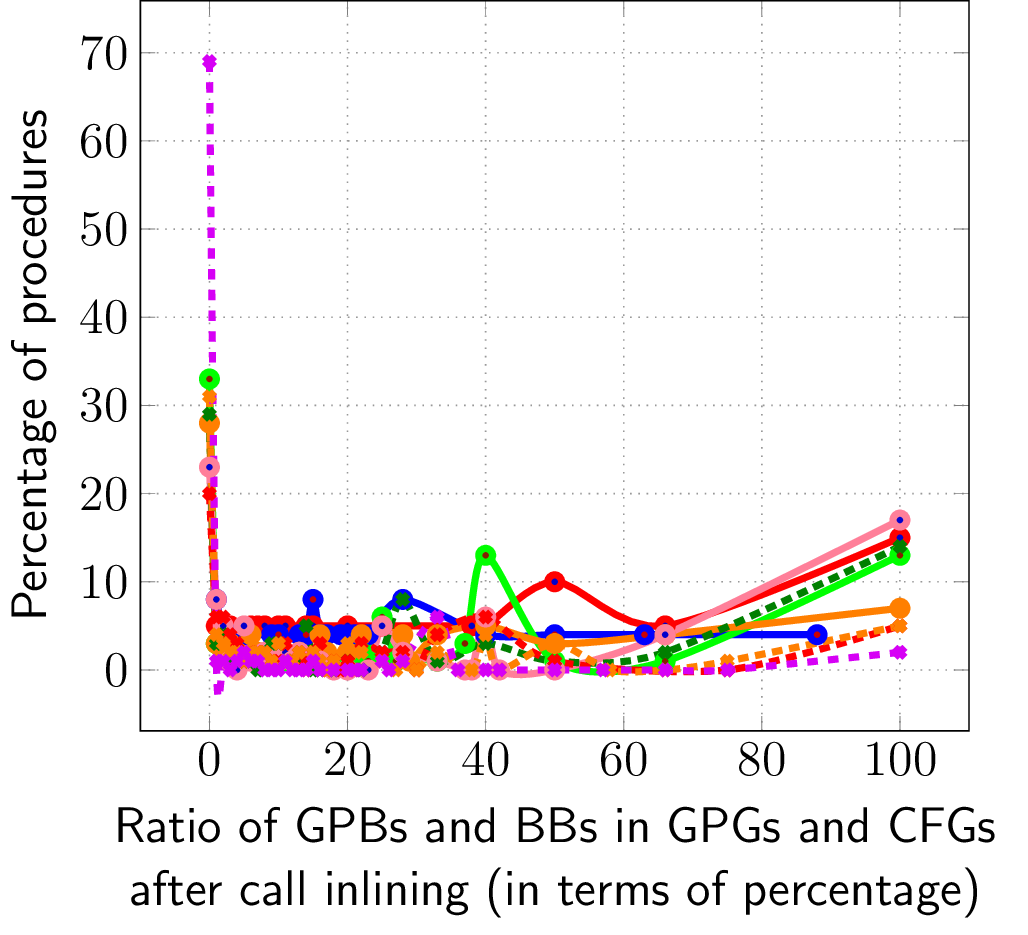}
\;\;
&
\includegraphics[width=7.3cm,height=6cm]{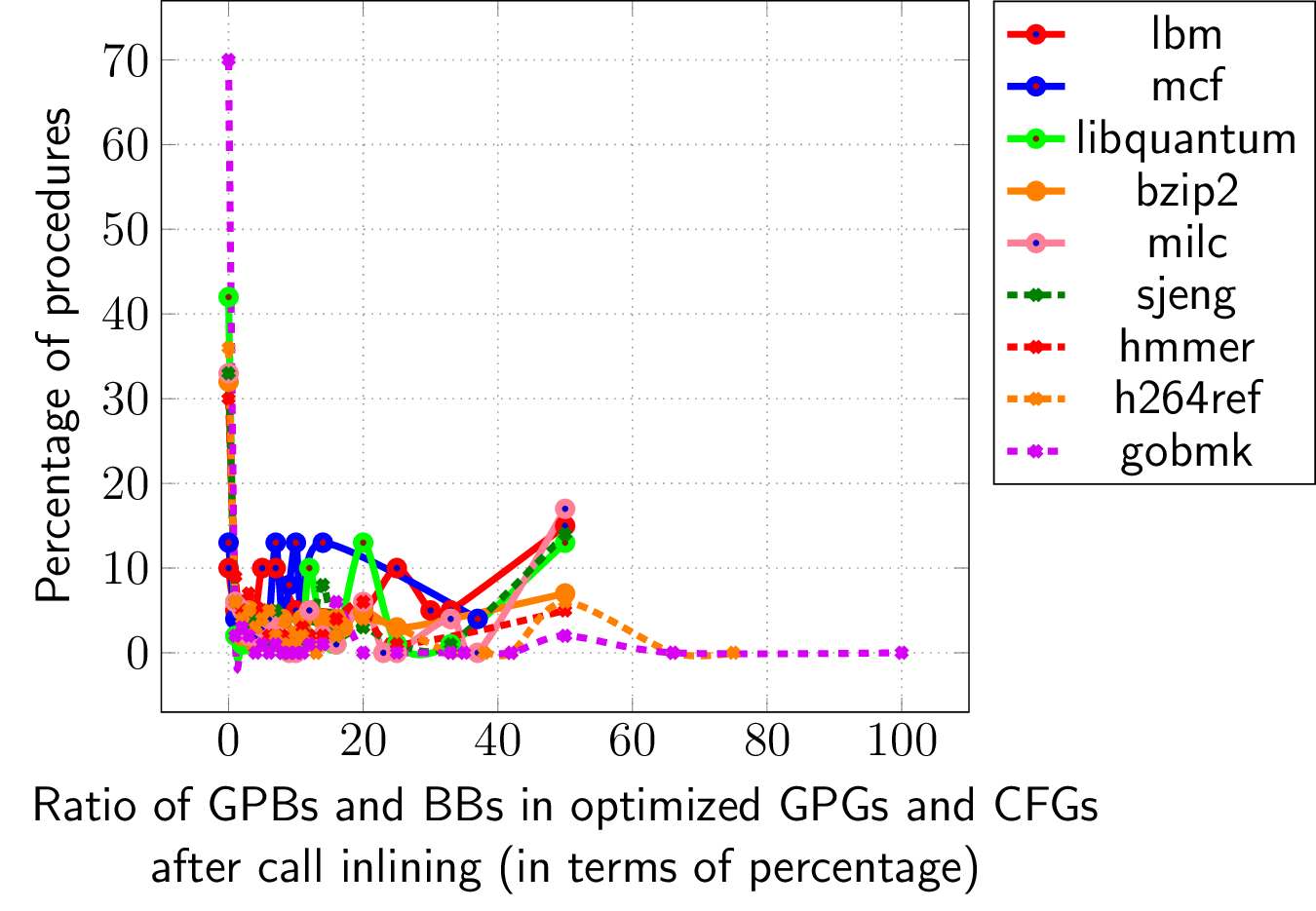}
\\ \hline
\multicolumn{2}{c}{\rule{0em}{17.5em}\includegraphics[width=7.5cm,height=6cm]{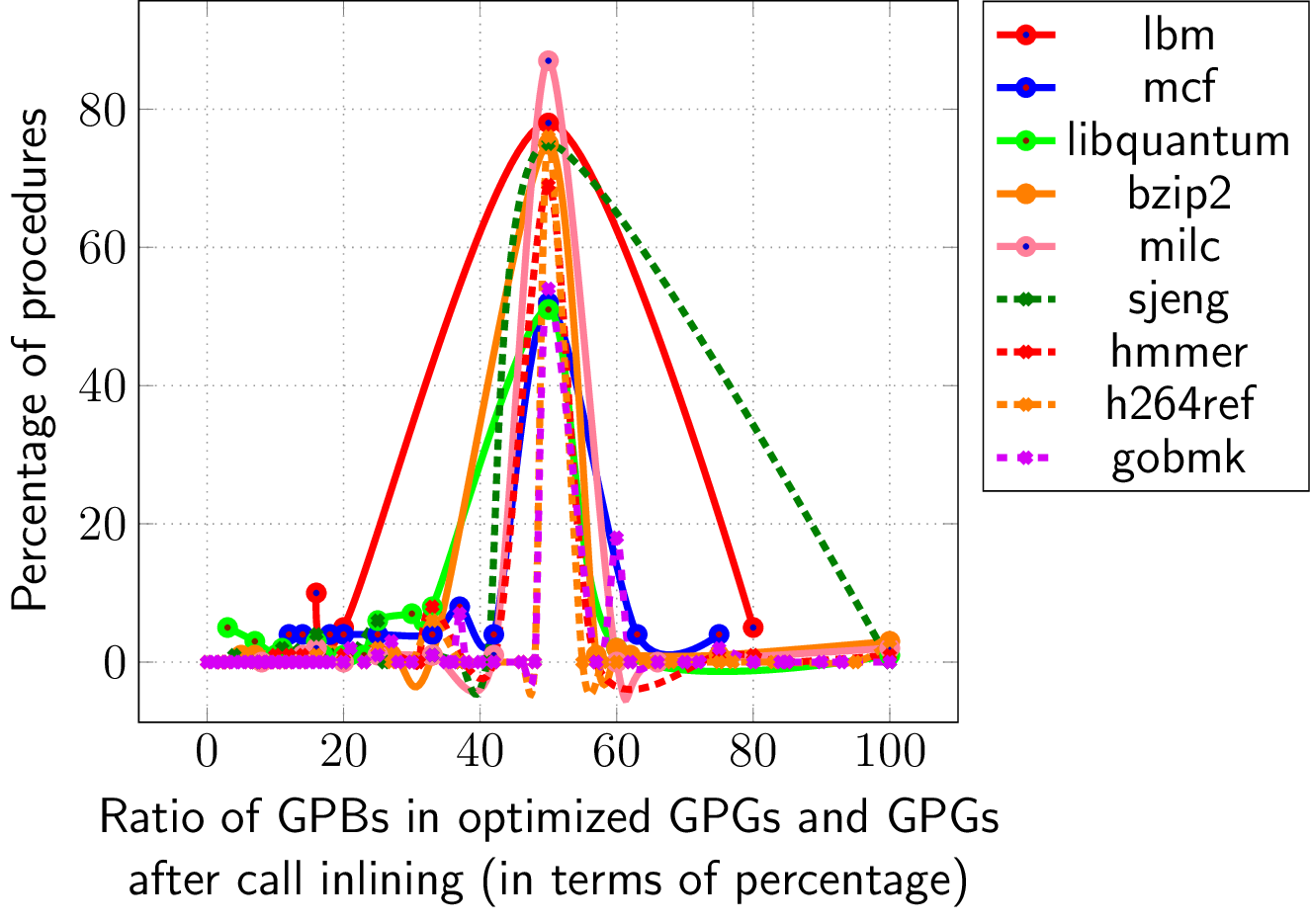}}
\\ \hline
\end{tabular}
\caption{Size of \gpgs relative to the size of corresponding procedures in terms of \gpbs and basic blocks.}
\label{plot:cmp-gpbs}
\end{figure}

\begin{figure}[t]
\begin{tabular}{c|c}
\includegraphics[width=5.8cm,height=6cm]{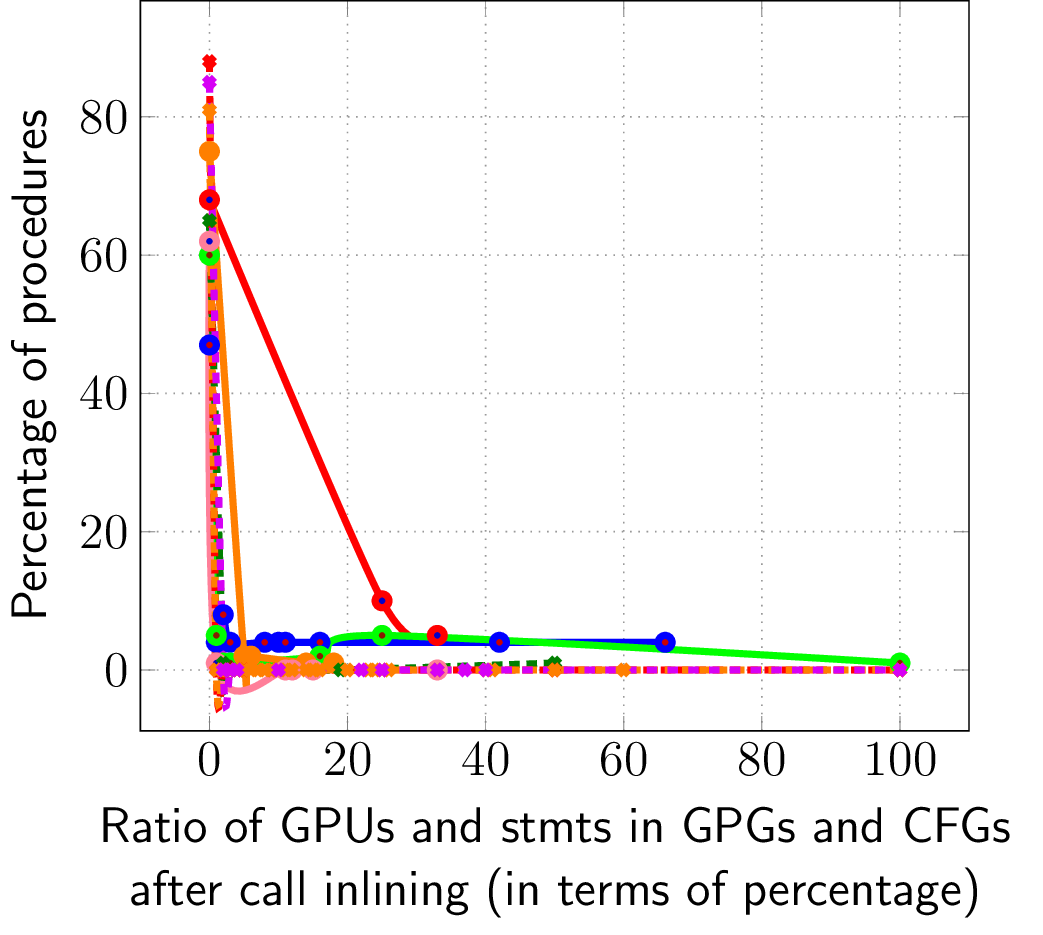}
\;\;
&
\includegraphics[width=7.3cm,height=6cm]{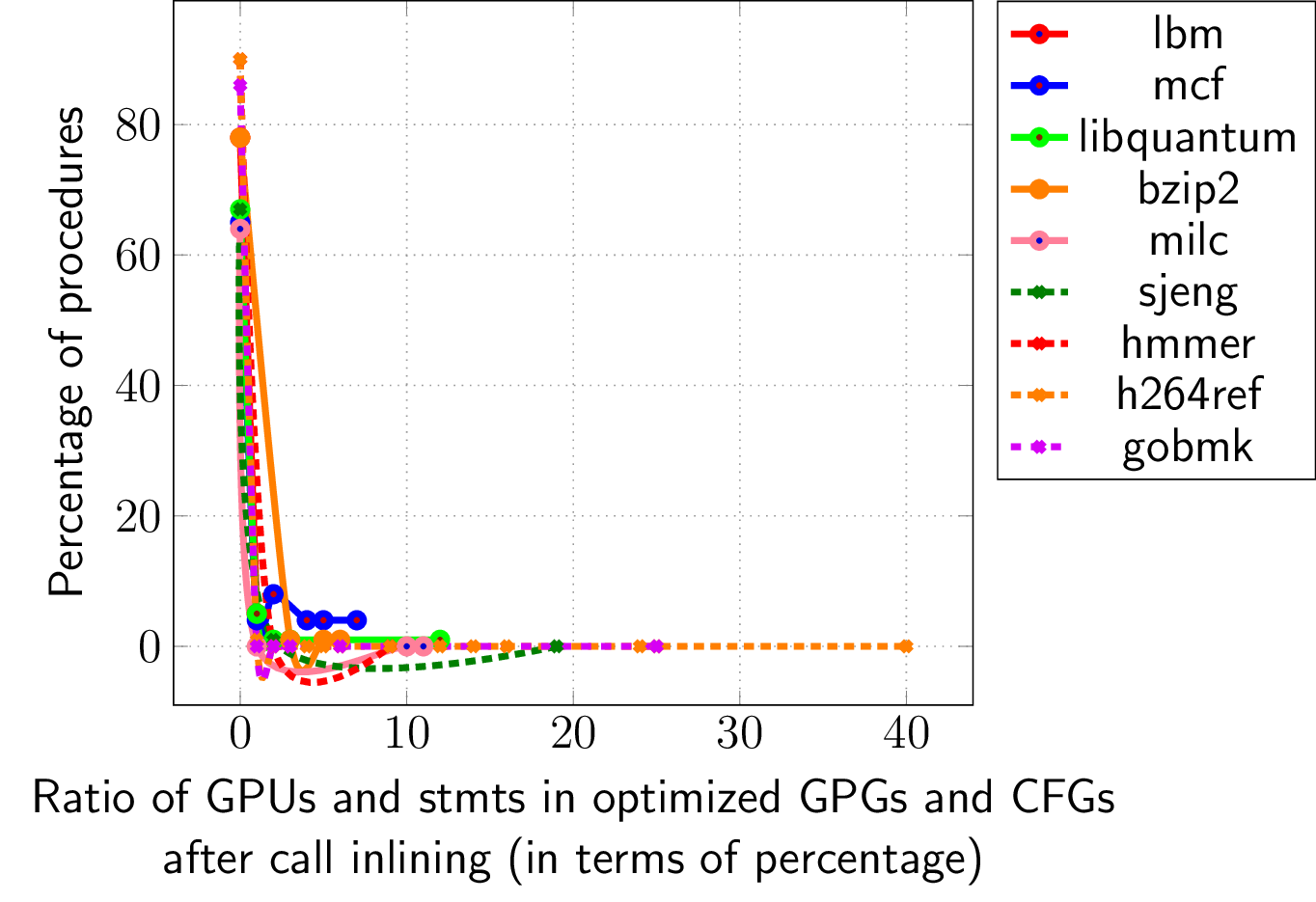}
\\ \hline
\multicolumn{2}{c}{\rule{0em}{17.5em}\includegraphics[width=7.5cm,height=6cm]{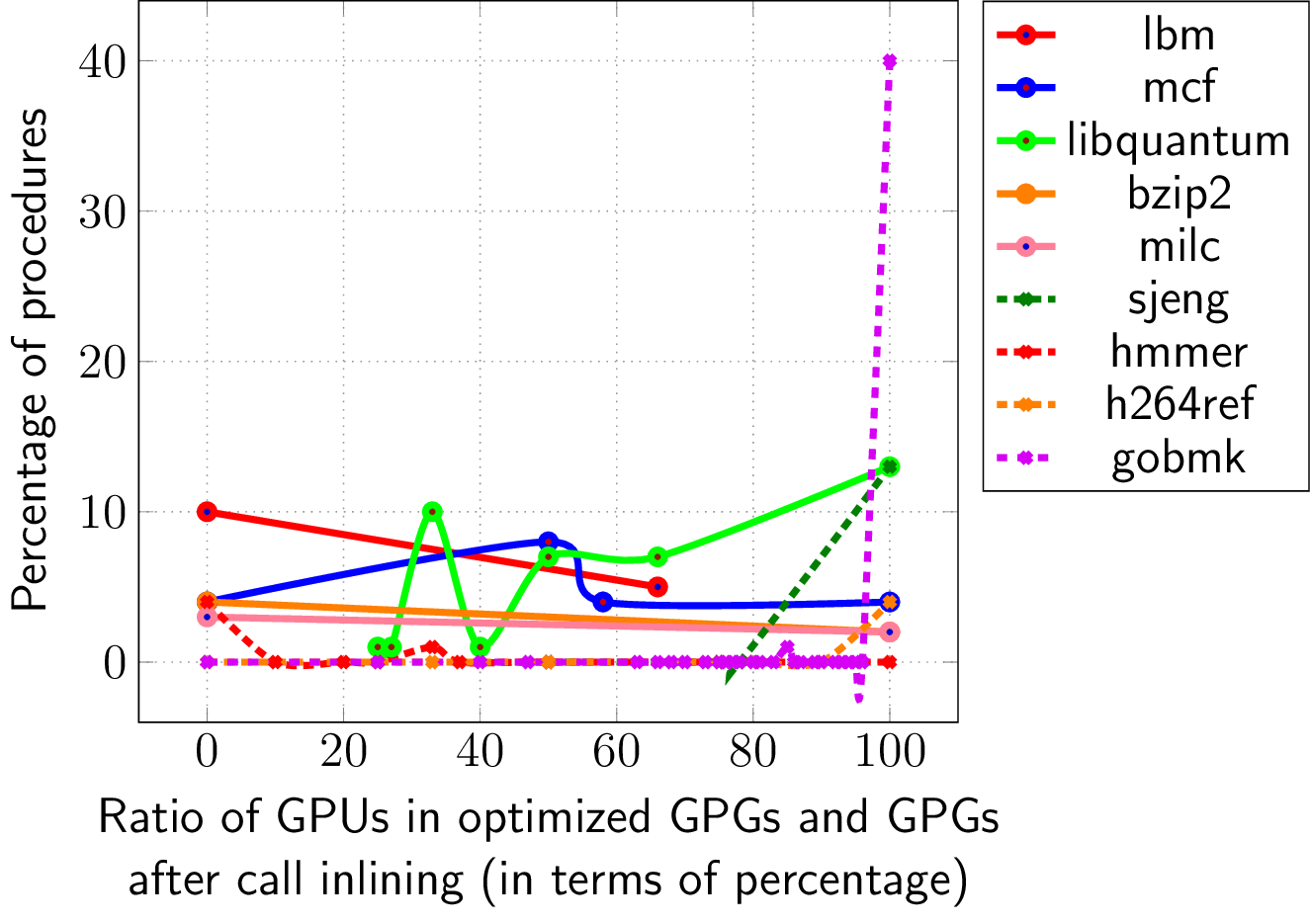}}
\\ \hline
\end{tabular}
\caption{Size of \gpgs relative to the size of procedures in terms of \gpus and pointer assignments.}
\label{plot:cmp-gpus}
\end{figure}

\begin{figure}[t]
\begin{tabular}{c|c}
\includegraphics[width=5.8cm,height=6cm]{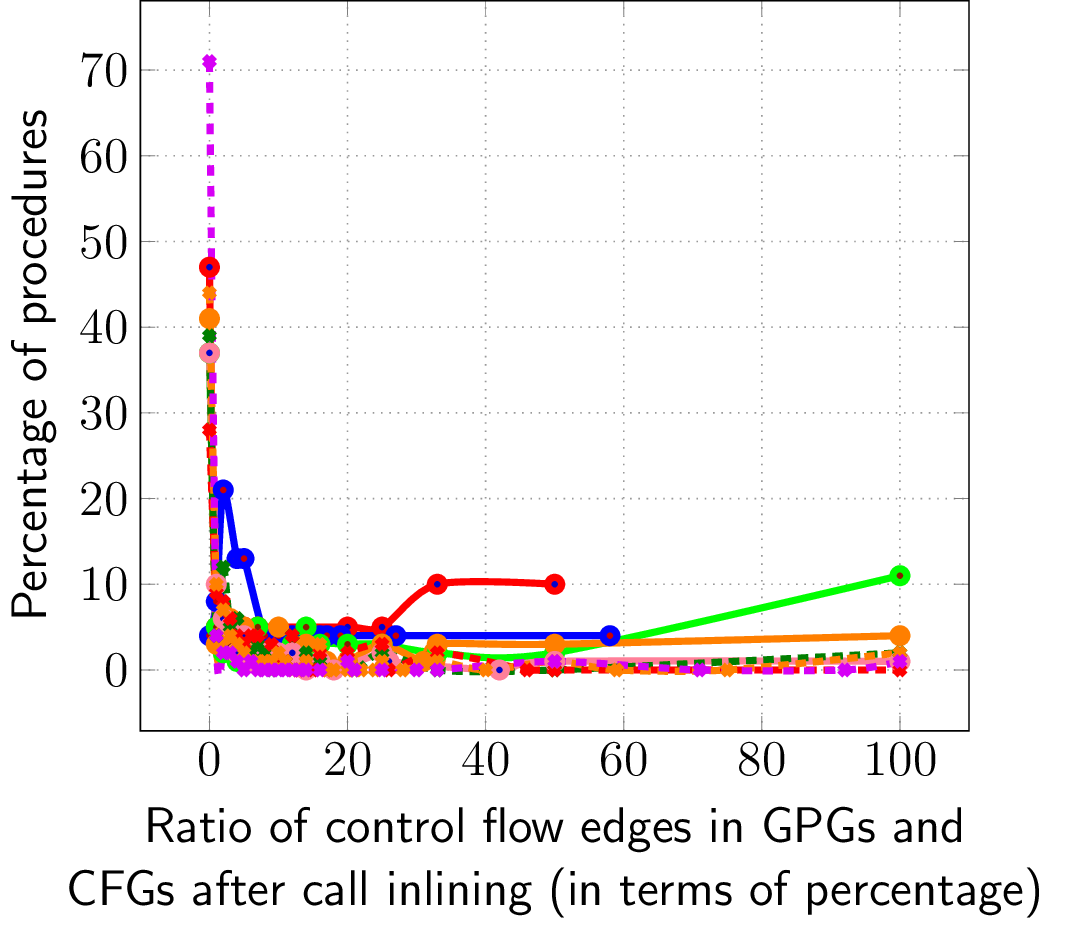}
\;\;
&
\includegraphics[width=7.3cm,height=6cm]{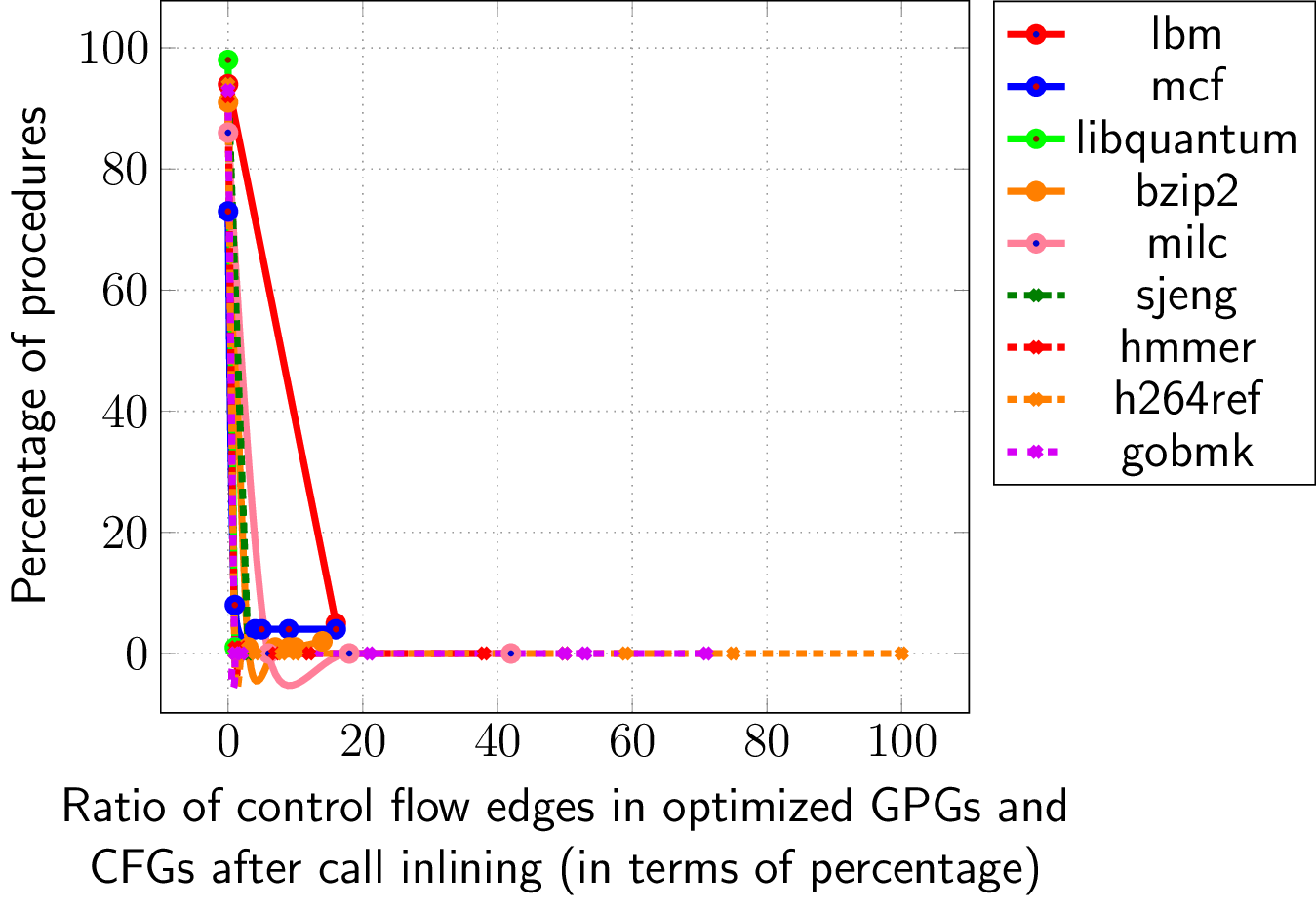}
\\ \hline
\multicolumn{2}{c}{\rule{0em}{17.5em}\includegraphics[width=7.5cm,height=6cm]{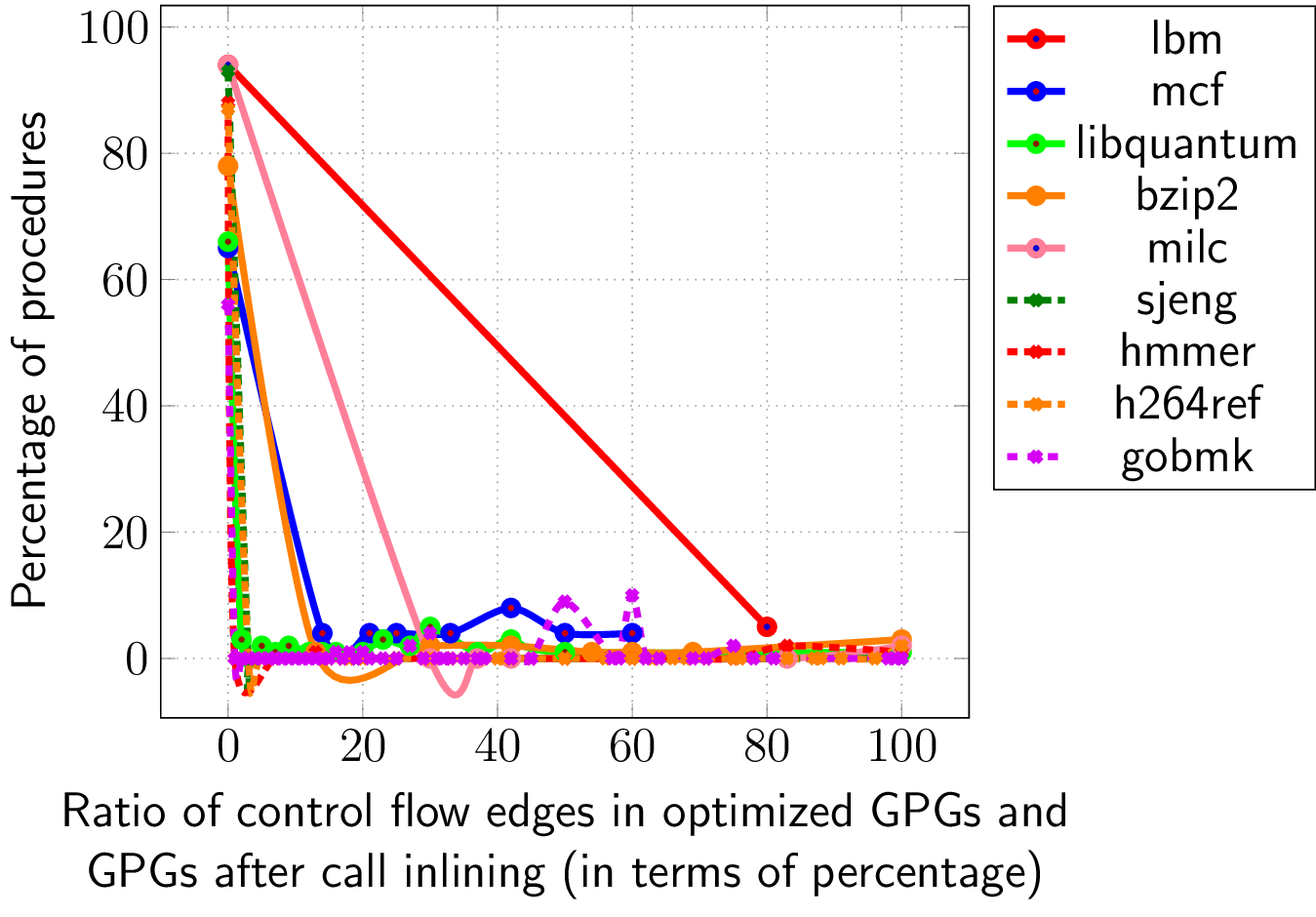}}
\\ \hline
\end{tabular}
\caption{Size of \gpgs relative to the size of corresponding procedures in terms of control flow edges.}
\label{plot:cmp-cf-edges}
\end{figure}

\subsection{Observations}
\label{sec:observations.meas}

We describe our observations about the sizes of \gpgs, \gpg optimizations, and performance of the analysis. Observations
related to the time measurements are presented in the end. Section~\ref{sec:discussion.meas} discusses these
observations by analyzing them.

\subsubsection{Effectiveness of Redundancy Elimination Optimizations} We observe that: %%This data is presented in Figure~\ref{plot:dead-coal}.
	\begin{enumerate}[(a)]
      \item The percentage of dead \gpus is very small and the dead \gpu elimination optimization 
            is the least effective of all the optimizations. Also, this optimization requires very little time compared to 
            other optimizations (see Figure~\ref{plot:time-meas-ptinfo}). Hence, disabling the optimization will neither improve 
            the efficiency or scalability of the analysis nor will it affect the compactness of the \gpgs. 

      \item The transformations performed by call inlining, strength reduction, and dead \gpu  elimination create empty \gpbs 
	    which are removed by empty \gpb elimination.  For most procedures, 0\%-5\% or close to 50\% of \gpbs are empty.

      \item The last optimization among the redundancy elimination optimizations, coalesces the
            adjacent \gpbs that do not require control flow between them. In our experience,
		many benchmarks had some very large \gpgs in the presence of
		recursion. \gpgs for recursive procedures are constructed by repeated 
		inlinings of recursive calls.
		Coalescing was most effective for such procedures. Once these \gpgs 
		were optimized, the \gpgs of the caller procedures did not have much scope for 
		coalescing. In other words, coalescing did not cause uniform reduction across all
		\gpgs but helped in the most critical \gpgs.	
		Hence we observe 
	    a reduction of 20\% to 50\% of \gpbs for some but not majority of procedures. 

		Even if coalescing did not reduce the number of \gpbs uniformly, it eliminated
		almost all back edges as shown in fourth 
            plot in Figure~\ref{plot:dead-coal}. This is significant because most of the inlined \gpgs are acyclic 
            and hence analyzing the \gpgs of the callers does not require additional iterations in a fixed-point
	    computation.
	\end{enumerate}

\subsubsection{Goodness of Procedure Summaries} This data is presented in Tables~\ref{fig:stats},~\ref{tab:stats2}, and Figure~\ref{plot:stats1}.
      We use the following goodness metrics on procedure summaries:
     \begin{enumerate}[(a)]
     \item Reusability. The number of calls to a procedure is a measure for the reusability of its summary. The construction of
           a procedure summary is meaningful only if it is use multiple times. 
           From column $E$ in Table~\ref{fig:stats}, it is clear that most 
           procedures are called from many call sites. This indicates a high reusability of procedure summaries. 
     \item Compactness of a procedure summary. 
%%In the worst case, a procedure summary may be same as the
%%procedure. In such a case, the application of a procedure summary at the call 
%%sites in its callers is 
%%meaningless because it is as good as visiting the procedure multiple 
%%times which is similar to a top-down approach. 
For scalability of a bottom-up approach, a procedure summary should be as compact as possible. 
           Figure~\ref{plot:stats1} and Table~\ref{tab:stats2} show that the procedure summaries are indeed small in terms of number of \gpbs and \gpus. 
           \gpgs for a large number of procedures have 0 \gpus because they do not manipulate global 
           pointers (and thereby represent the identity flow function).
           Further, the majority of \gpgs have 1 to 3 \gpbs.

	   Note that this is an absolute size of \gpgs. Observations about the relative size of \gpgs with respect to their \cfgs
           are presented in Section~\ref{sec:relative.size.meas} below.
     \item Percentage of context-independent information. A procedure summary is very useful if it contains high percentage of 
           context-independent information. We observe that the number of procedures with a high amount of 
           context-independent information is larger in the larger benchmarks.  Thus, a bottom-up approach is particularly
           useful for large programs.
\end{enumerate}

\begin{table}[t]
\small
\centering
\renewcommand{\tabcolsep}{2pt}
\begin{tabular}{|l|c|c|c|c|c|c|c|c|}
\hline 
Program 
& \begin{tabular}{@{}c@{}} \# of Proc. \\ which \\ have \\ 0 \gpus \end{tabular}
& \begin{tabular}{@{}c@{}} \# of Proc. \\ which \\  have \\ $\mtsym_{\top}$ as \\ \gpg \end{tabular}
& \begin{tabular}{@{}c@{}} \# of Proc. \\ in which  \\ back edges \\ are present \\ in a CFG \end{tabular}
& \begin{tabular}{@{}c@{}} \# of Proc. \\ in which \\ back edges \\ are present \\ in a \gpg \end{tabular}
& \begin{tabular}{@{}c@{}} \\ Exported \\ Definitions \\ \end{tabular}
& \begin{tabular}{@{}c@{}} \\ Imported \\ Uses \\ \end{tabular}
& \begin{tabular}{@{}c@{}} \\ \# Queued \\ \gpus \\ \end{tabular}
& \begin{tabular}{@{}c@{}} \\ \# Soundness \\ Alerts \\ \end{tabular}
\\ \hline \hline
lbm & 15 & 0 & 10 & 0 & 1.68 & 16.63 & 0 & 0
 \\ \hline
mcf & 12 & 0 & 20 & 1 & 12.30 & 29.26 & 117 & 0
 \\ \hline
libquantum & 38 & 0 & 36 & 0 & 1.54 & 1.89 & 0 & 0
 \\ \hline
bzip2 & 78 & 8 & 43 & 1 & 1.21 & 17.37 & 0 & 0
 \\ \hline
milc & 184 & 3 & 94 & 0 & 0.70 & 6.14 & 0 & 0
 \\ \hline
sjeng & 101 & 2 & 65 & 0 & 0.81 & 1.77 & 0 & 0
 \\ \hline
hmmer & 242 & 5 & 153 & 0 & 2.26 & 13.02 & 19 & 0
 \\ \hline
h264ref & 434 & 3 & 308 & 5 & 1.60 & 26.75 & 13 & 0
 \\ \hline
gobmk & 1436 & 2 & 464 & 8 & 0.39 & 1.36 & 6 & 0
 \\ \hline

\end{tabular}
\caption{Miscellaneous data about the \gpgs.}
\label{tab:stats2}
\end{table}

\begin{figure}[t]
\begin{tabular}{c|c}
\includegraphics[width=6.5cm,height=6cm]{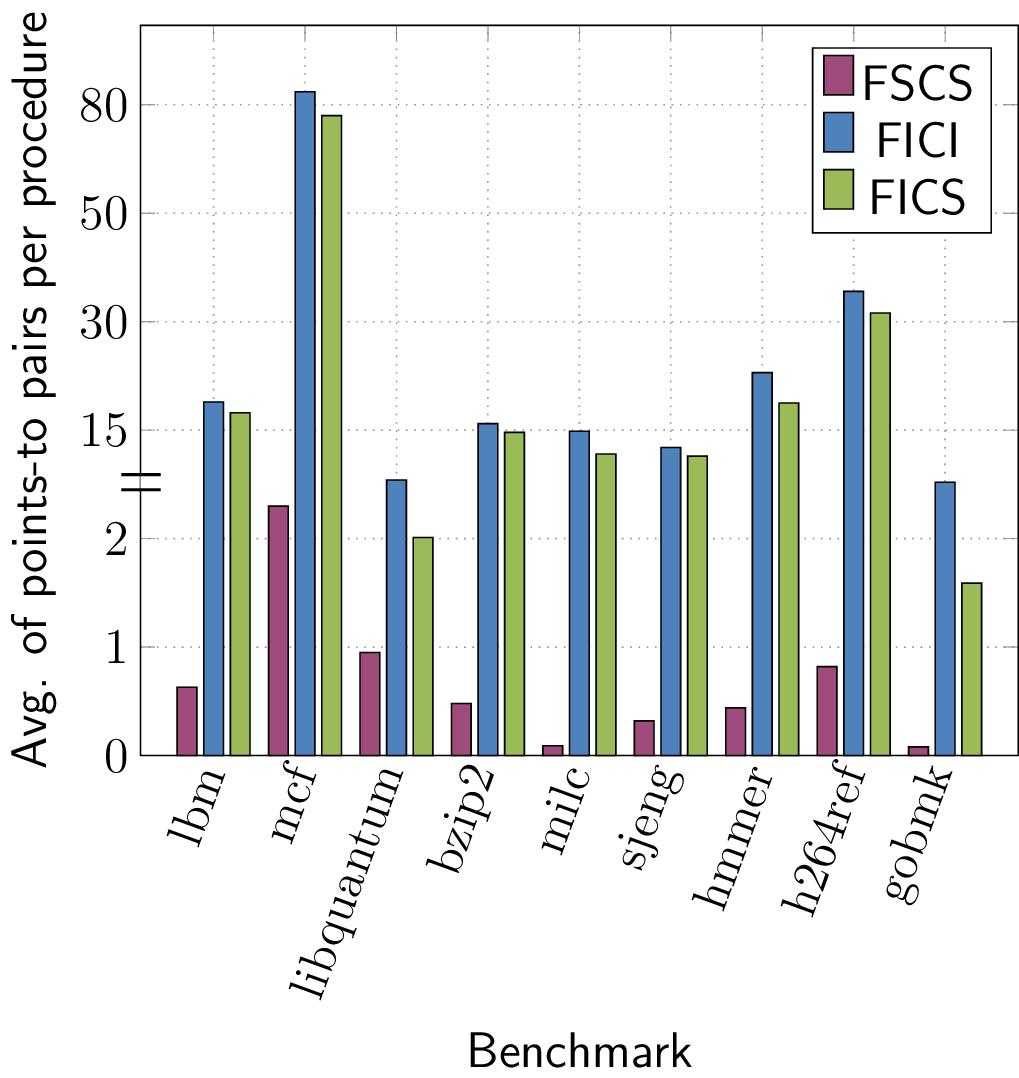}
\;\;
&
\includegraphics[width=6.5cm,height=6cm]{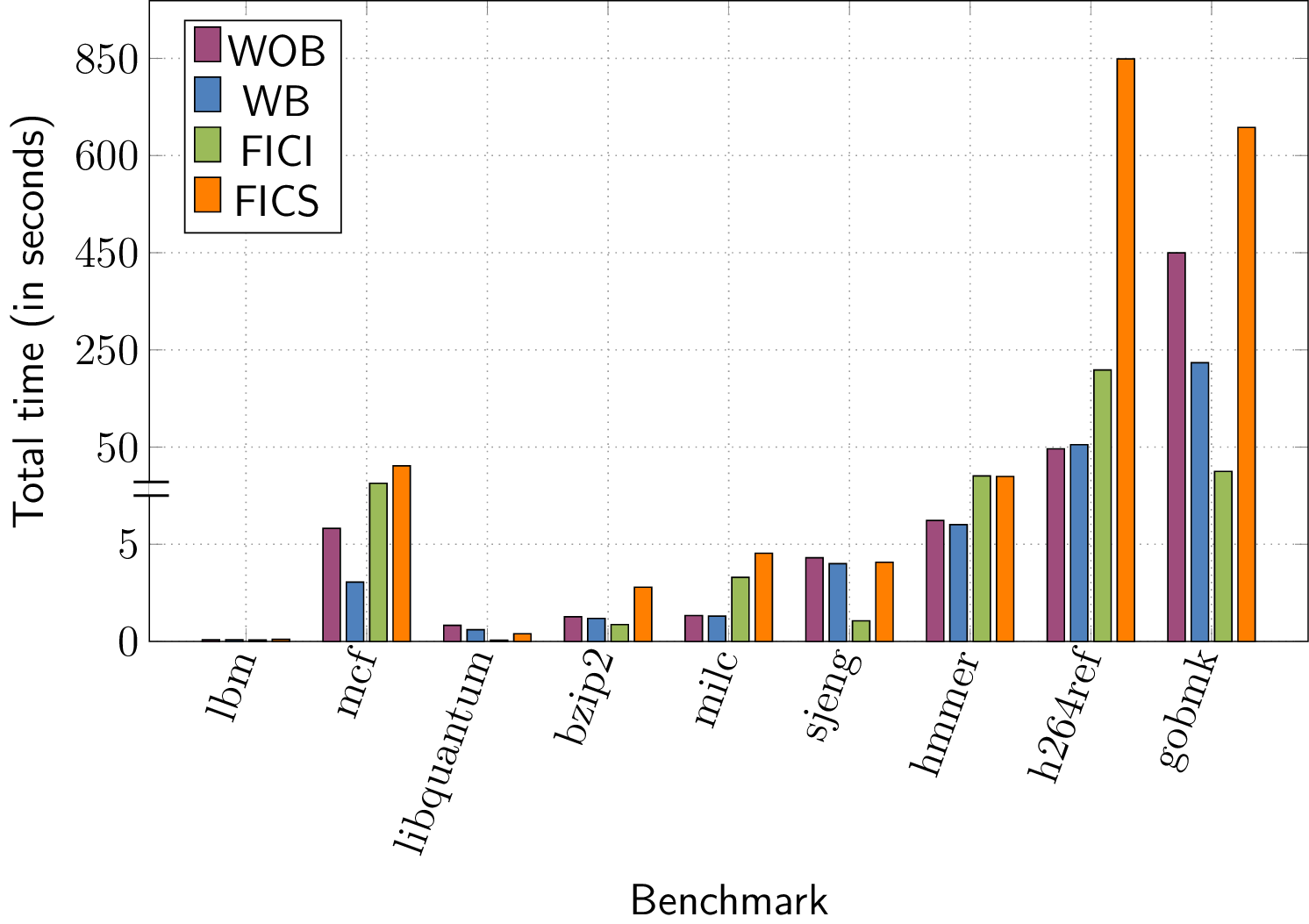}
\\ \hline
\rule{0em}{17.5em}
\includegraphics[width=6.5cm,height=6cm]{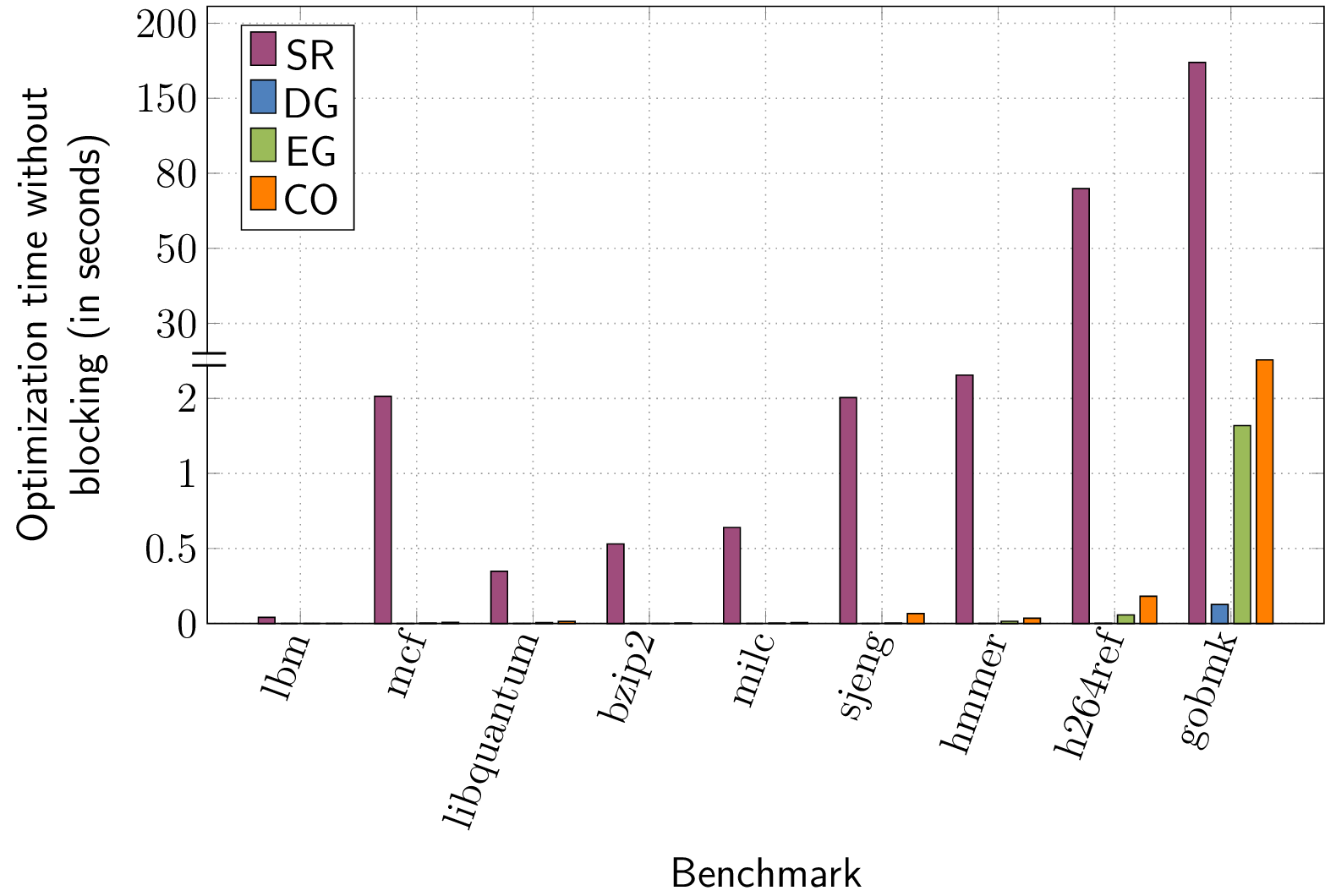}
\;\;
&
\includegraphics[width=6.5cm,height=6cm]{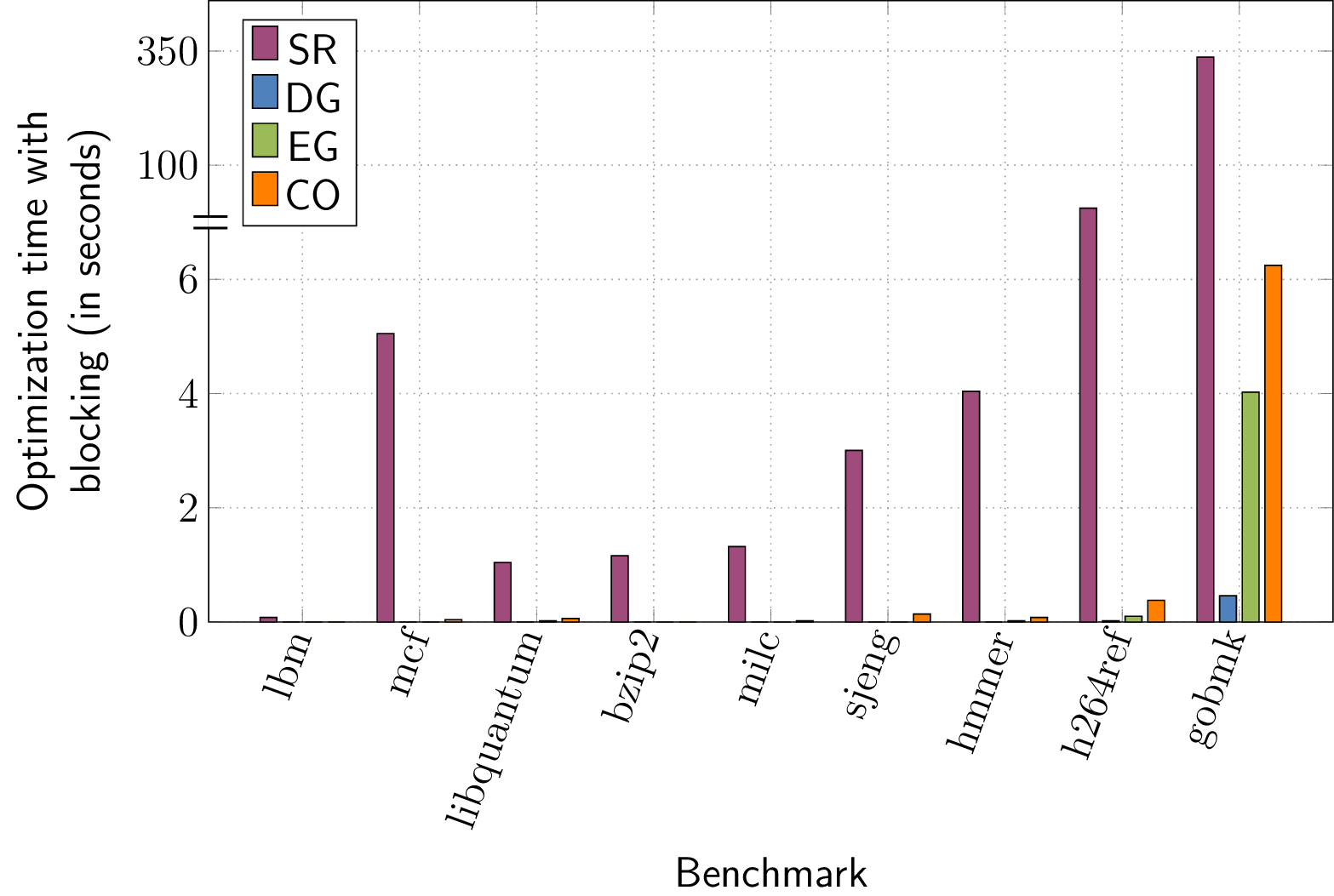}
\\ \hline
\end{tabular}
\caption{Final points-to information measurements (first plot) and time measurements (the remaining three plots). FSCS (flow- and context-sensitive), FICI (flow- and context-insensitive), FICS (flow-insensitive and context-sensitive), WOB (our analysis without blocking), WB (our analysis with blocking), SR (strength reduction optimization), DG (dead \gpu elimination), EG (empty \gpb elimination), CO (coalescing). The time taken by dead \gpu elimination, empty \gpb elimination, and coalescing 
is negligible for small benchmarks and hence the corresponding bars are not visible.}
\label{plot:time-meas-ptinfo}
\end{figure}

\subsubsection{Relative Size of \gpgs with respect to the Size of Corresponding Procedures} 
\label{sec:relative.size.meas}
%%	Recall that the \gpbs in the \gpgs are analogous to the basic blocks in the \cfgs, the \gpus are 
%%        analogous to the pointer assignments in the basic blocks\footnote{Except that the \gpus in a \gpb 
%%        do not have a control flow order between them.}, and the control flow between
%%       \gpbs is analogous to the control flow between basic blocks.
%%       We have computed the ratio of \gpbs to basic blocks (Figure~\ref{plot:cmp-gpbs}),
%%       the ratio of \gpus to the pointer assignments (Figure~\ref{plot:cmp-gpus}), and 
%%       the ratio of control flow edges in \gpgs to the control flow edges in the \cfg
%%	(Figure~\ref{plot:cmp-cf-edges}).
       For an exhaustive study, we compare three representations of a procedure with each other:
	\begin{inparaenum}[(I)] 
	\item the \cfg of a procedure, 
	\item the initial \gpg obtained after call inlining, and 	
	\item the final optimized \gpg. 
	\end{inparaenum}
	Since \gpgs have callee \gpgs inlined within them,
        for a fair comparison, the \cfg size must be counted by accumulating the sizes of the \cfgs of the callee procedures.
       This is easy for non-recursive procedures. For recursive procedures, we accumulate the size of a \cfg
       as many times as the number of inlinings of the corresponding \gpg (Section~\ref{sec:handling_recur}).
       Further, the number of statements in a \cfg is measured only in terms of the pointer 
       assignments.

       \begin{enumerate}[(a)]
       \item The first plot in these figures gives the size of the initial \gpg (i.e. II) relative to that of the
             corresponding \cfg (i.e. I).  It is easy to that the reduction is immense: a large number of initial \gpgs are in 
             the range 0\%-20\% of the corresponding \cfgs.
       \item The second plot in these figures gives the size of the optimized \gpg (i.e. III) relative to that of the
             corresponding \cfg (i.e. I).  
             The number of procedures in the range of 0\%-20\% is larger here than in the first plot 
             indicating more reduction because of optimizations.
       \item The third plot in these figures gives the size of the optimized \gpg (i.e. III) relative to that of the
             initial \gpg (i.e. I). Here the distribution of procedures is different for \gpbs, \gpus, and control flow edges.
             In the case of \gpbs, the reduction factor is 50\%.
%%		concentrated in the range of 20\%-60\%. 
		For \gpus, the reduction varies widely. 
%%		it takes two extremes: either large reduction (and many procedures in the 
%%             range 0\%-20\%) or small reduction (and many procedures in the range 80\%-100\%). 
	     The largest reduction is found for control flow: a large number of procedures fall in the range 
             0\%-20\%. The number of procedures in this range is larger than in the case of \gpbs or \gpus indicating that 
             the control flow is optimized the most.
       \item As a special case of control flow reduction, we have measured the effect of our optimizations on back edges. 
             This is because the presence of back edges increases the number of iterations required for fixed-point computation
             in an analysis. If a procedure summary needs to encode control flow, it is desirable to eliminate back edges 
             to the extent possible. The data in Table~\ref{tab:stats2} shows that
	     most of the \gpgs are acyclic in spite of the fact that the number of procedures with back edges in \cfg is 
             large.
\end{enumerate}

\subsubsection{Final Points-to Information}
We compared the amount of points-to information computed by our approach with flow- and context-insensitive (FICI) and
flow-insensitive and context-sensitive (FICS) methods (first plot of Figure~\ref{plot:time-meas-ptinfo} and Table~\ref{tab:pt-info}). For this purpose,
we computed number of points-to pairs per procedure in all the three approaches by dividing the total number
of unique points-to pairs across all procedures by the total number of procedures. Predictably, this number is smallest for our analysis (FSCS) and
largest for FICI method.

%\begin{landscape}
\begin{table}[t]
\centering
\renewcommand{\tabcolsep}{2pt}
\begin{tabular}{|l|c|c|c|c|c|c|c|}
\hline 
\multirow{3}{*}{
		\begin{tabular}{@{}c@{}} \\ Program \\ \end{tabular}
		}
&
	\multirow{3}{*}{
		\begin{tabular}{@{}c@{}} \\ \# of Proc. \\ \end{tabular}
		}
	& 
	\multirow{3}{*}{
		\begin{tabular}{@{}c@{}} \\ \# of Stmts. \\ \end{tabular}
		}
	& 
	\multicolumn{3}{c|}{
	\renewcommand{\arraystretch}{.8}%
		\begin{tabular}{@{}c@{}} 
		\rule{0em}{.9em}
			FSCS
		\end{tabular}}
	&
	\multicolumn{1}{c|}{
	\renewcommand{\arraystretch}{.8}%
		\begin{tabular}{@{}c@{}} 
		\rule{0em}{.9em}
			FICI
		\end{tabular}}
	&
	\multicolumn{1}{c|}{
	\renewcommand{\arraystretch}{.8}%
		\begin{tabular}{@{}c@{}} 
		\rule{0em}{.9em}
			FICS
		\end{tabular}}
\\ \cline{4-8}
\rule{0em}{1em} 
	& & &
	\multicolumn{1}{c|}{
	\renewcommand{\arraystretch}{.8}%
		\begin{tabular}{@{}c@{}} 
		\rule{0em}{.9em}
			FS
		\end{tabular}}
	&
	\multicolumn{1}{c|}{
	\renewcommand{\arraystretch}{.8}%
		\begin{tabular}{@{}c@{}} 
		\rule{0em}{.9em}
			FI
		\end{tabular}}
	& 
	\multicolumn{1}{c|}{
	\renewcommand{\arraystretch}{.8}%
		\begin{tabular}{@{}c@{}} 
		\rule{0em}{.9em}
			FS+FI
		\end{tabular}}
	& &
%%	\multicolumn{2}{c|}{
%%	\renewcommand{\arraystretch}{.8}%
%%		\begin{tabular}{@{}c@{}} 
%%		\rule{0em}{.9em}
%%			
%%		\end{tabular}}
\\ \cline{4-8}
	& &  
%%	& \begin{tabular}{@{}c@{}} \# of \\ pointees \end{tabular} 
	& \begin{tabular}{@{}c@{}} Avg \\ (per stmt) \end{tabular} 
%%	& \begin{tabular}{@{}c@{}} \# of \\ pointees \end{tabular} 
	& \begin{tabular}{@{}c@{}} Avg \\ (per proc) \end{tabular} 
	& \begin{tabular}{@{}c@{}} Avg \\ (per proc) \end{tabular} 
%%	& \begin{tabular}{@{}c@{}} \# of \\ pointees \end{tabular} 
	& \begin{tabular}{@{}c@{}} Avg \\ (per proc) \end{tabular} 
%%	& \begin{tabular}{@{}c@{}} \# of \\ pointees \end{tabular} 
	& \begin{tabular}{@{}c@{}} Avg \\ (per proc) \end{tabular} 
\\ \hline\hline
lbm & 19 & 367 & 1.99 & 0.79 & 0.63 & 19.26 & 17.11
 \\ \hline
mcf & 23 & 484 & 4.12 & 9.30 & 2.30 & 82.13 & 77.39
 \\ \hline
libquantum & 80 & 342 & 0.58 & 0.57 & 0.95 & 3.46 & 2.01
 \\ \hline
bzip2 & 89 & 1645 & 2.18 & 0.65 & 0.48 & 14.72 & 12.96
 \\ \hline
milc & 196 & 2504 & 1.18 & 3.10 & 0.09 & 13.21 & 8.71
 \\ \hline
sjeng & 133 & 684 & 1.44 & 1.83 & 0.32 & 10.04 & 8.17
 \\ \hline
hmmer & 275 & 6719 & 1.28 & 1.14 & 0.44 & 25.12 & 19.01
 \\ \hline
h264ref & 566 & 17253 & 2.35 & 12.02 & 0.82 & 35.04 & 30.75
 \\ \hline
gobmk & 2699 & 10557 & 0.74 & 6.36 & 0.08 & 2.95 & 1.59
 \\ \hline

\end{tabular}
\caption{Final points-to information. FSCS (flow- and context-sensitive), FICI (flow- and context-insensitive), FICS (flow-insensitive and context-sensitive).}
\label{tab:pt-info}
\end{table}
%\end{landscape}

\subsubsection{Time measurements}

We have measured the overall time as well as the time taken by each of the 
optimizations (Figure~\ref{plot:time-meas-ptinfo}). We have also measured the time taken by the FICI and FICS variants of points-to analysis.
Our observations are:

\begin{enumerate}[(a)]
\item Our analysis takes less than 8 minutes on gobmk.445  which is a large benchmark with 158 kLoC. 
      Our current implementation does not scale beyond that.
\item Strength reduction is the most expensive optimization followed by coalescing which is the most 
      expensive among the redundancy elimination optimizations.
\item We introduced reaching \gpus analysis with blocking to ensure soundness of strength reduction so that
      a barrier \gpu does not cause a side-effect invalidating strength reduction. However, our intuition
      was that very few of us write programs where a pointer is manipulated in such a manner. Hence we
      identified possible soundness alerts. The soundness alerts 
      arise when a \gpu whose composition was postponed, is updated by a \gpu within the same 
      \gpg after inlining in a caller \gpg. This is 
	identified by checking 
      if a \gpu in the set \queued of a \gpg is killed by 
	the \gpu of the same \gpg when it is inlined in a caller.

      We also measured the number of \gpus that were queued (i.e. not used as producer \gpus).
      Our measurements show that the number of \gpus in the \queued set is relatively small (see Table~\ref{tab:stats2}). 
      We did not find a single instance of a soundness alert that was valid; we did find a very 
      small number of false positives that were manually examined and rejected.

\item FICI variant is consistently faster than the FICS variant, and faster than FSCS in 
      most programs.  Further, FSCS is faster than FICS in most cases.

\end{enumerate}

\subsection{Discussion: Lessons From Our Empirical Measurements}
\label{sec:discussion.meas}

Our experiments and empirical data leads us to some important learnings as described below:

\begin{enumerate}
\item The real killer of scalability in program analysis is not the amount of data
      but the amount of control flow that it may be subjected to in search of precision.
\item For scalability, the bottom-up summaries must be kept as small as possible at each stage.
      %In particular, an exponential increase in size due to transitive call inlining is the biggest threat
      %to a bottom-up method. 
\item Some amount of top-down flow is very useful for achieving scalability.
\item Type-based non-aliasing aids  scalability significantly. 
\item The indirect effects for which we devised blocking to postpone \gpu compositions are extremely rare in
      practical programs. We did not find a single instance in our benchmarks.
\item Not all information is flow-sensitive. 
%%It is useful to distinguish between the two kinds of information.
\end{enumerate}

We learnt these lessons the hard way in the situations described in the rest of this section.

\subsubsection{Handling Recursion}
       In our first attempt of handling recursion, we converted indirect recursion to self recursion, and repeatedly
       inlined the recursive calls to optimize them.
%% until the effects of the \gpgs kept on changing.
       This failed because in some cases, the size of \gpg after inlining calls became too big and
       our analyses and optimizations did not scale. Hence, instead of first creating a naively large
       \gpg and then optimizing it to bring down the size, we decided to keep the \gpgs small at every stage
        by successive refinements of mutually recursive \gpgs starting from $\mtsym_{\top}$.

\subsubsection{Handling Large Size of Context-Dependent Information}
\label{sec:large-con-dep-info}

        Some \gpgs had a large amount of context-dependent information (i.e. \gpus with upwards-exposed 
        versions of variables) and the \gpgs could not be optimized much. This caused the size of
        the caller \gpgs to grow significantly, threatening the scalability of our analysis.
Hence we devised a heuristic threshold beyond which the procedure summary will be inlined as
        a symbolic $\mtsym_{\top}$ \gpg with an additional feature that it carries with it in a single \gpb,
        all context-dependent \gpus (i.e., the \gpus that have upwards-exposed versions of variables after optimizations).
        This keeps the size of the caller \gpg small and at the same time, allows reduction of the
        context-dependent \gpus. Once all \gpus are reduced to classical points-to edge, we effectively get the procedure summary
        of the original callee procedure for that call chain. Since the reduction of context-dependent
        \gpus is different for different calling contexts, the process needs to be repeated for each call chain.
        This is similar to the top-down approach where we analyze a procedure multiple times.
%%	 depending on the information that flows from its callers.

        Note that, in our implementation, we discovered very few cases (and only in large benchmarks) where the 
        threshold actually exceeded.\footnote{We used a threshold of 80\% context-dependent \gpus in a \gpg containing more than 10 \gpus.
		Thus, 8 context-dependent \gpus
         from a total of 11 \gpus was below our threshold as was 9 context-dependent \gpus from a total of 9 \gpus.}
        The number of call chains that required multiple traversals are in single digits and they are not very long. The 
        important point to note is that we got the desired 
        scalability only when we introduced this small twist.

\subsubsection{Handling Function Pointers}
\label{sec:function.pointers.meas}

         Function pointers used in a procedure but defined in its callers is another case where we had to inline unoptimized 
         \gpgs in the callers because the \gpgs of the procedure's callees were not known and hence their flow function was
         $\mtsym_{\top}$. This hampered scalability.
         Since our primary goal was to evaluate the effectiveness of our optimizations, our current
         implementation handles only locally defined function pointers (Section~\ref{sec:handling_fp})
	 Our implementation can be easily extended to handle function pointers defined in the calling contexts.
	 We can handle such function pointers by using
         a symbolic $\mtsym_{\top}$ \gpg and introducing a small touch of top-down analysis as was done above when handling
         a large number of context-dependent \gpus. We leave this as future work.

\subsubsection{Handling Arrays and SSA Form}

	 Pointers to arrays were weakly updated, hence we realized early on that maintaining this information flow
         sensitively prohibited scalability. This was particularly true for large arrays with static initializations.
         Similarly, \gpus involving SSA versions of variables were not required to be maintained flow sensitively.
         This allowed us to reduce the propagation of data across control flow without any loss in precision.

\subsubsection{Making Coalescing More Effective}

Unlike dead \gpu elimination, coalescing proved to be a very significant optimization for boosting the scalability of the analysis. 
The points-to analysis failed to scale in the absence of this optimization. However, this optimization was effective (i.e. coalesced many
\gpbs) only when we brought in the concept of types. In cases where
the data dependence between the \gpus was unknown because of the dependency on the context information, we used 
type-based non-aliasing to enable coalescing.

\subsubsection{Estimating the Number of Context-Dependent Summaries}

Constructing context-dependent procedure summaries (i.e. partial transfer functions) using the 
aliases or points-to information from 
calling contexts obviates the need of control flow. Since control flow is the real bottleneck as per our findings,
we computed the number of aliases after computing the final points-to information
to estimate the number of context-dependent summaries that may be required for real program.
This number (column $F$ in Table~\ref{fig:stats}) is large suggesting that it is undesirable to construct
multiple PTFs for a procedure using the aliases from the calling contexts.

\begin{figure}[t]
\newrgbcolor{newblue}{.6 0.8 1}
\newrgbcolor{newpink}{1 .48 .5}

\begin{pspicture}(0,0)(138,75)
%\psframe(0,0)(138,75)
\newcommand{\rA}{16}

%%{\psset{hatchwidth=1.pt,linewidth=.1}
%%\putnode{n1}{origin}{28}{45}{\psframebox[linecolor=newblue,framearc=.1,fillstyle=hlines,
%%			hatchangle=-45,hatchcolor=newblue,hatchsep=.75pt]{\rule{0mm}{18mm}\rule{45mm}{0mm}}}
%%\putnode{n2}{n1}{0}{-22}{\psframebox[linecolor=newblue,framearc=.1,fillstyle=hlines, 
%%			hatchcolor=newblue,hatchsep=.75pt]{\rule{0mm}{18mm}\rule{45mm}{0mm}}}
%%\putnode{n3}{n1}{50}{-11}{\psframebox[linecolor=newpink,framearc=.1,fillstyle=vlines, hatchangle=0, 
%%			hatchcolor=newpink,hatchsep=.75pt]{\rule{0mm}{18mm}\rule{45mm}{0mm}}}
%%\putnode{n4}{n3}{0}{22}{\psframebox[linecolor=newpink,framearc=.1,fillstyle=vlines, hatchangle=-45,
%%			hatchcolor=newpink,hatchsep=.75pt]{\rule{0mm}{18mm}\rule{45mm}{0mm}}}
%%\putnode{n5}{n3}{0}{-22}{\psframebox[linecolor=newpink,framearc=.1,fillstyle=vlines, 
%%			hatchangle=45,hatchcolor=newpink,hatchsep=.75pt]{\rule{0mm}{18mm}\rule{45mm}{0mm}}}
%%}

\putnode{n1}{origin}{67}{55}{\psframebox[fillcolor=lightgray,framearc=.1, fillstyle=solid,framesep=2.5,
	     linestyle=none]{\makebox[30mm]{\sf\begin{tabular}{@{}c@{}}Approximations of \\ data dependence\end{tabular}}}}

\putnode{n0}{n1}{-47}{-11}{\psframebox[fillcolor=lightgray,framearc=.03, fillstyle=solid,framesep=2.5,
	     linestyle=none]{\makebox[30mm]{\sf\begin{tabular}{@{}l@{}}
				Data handling
				\\ 
				\\
				Control flow 
				\\ 
				\\
				Higher order features
				\end{tabular}}}}

\putnode{n2}{n1}{0}{-22}{\psframebox[fillcolor=lightgray,framearc=.1, fillstyle=solid,framesep=2.5,
	     linestyle=none]{\makebox[30mm]{\sffamily\begin{tabular}{@{}c@{}}Data \\ abstractions\end{tabular}}}}

\putnode{n3}{n1}{47}{0}{\psframebox[fillcolor=lightgray,framearc=.1, fillstyle=solid,framesep=2.5,
	     linestyle=none]{\makebox[30mm]{\sf\begin{tabular}{@{}c@{}}Specialized \\ data structures\end{tabular}}}}

\putnode{n4}{n3}{0}{-22}{\psframebox[fillcolor=lightgray,framearc=.1, fillstyle=solid,framesep=2.5,
	     linestyle=none]{\makebox[30mm]{\sf\begin{tabular}{@{}c@{}}Relevant points-to \\ information\end{tabular}}}}

\putnode{n5}{n4}{0}{-24}{\psframebox[fillcolor=lightgray,framearc=.1, fillstyle=solid,framesep=2.5,
	     linestyle=none]{\makebox[30mm]{\sf\begin{tabular}{@{}c@{}}Order of computing \\ points-to information\end{tabular}}}}

\putnode{w}{n0}{0}{26}{\sf\begin{tabular}{c}
	     Language features 
	     \end{tabular}}
\putnode{w}{n1}{0}{15}{\sf\begin{tabular}{c}
	     Analysis features primarily \\ influencing precision
	     \end{tabular}}
\putnode{w}{n3}{0}{15}{\sf\begin{tabular}{c}
		Analysis features primarily \\ influencing  efficiency/scalability
		\end{tabular}}
\putnode{a}{n1}{22}{19}{}
\putnode{b}{n1}{22}{-54}{}
\ncline[linestyle=dashed, dash=.6 .6]{a}{b}
\putnode{a}{n1}{-24}{19}{}
\putnode{b}{n1}{-24}{-54}{}
\ncline[linestyle=dashed, dash=.6 .6]{a}{b}
\ncline[doubleline=true]{<->}{n4}{n5}
\ncline[doubleline=true]{->}{n4}{n3}
\ncline[doubleline=true]{->}{n2}{n1}
\ncline[doubleline=true]{->}{n1}{n3}
\ncline[doubleline=true]{->}{n0}{n1}
\ncline[doubleline=true]{->}{n0}{n2}

%%\putnode{w}{n0}{26}{-42}{
%%\begin{tabular}{|c|}
%%\hline
%%Examples of language features \\ \hline\hline
%%\rule[-2.1em]{0em}{4.7em}%
%%\begin{minipage}{88mm}
%%\raggedright\renewcommand{\baselinestretch}{.9}\small\normalsize%
%%addressof ($\&$) operator, type casts,
%%pointer arithmetic, function pointers,
%%unions, 
%%dynamic memory allocation,
%%receiver objects of calls, virtual calls, container objects, 
%%reflection, higher order (such as \emph{eval} in Javascript), concurrency 
%%\end{minipage}
%%\\ \hline
%%\end{tabular}
%%}
\end{pspicture}

\bigskip

{
\small\centering
\setlength{\tabcolsep}{3pt}
\begin{tabular}{|c|l|l|}
\hline
	\multicolumn{2}{|c|}{
	\rule[-.4em]{0em}{1.4em}%
		\sf Feature}
		& \multicolumn{1}{c|}{\sf Examples}
	\\ \hline\hline
\multirow{3}{*}{%
\rotatebox{90}{\sf Language\,\;}} 
&
\multirow{2}{*}{
Data handling} 
		&
		Addressof ($\&$) operator, type casts,
		unions,
		dynamic memory 
		\\
&
		& allocation,
		pointer arithmetic, 
		container objects
	\\ \cline{2-3}
&
Control flow 
		& Function pointers,
		receiver objects of calls, virtual calls, concurrency 
	\\ \cline{2-3}
&
Higher order features
		& Reflection, 
			\emph{eval} in Javascript
	\\ \hline
\multirow{9}{*}{%
\rotatebox{90}{\sf Analysis}} 
&
\rule[-.9em]{0em}{2.4em}%
\renewcommand{\arraystretch}{.9}%
\begin{tabular}{@{}l@{}}
Approximations of \\ data dependence 
\end{tabular}
			&  Path-sensitivity, flow-sensitivity, context-sensitivity, 
				 SSA form 
	\\ \cline{2-3}
&
\rule[-1.5em]{0em}{3.5em}%
Data abstractions &  
\renewcommand{\arraystretch}{.9}%
		\begin{tabular}{@{}l@{}}
			Field-sensitivity, object-sensitivity, allocation-site-based or 
			\\
			type-based abstraction of heap, heap cloning, summarized access 
			\\
			paths, summarization of aggregates 
			\end{tabular}
	\\ \cline{2-3}
&
\rule[-.9em]{0em}{2.4em}%
\renewcommand{\arraystretch}{.9}%
\begin{tabular}{@{}l@{}}
Relevant points-to \\ information 
\end{tabular}
		&
\renewcommand{\arraystretch}{.9}%
		\begin{tabular}{@{}l@{}}
			All pointers (exhaustive analysis), 
			relevant pointers in incremental, 
			\\	
			demand-driven, staged, level-by-level, or liveness-based
			analyses 
			\end{tabular}
	\\ \cline{2-3}
&
\rule[-.9em]{0em}{2.4em}%
\renewcommand{\arraystretch}{.9}%
\begin{tabular}{@{}l@{}}
Order of computing \\ points-to information 
\end{tabular}
			&  
\renewcommand{\arraystretch}{.9}%
		\begin{tabular}{@{}l@{}}
			Governed by relevance of pointers, or by algorithmic features 
			\\ 
			(e.g. top-down, bottom-up,
			parallel, or randomized algorithms)
			\end{tabular}
	\\ \cline{2-3}
&
\rule[-.9em]{0em}{2.4em}%
\renewcommand{\arraystretch}{.9}%
\begin{tabular}{@{}l@{}}
Specialized data \\ structures 
\end{tabular}
			& 
		\renewcommand{\arraystretch}{.9}%
		\begin{tabular}{@{}l@{}}
		BDDs, bloom filters, disjoint sets (for union-find), points-to graphs
			\\ 
		with placeholders, \gpgs
			\end{tabular}
	\\ \hline
\end{tabular}
}

\caption{Language and analysis features affecting the precision, efficiency, and scalability of points-to analyses.
	An arrow from feature A to feature B indicates that feature A influences feature B. 
The features influencing precision, influence efficiency and scalability indirectly.}
\label{fig:big-pic-pta}
\end{figure}

\section{Related Work: The Big Picture}
\label{sec:big-picture-pta}

Many investigations reported in the literature have described the popular points-to analysis methods
and have presented a comparative study of the methods with respect to scalability and 
precision~\cite{DBLP:conf/sas/HindP98, Hind:2000:PAI:347324.348916, PGL-014,DBLP:journals/toplas/Staiger-Stohr13,
lhotak_et_al:DR:2013:4169,Kanvar:2016:HAS:2966278.2931098}.
Instead of discussing these methods, we devise a metric 
of features that influence the precision and efficiency/scalability of points-to analysis. This metric 
can be used for identifying important characteristic of any points-to analysis at an abstract level.

\subsection{Factors Influencing the Precision, Efficiency, and Scalability of Points-to Analysis}

Figure~\ref{fig:big-pic-pta} presents our metric. At the top level, we have language features and analysis features. 
The analysis features have been divided further based on whether their primary influence is on the precision or
efficiency/scalability of points-to analysis. The categorization of language features is obvious. Here we describe
our categorization of analysis features.

\subsubsection{Features Influencing Precision}
Two important sources of imprecision in an analysis are approximation of data dependence and abstraction of data.
\begin{itemize}
	\item \emph{Approximations of data dependence.}
        The approaches that compromise on control flow by using flow-insensitivity or context-insensitivity
        over-approximate the control flow: flow-insensitivity effectively creates a complete graph out of a control flow graph
        whereas context-insensitivity treats call and returns as simple goto statements as far as the control transfer between procedures is
        concerned. 

	Observe that control flow in imperative languages is a proxy for implicit data dependence.  As a consequence, an 
	over-approximation of control flow amounts to over-approximation of data dependence. 
	In other words, control flow over-approximation may introduce spurious data dependences between pointer assignments
        that may have not existed if the analysis respected the control flow. This causes
	imprecision. 
	
	Note that SSA form also discards control flow but it avoids over-approximation in data dependences by creating use-def
        chains in the form of SSA edges.

	\item \emph{Data abstractions.}
        An abstract location usually represents a set of concrete locations. An over-approximation of this set of locations
	leads to spurious data dependences causing imprecision in points-to analysis. 
\end{itemize}

\subsubsection{Features Influencing Efficiency and Scalability}
Different methods use different techniques to achieve scalability. We characterize them based on the following three criteria:
\begin{itemize}
\item \emph{Relevant points-to information.}
      Many methods choose to compute a specific kind of points-to information which is then used to compute further points-to
      information. For example, staged points-to analyses begin with conservative points-to information which is then made more
      precise. Similarly, some methods begin by computing points-to information for top-level pointers whose indirections
      are then eliminated. This uncovers a different set of pointers as top-level pointers whose points-to information is then 
      computed.

\item \emph{Order of computing points-to information.}
      Most methods order computations based on relevant points-to information which may also be defined in terms of a
      chosen order of traversal over the call graph (eg. top-down or bottom-up).

\item \emph{Specialized data structures.}
      A method may use specialized data structures for encoding information efficiently (e.g. BDDs or \gpus and \gpgs) or may
      use them for modelling relevant points-to information (e.g. use of placeholders to model accesses of unknown pointees in a
      bottom-up method).
\end{itemize}

\subsubsection{Interaction between the Features}

In this section we explain the interaction between the features indicated by the
arrows in Figure~\ref{fig:big-pic-pta}.

\begin{itemize}
\item {\em Data abstraction influences approximations of data dependence.} 
	An abstract location may be over-approximated to represent a larger
	set of concrete locations in many situations such as in 
	field-insensitivity, type-based abstraction, allocation site-based abstraction. This
over-approximation creates spurious data dependence
	between the concrete locations represented by the abstract location. 
\item {\em Approximation of data dependence influences the choice of efficient data structures.} Some flow-insensitive methods 
	use disjoint sets for efficient union-find algorithms. Several methods use BDDs for 
	scaling context-sensitive analyses.
\item {\em Relevant points-to information affects the choice of data structures.} 
	Points-to information is stored in the form of graphs, points-to 
	pairs, or BDDs for top-down approaches. For bottom-up approaches, points-to information is computed using procedure summaries
		that use placeholders or \gpus.
\item {\em Relevant points-to information and order of computing influence each other mutually.} 
		In level-by-level analysis~\cite{Yu:2010:LLM:1772954.1772985}, points-to information is computed one level at a time. The relevant information to be computed at a 
		given level requires points-to information computed by the higher levels. Thus, in this case the relevance of points-to information
		influences the order of computation. In LFCPA~\cite{lfcpa} 
		only the live pointers are relevant. Thus, points-to information is computed only when the liveness of pointers is generated.
		Thus, the generation of liveness information influences the relevant points-to information to be computed.
\end{itemize}

\subsubsection{Our Work in the Context of Big Picture of Points-to Analysis}

\gpg-based points-to analysis preserves data dependence by being flow- and context-sensitive. It is path-insensitive and uses 
SSA form for top-level local variables.
        Unlike the approaches that over-approximate control flow indiscriminately, 
        we discard control flow as much as possible but only when there is a guarantee that it does not over-approximate data dependence.

Our analysis is field-sensitive. It over-approximates arrays by treating all its elements alike. We use allocation-site-based abstraction for representing heap
locations and use $k$-limiting for summarizing the unbounded accesses of heap where allocation sites are not known.

Like every bottom-up approach, points-to information is computed when all the information is available in the context. 
Our analysis computes 
points-to information for all pointers.

\subsection{Approaches of Constructing Procedure Summaries}

We restrict our description of related work to bottom-up approaches.
We begin with the two broad categories of approaches introduced in 
Section~\ref{sec:limitations-past-work}.

\subsubsection{MTF Approach}

In this approach~\cite{ptf,summ2,Yu:2010:LLM:1772954.1772985,Kahlon:2008:BTS:1375581.1375613}, 
control flow is not required to be recorded between memory updates. This is because
the data dependency between memory updates (even the ones which access unknown pointers) is known by using either the alias information
or the points-to information from the calling context.
These approaches construct symbolic procedure summaries. This involves computing 
preconditions and corresponding postconditions (in terms of aliases or points-to information).
A calling context is matched against a precondition
and the corresponding postcondition gives the result.

Level-by-level analysis~\cite{Yu:2010:LLM:1772954.1772985} constructs a procedure summary with multiple interprocedural conditions.
It matches the calling context with these conditions and chooses the appropriate summary for the given context. 
This method partitions the pointer variables in a program into different levels based on the Steensgaard's points-to graph for the program.
It constructs a procedure summary for each level (starting with the highest level) and uses the points-to information from the previous level.
This method constructs interprocedural def-use chains by using extended SSA form. 
When used in conjunction with conditions based on points-to information from calling contexts, the chains become context 
sensitive.

The scalability of these approaches depends on the 
number of aliases/points-to pairs in the calling contexts, which could be large.
%%Also, the constructed procedure summaries depend on the calling context for alias information or points-to information (for interprocedural conditions 
%%in level-by-level analysis~\cite{Yu:2010:LLM:1772954.1772985}) and hence are not context-independent.
Thus, this approach may not be useful for constructing summaries for library functions which 
have to be analyzed without the benefit of different calling contexts.
Saturn~\cite{Saturn} creates sound summaries but they may not be precise across applications 
because of their dependence on
context information.

Relevant context inference~\cite{Chatterjee:1999:RCI:292540.292554} constructs a procedure summary
by inferring the relevant potential aliasing between unknown pointees that are accessed in the procedure.
Although, it does not use the information from the context, 
it has multiple versions of the summary depending
on the alias and the type context. This analysis could be inefficient if the inferred possibilities of aliases and types do not actually occur in the program.
It also over-approximates the alias and the type context as an optimization thereby being 
only partially context-sensitive.

\subsubsection{STF Approach}

This approach
does not make any assumptions about the calling contexts~\cite{value.graph,summ1,Shang:2012:ODS:2259016.2259050,purity1,Whaley}
but constructs large procedure summaries
causing inefficiency in fixed-point computation at the intraprocedural
level. It introduces separate placeholders for every distinct access of a pointee (Section~\ref{sec:limitations-past-work}).
Also, the data dependence is not known in the case of indirect accesses of unknown pointees 
and hence control flow is required for constructing the summary for a flow-sensitive 
points-to analysis. However, these methods do not record control flow between memory updates in the summaries so
constructed.
Thus, in order to ensure soundness, the procedure summaries do not assume any ordering between the
memory updates and are effectively applied 
flow-insensitively even though they are constructed flow-sensitively.
This introduces imprecision 
by prohibiting killing of points-to information. However, it may not have much adverse impact on
programs written in Java because 
all local variables in Java have SSA versions, thanks to the absence of
indirect assignments to variables (there is no addressof operator).
Besides, there are few static variables in Java programs and absence of kill for them may not matter much;
the points-to relations of heap locations are not killed in any case.

Note that the MTF approach is precise even though no control flow in the procedure summaries is recorded because
the information from calling context obviates the need for control flow.

\subsubsection{The Hybrid Approach}

Hybrid approaches use customized summaries and
combine the top-down and bottom-up analyses to construct summaries~\cite{summ2}.
This choice is controlled by
the number of times a procedure is called. If this number 
exceeds a fixed threshold, a summary is constructed using the information of the calling contexts
that have been recorded for that procedure. A new calling context may 
lead to generating a new precondition and hence a new summary.
If the threshold is set to zero, then a summary is constructed for every procedure and hence we have a pure bottom-up approach.
If the threshold is set to a very large number, then we have a pure top-down approach and no procedure summary is constructed.

Additionally, we can set a threshold on the size of procedure summary or the percentage of context-dependent information in the summary or a combination of these
choices. 
In our implementation, we have used the percentage of context-dependent information as a threshold---when a procedure has a
significant amount of context-dependent information, it is better to introduce a small touch of top-down analysis
(Section~\ref{sec:large-con-dep-info}).
If this threshold is set to 0\%, our method becomes purely bottom-up approach; if it is set to 100\%, our method
becomes a top-down approach.

\section{Conclusions and Future Work}
\label{sec:conclusions}

%%We thus conclude that a pure top-down approach or a pure bottom-up approach is not sufficient for achieving scalability without introducing imprecision.
%%The need for scalability and precision demands an hybrid approach which explores the pros and cons of each of the approaches. A good study of the nature
%%of procedures and their interaction with other procedures is needed to classify them as a top-down procedure or a bottom-up procedure. Meaning, the procedure
%%is either visited in a top-down approach fashion or a corresponding procedure summary is constructed which is later applied at the call sites in all its callers.
%%A detailed study of the heuristics is required which decides the threshold for classifying the procedures in these two categories.

Constructing compact procedure summaries
for flow- and context-sensitive points-to
analysis seems hard because it 
\begin{enumerate}[(a)]
\item needs to model the accesses of pointees defined in callers
      without examining their code, 
\item needs to preserve data dependence between memory updates, and
\item needs to incorporate the effect of the summaries of the callee procedures transitively.   

%by either using
%	knowledge of aliases from calling contexts or by maintaining control flow 
%	between the memory updates.
\end{enumerate}
The first issue has been handled by modelling 
accesses of unknown pointees using placeholders. However, it may require
a large number of placeholders. The second issue has been handled by
constructing multiple versions of a procedure summary for different aliases in the calling 
contexts.  The third issue can only be handled by inlining the summaries of the callees. However,
it can increase the size of a summary exponentially 
thereby hampering the
scalability of analysis.

We have handled the first issue by 
proposing the concept of 
generalized points-to updates (\gpus) which track indirection levels.
Simple arithmetic on indirection levels
allows composition of \gpus to create
new \gpus with smaller indirection levels;
this reduces them progressively 
to classical points-to edges.

In order to handle the second issue,
we maintain control flow within a \gpg and perform optimizations of
strength reduction and redundancy elimination. Together, these optimizations
reduce the indirection levels of \gpus, eliminate data dependences between \gpus,
and minimize control flow significantly. 
These optimizations also mitigate the impact of the third issue.

In order to achieve the above, we have 
devised novel data flow analyses such as reaching \gpus analysis (with and without blocking)
and coalescing analysis which is a bidirectional analysis.
Interleaved call inlining and strength reduction of \gpgs facilitated a novel optimization that
computes flow- and context-sensitive points-to information in the first phase of a bottom-up
approach. This obviates the need for the second phase.

Our measurements on SPEC benchmarks show that \gpgs are small enough to
scale { fully} flow- and context-sensitive { exhaustive} points-to analysis to C programs
as large as 158 kLoC.
Two important takeaways from our empirical evaluation are:
\begin{enumerate}[(a)]
\item Flow- and context-sensitive points-to information is small and sparse. 
\item The real killer of scalability in program analysis is not the amount of data
      but the amount of control flow that it may be subjected to in search of precision.
      Our analysis scales because it minimizes the control flow significantly.
\end{enumerate}
Our empirical measurements show that most of the \gpgs are acyclic even if they
represent procedures that have loops or are recursive.

As a possible direction of future work, it would be useful to explore the possibility 
of  scaling the implementation to larger programs; we suspect that this would be
centered around examining the control flow in the \gpgs and optimizing it still further.
Besides, it would be interesting to explore the 
possibility of restricting \gpg construction to live pointer variables~\cite{lfcpa} for scalability.
It would also be useful to extend the scope of the implementation to C++ and Java programs.

The concept of \gpg provides a useful 
abstraction of memory and memory transformers involving pointers by directly modelling
%%The way matrices represent values as well as transformations, 
%%\gpgs represent memory as well as memory transformers defined in 
load, store, and copy of memory addresses. 
Any client program analysis
   that uses these operations { may be able to} use \gpgs{} { by} combining them 
   with { the original abstractions of the analysis. This direction can also be explored in future.}

\begin{acks}
Pritam Gharat is partially supported by a TCS Research Fellowship.
\end{acks}

%%\change{}{
%%\section*{Acknowledgments}
%%We would like to thank the anonymous reviewers of the SAS~2016 conference
%%for their comments on an initial version of this paper which appeared there~\cite{gpg.sas.16}. 
%%The current paper has
%%benefited from the feedback of many people including Supratik
%%Chakraborty, Sriram Srinivasan, Amitabha Sanyal, Supratim Biswas, and Venkatesh
%%Chopella. The seeds of \gpgs were explored in a very different form in
%%the Master's thesis of Shubhangi Agrawal in 2010.
%%}

\nocite{DBLP:conf/sas/NystromKH04}

\bibliographystyle{ACM-Reference-Format}
\bibliography{toplas-gpgpta}

\end{document}